\documentclass[3p,times]{elsarticle}

\usepackage{amssymb}
\usepackage{amsmath}
\usepackage{amsthm}

\usepackage{float}
\usepackage[pdfborder={0 0 0}, colorlinks=true, linkcolor=blue, citecolor=red]{hyperref}
\hypersetup{pdfborder=0 0 0}
\usepackage{algorithm}
\usepackage{algpseudocode}
\usepackage{mathrsfs}
\usepackage{subfigure}
\usepackage{multirow}
\usepackage{wrapfig}
\usepackage{siunitx} 

\usepackage{leftidx}
\usepackage[latin1]{inputenc}
\usepackage{ulem}

\PassOptionsToPackage{hidelinks}{hyperref}

\usepackage{IEEEtrantools}

\usepackage{lineno,hyperref}
\modulolinenumbers[5]

\usepackage[figuresright]{rotating}

\def\e{{\epsilon}}

\def\norm#1{\|#1\|} 

\newcommand{\beq} {\begin{equation}}
\newcommand{\eeq} {\end{equation}}
\newcommand{\bdm} {\begin{displaymath}}
\newcommand{\edm} {\end{displaymath}}
\newcommand{\bit}{\begin{itemize}}
\newcommand{\eit}{\end{itemize}}
\newcommand{\bde}{\begin{description}}
\newcommand{\ede}{\end{description}}
\newcommand{\bce}{\begin{center}}
\newcommand{\ece}{\end{center}}
\newcommand{\ben} {\begin{enumerate}}
\newcommand{\een} {\end{enumerate}}
\newcommand{\bea} {\begin{eqnarray}}
\newcommand{\eea} {\end{eqnarray}}
\newcommand{\barr} {\begin{array}}
\newcommand{\earr} {\end{array}}
\newcommand{\bean} {\begin{eqnarray*}}
\newcommand{\eean} {\end{eqnarray*}}
\newcommand{\edoc} {\end{document}}

\newcommand*\ddiff{\mathop{}\!\mathrm{d}}

\newcommand{\dd}[2] {\ensuremath{\frac{\partial {#1}}{\partial {#2}}}}
\newcommand{\DD}[2] {\ensuremath{\frac{\ddiff {#1}}{\ddiff {#2}}}}
\newcommand{\Grad} {\ensuremath{\nabla}}

\newcommand{\Div} {\ensuremath{\nabla\cdot}}

\newcommand{\eval}[2][\right]{\relax
  \ifx#1\right\relax \left.\fi#2#1\rvert}

\providecommand{\norm}[1]{\left\hspace{-0.5ex}\lVert#1\right\hspace{-0.5ex}\rVert}

\newcommand{\Nel} {\ensuremath{{N_\text{el}}}}

\newcommand{\half} {\ensuremath{\frac{1}{2}}}

\newcommand{\mc}[1]{\mathcal{#1}}

\newcommand{\mb}[1]{\mathbf{#1}}

\newcommand{\nor}[1]{\left\vert\kern-0.35ex\left\vert #1 \right\vert\kern-0.35ex\right\vert}
\newcommand{\norL}[1]{\left\vert\kern-0.5ex\left\vert #1 \right\vert\kern-0.5ex\right\vert}
\newcommand{\nort}[1]{{\left\vert\kern-0.35ex\left\vert\kern-0.35ex\left\vert #1 
    \right\vert\kern-0.35ex\right\vert\kern-0.35ex\right\vert}}
\newcommand{\snor}[1]{\left| #1 \right|}
\newcommand{\LRp}[1]{\left( #1 \right)}
\newcommand{\LRs}[1]{\left[ #1 \right]}
\newcommand{\LRa}[1]{\left< #1 \right>}
\newcommand{\LRc}[1]{\left\{ #1 \right\}}

\newcommand{\jumpL}[1] {\ensuremath{\LRs{\hspace{-1ex}\LRs{#1}\hspace{-1ex}}}}

\newcommand{\average}[1] {\ensuremath{\LRc{\hspace{-0.9ex}\LRc{#1}\hspace{-0.9ex}}}}
\newcommand{\averageM}[1] {\ensuremath{\LRc{\hspace{-1.ex}\LRc{#1}\hspace{-1.ex}}}}

\newcounter{tablecell}

\newcommand{\figlab}[1]{\label{fig:#1}}
\newcommand{\eqnlab}[1]{\label{eq:#1}}

\newcommand{\tablab}[1]{\label{tab:#1}}

\newcommand{\figref}[1]{\ref{fig:#1}}

\newcommand{\eqnref}[1]{\eqref{eq:#1}}

\newcommand{\seclab}[1]{\label{sect:#1}}
\newcommand{\secref}[1]{\ref{sect:#1}}
\newcommand{\tabref}[1]{\ref{tab:#1}}

\renewcommand{\u}{u}

\renewcommand{\e}{e}

\newcommand{\ub}{\mb{\u}}

\newcommand{\R}{{\mathbb{R}}}

\newcommand{\Poly}{{\mc{P}}}

\newcommand{\Vb}{{\mb{V}}}
\newcommand{\V}{{V}}

\newcommand{\Vhp}[1]{{\V_h^#1}}

\newcommand{\Vbhp}[1]{{\Vb_h^#1}}

\newcommand{\Rb}{\mb{R}}

\newcommand{\n}{\mb{n}}
\newcommand{\nm}{\n^-}
\newcommand{\np}{\n^+}

\newcommand{\ubh}{{\hat{\ub}}}

\renewcommand{\sb}{\mb{s}}

\newcommand{\projL}{{\Pmat^L}}

\newcommand{\Emat}{{\bf{E}}}

\newcommand{\Tmesh}{\mc{T}}
\newcommand{\pTmesh}{\partial\mc{T}}
\newcommand{\Tmeshh}{{{\Tmesh}_h}}

\newcommand{\Nfield} {\ensuremath{{N_\text{field}}}}

\newcommand{\Neighborh}{{{\mc{N}}_h}}

\newcommand{\Ub}{{\bf U}}

\newcommand{\dtt}{\triangle t}

\newcommand{\dt}{{\triangle t}}
\newcommand{\dx}{{\triangle x}}

\newcommand{\Qmat}{{\bf{Q}}}

\newcommand{\pres}{{{p}}}

\newcommand{\Ical}{\mathcal{I}}

\newcommand{\Sb}{{\bf S }}

\renewcommand{\u}{{u}}

\newcommand{\strain}{{\boldsymbol{\epsilon}}}
\newcommand{\vort}{{\boldsymbol{\omega}}}

\newcommand{\Mmat}{{\bf{M}}}

\newcommand{\Bmat}{{\bf{B}}}
\newcommand{\Pmat}{{\bf{P}}}
\newcommand{\Cmat}{{\bf{C}}}

\newcommand{\Elem}{{I}}

\usepackage{scalerel,stackengine}
\stackMath
\newcommand\reallywidehat[1]{%
\savestack{\tmpbox}{\stretchto{%
  \scaleto{%
    \scalerel*[\widthof{\ensuremath{#1}}]{\kern-.6pt\bigwedge\kern-.6pt}%
    {\rule[-\textheight/2]{1ex}{\textheight}}
  }{\textheight}%
}{0.5ex}}%
\stackon[1pt]{#1}{\tmpbox}%
}

\newtheorem{corollary}{Corollary}
\newtheorem{proposition}{Proposition}

\newtheorem{remark}{Remark}

\usepackage{color}
\usepackage{soul,xargs}
\usepackage[pdftex,dvipsnames]{xcolor}

\usepackage[colorinlistoftodos,prependcaption,textsize=tiny]{todonotes}

\newcommandx{\question}[2][1=]{\todo[linecolor=red,backgroundcolor=red!25,bordercolor=red,#1]{#2}}
\newcommandx{\change}[2][1=]{\todo[linecolor=blue,backgroundcolor=blue!25,bordercolor=blue,#1]{#2}}
\newcommandx{\add}[2][1=]{\todo[linecolor=OliveGreen,backgroundcolor=OliveGreen!25,bordercolor=OliveGreen,#1]{#2}}
\newcommandx{\improve}[2][1=]{\todo[linecolor=Plum,backgroundcolor=Plum!25,bordercolor=Plum,#1]{#2}}
\newcommandx{\thiswillnotshow}[2][1=]{\todo[disable,#1]{#2}}
\newcommandx{\remove}[2][1=]{\todo[linecolor=yelllow,backgroundcolor=yellow!10,bordercolor=red,#1]{#2}}

\begin{document}

\begin{frontmatter}




\title{
Differentiable DG with Neural Operator Source Term Correction
}


\author[AddrKUS]{Shinhoo Kang\corref{mycorrespondingauthor}}
\cortext[mycorrespondingauthor]{Corresponding author}
\ead{shinkang@korea.ac.kr}
\address[AddrKUS]{Korea University, Sejong 30019, KR.}
\address[AddrANL]{Mathematics and Computer Science Division, Argonne National Laboratory, Lemont, IL 60439, USA.}

\author[AddrANL]{Emil M. Constantinescu} 
\ead{emconsta@anl.gov}

\begin{abstract} 
Computational advances have fundamentally transformed the landscape of numerical simulations, enabling unprecedented levels of complexity and precision in modeling physical phenomena. While these high-fidelity simulations offer invaluable insights for scientific discovery and problem solving, they impose substantial computational requirements. Consequently, low-fidelity models augmented with subgrid-scale parameterizations are employed to achieve computational feasibility.
We introduce an end-to-end differentiable framework for solving the compressible Navier--Stokes equations. This integrated approach combines a differentiable discontinuous Galerkin (DG) solver with a neural network source term. Through the implementation of neural ordinary differential equations (NODEs) for network parameter optimization, our methodology ensures continuous interaction with the governing equations throughout the training process. We refer to this approach as NODE-DG.
This hybrid approach combines the accuracy of numerical methods with the efficiency of machine learning, offering the following key advantages: (1) improved accuracy of low-order DG approximations by capturing subgrid-scale dynamics; (2) robustness against nonuniform or missing temporal data; (3) elimination of operator-splitting errors; 
(3) total mass conservation; and (4) a continuous-in-time operator that enables variable time step predictions, which accelerate projected high-order DG simulations.
We demonstrate the performance of the proposed framework through two examples: two-dimensional Kelvin--Helmholtz instability and three-dimensional Taylor--Green vortex examples.

  
\end{abstract}

\begin{keyword}
neural ordinary differential equations, 
machine learning, acceleration, 
Navier--Stokes equations,
discontinuous Galerkin,
differentiable programming

\end{keyword}

\end{frontmatter}




\section{Introduction}

Computational fluid dynamics has become the cornerstone of fluid mechanics, transforming how researchers and engineers address complex fluid flow challenges. It is widely used in applications ranging from predicting large-scale weather phenomena such as hurricanes, cyclones, and jet streams to modeling blood flows, wildfires, jet engines, turbines, scramjets, and more. This progress has been fueled by advancements in high-performance computing and sophisticated numerical methods, such as finite volume methods, spectral methods, and discontinuous Galerkin (DG) methods. 
Among these, DG methods have received significant attention, owing to their high-order accuracy, flexibility for handling complex geometry, hp-adaptivity, and efficiency on modern computing architectures \cite{cockburn2000development,arnold2002unified,demkowicz2006computing}.
DG methods blend the strengths of finite volume and finite element approaches, as they approximate a global solution using a finite set of local functions. Communications between elements are restricted to neighboring elements through numerical fluxes at the element boundaries. This localized structure makes DG methods highly suitable for parallel computing.
 
Despite these advancements, perfectly capturing real-world phenomena in numerical models remains a significant challenge due to the multiscale nature of these systems and the complexity of their underlying physics. Moreover, resolving all scales is computationally prohibitive. 
Consequently, physical modeling is typically carried out on a coarse grid, supplemented by appropriate physical parameterization as a source term. In general, a numerical model incorporating parameterization can be expressed as 
\begin{align}
\DD{u}{t} = R(u) + S(u),
\eqnlab{goveq-with-source}
\end{align}
where $u$ denotes the resolved-scale state vector; $R(u)$ represents the spatially discretized fluid dynamic equations; and $S(u)$ corresponds to the source term provided by the physical parameterization.

Physical parameterization approximates physical processes that occur beyond the resolution of the numerical model and are not directly captured by its governing equations.
Examples include radiation, cloud formation, heat transfer, evaporation, and other physical interactions in weather models. Turbulence modeling is another example of physical parameterization, accounting for the effect of unresolved small-scale structures that influence the resolved scales~\cite{sanderse2024scientific}. For instance, large eddy simulation resolves large-scale turbulent motions that carry the majority of the flow energy, while modeling the subgrid scales (SGS) that have a relatively minor effect on the mean flow \cite{sagaut2005filtering}. 
Various turbulence models have been proposed to represent SGS dynamics, with the static Smagorinsky model \cite{smagorinsky1963general} and the dynamic Smagorinsky model \cite{germano1991dynamic} being among the most widely used. 
%

The development of physical parameterization is inherently challenging because of the multiscale, nonlinear, and often poorly understood nature of many physical processes. As a result, parameterizations often rely on empirical or semi-empirical relationships that may not be universally applicable. 
For example, Smagorinsky models estimate the dissipation of SGS energy but often perform poorly for certain flows \cite{piomelli1991large,liu1994properties}.
The reason is primarily that they assume that eddy viscosity is always purely dissipative, failing to account for energy transfer from smaller scales to larger scales (backscatter) \cite{meneveau2000scale, germano1991dynamic}.
Furthermore, many parameterization schemes include tuning parameters that require calibration. This calibration is crucial to ensure that the parameterizations accurately represent the underlying physical processes under the specific conditions of a simulation \cite{schmidt2017practice,couvreux2021process}.
 These challenges have prompted the adoption of alternative approaches, such as machine learning--based parameterizations, to address these limitations more effectively.

Machine learning (ML) methods have successfully demonstrated their capability for physical parameterization and parameter tuning, overcoming many limitations of traditional approaches. By utilizing observational data or high-resolution simulations, ML models can represent complex physical processes more accurately, automate parameter tuning, and accelerate the development of robust simulation frameworks across a wide range of scientific and engineering applications. For instance, Schneider et al.~\cite{schneider2017earth} proposed calibrating free parameters in the two-scale Lorenz 96 model using high-resolution simulations.
Rasp et al.~\cite{rasp2018deep} learned  subgrid processes in climate models from cloud-resolving model data.
Couvreux et al.~\cite{couvreux2021process} 
 used Gaussian process-based methods from uncertainty quantification to calibrate the model-free parameters. Tsai et al.~\cite{tsai2021calibration} showed that in a hydrological land surface model, gradient-based calibration schemes that adjust all parameters become increasingly effective as data availability grows. Beck et al.~\cite{beck2019deep} proposed data-driven closure through supervised learning by constructing a mapping from direct numerical simulation data and large eddy simulation data. 

Moreover, ML-augmented physical parameterization has shown potential for improving the accuracy of coarse-grid solvers.
Fabra et al.~\cite{fabra2022finite} introduced an ML-based correction term 
derived from fine-scale solutions to enhance coarse-scale models for the wave equations. 
Similarly, Lara and Ferrer~\cite{de2022accelerating,de2023accelerating} 
proposed improving a low-order DG solver by incorporating a parameterized source term 
for Burgers' equations and compressible Navier--Stokes equations. 
Their approach was tested using the turbulent Taylor--Green vortex case 
under various Reynolds numbers (30, 200, and 1600) and 
for laminar, transitional, and turbulent regimes, 
demonstrating its effectiveness across different flow conditions. 
Both Fabra et al.~\cite{fabra2022finite} and Lara and Ferrer~\cite{de2022accelerating,de2023accelerating} demonstrated that the correction term bridges the gap between projected high-order and low-order solutions. By employing supervised learning to construct the corrective term 
\footnote{
Fabra et al.
utilized an iterative approach, while Lara and Ferrer
applied a post-time-step correction method.
},
they successfully improved the speed and accuracy of low-order simulations. 
However, this approach relies on using uniform time steps for training the corrective term and for making predictions. Consequently, nonuniform observations or missing data may lead to poor results, limiting the potential of ML-augmented physical parameterization.
This limitation becomes critical when high-resolution simulations or observational datasets are temporally sparse or do not align with the time scales of a coarse-grid solver. Furthermore, the discrete correction step in ~\cite{de2022accelerating,de2023accelerating} relies on split time-stepping methods, which can introduce significant operator splitting errors.


 To address the limitations related to the training-prediction step-size restrictions,
 Kang and Constantinescu~\cite{kang2023learning} generalized
 the discrete correction method~\cite{fabra2022finite,de2022accelerating,de2023accelerating} to a continuous correction method within the context of DG spatial discretization. 
 The key idea involves replacing the source term in \eqnref{goveq-with-source} with a neural network,
 applied continuously in time, as expressed by
 \begin{align}
  \DD{u}{t} = R(u) + S_\theta(u),
  \eqnlab{goveq-with-nnsource}
  \end{align}
where $S_\theta(u)$ is a neural network source term. 
The neural network parameters are then trained by using neural ordinary differential equations (NODEs)~\cite{chen2018neural}.
Equation \eqnref{goveq-with-nnsource} represents a differentiable hybrid model 
that integrates scientific knowledge with machine learning techniques \cite{rackauckas2020universal}.
This hybrid approach is particularly appealing because the neural network parameters are optimized through gradient-based methods while interacting with the governing equations, offering enhanced accuracy, computational efficiency, and interpretability compared with conventional black-box machine learning models or traditional numerical models.
For instance,
Kochkov et al.~\cite{kochkov2024neural} demonstrated that
 an ML component can successfully replace or correct the traditional physical parameterizations 
 in a general circulation model. 
 Their hybrid model provides computational savings by orders of magnitude while maintaining comparable accuracy to that of conventional models. 
 Similarly, Shankar et al.~\cite{shankar2023differentiable2} showed that the hybrid approach enhanced the accuracy of turbulence closure compared with conventional methods. 
 Additionally, Kang and Constantinescu~\cite{kang2023learning} 
showed that the hybrid approach could enhance the performance of low-order DG solvers for the one-dimensional convection-diffusion equation and viscous Burgers' equations.
 

%
Building on our previous work~\cite{kang2023learning}, this study introduces an end-to-end differentiable NODE-DG framework based on source term correction for solving the compressible Navier--Stokes equations. We develop a differentiable DG solver for compressible Navier--Stokes equations on a structured mesh, incorporating a neural network source term as defined in \eqnref{goveq-with-nnsource}, and train the network parameters using NODEs.
The NODE-DG framework offers several key advantages:
(1) it improves the accuracy of low-order DG approximations by capturing subgrid-scale dynamics;
(2) it can effectively handle nonuniform or missing temporal data; 
(3) it eliminates operator splitting errors; 
(4) it preserves total mass; and 
(5) it learns a continuous-in-time operator,
enabling predictions with variable step sizes. This approach significantly accelerates projected high-order DG simulations.

The paper is organized as follows. 
In Section 
\secref{problem} we review nodal DG methods and the discrete corrective forcing approach, and we introduce a projection operator.
In Section \secref{method} we present the NODE-DG framework.
Section \secref{NumericalResults} demonstrates the performance of our proposed methodology using the examples of the Kelvin--Helmholtz instability and the Taylor--Green vortex. 
In Section \secref{Conclusion} we summarize our current approach and results, discuss its limitations, and outline possible directions for further research.

\section{Preliminaries}
\seclab{problem}

\subsection{Discontinuous Galerkin method }
We briefly review nodal discontinuous Galerkin methods \cite{hesthaven2007nodal,kopriva2009implementing}. 
For clear exposition we focus on a one-dimensional system of a hyperbolic equation, 
\begin{align} 
\dd{u}{t} + \dd{f(u)}{x}= 0, \text{ on } \Omega \times \LRs{0,T},
\eqnlab{hyperboliceq}
\end{align}
where $u,f \in\R$ are conservative variable and physical flux and $(x,t)\in\Omega\times[0,T]$.
We partition the domain $\Omega$ into $\Nel$ non-overlapping elements $\Elem_i=\LRs{x_{i-\half},x_{i+\half}}$ for $i=1,2,\cdots,\Nel$,
and define the mesh $\Tmeshh := \cup_{i=1}^\Nel \Elem_i$ 
by a finite collection of the elements $\Elem_i$. 
Here, $x_\half=x_{\min}$, 
$x_{\Nel+\half}=x_{\max}$, and $h := \max_{j\in
\LRc{1,\hdots,\Nel}}\dx^j$, where 
$\dx^j=x_{j+\half}-x_{j-\half}$ is the size of the element $j$.
We denote the boundary of element $\Elem$ by $\partial \Elem$.
We let $\pTmesh_h := \LRc{\partial I:\Elem \in \Tmesh_h}$ be the collection of the boundaries of all elements.
For two neighboring elements $\Elem^+$ and $\Elem^-$ that share an interior
interface $\e = \Elem^+ \cap \Elem^-$, we denote by $q^\pm$ the trace of their
solutions on $\e$. 
We define $\nm$ as the unit outward normal vector on
the boundary $\partial \Elem^-$ of element $\Elem^-$, and we define $\np = -\nm$ as the unit outward
normal of a neighboring element $\Elem^+$ on $\e$.
On the interior face $\e$ we define the mean/average operator $\average{\bf v}$, where $\bf v$ is
either a scalar or a vector quantity, as
$\averageM{{\bf v}}:=\LRp{{\bf v}^- + {\bf v}^+}/2$, 
and the jump operator $\jumpL{\bf v} := {\bf v}^+ \cdot \np + {\bf v}^- \cdot \nm $.
Let $\Poly^{N}\LRp{D}$ denote the space of polynomials of degree at
most $N$ on a domain $D$. Next, we introduce discontinuous
piecewise polynomial spaces for scalars and vectors of size $k$ as
\begin{align*}
\Vhp{N}\LRp{\Tmeshh} &:= \LRc{v \in L^2\LRp{\Tmeshh}:
  \eval{v}_{\Elem} \in \Poly^N\LRp{\Elem}, \forall \Elem \in \Tmeshh}, \\
  \Vbhp{N}\LRp{\Tmeshh} &:= \LRc{{\bf v} \in \LRs{L^2\LRp{\Tmeshh}}^k:
  \eval{{\bf v}}_{\Elem} \in \LRs{\Poly^N\LRp{\Elem}}^k, \forall \Elem \in \Tmeshh},
\end{align*}
and elementwise polynomial spaces for scalars and vectors as 
\begin{align*}
\Vhp{N}\LRp{\Elem} := \Poly^N\LRp{\Elem} 
\text{ and } 
\Vbhp{N}\LRp{\Elem} := \LRs{\Poly^N\LRp{\Elem}}^k.
\end{align*}
We define $\LRp{\cdot,\cdot}_\Elem$ as the $L^2$-inner product on an
element $\Elem$ and $\LRa{\cdot,\cdot}_{\partial \Elem}$ as the
$L^2$-inner product on the element boundary $\partial \Elem$. 
We define the associated norm as $\norm{ \cdot }_{DG}:= \LRp{ \sum_{\Elem \in \Tmesh_h} \norm{ \cdot }_{\Elem}^2 }^\half$, where $\norm{ \cdot }_{\Elem}=\LRp{\cdot,\cdot}_\Elem^\half$.

Each element $\Elem$ is the image of the reference element $\Elem^*=[-1,1]$ by an affine map $x=\phi^\Elem(\xi)$. $\snor{J^\Elem}=\det \LRp{\dd{x}{\xi}}=\frac{\dx^\Elem}{2}$ is the determinant of a Jacobian of $\phi^\Elem$. 
On each $\Elem$, a local solution $u \in \Vhp{N}\LRp{I}$ is approximated by a linear combination of Lagrange basis functions $\ell_j(\xi)=\Pi_{i=0,i\ne j}^N \frac{\xi - \xi_i}{\xi_j - \xi_i}$,
$$
u(\xi,t) = \sum_{j=0}^N u_j \ell_{j}(\xi), 
$$
where $\xi_j$ is the Legendre--Gauss--Lobatto (LGL) points \cite{kopriva2009implementing} and $u_j:=u(\xi_j,t)$ is the nodal value of $u$ at $\xi_j$.

The DG weak formulation of \eqnref{hyperboliceq} yields the following: Seek $u \in \Vhp{N}\LRp{\Elem}$ such that 
\begin{align} 
\snor{J^I}\Mmat_{ij} \DD{u_j}{t} 
- \Qmat^\top_{ij} f_j
+ \LRp{\Emat^\top \Bmat}_{ij} f^*_j= 0
\eqnlab{matrixform-gov}
\end{align}
holds for each element $\Elem \in \Tmeshh$ and $i,j=0,1,\cdots,N$.
Here, $f^*_j$ is a numerical flux; $\Mmat_{ij}=\LRp{\ell_i(\xi),\ell_j(\xi)}_{\Elem^*}$ is a mass matrix;
$\Qmat_{ij}=\LRp{\ell_i(\xi),\dd{\ell_j(\xi)} {\xi}}_{\Elem^*}$ is an integrated differentiation matrix; 
$$
\Emat=\begin{pmatrix} 
\ell_0(-1)\\
\ell_N(1)
\end{pmatrix}, \text{ and } \Bmat=\begin{pmatrix}
-1 & 0 \\ 
0 & 1
\end{pmatrix}.$$

Now, we rewrite the DG weak formulation in \eqnref{matrixform-gov} as an explicit ordinary differential equation,
\begin{align}
  \eqnlab{gov-ode} 
  \DD{u}{t} = 
  \snor{J^\Elem}^{-1}\Mmat^{-1}\LRp{\Qmat^\top f-\Emat^\top \Bmat f^*}=: R (u),
\end{align}
where we abuse notation and interpret $u,f,f^*$ as spatially discrete quantities evaluated at $\xi_0,\xi_1,\cdots,\xi_N$.





\subsection{Elementwise projection operators}

We let $u^{H} \in \Vhp{H}\LRp{\Elem}$ and $u^{L} \in \Vhp{L}\LRp{\Elem}$ be the high- and the low-order DG approximations 
\footnote{
For simplicity, we omit the term ``DG'' from   ``DG approximations'' from now on.
}
($H>L$) on $\Elem$, respectively.  
We define a low-order projection operator, $\mathscr{P}^L$, which projects the high-order approximation $u^H$ into the low-order space, ensuring $\mathscr{P}^L (u^H)\in \Vhp{L}(\Elem)$.
%
%
To implement this, we begin with the definition of the $L^2$ projection on each element,
\begin{align}
\eqnlab{l2projection}
  \LRp{u^H - \mathscr{P}^L (u^H), v}_\Elem = 0, \quad \forall v \in \Vhp{L}(\Elem). 
\end{align}
Then, we expand both the high-order approximation and its projected counterpart to $u^H(\xi,t)=\sum_{j=0}^H\ell_j^H(\xi) u_j^H(t)$ and 
$\mathscr{P}^L (u^H)(\xi,t)=\sum_{j=0}^L\ell_j^L(\xi) \LRp{\mathscr{P}^L (u^H)}_j(t)$ 
in \eqnref{l2projection}.
Setting $v=\ell_i^L(\xi)$ for $i=0,\cdots,L$ in \eqnref{l2projection}, we construct the low-order projection matrix $\Pmat^L \in \R^{(L+1)\times(H+1)} $ such that 
\begin{align}
  \LRp{\mathscr{P}^L (u^H)}_i = \underbrace{\LRp{\Mmat^L}^{-1}_{ij}\Cmat_{jk}}_{\Pmat^L_{ik}} u^H_k
  \eqnlab{l2proj-high2low-op}
\end{align}
where 
$\Mmat^L_{ij}=\LRp{\ell_i^L(\xi),\ell^L_j(\xi)}_{\Elem^*}$ and $\Cmat_{ij}=\LRp{\ell_i^L(\xi),\ell^H_j(\xi)}_{\Elem^*}$.

\subsection{Neural ODEs}

A NODE \citep{chen2018neural} is a continuous-depth neural network that leverages ordinary differential equations to capture the underlying dynamics of input-output datastreams.
In a NODE, the functional $R$ in \eqnref{gov-ode} is approximated by a neural network model $R_\theta$ with learnable parameters $\theta$: 
\begin{align}
\eqnlab{node0}
    \DD{u(t)}{t} = R_\theta(t,u(t)).
\end{align}
By solving the initial value problem \eqnref{node0}, 
we can evaluate the state $u$ at any target time $\tau$, 
\begin{align}
\eqnlab{node0-integral}
    u(\tau) &= u(0) + \int_{0}^{\tau} R_\theta(\iota,u(\iota)) d\iota. 
\end{align}
Dividing the time domain into $N_t$ subintervals from the initial time $t_0=0$ to the final time $t_{N_t}=\tau$, we can write \eqnref{node0-integral} as
\begin{align}
\eqnlab{node0-integral-Nt}
    u(\tau) &= u(0) + \sum_{n=1}^{N_t}\int_{t_{n-1}}^{t_n} R_\theta(\iota,u(\iota)) d\iota. 
\end{align}
We note that \eqnref{node0-integral-Nt} employs a single neural network architecture that is independent of the time horizon.
Moreover, 
within subintervals, a NODE does not save intermediate network parameters. Therefore, compared with a conventional neural network such as ResNet~\cite{he2016deep}, which requires $N_t$ separate network architectures to compute $u(\tau)$, 
a NODE offers greater memory efficiency~\cite{gupta2022galaxy}
.

A NODE can effectively manage nonuniform samples or missing data in time series by allowing flexible selection of time subintervals ~\cite{rubanova2019latent}. Additionally, incorporating backpropagation \cite{paszke2019pytorch} or adjoint methods \cite{pontryagin1987mathematical,chen2018neural,liang2021stiffness,zhang2022memory} within the ODE solver facilitates the training of NODEs, enabling the neural network to dynamically adjust its parameters and architecture in response to input data. This adaptability makes NODEs particularly well suited for modeling time series data
.
Numerous encouraging results have been reported in the literature.  
Zhuang et al.~\cite{zhuang2020adaptive} applied NODEs for image classification on the CIFAR10 dataset. 
Rackauckas et al.~\cite{rackauckas2020universal} identified
the missing terms in the Lotka--Volterra system. 
Huang et al.~\cite{huang2022accelerating} addressed temporal discretization error caused by coarse time step sizes. 
Lee and Parish~\cite{lee2021parameterized} extended NODEs to learn latent dynamics of parameterized ODEs including Burgers' equation, Euler equations, and reacting flows.  
Maulik et al.~\cite{maulik2020time} investigated the learning latent-space representations for dynamical equations, focusing on the viscous Burgers' equation. 

While NODEs offer various advantages over traditional networks, they also have some limitations. 
One of these  is the computational demand of training NODEs. 
Grathwohl et al.~\cite{grathwohl2018ffjord} observed that the number of function evaluations required for integrating the dynamics can become excessively costly during training.
Additionally, NODEs are not robust against adversarial attack~\cite{anumasa2021improving} and lack a direct mechanism for assimilating data that arrives at a later time \cite{kidger2020neural}
.
To address these shortcomings, researchers have proposed several NODE variants~\cite{djeumou2022taylor,anumasa2021improving,kidger2020neural}. 
In this study, however, we focus on standard NODEs as a proof of concept.

\section{A differentiable NODE-DG framework}
\seclab{method}

In this section we introduce a novel continuous-in-time hybrid method, referred to as a differentiable NODE-DG framework, which integrates traditional DG methods with machine learning methods to enhance the performance of classical low-order DG discretizations.
%
%
Our approach can be viewed as a generalization of the discrete corrective forcing method~\cite{de2022accelerating,de2023accelerating}.  Section \secref{discrete-correction} provides a brief overview of that method, and Section \secref{continuous-correction} introduces the NODE-DG framework.


\subsection{Brief review of the discrete corrective forcing approach}
\seclab{discrete-correction}


The high-order solution $u^H$ in \eqnref{gov-ode}, 
\begin{align}
    \eqnlab{gov-ode-high}
    \DD{u^H}{t} = R^H(u^H), 
\end{align} 
provides high-order accuracy, but it comes with expensive computational cost at high resolution due to its large degrees of freedom \cite{de2023accelerating}. 
In contrast, the low-order solution $u^L$ in \eqnref{gov-ode}, 
\begin{align}
    \eqnlab{gov-ode-low}
    \DD{u^L}{t} = R^L(u^L),
\end{align}
is faster but less accurate. 
We apply the low-order projection matrix $\Pmat^L$ to \eqnref{gov-ode-high}, which yields 
\begin{align}
    \eqnlab{gov-ode-filtered}
    \DD{\Pmat^L u^H}{t} = \Pmat^L R^H(u^H).
\end{align}
Both $u^L$ in \eqnref{gov-ode-low} and $\Pmat^L u^H$ in \eqnref{gov-ode-filtered} belong to the same low-order space $\Vhp{L}(\Elem)$, but their tendency terms are not the same \cite{de2022accelerating};
that is, 
$\DD{\Pmat^L u^H}{t} \ne \DD{u^L}{t}$. 
Given that $\Pmat^L u^H$ is considered optimal in $\Vhp{L}(I)$, can machine learning methods be utilized to improve the accuracy of the low-order solution? 
 
Lara and Ferrer  \cite{de2022accelerating,de2023accelerating} presented 
the discrete corrective forcing approach to improve the performance of low-order DG solvers. An $s$-stage explicit Runge--Kutta (ERK) schemes with a uniform time step size is used to integrate the coarse-grid operator $R^L$ in \eqnref{gov-ode-low},  
\begin{subequations}
\begin{align}
    U_{n,i} &= u_n^L + \dt^L\sum_{j=1}^{i-1} a_{ij} R^L_j, \quad i=1,2,\cdots,s,\\
  u_{n+1}^L &= u_n^L + \dt^L \sum_{i=1}^s b_i R^L_i,
\end{align}
\eqnlab{erk-low}
\end{subequations} 
where $R_i^L:=R^L(t_n + c_i\dt,U_{n,i})$; 
$u^L_0=u^L(0)$; and 
$a_{ij}$, $b_i$, and $c_i$ are scalar coefficients for $s$-stage ERK methods. Here, $\dt^L$ is a coarse time step size.
The results $u^L_{n+1}$ in \eqnref{erk-low} are then corrected by incorporating the neural network source term $S_\theta$ using a forward Euler method, 
\begin{subequations}
\begin{align}
    \tilde{u}_{n+1} &= u_{n+1}^L + \dt^L
    S_{\theta}
    (u_n^L),\\
  u_{n+1}^L &= \tilde{u}_{n+1}.
\end{align}
\eqnlab{discrete-corrective-forcing}
\end{subequations}
Here, $\tilde{u}$ is the corrected approximation.

This two-step approach in \eqnref{erk-low} and \eqnref{discrete-corrective-forcing} can be seen as a temporal discretization of sequential operator-splitting step \cite{sportisse2000analysis} for \eqnref{goveq-with-nnsource}, 
which solves two subproblems sequentially on subintervals $\LRs{t_n,t_{n+1}}$, 
\begin{subequations}
\begin{align}
    \DD{u^L}{t} &= R^L (u^L),  \text{ with } u^L(t_n) = \tilde{u}_n,\\
    \DD{\tilde{u}}{t} &= S_\theta(\tilde{u}), \text{ with } \tilde{u}(t_n) = u^L_{n+1}, 
\end{align}
\eqnlab{sequential-operator-spliting}
\end{subequations}
for $n=0,1,\cdots,N_t-1$ and $\tilde{u}_0=u^L_0$.
This approach introduces errors in time because the correction is formally integrated by using the first-order forward Euler and the first-order Strang-type temporal splitting resulting in a first order scheme.
\footnote{
This approach can also be viewed as a partitioned method in time. The large scale is solved by using a high-order Runge--Kutta, while the correction is discretized with the first-order forward Euler. Moreover, the coupling between the two integrators is also first order at a continuous level. This is contrasted in the next section with a tightly coupled integration that simultaneously advances the source and the large scale.
}
To distinguish between the corrective forcing terms in \eqnref{goveq-with-nnsource} and \eqnref{discrete-corrective-forcing}, we denote the latter as $\tilde{S}_{\theta}$ and refer to it as the \textit{discrete corrective forcing} term.


According to the procedure in \cite{de2022accelerating,de2023accelerating}, we train the discrete corrective forcing term as follows. 
First, we generate the projected high-order solution by applying the low-order projection matrix $\Pmat^L$ to the high-order approximations $u^H$. 
Second, 
we compute the output feature of $\tilde{S}_{\theta,n}$ at every time step $t_n$ by rearranging the forward Euler step in \eqnref{discrete-corrective-forcing} and replacing $\tilde{u}_{n+1}$ with $\Pmat^L u^H$, 
\begin{align}
\tilde{S}_{\theta,n} = \frac{\Pmat^L u^H_{n+1} - u^L_{n+1}}{\dt^L}. 
\eqnlab{discrete-correction}
\end{align}
Then we construct a nonlinear map $\tilde{S}_{\theta}$ from the input feature of $\Pmat^L u^H_n$ to the output feature of $\tilde{S}_{\theta,n}$ by using a feed-forward neural network.
  For simplicity we do not consider the dependency of the earlier projected solutions.

We note that the output feature of $\tilde{S}_{\theta,n}$ is produced using 
a uniform time step size $\triangle t^L$ and a specific time-stepping scheme. As a result, the discrete corrective forcing term $\tilde{S}_\theta$ is not compatible with variable step sizes for prediction or alternative time-stepping schemes. 
Moreover, $\tilde{S}_\theta$ does not represent the continuous operator in \eqnref{goveq-with-nnsource} or \eqnref{sequential-operator-spliting}; rather, it acts as a nonlinear mapping between finite sets. 
This limitation restricts the applicability of the discrete corrective forcing approach, since the neural network model may have difficulty generalizing to adaptive time-stepping methods, nonuniform data, or different time-stepping schemes. 
In our previous work ~\cite{kang2023learning}, we observed that the accuracy of the discrete corrective forcing approach was highly sensitive to the choice of time step size. 
 
\subsection{NODE-DG framework}
\seclab{continuous-correction}

To overcome the drawbacks of the discrete corrective forcing approach in Section \secref{discrete-correction}, we proposed to learn the continuous source dynamics through NODEs~\cite{kang2023learning} in the context of DG discretization. 
By interpreting the neural network--based source $S_\theta(u)$ in \eqnref{discrete-corrective-forcing} as a continuous operator, 
the neural network source term can be incorporated in a continuous-in-time form,  
\begin{align}
\eqnlab{eq-nnsource}
  \DD{\hat{u}^L}{t} &= R^L (\hat{u}^L) + S_\theta(\hat{u}^L), 
\end{align}
eliminating operator splitting errors by solving the system in a coupled fashion.
We integrate \eqnref{eq-nnsource} and obtain the corrected solution $\hat{u}(\tau)^L$ at $t=\tau$ by  
\begin{align}
\eqnlab{node-integral}
    \hat{u}^L(\tau) &= \hat{u}^L(0) + \int_{0}^{\tau}  R^L(\iota,\hat{u}^L(\iota)) + 
    S_\theta(\hat{u}^L(\iota)) \ddiff\iota,
\end{align}
where $\hat{u}^L$ is
the prediction of the low-order solution with the neural network source term $S_\theta$. Hereafter, we call \eqnref{eq-nnsource} and $\hat{u}^L$ the augmented system and the augmented solution, respectively. 
Since $S_\theta(u)$ is a continuous operator, 
various time-stepping methods can be employed for approximating the integral of \eqnref{node-integral}. In this work, however, we focus on $s$-stage explicit Runge--Kutta  schemes with uniform time step size $\dt$.
Applying ERK schemes to \eqnref{node-integral} yields
\begin{subequations}
\begin{align}
  \hat{U}_{n,i} &= \hat{u}^L_n + \dt\sum_{j=1}^{i-1} a_{ij} \LRp{\hat{R}^L_j + \hat{S}_{\theta_j} }, \quad i=1,2,\cdots,s,\\
  \hat{u}^L_{n+1} &= \hat{u}^L_n + \dt\sum_{i=1}^{s}b_i \LRp{\hat{R}^L_i+\hat{S}_{\theta_i}},
\end{align}
\eqnlab{node-erk}
\end{subequations}
where $\hat{R}_i:=R^L(t_n + c_i\dt,\hat{U}_{n,i})$; and 
$\hat{S}_{\theta_i}:=S_\theta(\hat{U}_{n,i})$.  
Given an initial condition $\hat{u}^L_0=\hat{u}^L(0)$, the $N_t$ step ERK time-stepping method generates $N_t$ discrete instances of a single trajectory, $\LRc{\hat{u}^L_1,\hat{u}^L_2,\cdots,\hat{u}^L_{N_t}}$.  

\begin{remark}
In this study we solve the 2D/3D compressible Navier--Stokes equations in \secref{cns-gov}. 
Our goal is to enhance the low-order solution accuracy by solving the augmented system,
\begin{align}
  \eqnlab{gov-ode-system} 
  \DD{\ubh^L}{t} &= \Rb^L (\ubh^L) + {\Sb_\theta}(\ubh^L),
\end{align}
where $\Rb^L$ is a low-order DG spatial discretization of \eqnref{cns-gov}; 
$\ub=\LRp{\rho,\rho \varphi_1, \cdots, \rho \varphi_d, \rho E}^\top$ is the vector of conservative variables in $d$-dimension; 
$\rho$ is the density; $\varphi_i$ is the velocity component in the $i$th coordinate direction; $\rho E$ is the total energy; $\ub^L \in \Vbhp{L}\LRp{\Elem}$ is the low-order approximation of $\ub$; and $\ubh^L \in \Vbhp{L}\LRp{\Elem}$ is the augmented solution of $\ub^L$.

\end{remark}

\subsection{Neural network source term architecture} 
For $n_i \in \mathbb{N}$, a 
$D$-layer
feed-forward neural network is the function that maps  an input $z^0 \in \R^{n_0}$ to an output $z^D \in \R^{n_D}$, which consists of $D$ network layers, 
\begin{align*}
    z^i &= \sigma^i \LRp{ W^i z^{i-1} + b^i}, 
    \quad \text{ for } i=1,2,\cdots,D. 
\end{align*}
Here, $\sigma^i$ is the rectified linear unit  \cite{nair2010rectified} activation functions for $i=1,2,\cdots,D\!-\!1$, and $\sigma^D$ is the linear activation function, with $W^i \in \R^{n_i\times n_{i-1}}$ and $b^i \in \R^{n_i}$ being $i$th weight matrix and basis vectors. The dimension of the neural network architecture is denoted by $\LRc{n_0,n_1,\cdots,n_D}$.


In a neural network, a convolution is a method
used to extract features by condensing an image into fewer pixels.
 This is achieved by sliding a convolutional kernel, 
 represented as an $n\times n$ matrix (for $n \in \mathbb{N}$), across the image for each pixel.
 Similarly, in this work 
 we devise a convolution kernel for $S_\theta$, 
 where a local$D$-layer feed-forward neural network 
 slides over the low-order solutions of all the elements and
  produces a local source approximation for each element:  
\begin{align*}
S_\theta: u^{\Neighborh(\Elem)} \rightarrow s \in \Vhp{L}(\Elem),
\end{align*}
for a scalar equation and 
\begin{align}
\eqnlab{nnsource-local-vector} 
\Sb_\theta: \ub^{\Neighborh(\Elem)} \rightarrow \sb \in \Vbhp{L}(\Elem)
\end{align}
for a system of equations. 
For the compressible Navier--Stokes equations, 
the input vector is $\ub=\LRp{\rho, u_1, \cdots, u_d, T}^\top$, 
and the output vector is $\sb=\LRp{s_\rho, s_{\rho u_1}, \cdots, s_{\rho u_d}, s_{\rho E}}^\top$. 
Here, $\Neighborh(\Elem)$ denotes the set of elements adjacent to $\Elem$, including $\Elem$ itself, and $u^\Neighborh(\Elem)$ refers to the local solutions on the neighbor elements $\Neighborh(\Elem)$. 
To simplify implementation, we focus on Cartesian grids.
Analogous to an $n\times n $ image kernel, 
the adjacent elements are defined by a $k_w\times k_w$ kernel in a two-dimensional domain,
 where $k_w$ is the kernel width (the number of elements along one axis).
 In a three-dimensional domain, the kernel extends to $k_w\times k_w\times k_w$.
\begin{figure}[h!t!b!]
  \centering
   \includegraphics[trim=0cm 0cm 0cm 0cm,clip=true,width=0.91\textwidth]{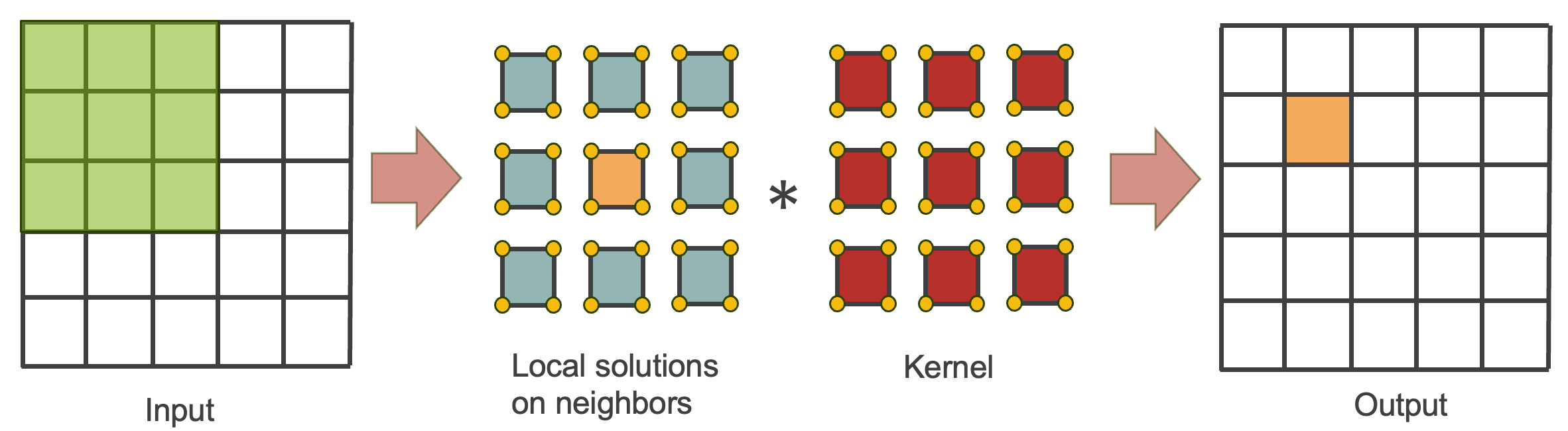}
  \caption{Schematic of a two-dimensional local neural network source for $N=1$ and a $3\times 3$ kernel.
  }
  \figlab{localkernel}
\end{figure}
An $N$th-order scalar solution $u$ within an element has $(N+1)^d$ degrees of freedom
 in a $d$-dimensional domain.
 Consequently, the input and output dimensions for the local neural source function are
  $n_0=(N+1)^dk_w^d\,\Nfield$ and $n_D=(N+1)^d\,\Nfield$, respectively,  
  where $\Nfield$ represents the number of conservative variables. 
%
%
 Figure \figref{localkernel} illustrates a schematic of a two-dimensional local neural network source for $N=1$ and a $3\times 3$ kernel, where the input and output dimensions are $n_0=36N_{field}$ and $n_D=4N_{field}$, respectively. 
That is, the neural network predicts four nodal values of $s_\rho, s_{\rho u}, s_{\rho v}$, and $s_{\rho E}$.

\subsection{Mass conserving neural network source terms}
\seclab{continuous-correction-massconservation}

The na\"{\i}ve neural network source term in \eqnref{nnsource-local-vector} improves the accuracy of low-order solver; however, it lacks any inherent structure for mass conservation, allowing it to behaves as either a sink or a source. Numerical instability typically arise when fundamental physical laws are violated. To address the issue and improve stability, we design a mass conserving neural network source term. 
The key idea is to impose the constraint $\LRp{s_\rho,1}_I=0$ to the mass conservation equation, which enforces the mean value of the density source term on each element to vanish. This constraint can easily be imposed in modal space. For simple exposition, we omit the subscript $\rho$ and consider one-dimensional scalar source $s$. We let $\tilde{\ell}_i(\xi)$ be the $i$th Legendre polynomial for $i=0,1,\cdots,N$, and define Vandermonde matrix as $\Vb_{ij}=\tilde{\ell}_j(\xi_i)$ for $i,j=0,1,\cdots,N$.
Since the zeroth modal value ($\tilde{u}_0$) represents the mean local mass, we set $\tilde{s}_0=0$ and only predict $\tilde{s}_j$ for $j=1,2,\cdots,N$.
The nodal values of the source term are then obtained by applying Vandermonde matrix to the modal values, 
\begin{align}
\eqnlab{mcnnsource} 
s_i = \sum_{j=0}^N \Vb_{ij} \tilde{s}_j,
\end{align}
for $i,j=0,1,\cdots,N$.
Due to the constraint $\tilde{s}_0=0$, the output dimension of the neural network source term is reduced by one, leading to $n_D=\LRp{(N+1)^d-1}\,\Nfield$.

\begin{proposition}\label{prop:cons:source}
The neural network source term in \eqnref{mcnnsource} with $\tilde{s}_0=0$ preserves the total mass. 
\end{proposition}

\begin{proof}
Consider the one dimensional mass conservation equation with a source term, 
$$
\dd{\rho}{t} + \dd{\rho u}{x} = s.
$$
To preserve the total mass over the domain $\Omega$, we need to show that $\LRp{s,1}_\Omega=0$.
Since $\LRp{s,1}_\Omega=\sum_{\Elem \in \Tmesh_h} \LRp{s,1}_\Elem $, it is sufficient to prove that $\LRp{s,1}_\Elem=0 $ for each element.
A local source is approximated by a linear combination of Lagrange basis functions, 
$$
s(\xi,t) = \sum_{i=0}^N \ell_{i}(\xi) s_i 
= \sum_{i=0}^N \ell_{i}(\xi) \LRp{\sum_{j=0}^N \tilde{\ell}_j(\xi_i) \tilde{s}_j}
= \sum_{j=0}^N \underbrace{\LRp{\sum_{i=0}^N \ell_{i}(\xi) \tilde{\ell}_j(\xi_i) }}_{\tilde{\ell}_j (\xi)}\tilde{s}_j, 
$$
where $s_i$ corresponds to the evaluation of a linear combination of Legendre polynomials at $\xi_i$. Reordering the summation yields $\tilde{\ell}_j (\xi)=\sum_{i=0}^N \Vb_{ji}^T \ell_i(\xi)$. 
Using this relation, we obtain 
\begin{align*}
    \LRp{s,1}_\Elem 
    = \sum_{i=0}^N \LRp{ \ell_i(\xi) s_i,1}_\Elem
    = \sum_{i=0}^N \LRp{ \sum_{j=0}^N\Vb^{-\top}_{ij}\tilde{\ell}_j(\xi) \sum_{k=0}^N\Vb_{ik}\tilde{s}_k ,1}_\Elem 
    = \sum_{j=0}^N \sum_{k=0}^N \LRp{\tilde{\ell}_j(\xi) \delta_{jk} \tilde{s}_k,1}_\Elem 
    = |J^\Elem|\sum_{j=0}^N \LRp{\tilde{\ell}_j(\xi) \tilde{s}_j,1}_{\Elem^*}.
\end{align*}
Because the zeroth Legendre polynomial $\tilde{\ell}_0(\xi)$ is 1, and  
Legendre polynomials are orthogonal with respect to the inner product on $\Elem^*$, it follows that 
\begin{align*}
    \LRp{s,1}_\Elem 
    = |J^\Elem|\sum_{j=0}^N \LRp{\tilde{\ell}_j(\xi),\tilde{\ell}_0(\xi)}_{\Elem^*}\tilde{s}_j
    = |J^\Elem|\LRp{\tilde{\ell}_0(\xi),\tilde{\ell}_0(\xi)}_{\Elem^*}\tilde{s}_0 = 0.
\end{align*}
Therefore, the total mass is preserved.
\end{proof}

Next we show that time stepping the augmented system will conserve linear invariants if conditions in Proposition \ref{prop:cons:source} are satisfied and if the time stepping applied to the original high- or low-order systems is conservative. We express the result just for mass conservation for simplicity.
\begin{corollary}\label{prop:rk-linear}
Consider the augmented semi-discrete system
in \eqnref{gov-ode-system},  
which is advanced by an $s$-stage Runge--Kutta (RK) method, as in  \eqnref{node-erk}.
Assume the following conditions hold:
\begin{itemize}
  \item[(i)] DG conservation of mass: For every state $\ub$, the density component of $\Rb_L$ satisfies
  \[
  \sum_{\Elem\in\mathcal T_h}\,(\Rb_L(\ub)_\rho,1)_\Elem=0
  \quad\text{(e.g., with periodic/no-flux boundary conditions).}
  \]
  \item[(ii)] Zero-mean source (Prop.~\ref{prop:cons:source}): For every state $\ub$, the density source from \eqnref{mcnnsource} obeys
  \[
  (\Sb_\theta(\ub)_\rho,1)_\Elem=0\quad\text{for each element }\Elem\in\mathcal T_h,
  \]
  which is enforced by the modal constraint $\tilde s_0=0$ in Proposition~\ref{prop:cons:source}.
\end{itemize}
Then the RK step preserves the discrete total mass:
\[
\sum_{I\in\mathcal T_h}(\rho_{n+1},1)_I=\sum_{I\in\mathcal T_h}(\rho_n,1)_I .
\]
Moreover, each RK stage preserves total mass as well.
\end{corollary}
\begin{proof}
Let $M(\ub):=\sum_{\Elem\in\mathcal T_h}(\rho,1)_\Elem$ denote the discrete total mass. For any RK scheme,
  \[
  \Ub_{n,i}=\ub_n+\Delta t\sum_{j=1}^s a_{ij}\Big(\Rb_L(\Ub_{n,j})+\Sb_\theta(\Ub_{n,j})\Big),\qquad i=1,\dots,s.
  \]
Using linearity of $M(\cdot)$,
  \[
  M(\Ub_{n,i})=M(\ub_n)+\Delta t\sum_{j=1}^s a_{ij}\Big( (\Rb_L(\Ub_{n,j})_\rho,1)_\Omega + (\Sb_\theta(\Ub_{n,j})_\rho,1)_\Omega \Big),
  \]
  where $(\cdot,\cdot)_\Omega:=\sum_{\Elem\in\mathcal T_h}(\cdot,\cdot)_\Elem$.
Assumption (i) gives $(\Rb_L(\cdot)_\rho,1)_\Omega=0$ for any argument; Assumption (ii) gives $(\Sb_\theta(\cdot)_\rho,1)_\Omega=0$ for any argument. Hence, each summand vanishes:
  \[
  M(\Ub_{n,i})=M(\ub_n)\qquad\text{for }i=1,\dots,s .
  \]

The step update is
  \[
  \ub_{n+1}=\ub_n+\Delta t\sum_{i=1}^s b_i\Big(\Rb_L(\Ub_{n,i})+\Sb_\theta(\Ub_{n,i})\Big).
  \]
  Applying $M(\cdot)$ and exploiting the stagewise orthogonality, we have
  \[
  M(\ub_{n+1})=M(\ub_n)+\Delta t\sum_{i=1}^s b_i\Big( (\Rb_L(\Ub_{n,i})_\rho,1)_\Omega + (\Sb_\theta(\Ub_{n,i})_\rho,1)_\Omega \Big)
  =M(\ub_n).
  \]
Therefore, $M(\ub_{n+1})=M(\ub_n)$ and for every stage we also have $M(\Ub_{n,i})=M(\ub_n)$.
\end{proof}
If the conditions of Proposition \ref{prop:cons:source} are satisfied for each conserved field individually, then mass, momentum, and energy are preserved. 

\subsection{Training of neural network source term}

The neural network source term in \eqnref{eq-nnsource} is trained along with the governing equations of the system.
This approach allows the network to learn while respecting the underlying physical laws. 
The approach results in models that are both accurate and physically interpretable,
combining data-driven insights with domain-specific knowledge.

Under the assumption that the projected high-order approximations are optimal,  
we generate a single trajectory of a high-order solution $u^H$ using $s$-stage ERK methods with a uniform time step size $\dt^H$ 
in order to obtain the training data. 
We then apply the low-order projection matrix in \eqnref{l2proj-high2low-op} to the trajectory,
 which yields the projected high-order solutions $\Pmat^Lu^H$.
In training, we randomly select $n$-batch instances from the trajectory and gather $m$ consecutive instances 
$\LRc{\LRp{\Pmat^L u^H}_{0}^{(i)},\LRp{\Pmat^Lu^H}_{1}^{(i)},\cdots,\LRp{\Pmat^Lu^H}_{m}^{(i)} }$ for $i=1,\cdots,n$.
Here, $(\Pmat^Lu^H)_{0}^{(i)}$ is the projected high-order solution at time $t=t_s$ for the $i$th sample.
Next, we predict the augmented solutions with $m$ time steps using \eqnref{node-integral}. 
%
%
Given an initial condition $\hat{u}^L_0=\LRp{\Pmat^L u^H}_{0}^{(i)}$, the $m$ step ERK time-stepping method generates $m$ discrete instances for the $i$th sample, $\LRc{\LRp{\hat{u}^L}_{1}^{(i)},\LRp{\hat{u}^L}_{2}^{(i)},\cdots,\LRp{\hat{u}^L}_{m}^{(i)} }$.
%
%
We then update the network parameters $\theta=\LRc{W^i,b^i}_{i=1}^D$ by minimizing the loss, 
\begin{align*}
  \mc{L}(\theta) =  \frac{1}{n m}\sum_{i=1}^n \sum_{j=1}^m \norm{\LRp{\Pmat^L u^H}_{j}^{(i)} - \LRp{\hat{u}^L}_{j}^{(i)}  }^2,  
\end{align*}
where $\norm{\cdot}^2$ is the mean squared error. 
For the compressible Navier--Stokes systems, the loss becomes
\begin{align}
 \eqnlab{loss-cns}
  \mc{L}(\theta) =  \frac{1}{n m}\sum_{i=1}^n \sum_{j=1}^m 
  \sum_{k=1}^\Nfield 
  \norm{\LRp{\Pmat^L u^H_k}_{j}^{(i)} - \LRp{\hat{u}^L_k}_{j}^{(i)}  }^2.  
\end{align}
Here, $\hat{u}^L_k$ and $u^H_k$ represent the $i$th conservative variables of $\hat{\ub}^L$ and $\ub^H$, respectively.

In addition, the loss function \eqnref{loss-cns} is augmented with an $L_2$ regularization term, with $\alpha$ serving as the tuning parameter,
\begin{align}
 \eqnlab{loss-cns-l2regularization}
  \mc{L}(\theta) =  \frac{1}{n m}\sum_{i=1}^n \sum_{j=1}^m 
  \sum_{k=1}^\Nfield 
  \norm{\LRp{\Pmat^L u^H_k}_{j}^{(i)} - \LRp{\hat{u}^L_k}_{j}^{(i)}  }^2 
  + \alpha \norm{\theta}^2.
\end{align}
Unless otherwise specified, we set $\alpha=0$ as the default in all numerical experiments. 

\subsection{Implementation}
We develop a differentiable solver for the compressible Navier--Stokes equations 
using the automatic differentiation Python package \texttt{JAX} \cite{jax2018github}. 
To implement our proposed approach, 
we leverage the \texttt{Optax} \cite{hessel2020optax} optimization library, 
the \texttt{Equinox} \cite{kidger2021equinox} neural network library, and the 
 \texttt{Diffrax} \cite{kidger2022neural} NODE package. 
 Built on \texttt{XLA} (Accelerated Linear Algebra), \texttt{JAX} delivers significant computational speedups.  

For the DG implementation, we use the Lax--Friedrichs flux $f^* = \averageM{f} + \half\max\LRp{\snor{\dd{f}{u}}}\jumpL{u}$ for inviscid flux and the local discontinuous Galerkin method \cite{cockburn1998local} for viscous flux in \eqnref{gov-ode}. 
To reduce the computational cost of inverting the mass matrix, 
we use an inexact mass matrix defined as $\Mmat_{ij}=\delta_{ij}w_i$, 
where $w_i$ are the LGL quadrature weights~\cite{kopriva2009implementing}.


\section{Numerical Results}
\seclab{NumericalResults}
 In this section we present numerical results to demonstrate the performance of the proposed approach for learning continuous source dynamics. 
 In all the examples, we have used structured meshes (e.g., quadrilateral elements for 2D and hexahedral elements for 3D) for spatial discretization and uniform time step size.  
We measure the $L^2$ relative error of $q$ by 
$$ \left\Vert q  - q_r   \right\Vert_{rel} := 
\frac{\left\Vert q  - q_r   \right\Vert_{DG} }{\left\Vert q_r \right\Vert_{DG} },$$
where $q_r$ is a reference solution.


\subsection{
Two-dimensional Kelvin--Helmholtz instability}

We consider the two-dimensional Kelvin--Helmholtz instability (KHI) on $\Omega=[-1,1]^2$ with doubly periodic boundary conditions.
KHI arises when a velocity difference exists across the interface of two fluids with different densities.
When the interface is subject to an initial perturbation, such as small waves, 
the disturbance grows exponentially, eventually forming Kelvin--Helmholtz rotors \cite{lecoanet2016validated}.
KHI plays a key role in the development of turbulence in the stratified fluids.
The initial condition is defined as in \cite{chan2020entropy}:
\begin{align*}
  \rho &= 1 + \half \LRp{ 
                            \frac{1}{1+\exp{\LRp{-\frac{(y-y_1)}{\beta^2}}}} 
                          - \frac{1}{1+\exp{\LRp{-\frac{(y-y_2)}{\beta^2}}}}
                          },\\
     u &= -0.5 + \LRp{ 
            \frac{1}{1+\exp{\LRp{-\frac{(y-y_1)}{\beta^2}}}} 
          - \frac{1}{1+\exp{\LRp{-\frac{(y-y_2)}{\beta^2}}}}
          },\\
     v &= \alpha \sin\LRp{2\pi x} \LRp{ \exp\LRp{-\frac{(y-y_1)^2}{\beta^2} } 
                                 + \exp\LRp{-\frac{(y-y_2)^2}{\beta^2} }  }, \\
  \pres &= \gamma^{-1},
\end{align*}
where $\gamma=1.4$, $y_1=-0.5$, $y_2=0.5$, $\alpha=0.1$, and $\beta=0.1$. 
We examine a fluid with a Prandtl number of $Pr=0.72$ and a Reynolds number of $Re=200$.

We generate the eighth-order approximation with the third-order ERK scheme 
and the time step size of $2\times 10^{-4}$ for $t=[0,3]$ over the mesh of $32\times32$ elements. 
Figure \figref{cns2d-khi-re200-ss-rH-t1234} shows the evolution of the density and the vorticity fields at $t={1,2,3}$. 
The density and vorticity fields are colored from $0.85$ to $1.7$ and from -0.85 to 0.85, respectively.  
Over time, 
the disturbance grows stronger, leading to the formation of vortices at $t=3$.  
We select the simulation period from $t=2$ to $t=3$ to demonstrate our approach. 
\begin{figure}[h!t!b!]
  \centering
  \subfigure[$\rho^H$ at $t=1$]{
    \figlab{cns2d-khi-re200-ss-rH-t1}
      \includegraphics[trim=20cm 8cm 20cm 8cm,clip=true,width=0.31\textwidth]{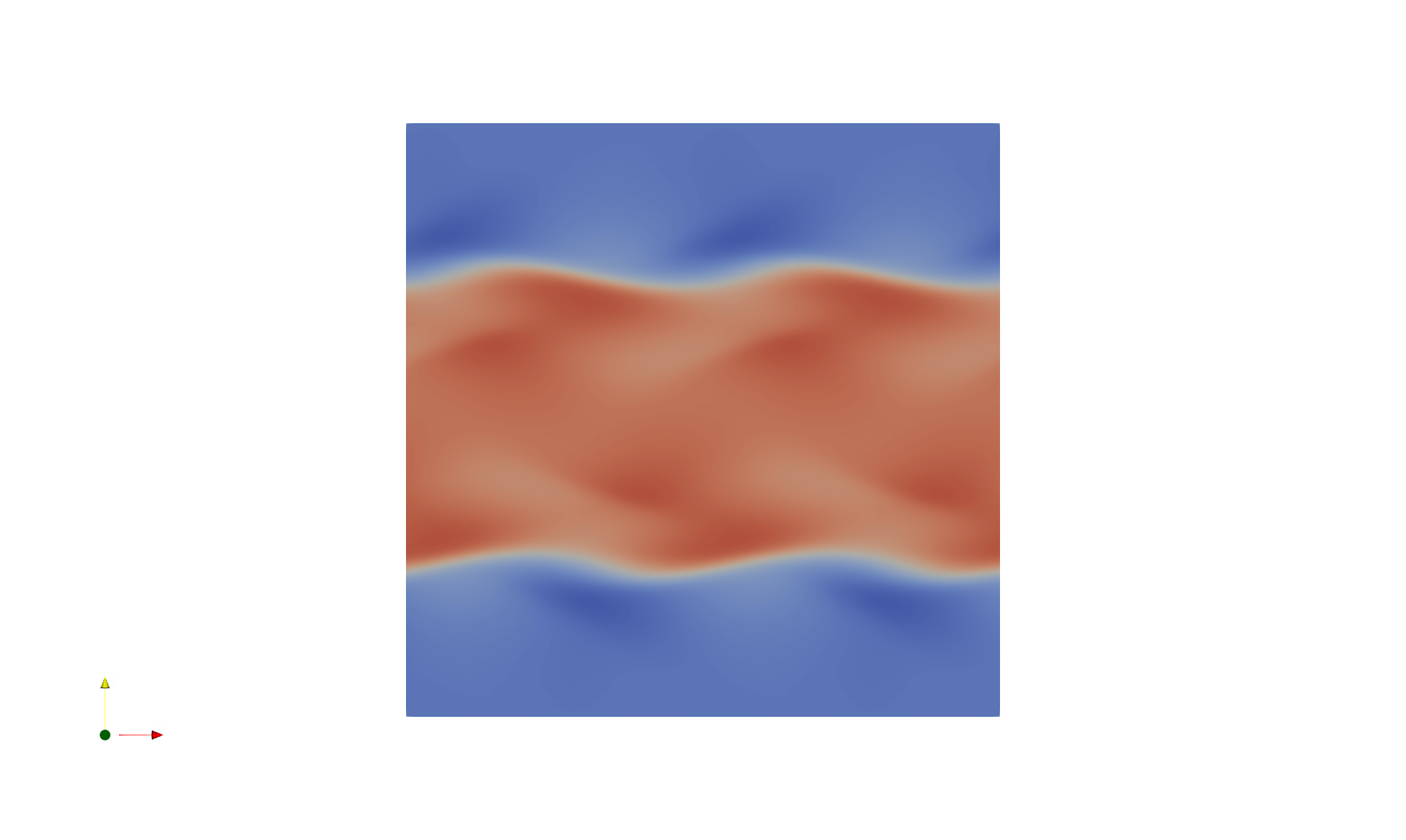}
  }
  \subfigure[$\rho^H$ at $t=2$]{
    \figlab{cns2d-khi-re200-ss-rH-t2}
      \includegraphics[trim=20cm 8cm 20cm 8cm,clip=true,width=0.31\textwidth]{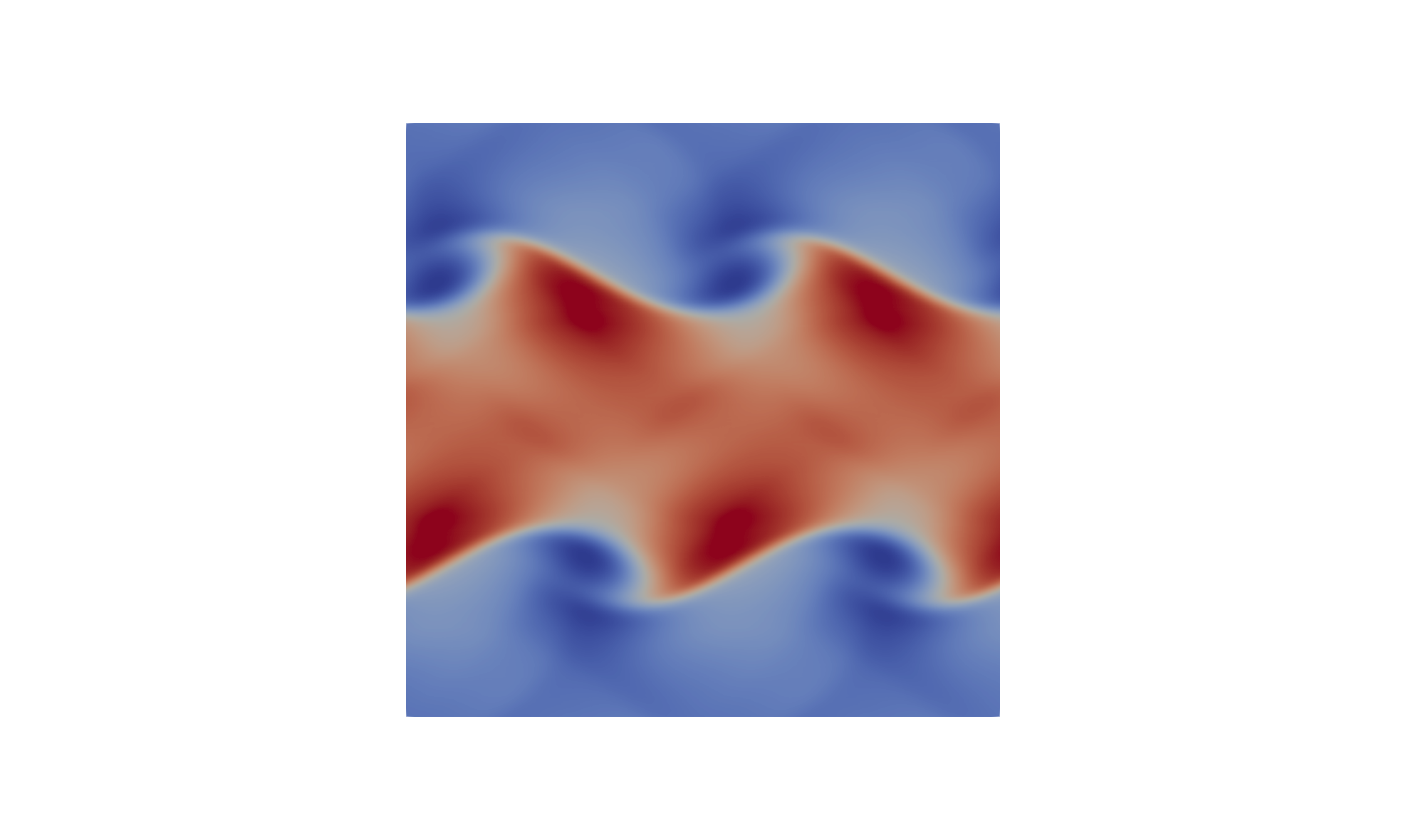}
  }
  \subfigure[$\rho^H$ at $t=3$]{
    \figlab{cns2d-khi-re200-ss-rH-t3}
      \includegraphics[trim=20cm 8cm 20cm 8cm,clip=true,width=0.31\textwidth]{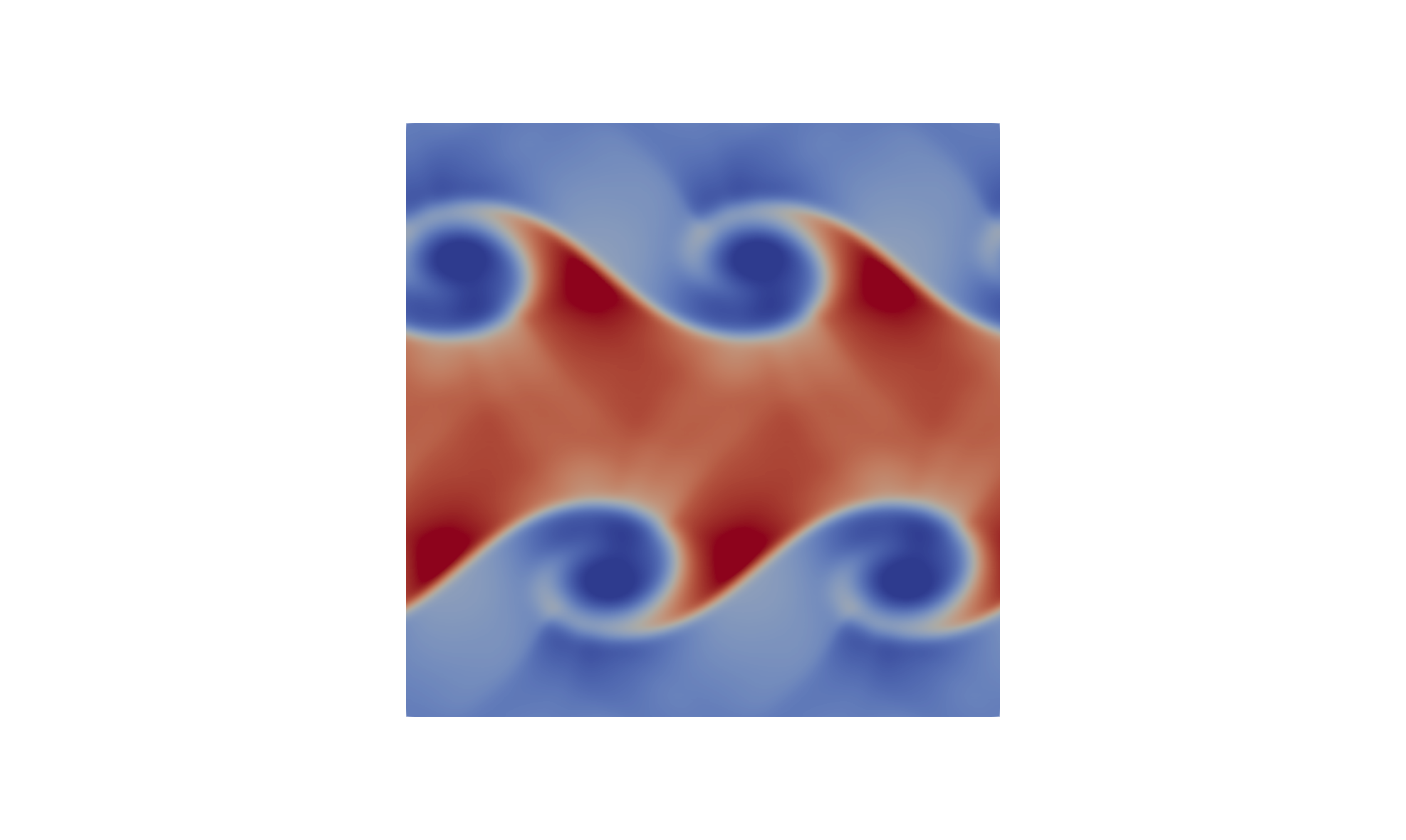}
  } \\
  \subfigure[$\vort^H$ at $t=1$]{
    \figlab{cns2d-khi-re200-ss-vortH-t1}
      \includegraphics[trim=20cm 8cm 20cm 8cm,clip=true,width=0.31\textwidth]{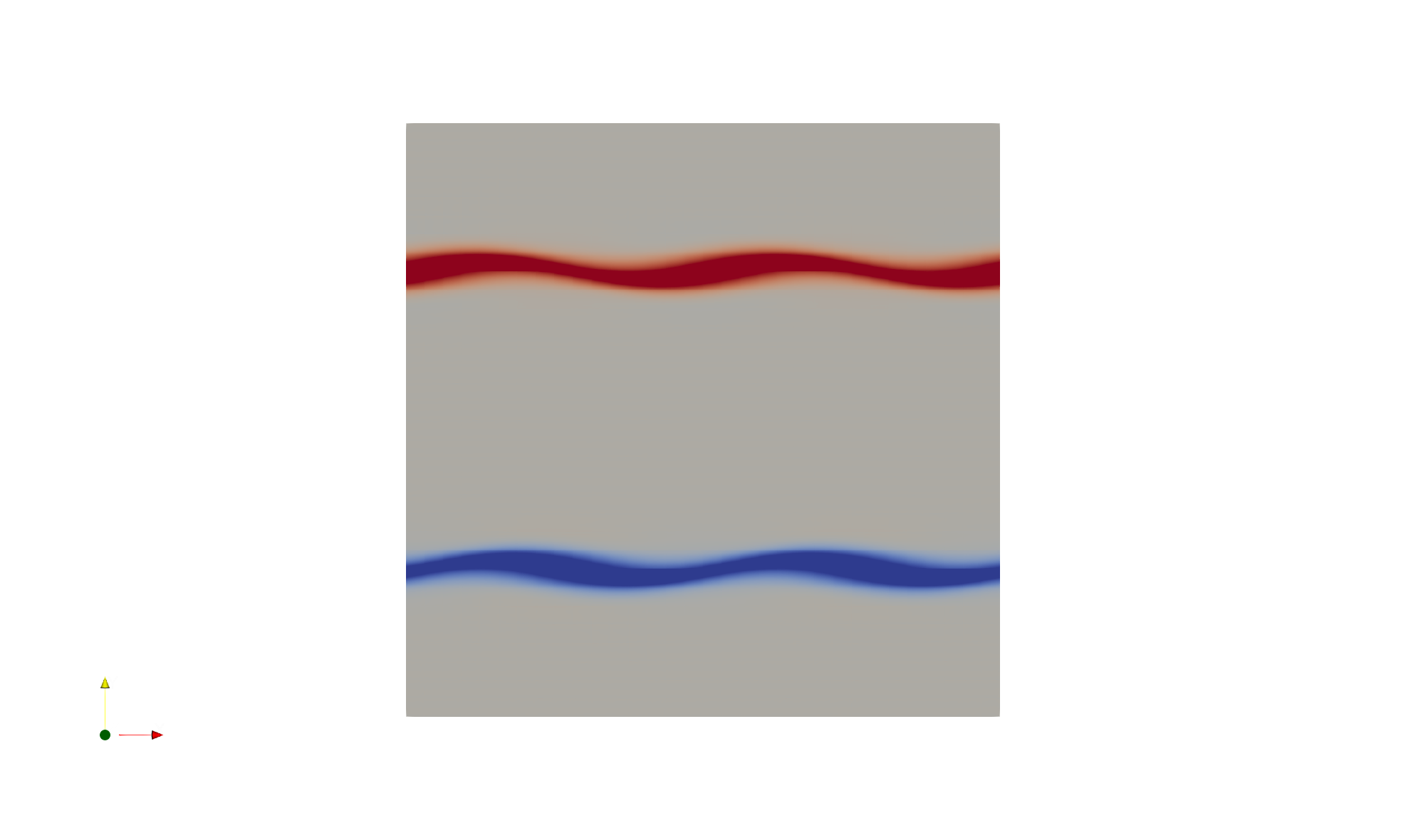}
  }
  \subfigure[$\vort^H$ at $t=2$]{
    \figlab{cns2d-khi-re200-ss-vortH-t2}
      \includegraphics[trim=20cm 8cm 20cm 8cm,clip=true,width=0.31\textwidth]{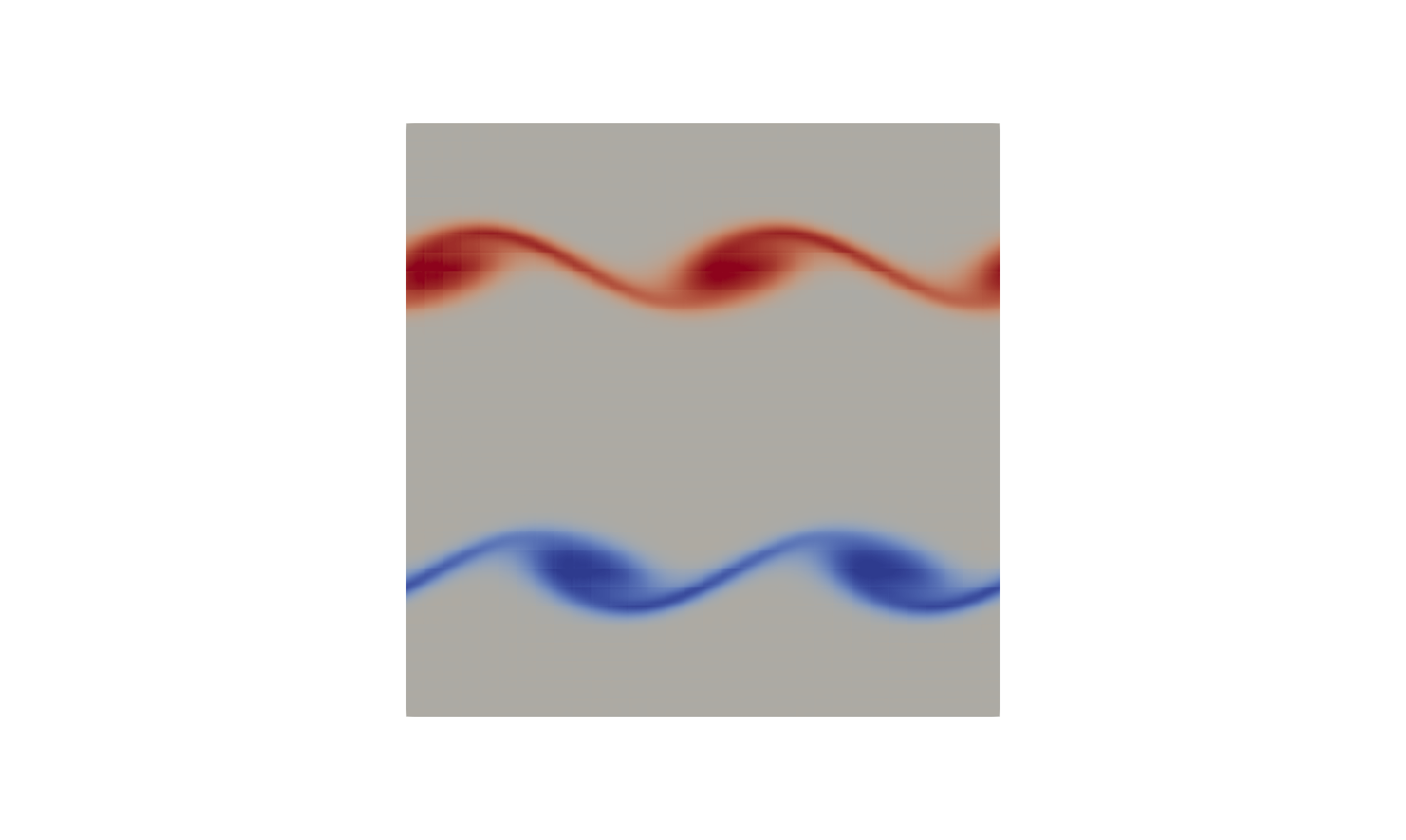}
  }
  \subfigure[$\vort^H$ at $t=3$]{
    \figlab{cns2d-khi-re200-ss-vortH-t3}
      \includegraphics[trim=20cm 8cm 20cm 8cm,clip=true,width=0.31\textwidth]{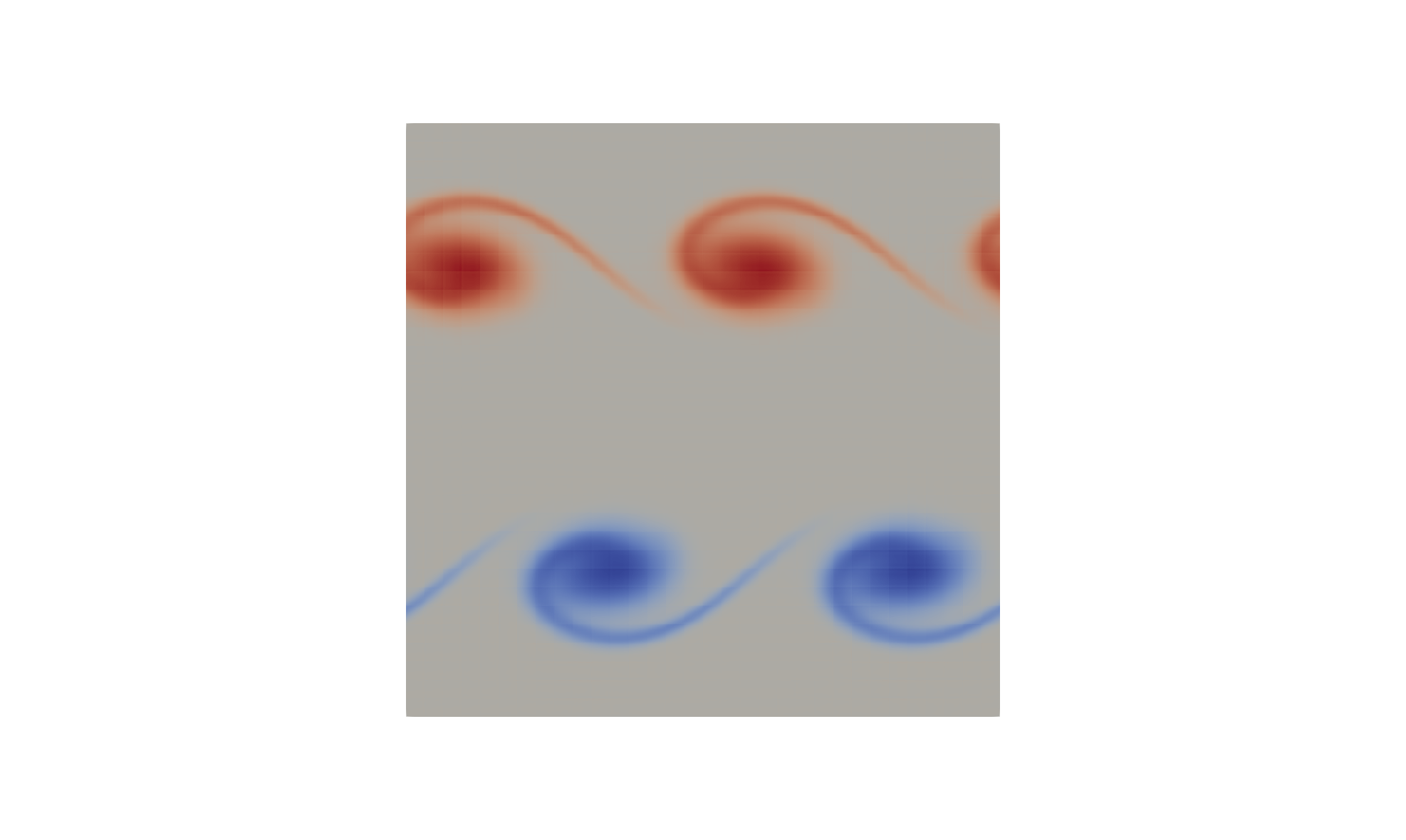}
  }
  \caption{2D Kelvin--Helmholtz instability: snapshots of the density and vorticity fields of high-order solution at $t=\LRc{1,2,3}$.
  }
  \figlab{cns2d-khi-re200-ss-rH-t1234}
\end{figure}

\begin{figure}[h!t!b!]
  \centering
  \subfigure[Projected density]{
    \figlab{cns2d-khi-re200-ss-t2-GrH-t3}
      \includegraphics[trim=20cm 8cm 20cm 8cm,clip=true,width=0.31\textwidth]{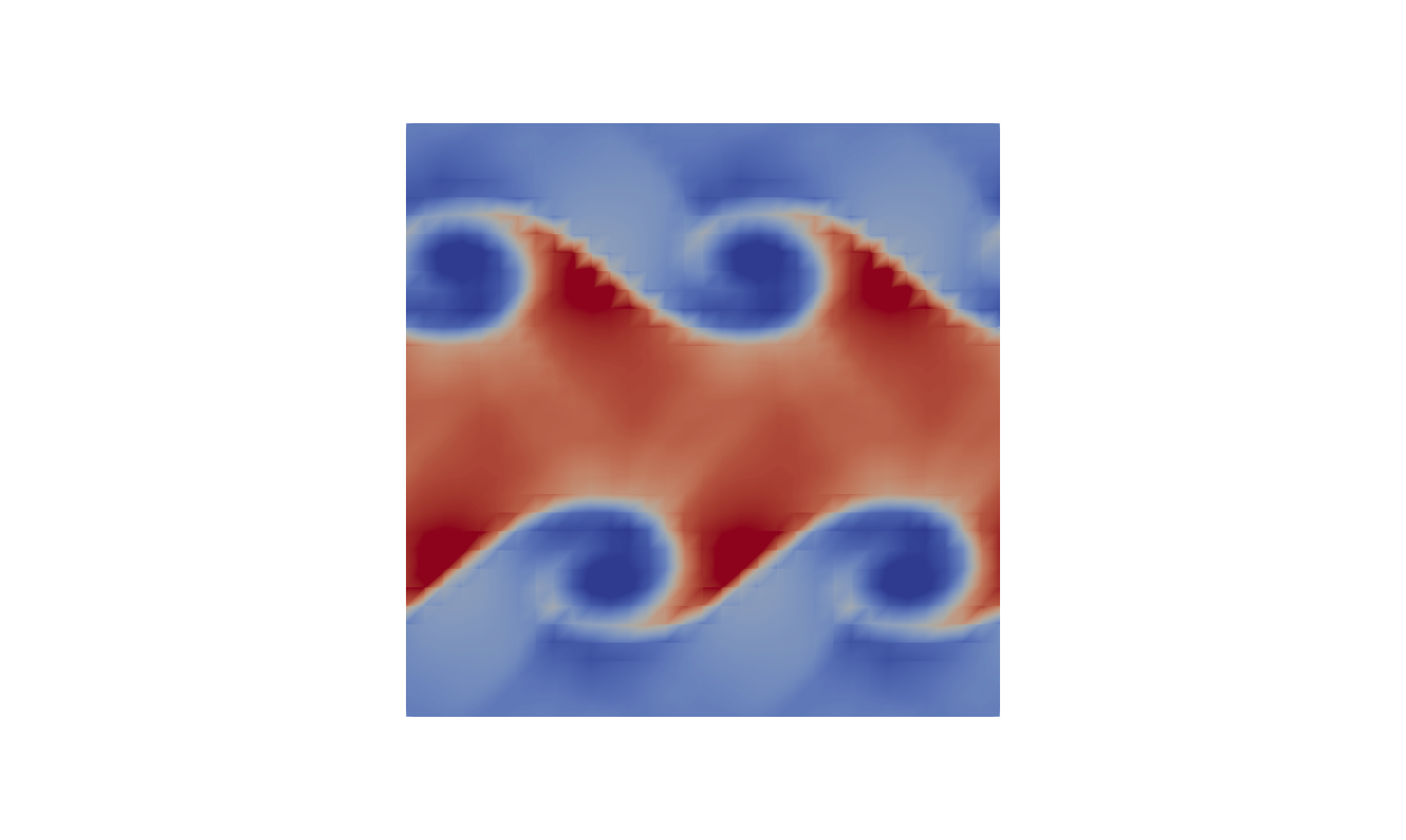}
  }
  \subfigure[Augmented density]{
    \figlab{cns2d-khi-re200-ss-t2-rLh-t3}
      \includegraphics[trim=20cm 8cm 20cm 8cm,clip=true,width=0.31\textwidth]{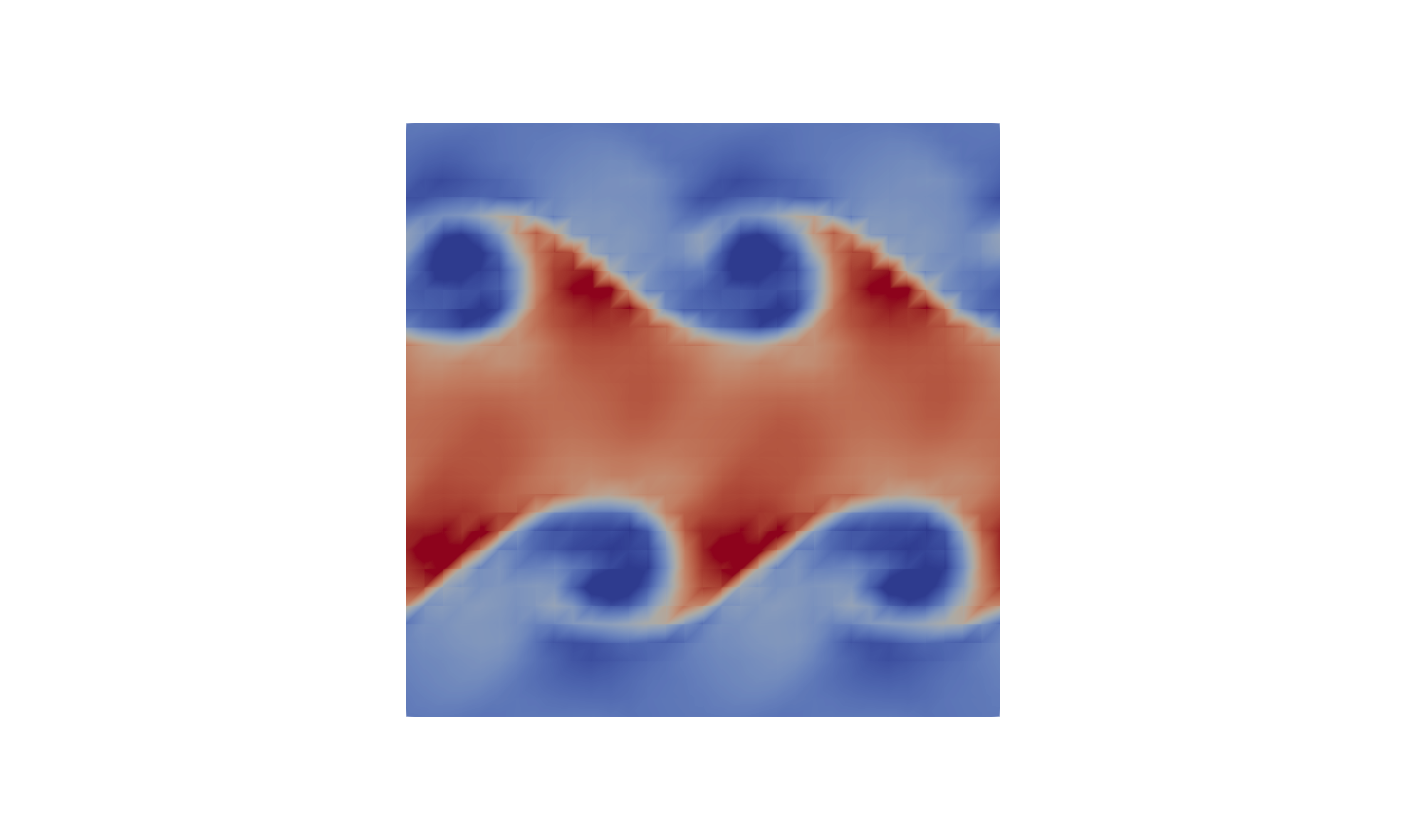}
  }
  \subfigure[Low-order density]{
    \figlab{cns2d-khi-re200-ss-t2-rL-t3}
      \includegraphics[trim=20cm 8cm 20cm 8cm,clip=true,width=0.31\textwidth]{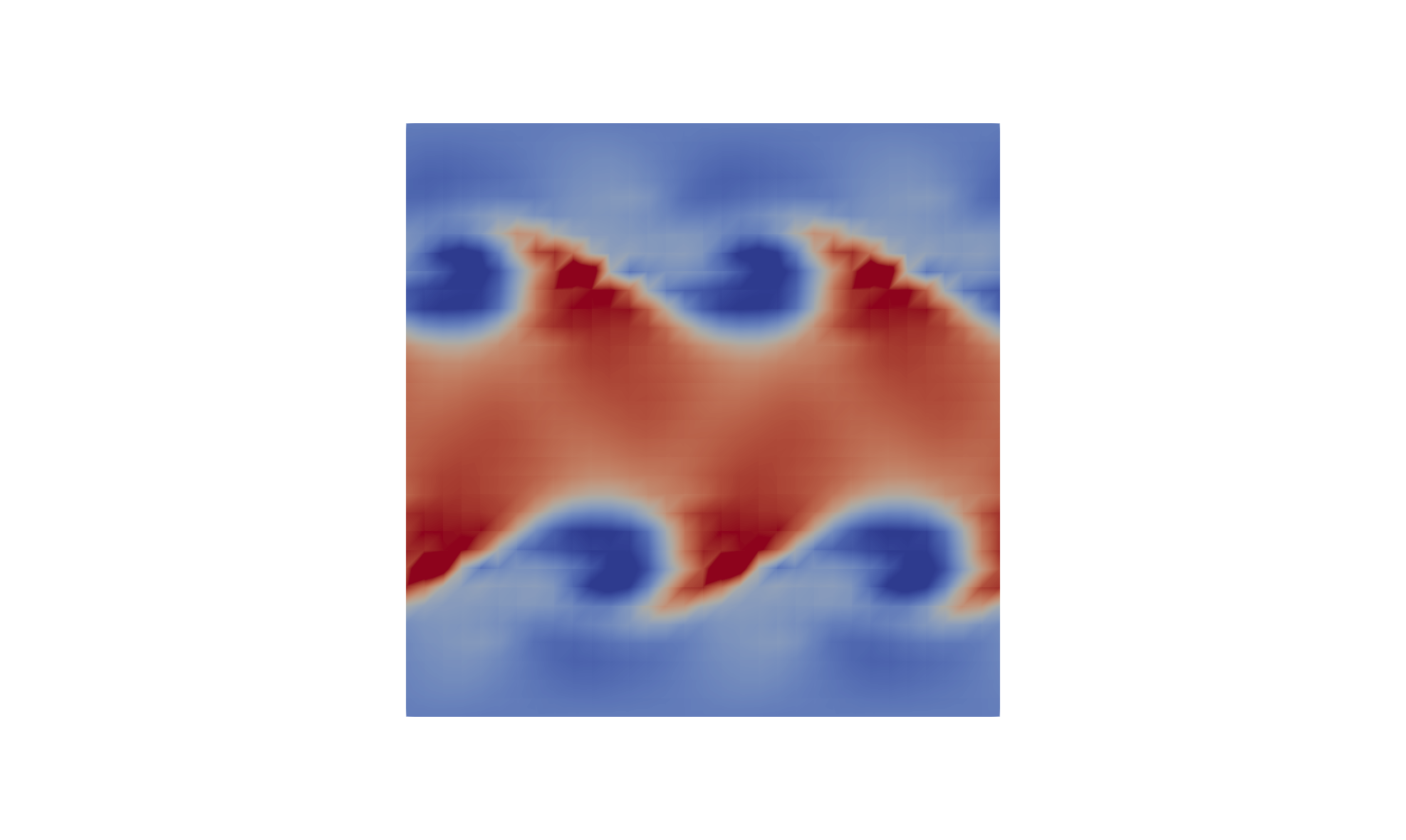}
  } \\
  \subfigure[Projected vorticity]{
    \figlab{cns2d-khi-re200-ss-GvortH-t3}
      \includegraphics[trim=20cm 8cm 20cm 8cm,clip=true,width=0.31\textwidth]{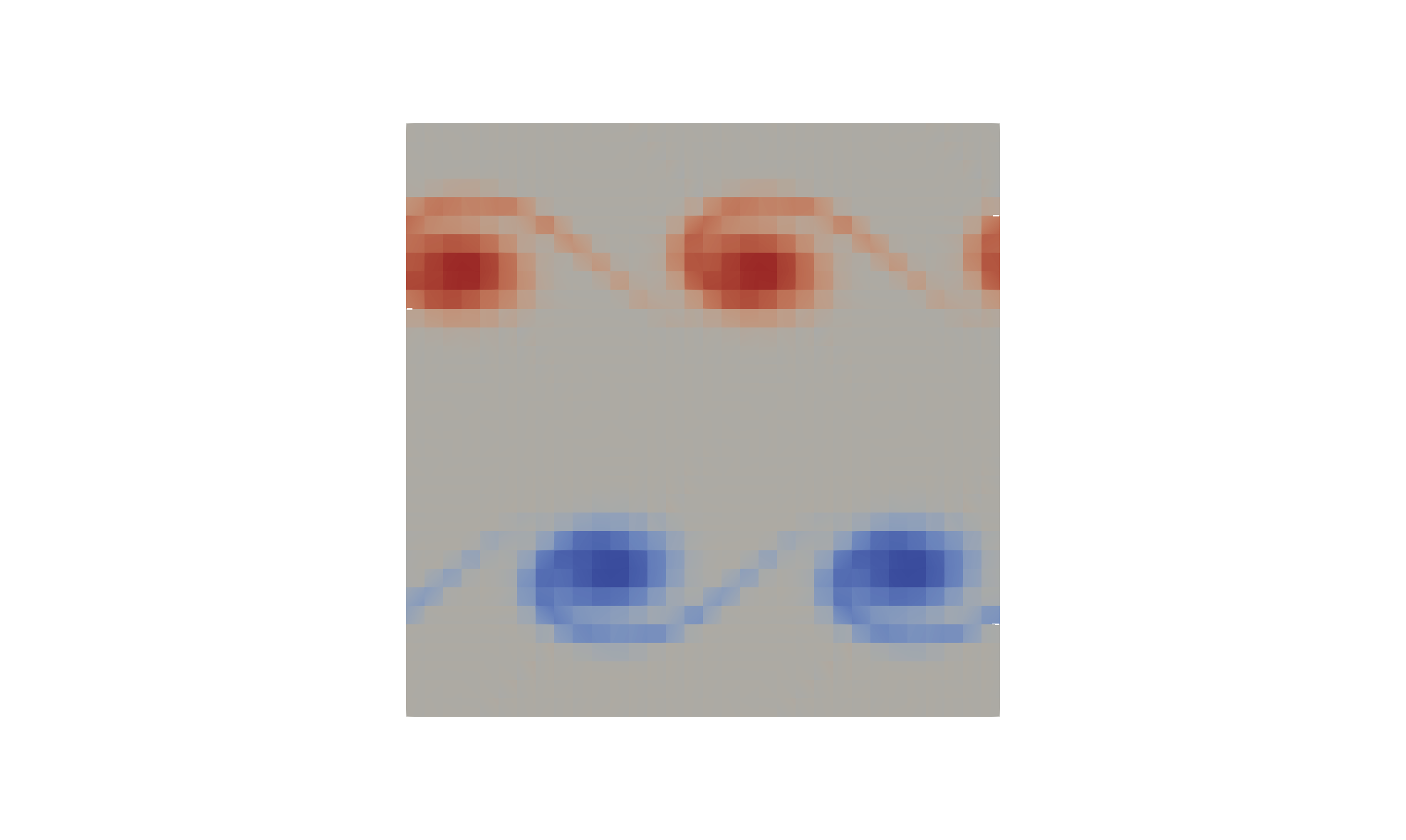}
  }
  \subfigure[Augmented vorticity]{
    \figlab{cns2d-khi-re200-ss-vortLh-t3}
      \includegraphics[trim=20cm 8cm 20cm 8cm,clip=true,width=0.31\textwidth]{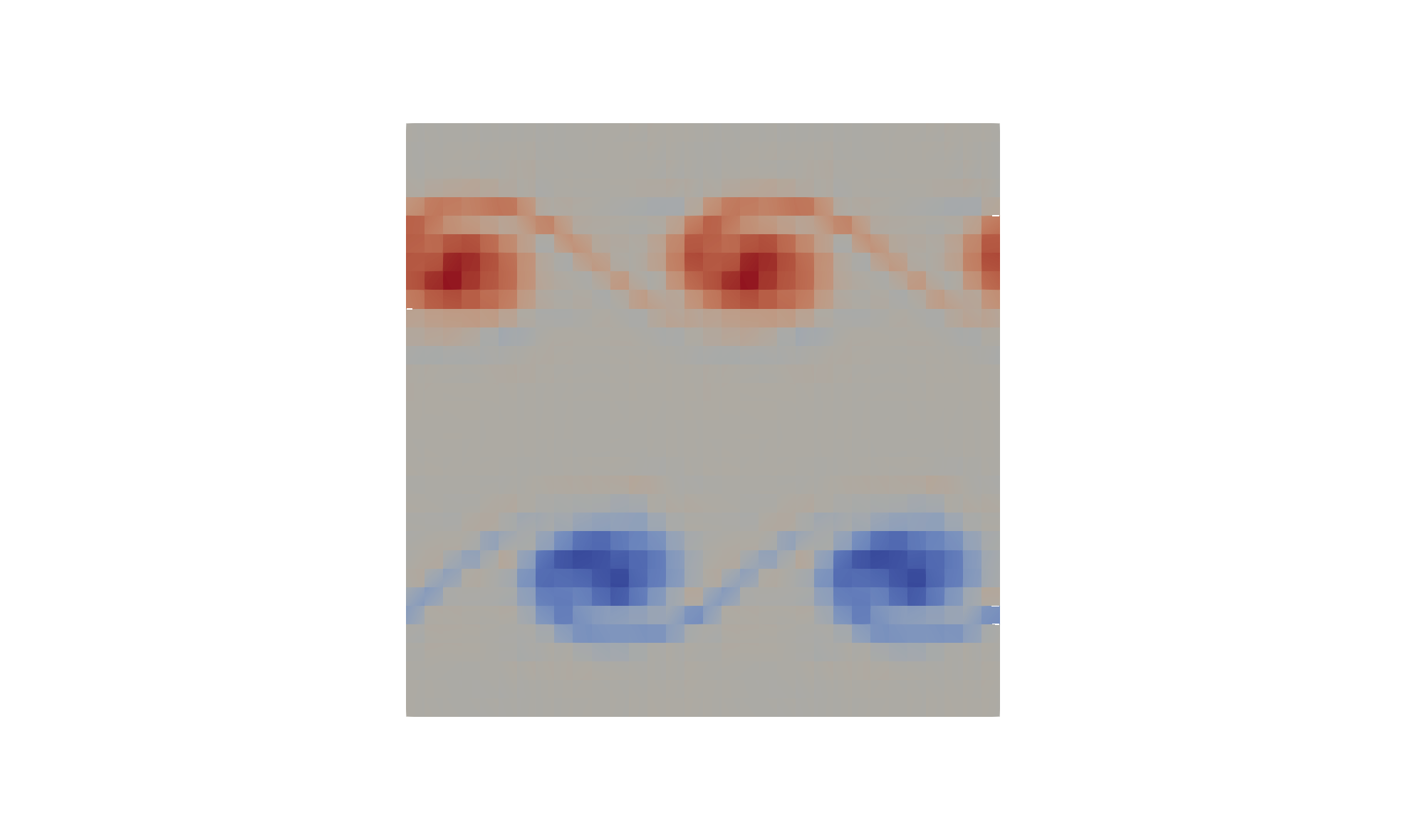}
  }
  \subfigure[Low-order vorticity]{
    \figlab{cns2d-khi-re200-ss-vorL-t3}
      \includegraphics[trim=20cm 8cm 20cm 8cm,clip=true,width=0.31\textwidth]{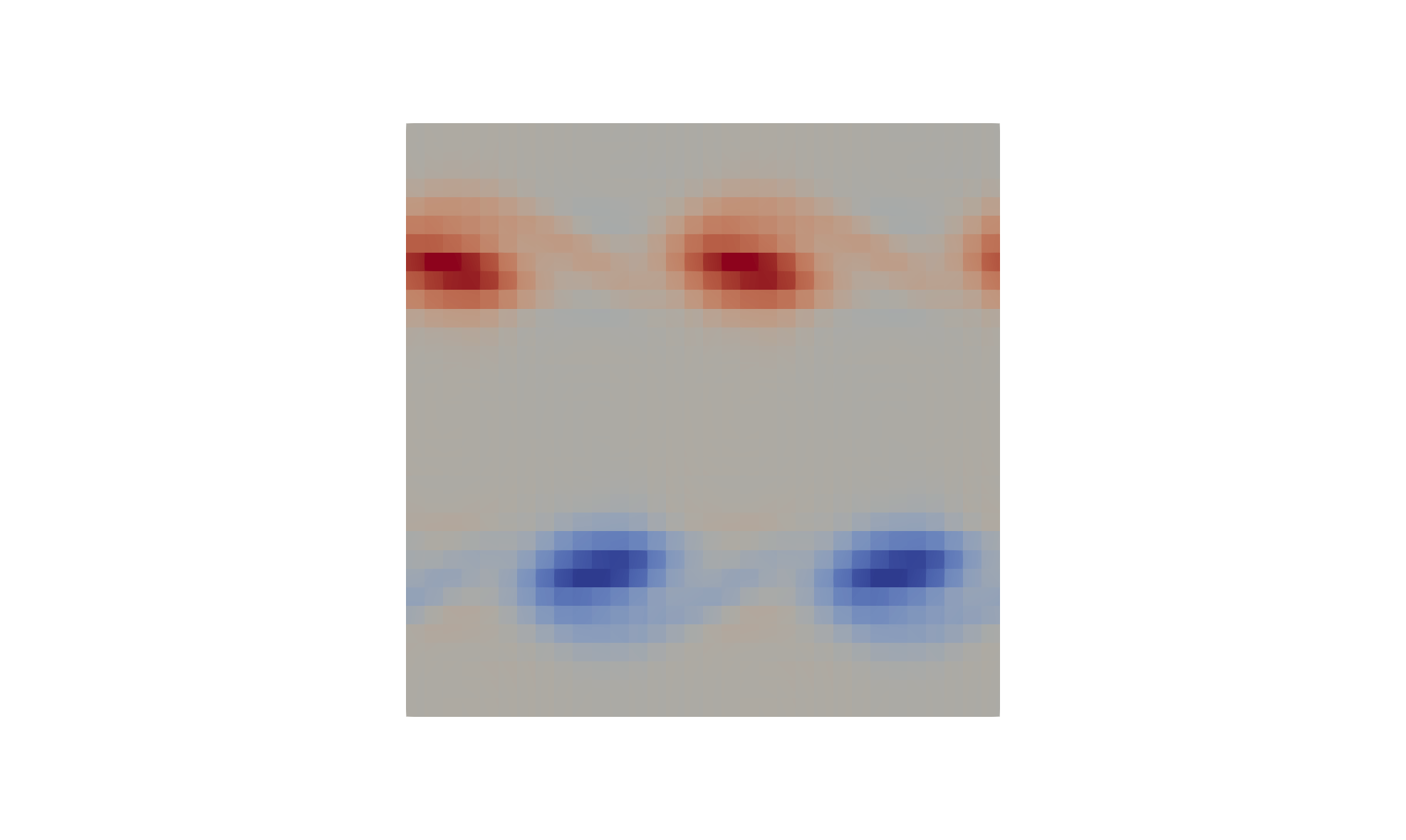}
  }
  \caption{2D Kelvin--Helmholtz instability: snapshots of the density and vorticity fields for the projected, augmented, and low-order approximations at $t=3$.
  }
  \figlab{cns2d-khi-re200-ss-from-t2-to-t3}
\end{figure}

To generate projected high-order approximations, 
we apply the projection matrix $\Pmat^L$ in \eqnref{l2proj-high2low-op} 
to the conservative variables $\ub^H$ on each element at every time step.
Figure \figref{cns2d-khi-re200-ss-t2-GrH-t3} 
displays the projected (filtered) solution of the density field at $t=3$, while  
Figure \figref{cns2d-khi-re200-ss-GvortH-t3} 
shows the vorticity field derived from the filtered conservative variables at $t=3$. 
Next, we divide the time series of $\projL\ub^H$ into the training data for $t=[2,2.7]$ and 
the test data for $t=[2.7,3]$. 
\footnote{
For convenience, $\projL\ub^H$ is implicitly understood as $\LRp{\projL\rho^H,\projL\rho u^H,\projL\rho v^H,\projL\rho E^H}$.
}

To train the neural network source term ${\Sb_\theta}$ in \eqnref{gov-ode-system},
 we randomly select four batch instances of $\projL \ub^H$ from the training data and integrate \eqnref{gov-ode-system} for $m$ time steps by using the fifth-order ERK method 
\cite{tsitouras2011runge} with a time step size of $\dt=2\times 10^{-3}$. The integration is performed on a mesh with the first-order ($N=1$) polynomial and $32\times32$ elements. 
Since the first-order DG method  has four nodal points per element and four conservative variables per node, the input and output layers of the neural network in \eqnref{nnsource-local-vector} have $n_0=16\times k_w^2$ and $n_D=16$ degrees of freedom, respectively.
The architecture of the neural network source is chosen as $\LRc{n_0,256,64,n_{D}}$. 
We use the AdaBelief \cite{zhuang2020adabelief} optimizer with a learning rate of $10^{-3}$ and train the network for $4,000$ epochs. 

For prediction, we use the projected solution $\projL\ub^H$ at $t=2$ as the initial condition 
and integrate it over $500$ time steps with $\dt=2\times 10^{-3}$,  
resulting in the final solution at $t=3$.
Figure \figref{cns2d-khi-re200-ss-from-t2-to-t3} 
shows the snapshots of the density and the vorticity fields for the projected solution (left), 
the augmented solution with $k_w=3$ and $m=11$ (middle), and the low-order solution (right) at $t=3$.
The results clearly show that the augmented solutions of the density ($\hat{\rho}$) and the vorticity ($\hat{\vort}$) 
are closer to the projected solutions of $\projL{\rho^H}$ and $\projL{\vort^H}$ 
than are the low-order solutions of $\rho^L$ and $\vort^L$, respectively.  
While the low-order solutions suffer from numerical dissipation, the augmented solutions effectively capture the vortical structure.

\subsubsection{
Incorporating Reynolds number}
\seclab{reynoldsnumber}

To illustrate the generalization of the NODE-DG approach and its use for parametric partial differential equations, we incorporate the Reynolds number as an input to the neural network source term
, ${\Sb_\theta}(\ubh^L; Re)$, increasing the input dimension of ${\Sb_\theta}$ to $n_0 = 16 \times k_w^2 + 1$. 
Similar to the case with $Re=200$, we generate projected simulation trajectories for $Re=\{100,300\}$. 
Using the dataset across Reynolds numbers $Re=\{100,200,300\}$, 
we randomly select four states from the trajectories and gather $m$ consecutive sequences.
The neural network source term is then trained following the same procedure as for $Re=200$ with $k_w=3$ and $m=11$. 
We use the AdaBelief optimizer with a learning rate of $5\times 10^{-4}$ and train the network for $2,000$ epochs. 
Figure \figref{cns2d-khi-rexxx2-density-from-t2-to-t3} 
and 
Figure \figref{cns2d-khi-rexxx2-vort-from-t2-to-t3}
display 
the snapshots of the density and vorticity fields for the projected (left), augmented (middle), and low-order (right) solutions at $t=3$, respectively.
In Figure \figref{cns2d-khi-rexxx2-density-from-t2-to-t3}, 
the augmented density fields more accurately capture the vortical structure compared with the low-order density fields. 
Similarly, in Figure \figref{cns2d-khi-rexxx2-vort-from-t2-to-t3}, the vortical structures become sharper 
when augmenting the neural network source term, as opposed to the low-order approximations. 
Augmenting the neural network source term improves the accuracy of low-order approximations. 
 We note that our approach successfully predicts for $Re=150$ and $Re=250$, which were not included in the training data.
\begin{figure}[h!t!b!]
  \centering
  \includegraphics[trim=20cm 8cm 20cm 8cm,clip=true,width=0.29\textwidth]{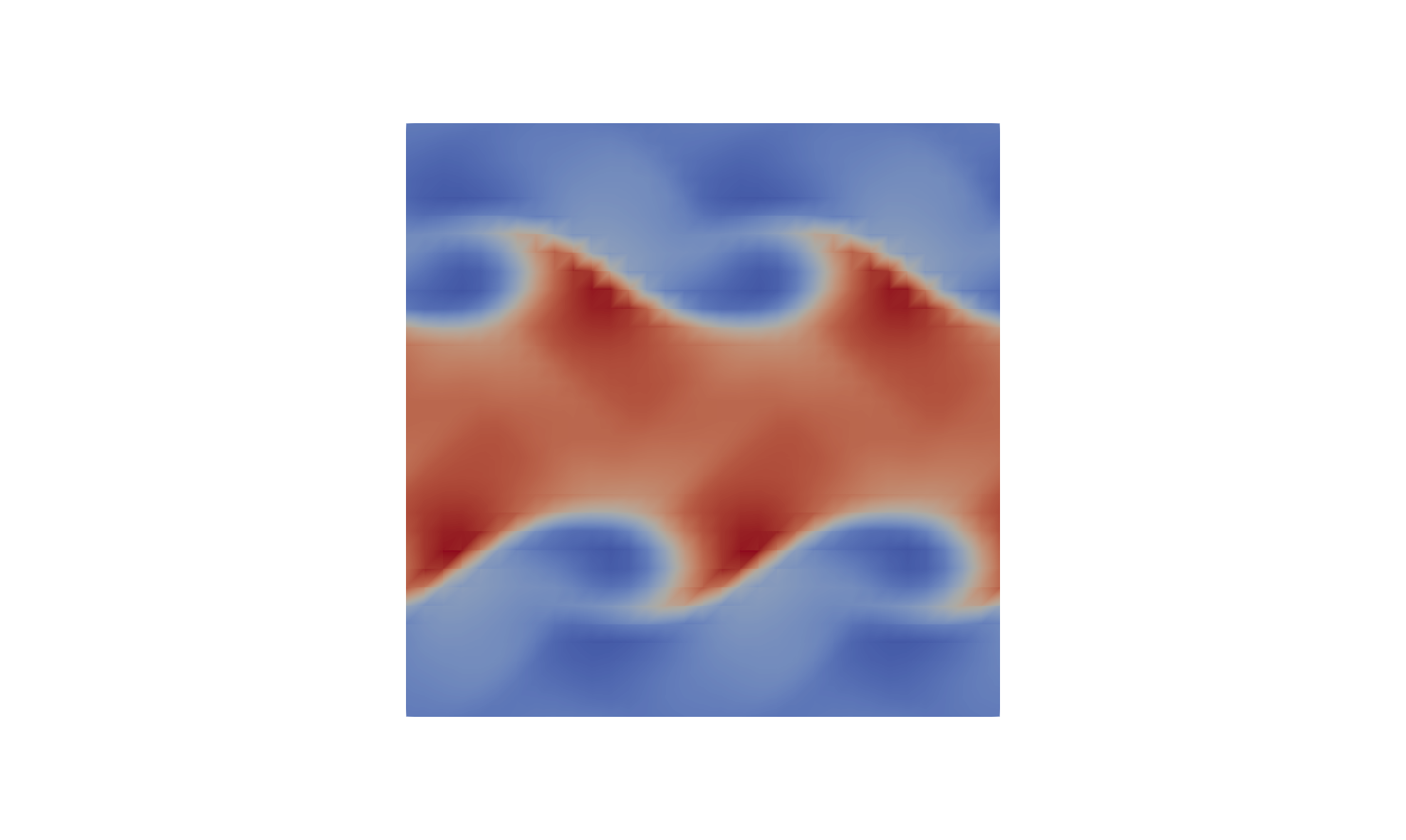}
  \includegraphics[trim=20cm 8cm 20cm 8cm,clip=true,width=0.29\textwidth]{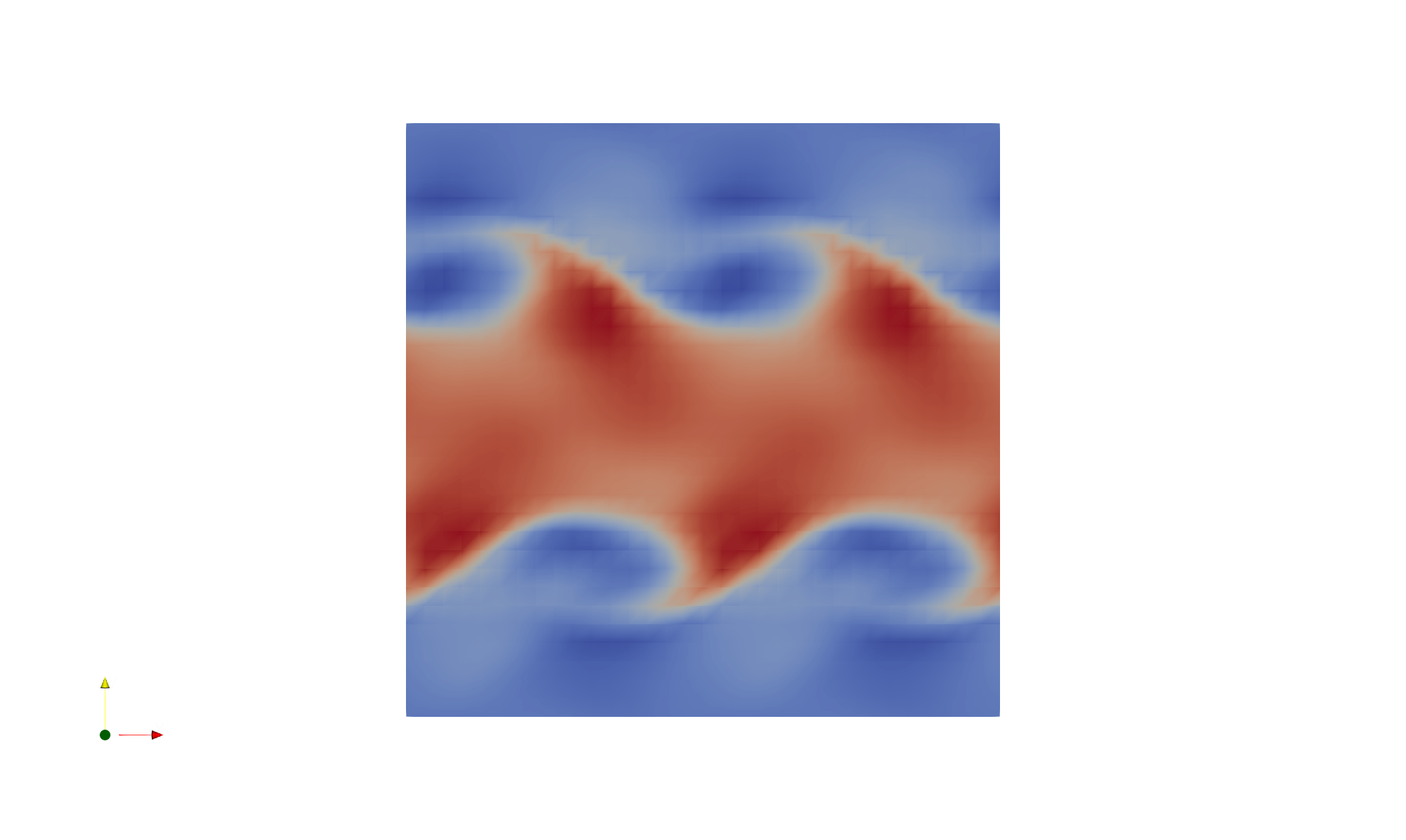}
  \includegraphics[trim=20cm 8cm 20cm 8cm,clip=true,width=0.29\textwidth]{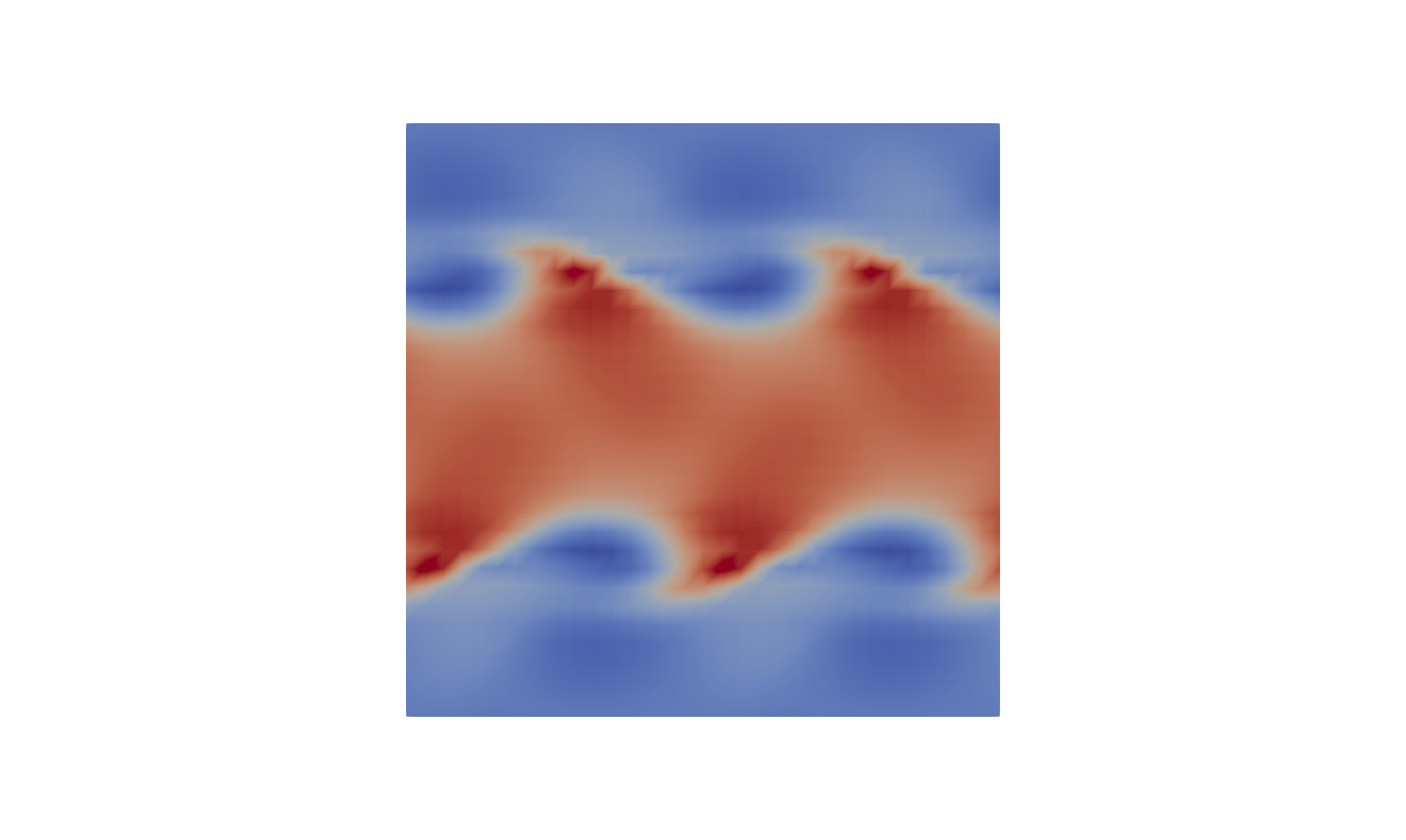}
  \\
  \includegraphics[trim=20cm 8cm 20cm 8cm,clip=true,width=0.29\textwidth]{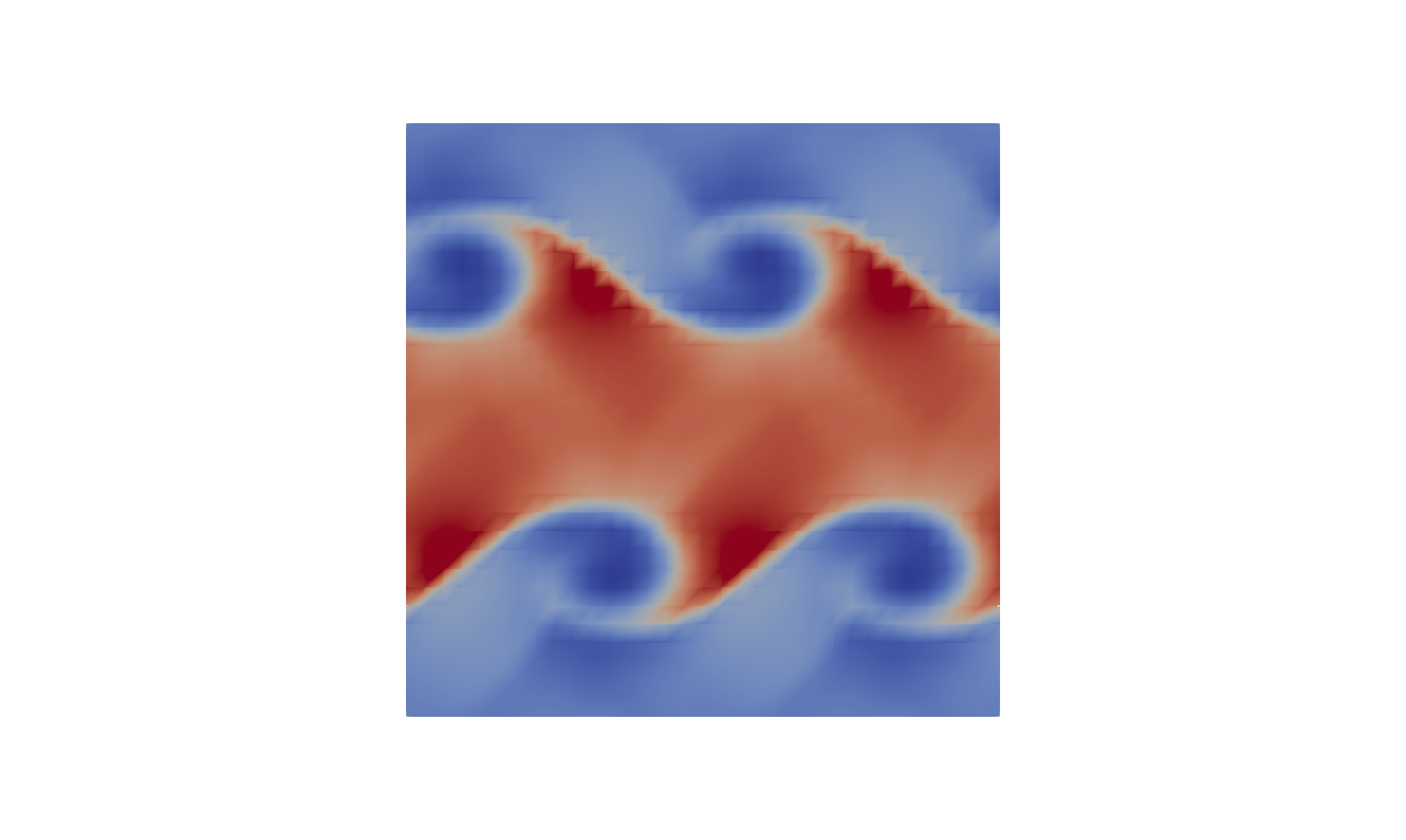}
  \includegraphics[trim=20cm 8cm 20cm 8cm,clip=true,width=0.29\textwidth]{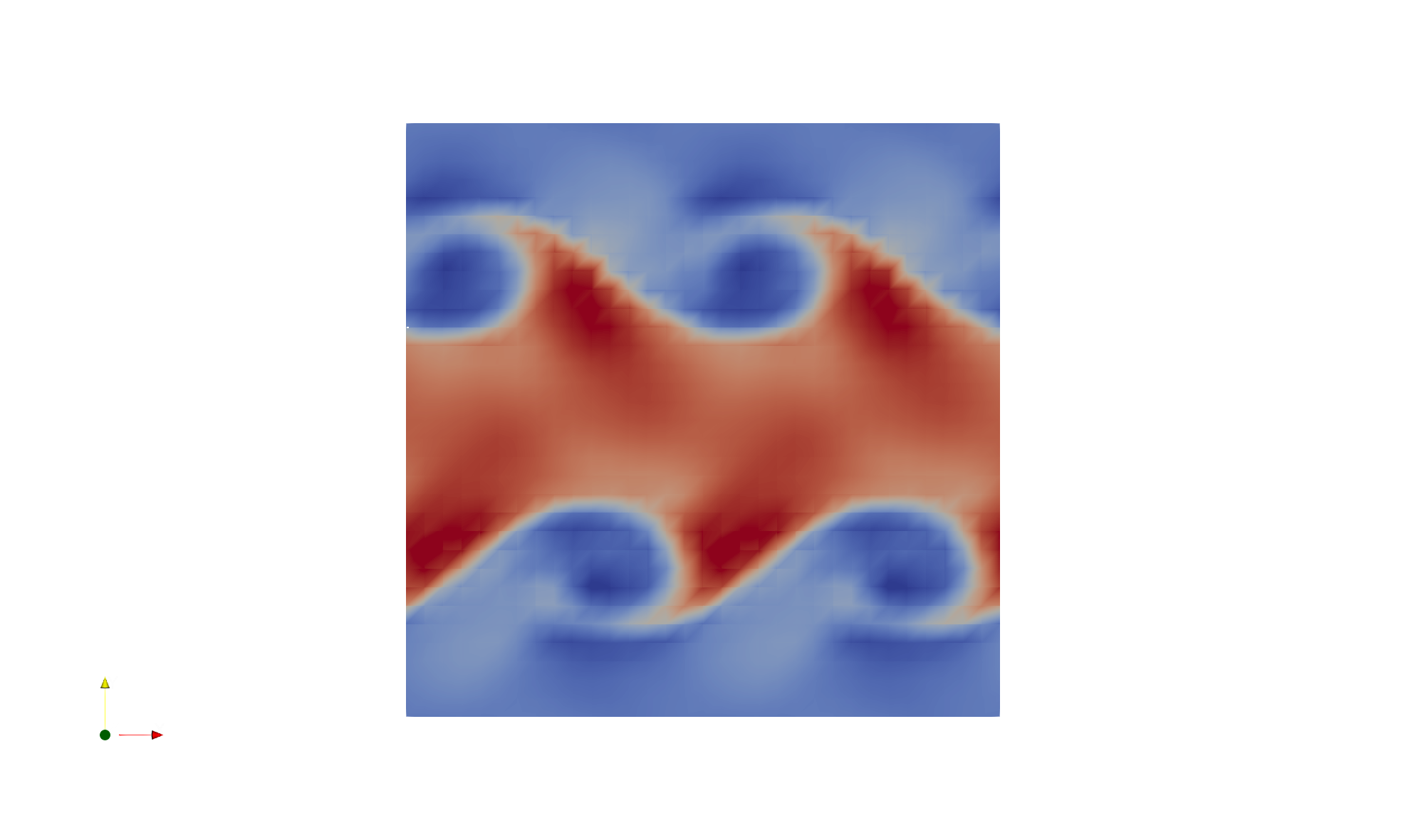}
  \includegraphics[trim=20cm 8cm 20cm 8cm,clip=true,width=0.29\textwidth]{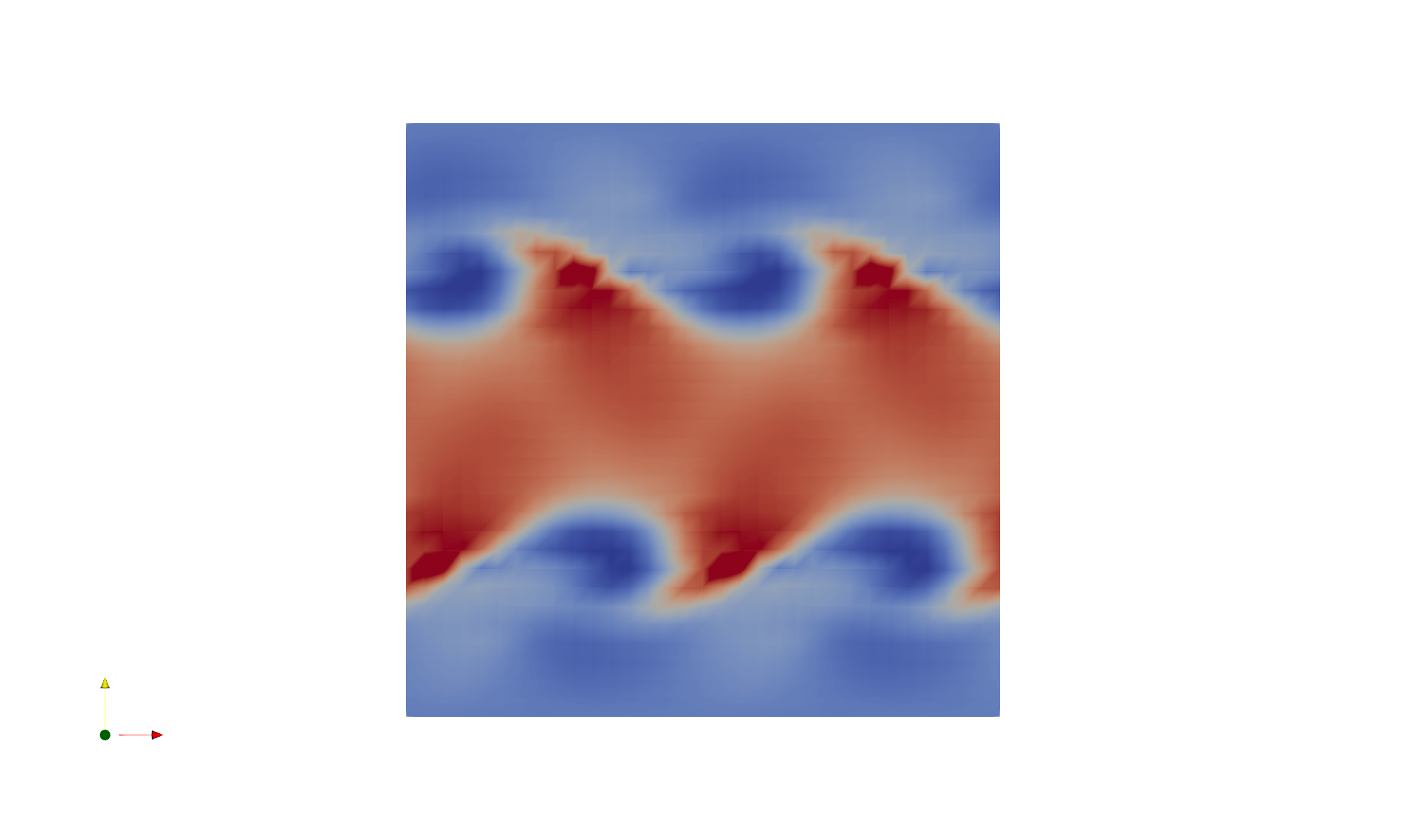}
  \\
  \includegraphics[trim=20cm 8cm 20cm 8cm,clip=true,width=0.29\textwidth]{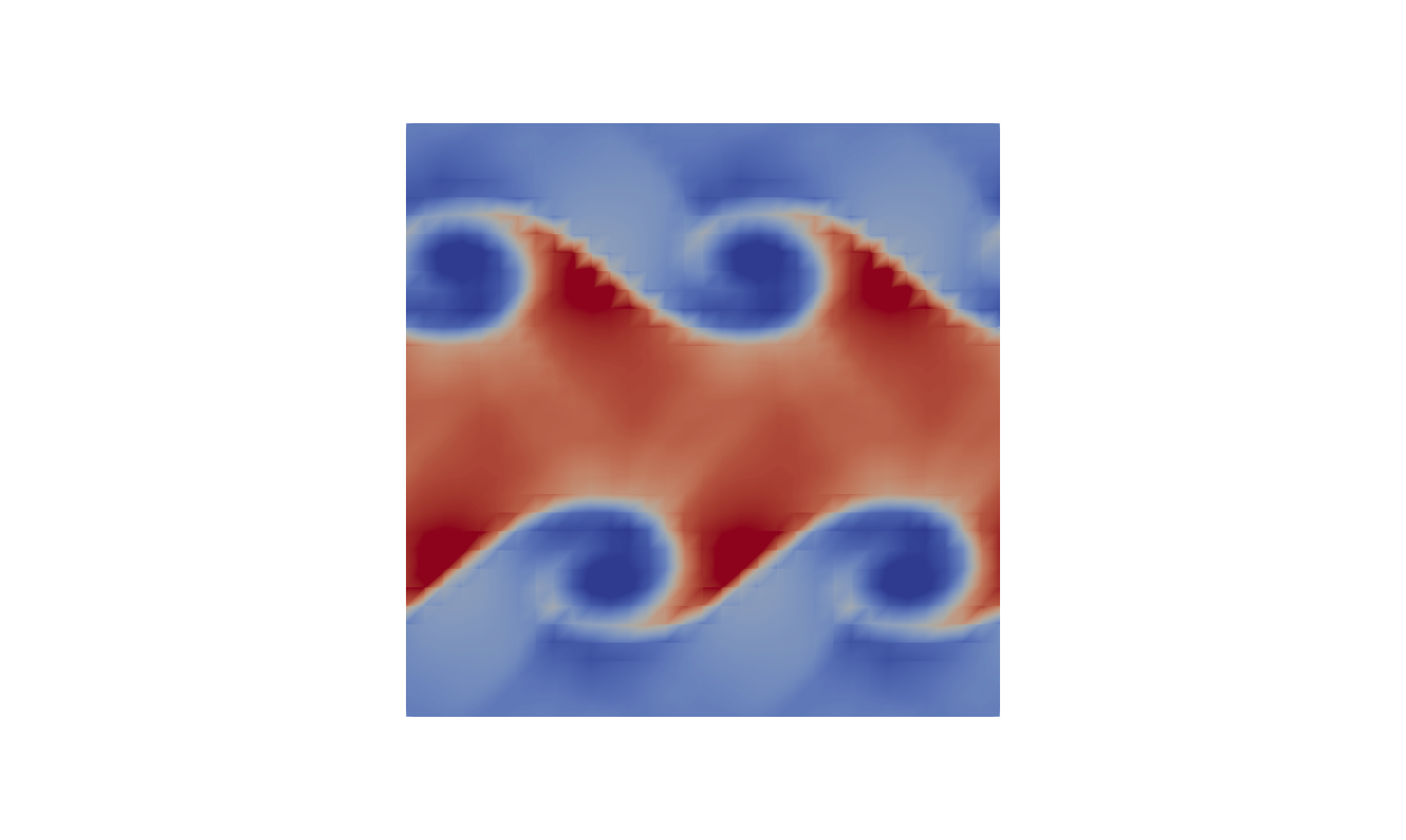}
  \includegraphics[trim=20cm 8cm 20cm 8cm,clip=true,width=0.29\textwidth]{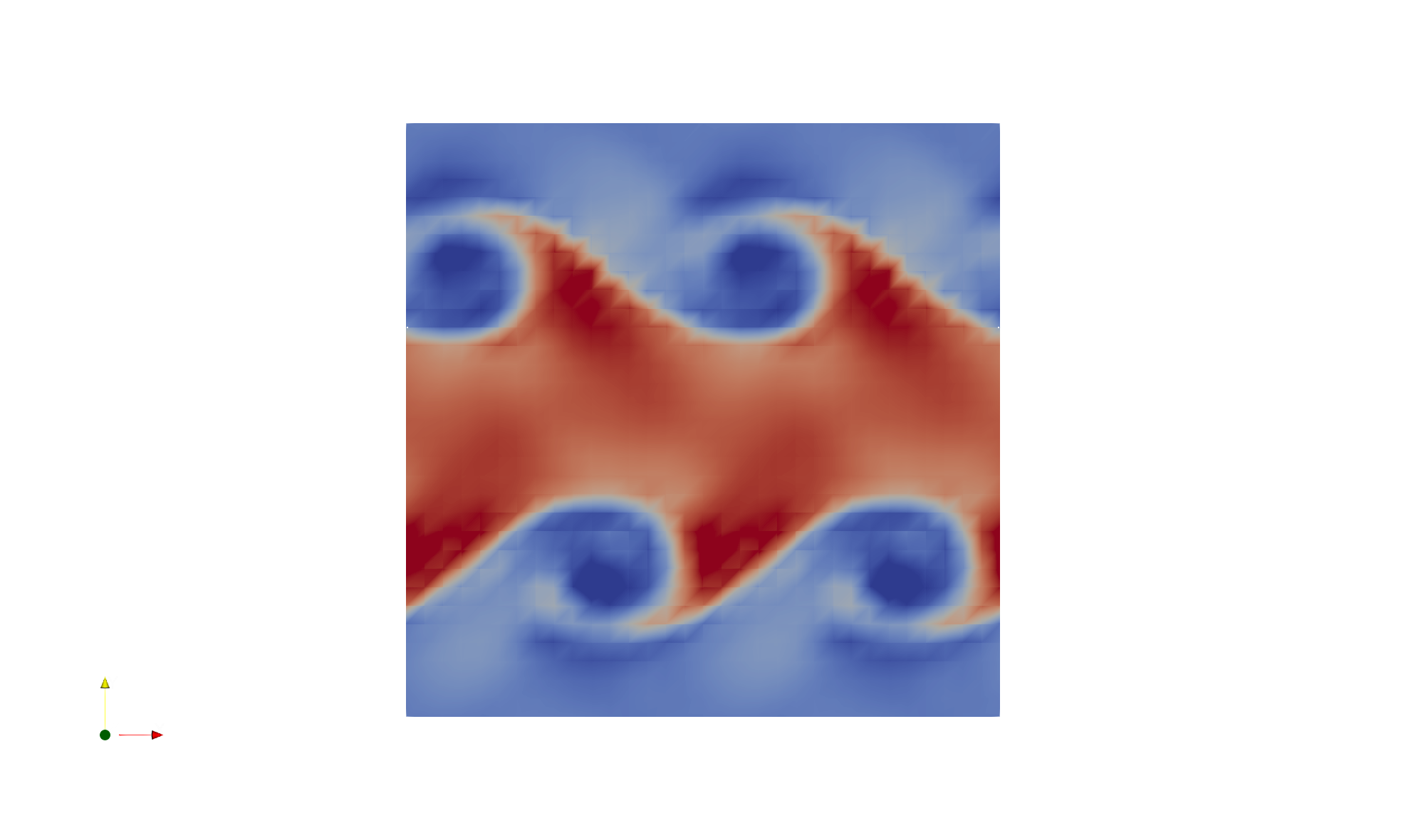}
  \includegraphics[trim=20cm 8cm 20cm 8cm,clip=true,width=0.29\textwidth]{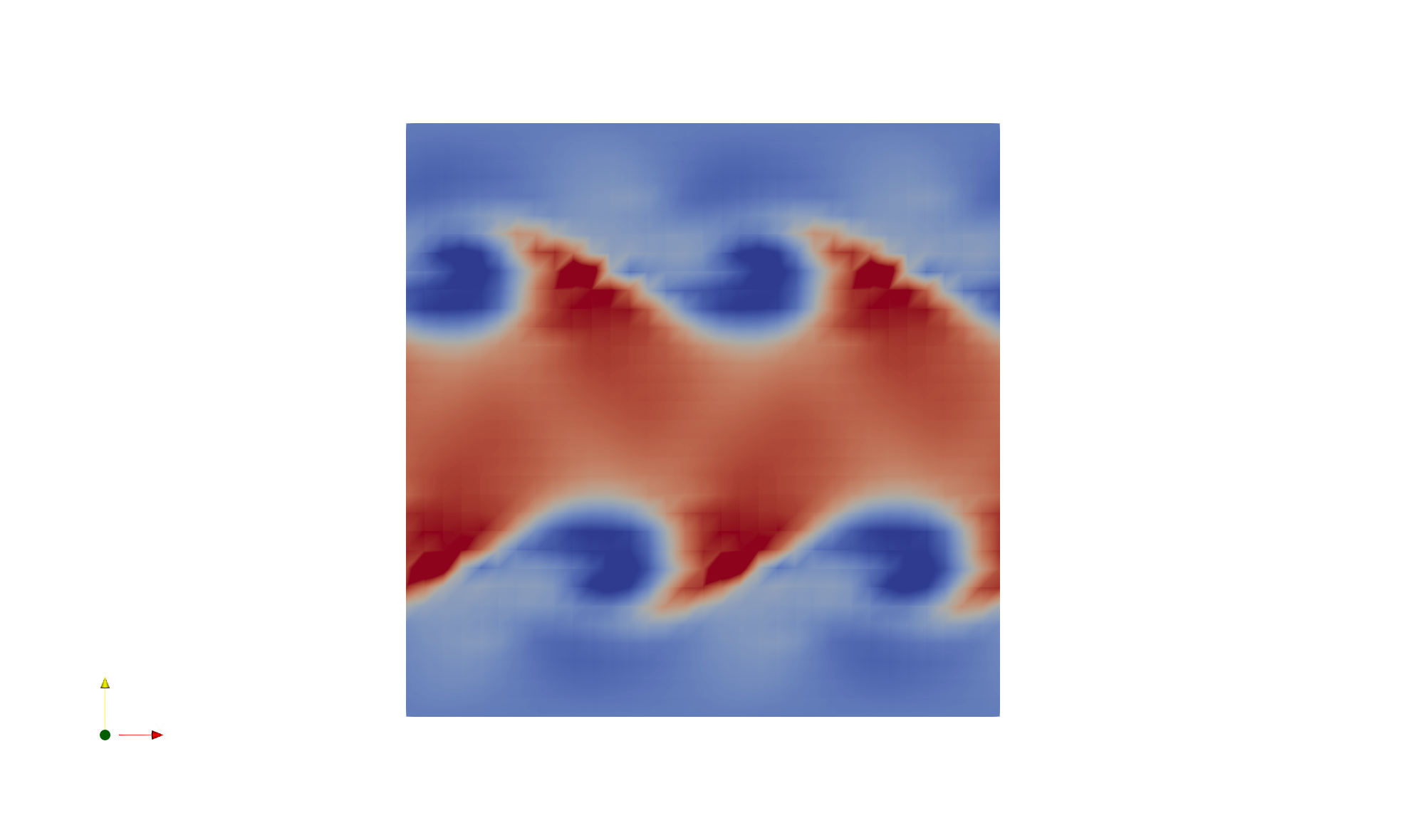}
  \\
  \includegraphics[trim=20cm 8cm 20cm 8cm,clip=true,width=0.29\textwidth]{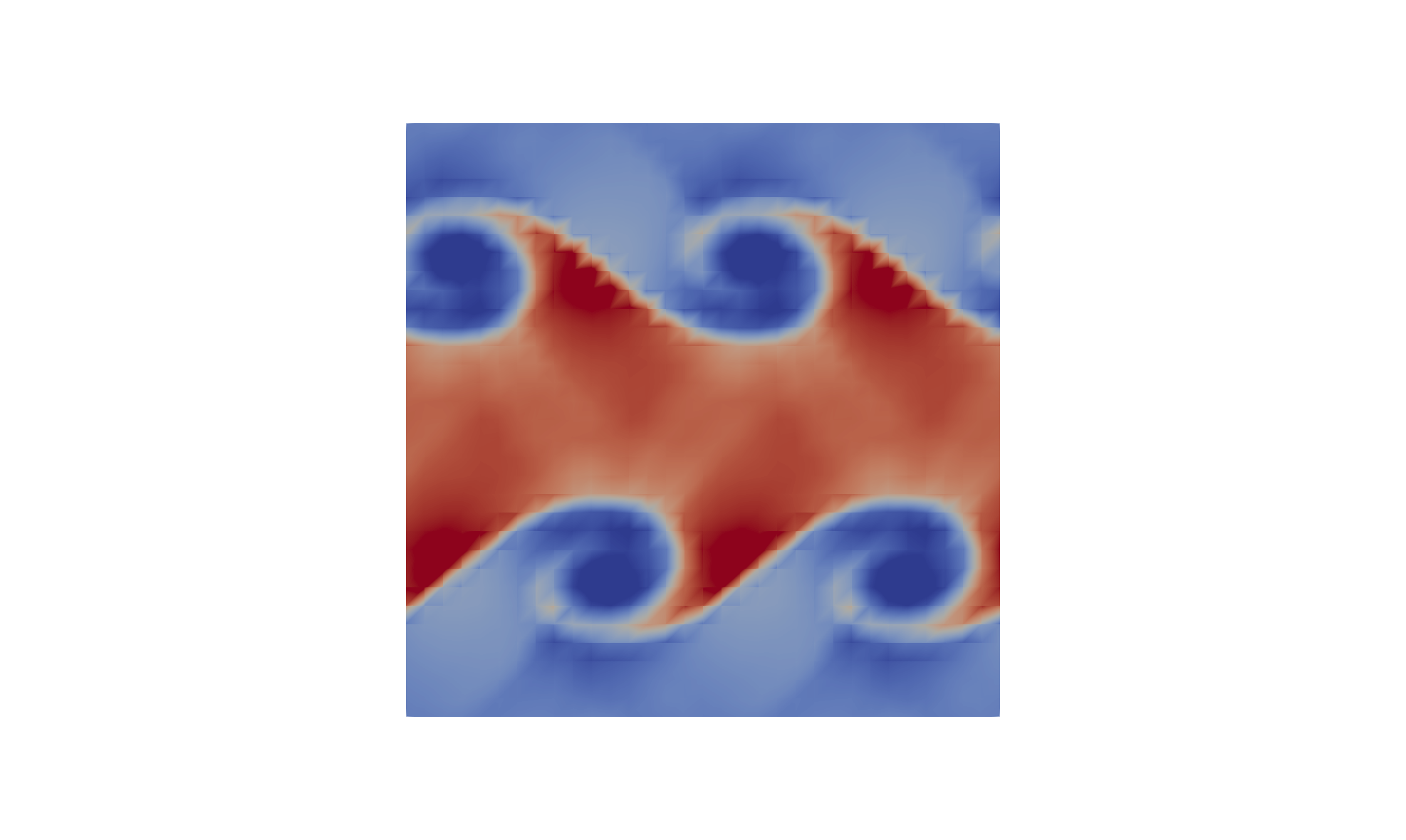}
  \includegraphics[trim=20cm 8cm 20cm 8cm,clip=true,width=0.29\textwidth]{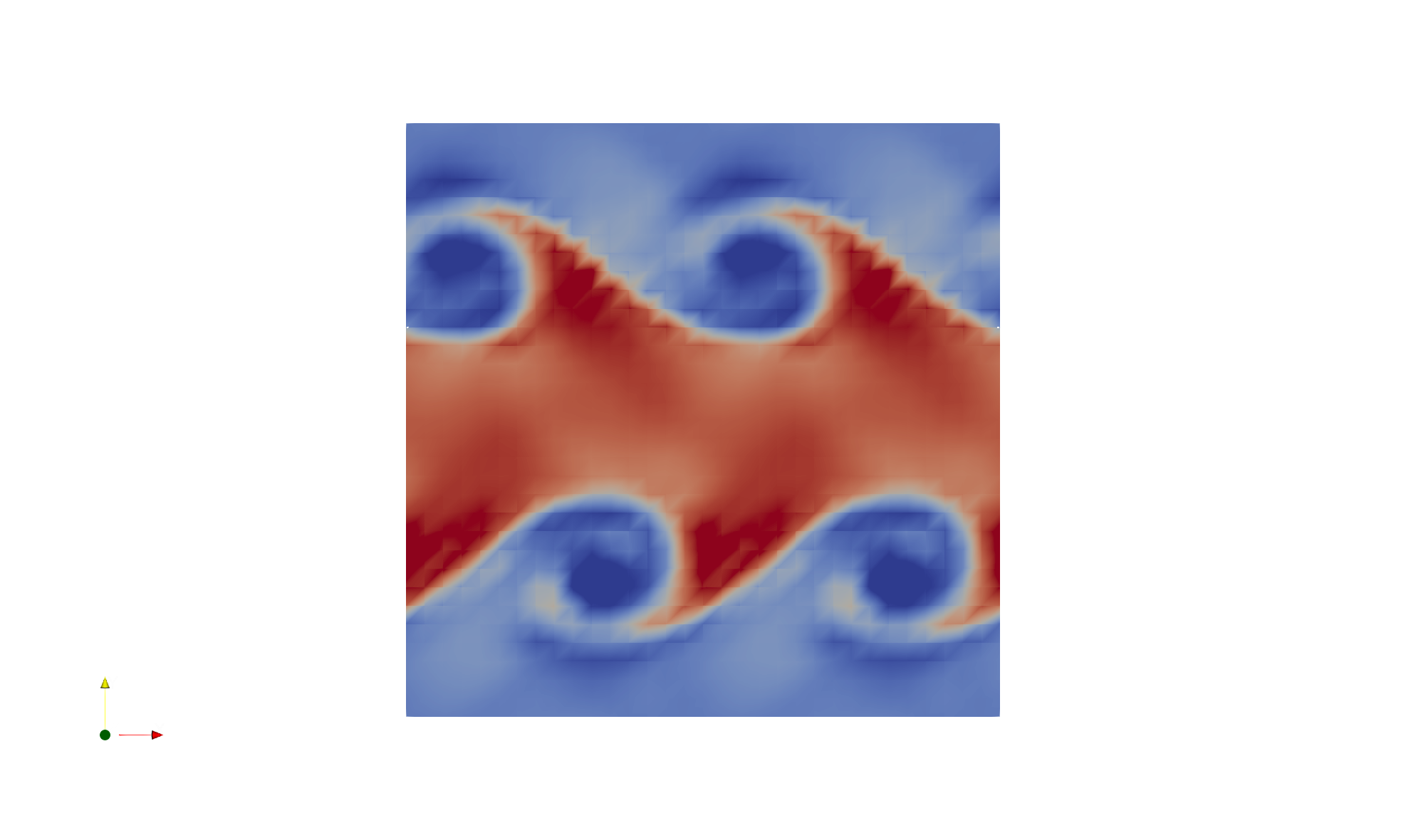}
  \includegraphics[trim=20cm 8cm 20cm 8cm,clip=true,width=0.29\textwidth]{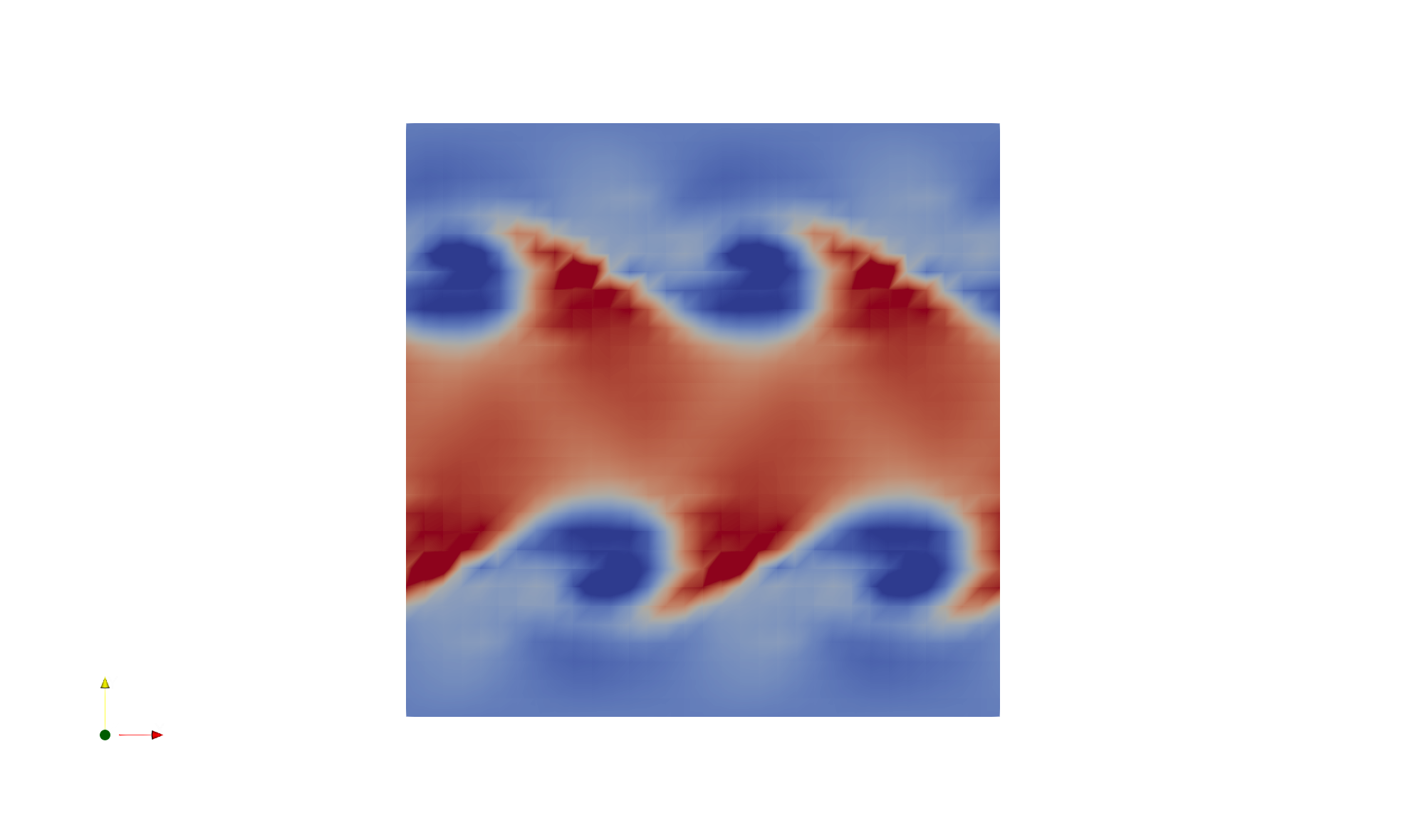}
  \\
  \includegraphics[trim=20cm 8cm 20cm 8cm,clip=true,width=0.29\textwidth]{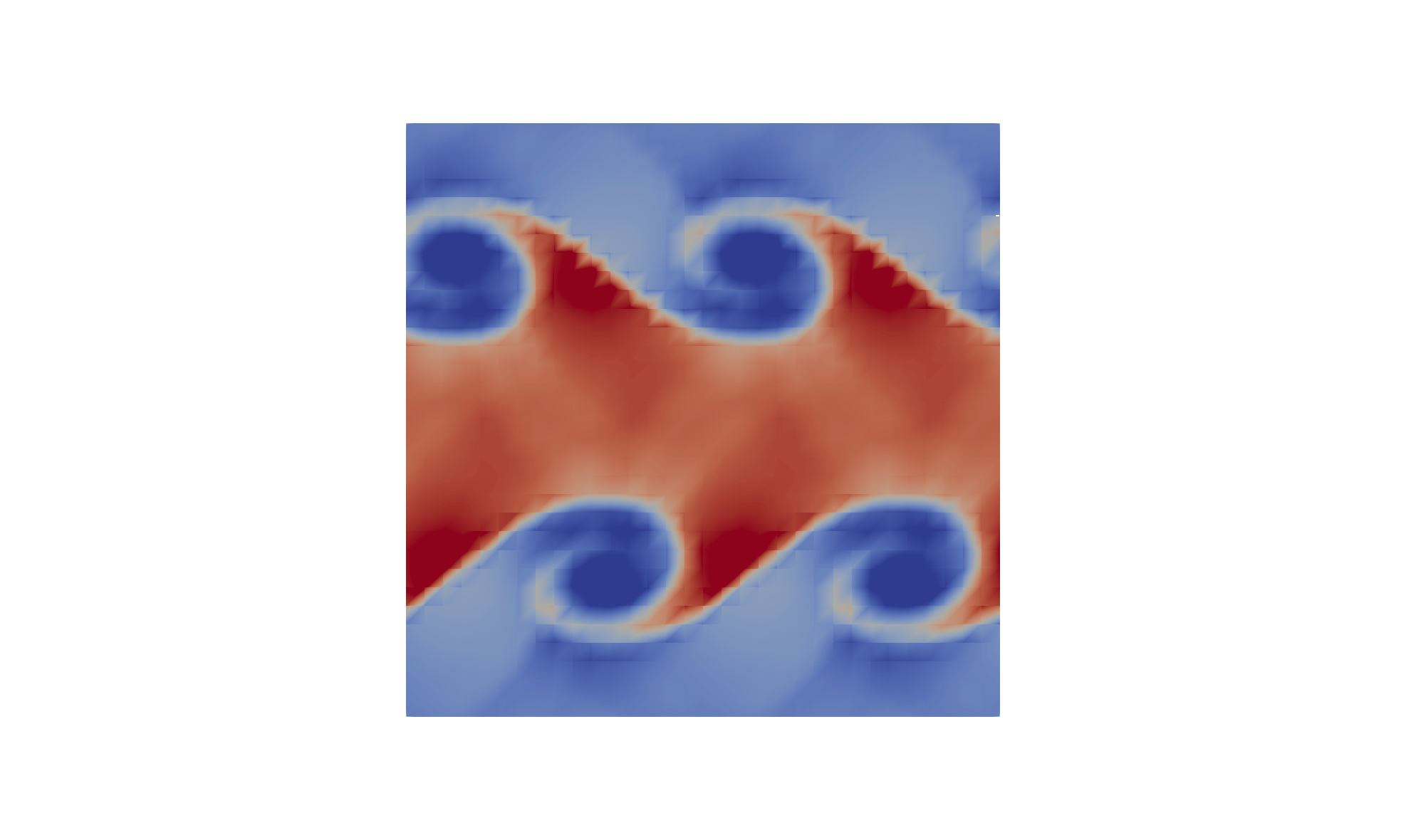}
  \includegraphics[trim=20cm 8cm 20cm 8cm,clip=true,width=0.29\textwidth]{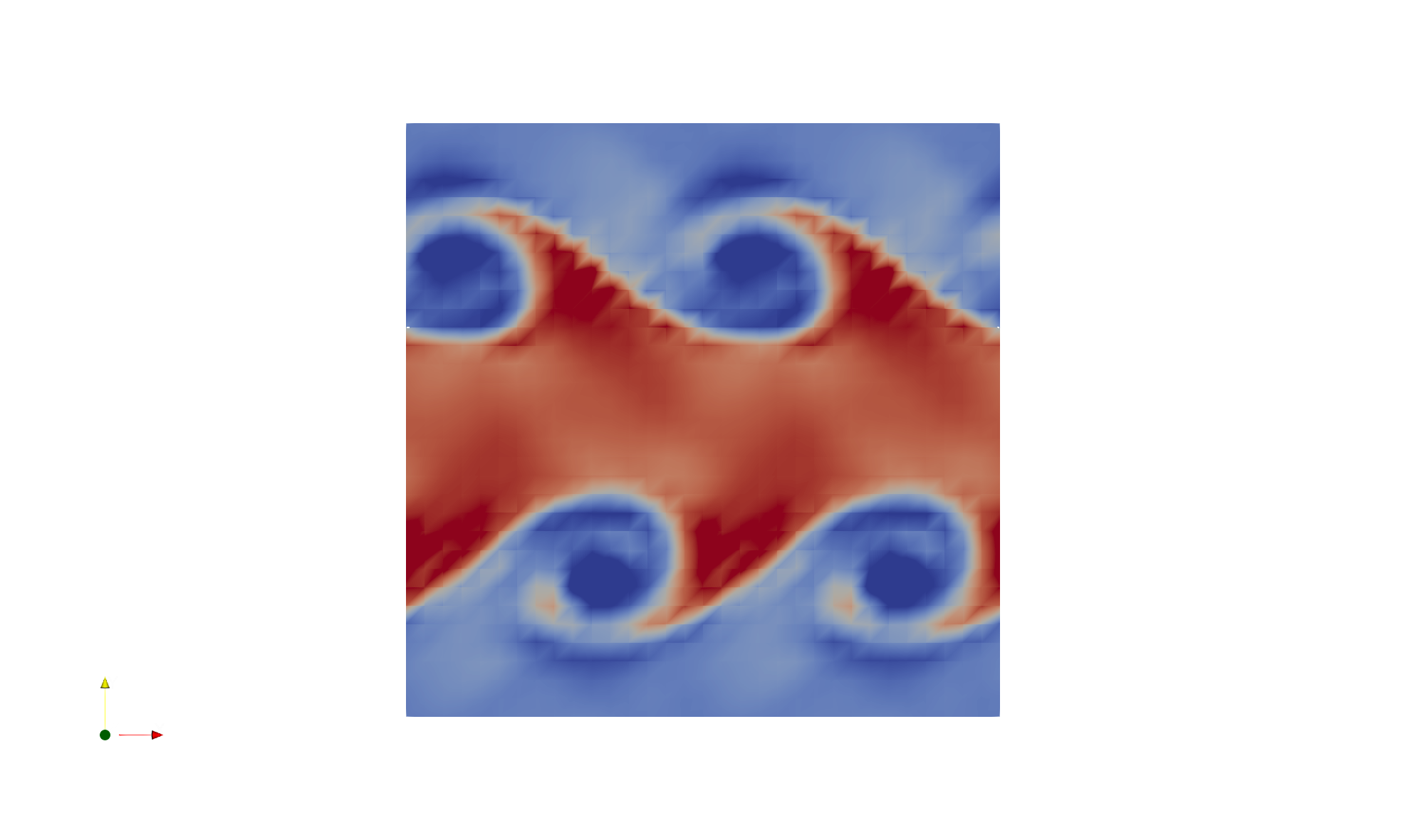}
  \includegraphics[trim=20cm 8cm 20cm 8cm,clip=true,width=0.29\textwidth]{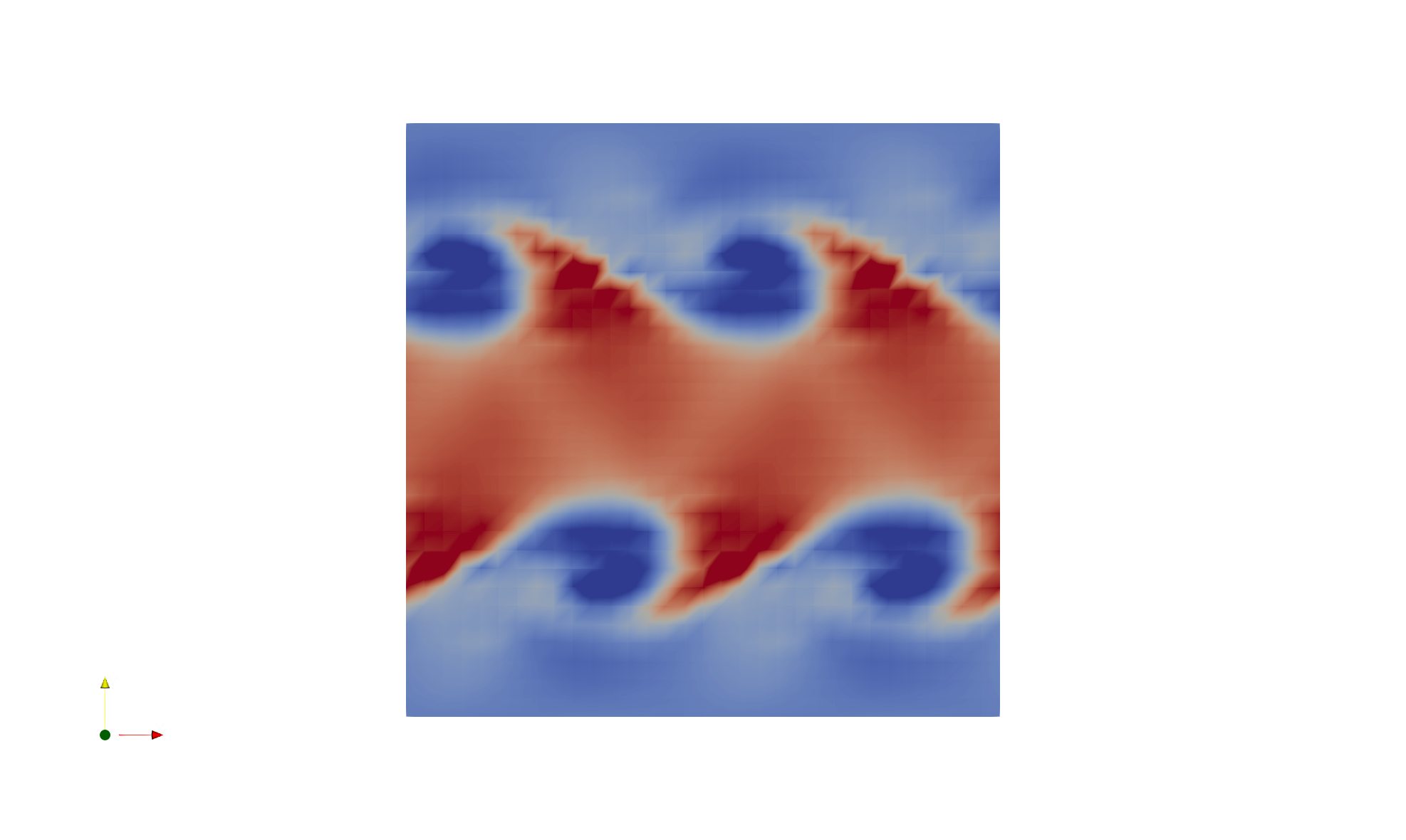}
  \caption{2D Kelvin--Helmholtz instability: snapshots of the density fields at $t=3$ for the projected solution (left), augmented solution (middle), and low-order solution (right).
  From top to bottom, results are shown for $Re=100$, $Re=150$, $Re=200$, $Re=250$, and $Re=300$. Our approach successfully predicts for $Re=150$ and $Re=250$, which were not included in the training data.
  }
  \figlab{cns2d-khi-rexxx2-density-from-t2-to-t3}
\end{figure}  

\begin{figure}[h!t!b!]
  \centering
  \includegraphics[trim=20cm 8cm 20cm 8cm,clip=true,width=0.29\textwidth]{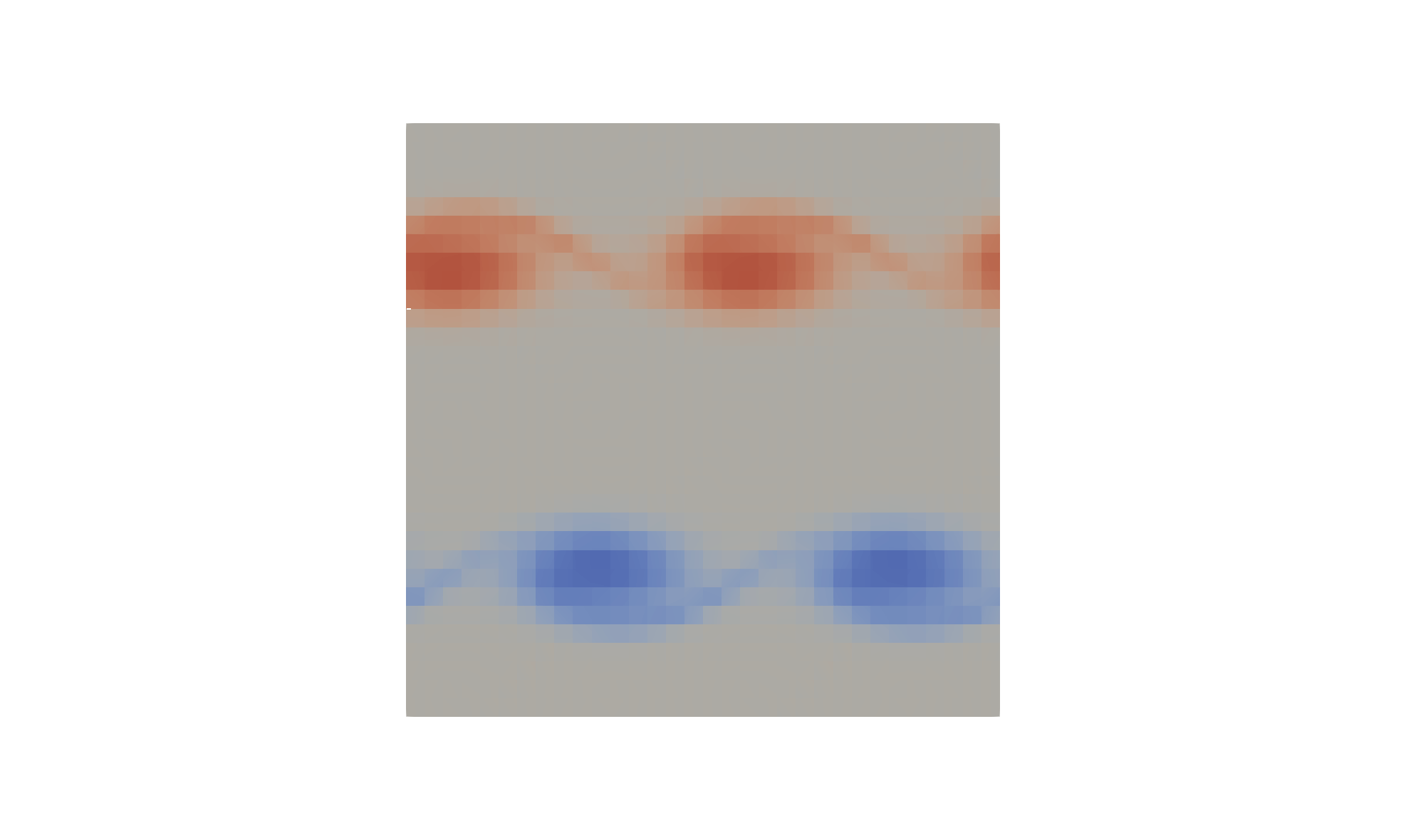}
  \includegraphics[trim=20cm 8cm 20cm 8cm,clip=true,width=0.29\textwidth]{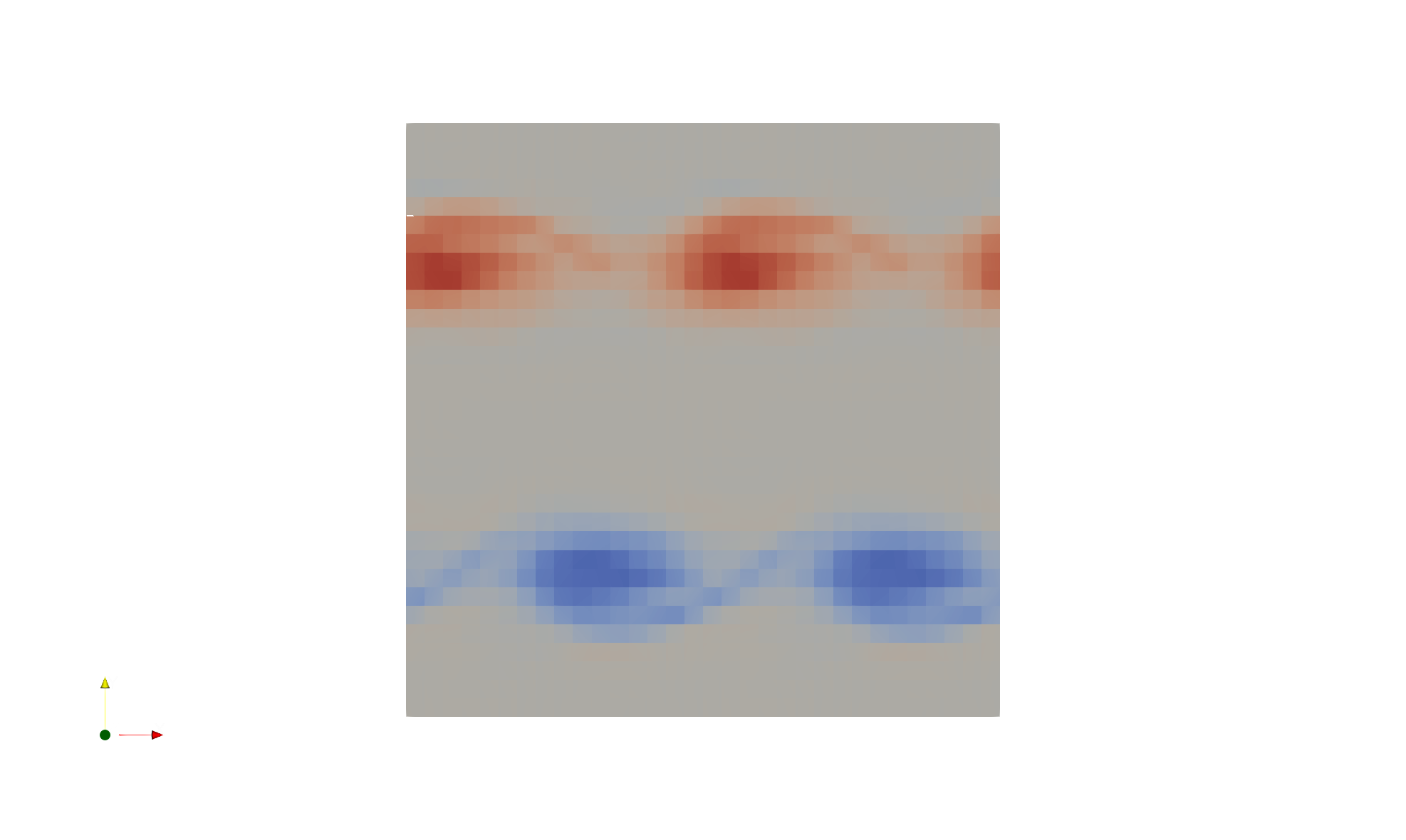}
  \includegraphics[trim=20cm 8cm 20cm 8cm,clip=true,width=0.29\textwidth]{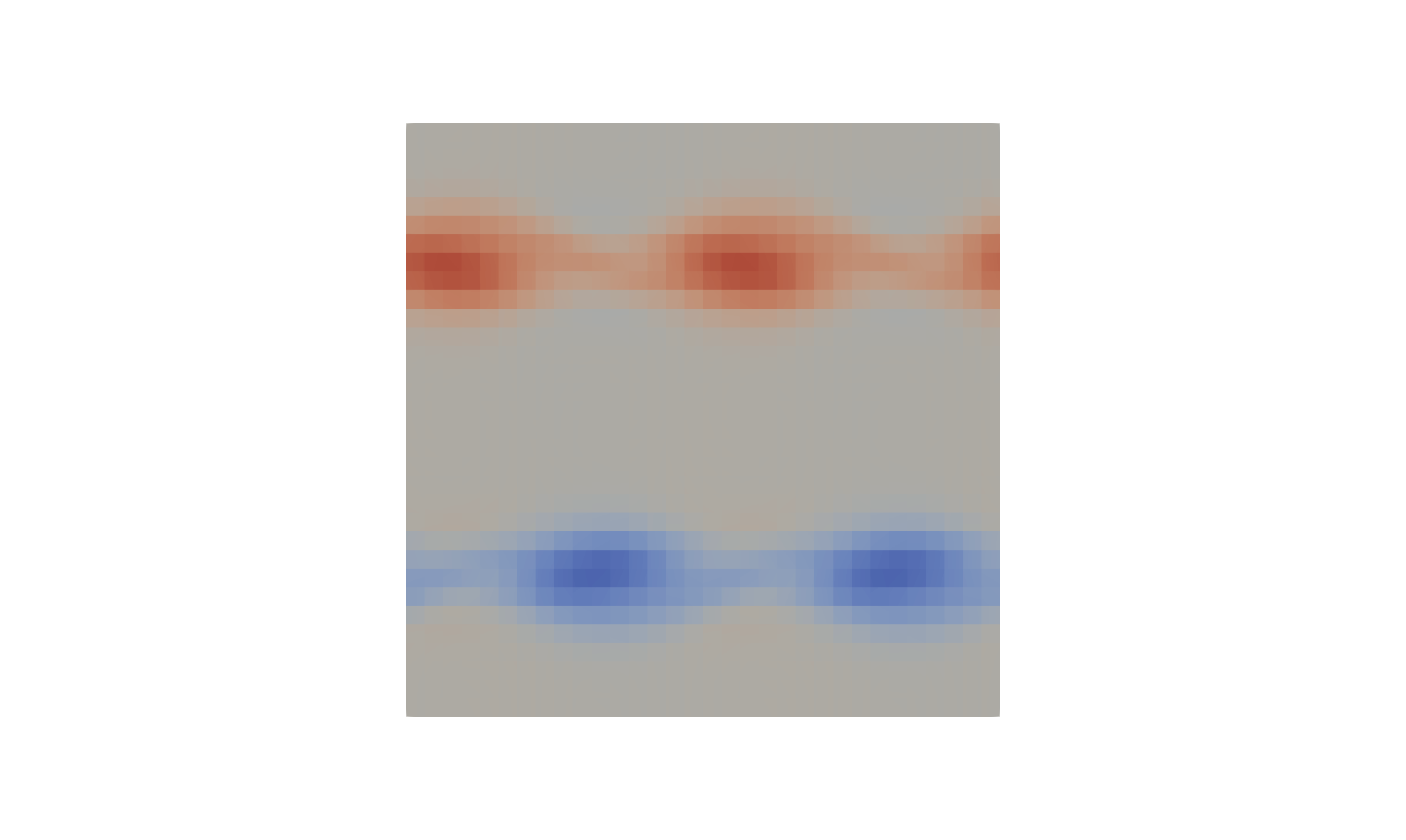}
  \\
  \includegraphics[trim=20cm 8cm 20cm 8cm,clip=true,width=0.29\textwidth]{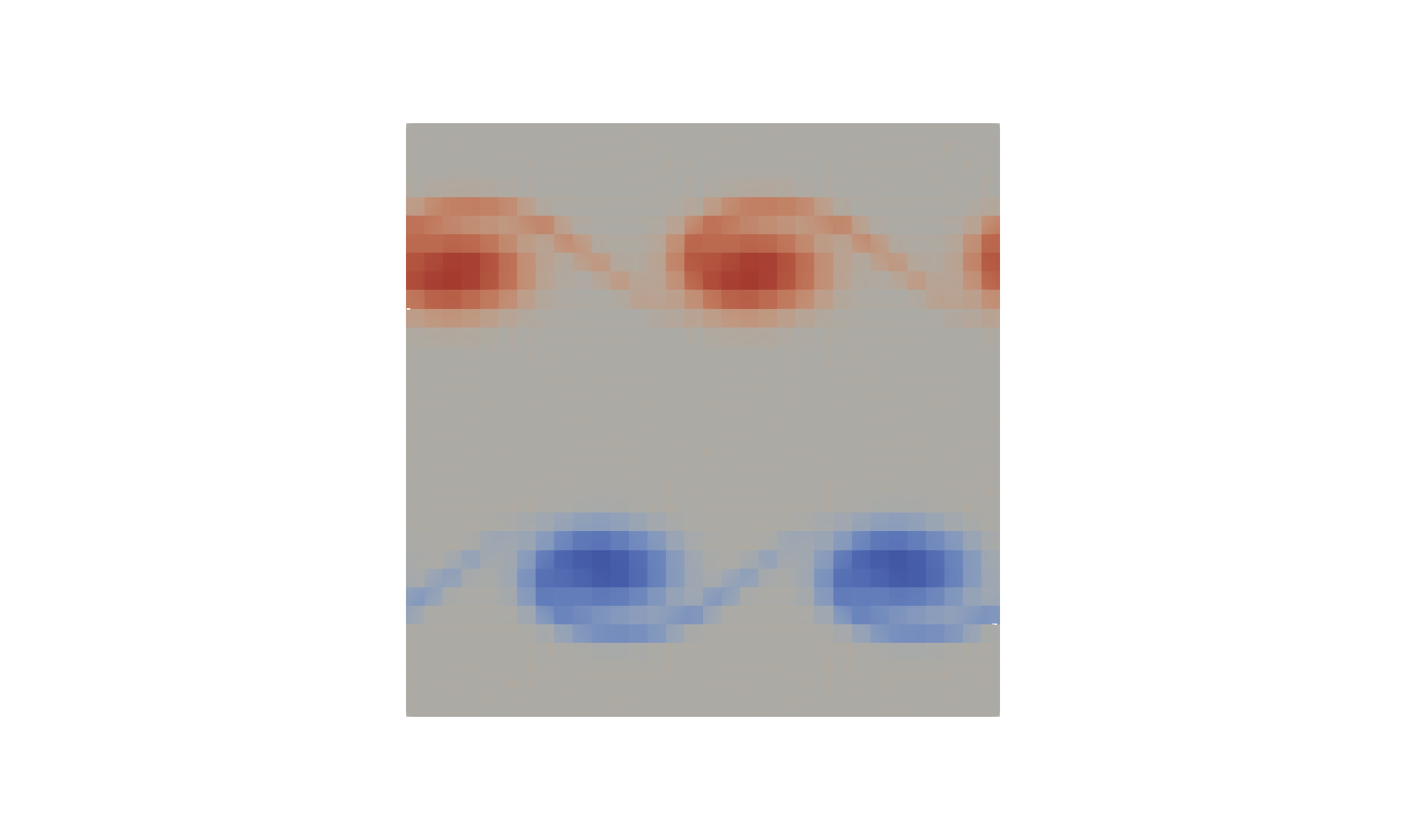}
  \includegraphics[trim=20cm 8cm 20cm 8cm,clip=true,width=0.29\textwidth]{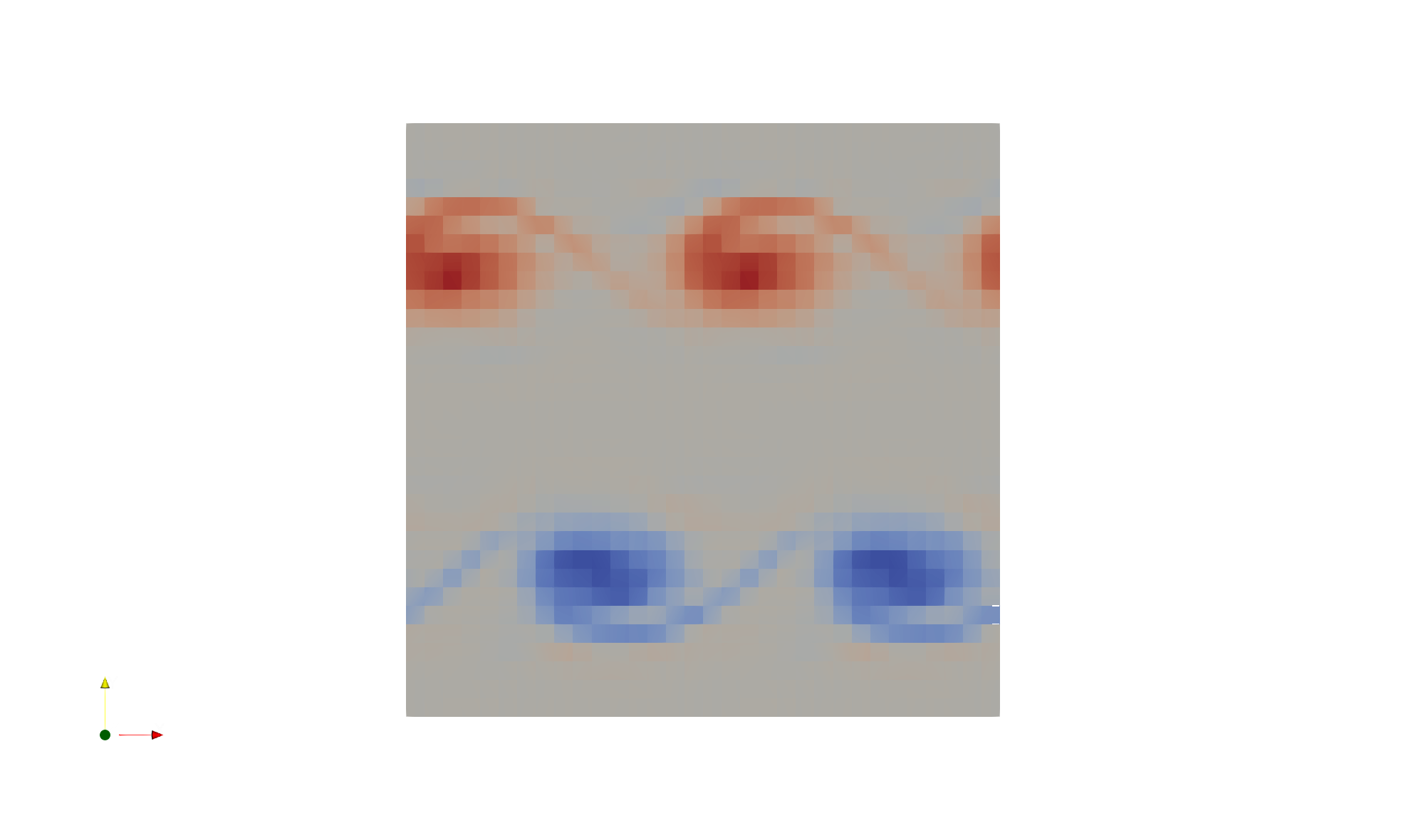}
  \includegraphics[trim=20cm 8cm 20cm 8cm,clip=true,width=0.29\textwidth]{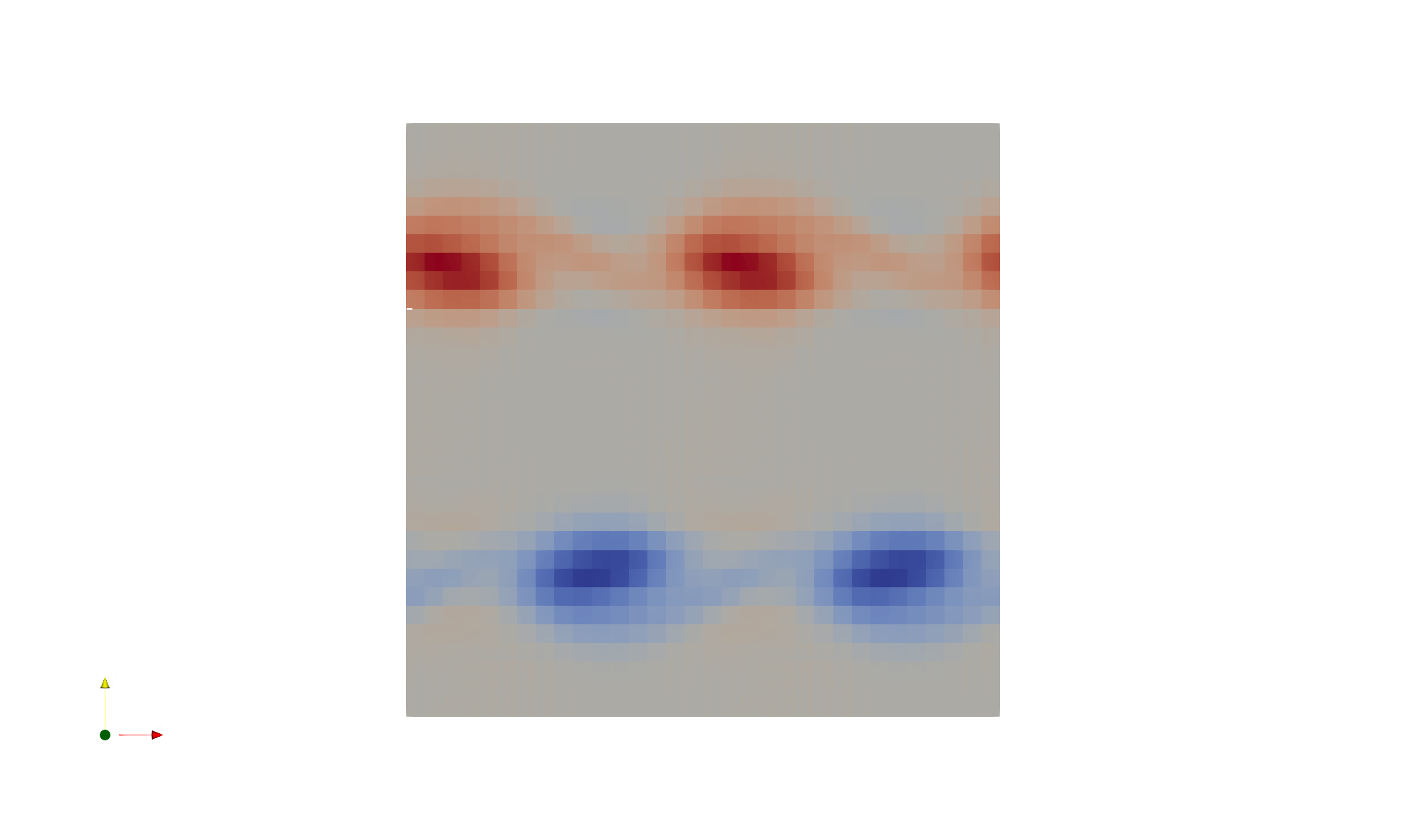}
  \\
  \includegraphics[trim=20cm 8cm 20cm 8cm,clip=true,width=0.29\textwidth]{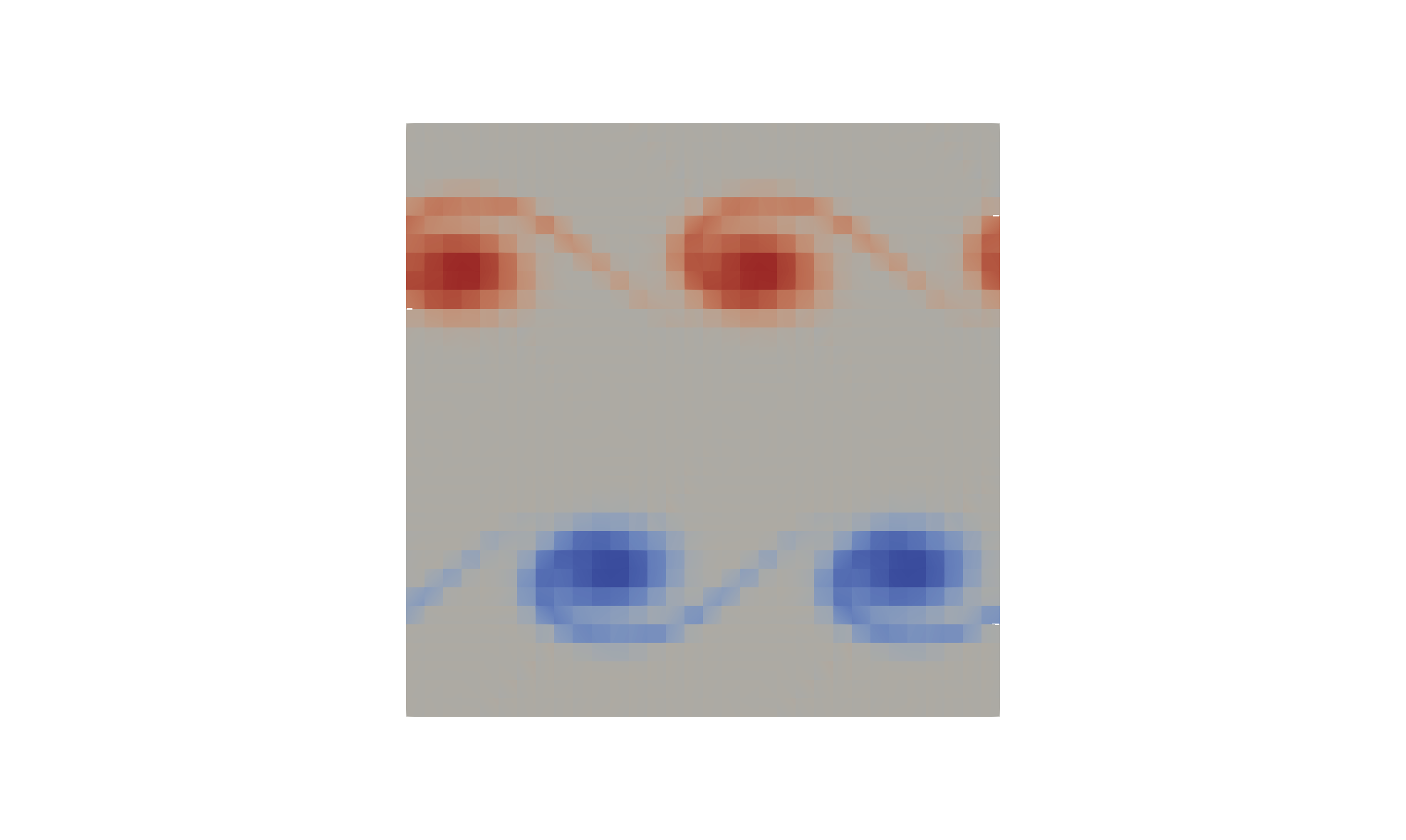}
  \includegraphics[trim=20cm 8cm 20cm 8cm,clip=true,width=0.29\textwidth]{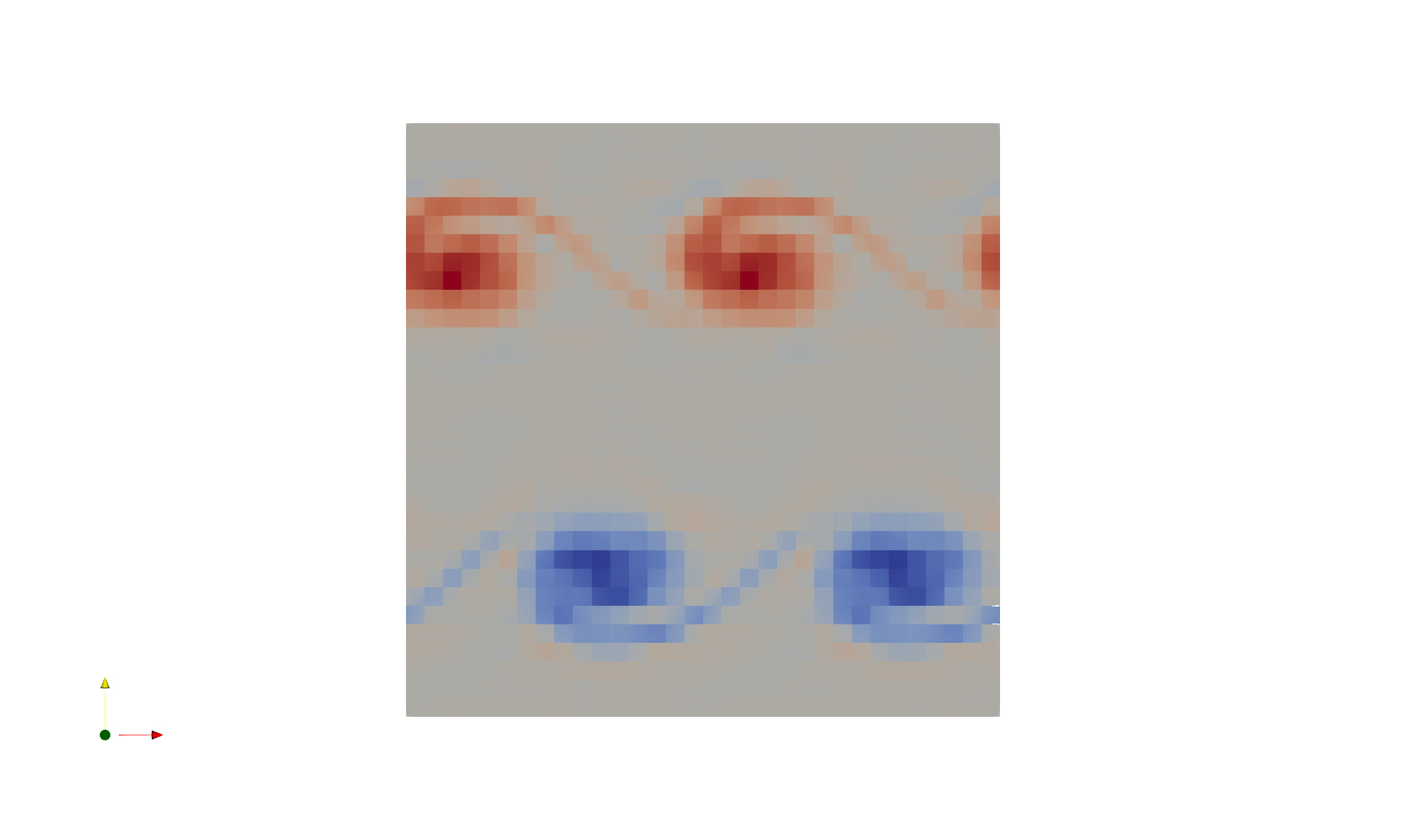}
  \includegraphics[trim=20cm 8cm 20cm 8cm,clip=true,width=0.29\textwidth]{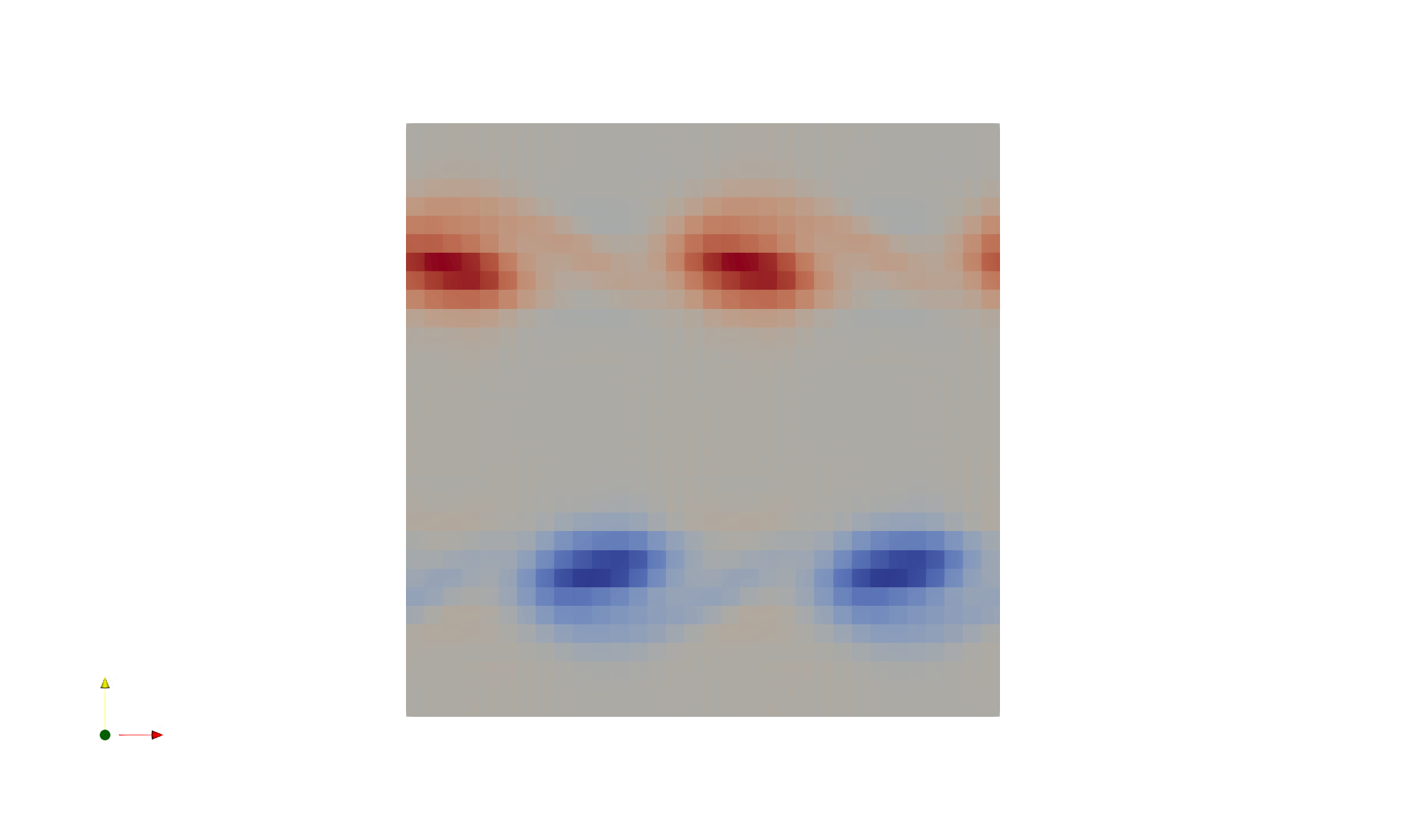}
  \\
  \includegraphics[trim=20cm 8cm 20cm 8cm,clip=true,width=0.29\textwidth]{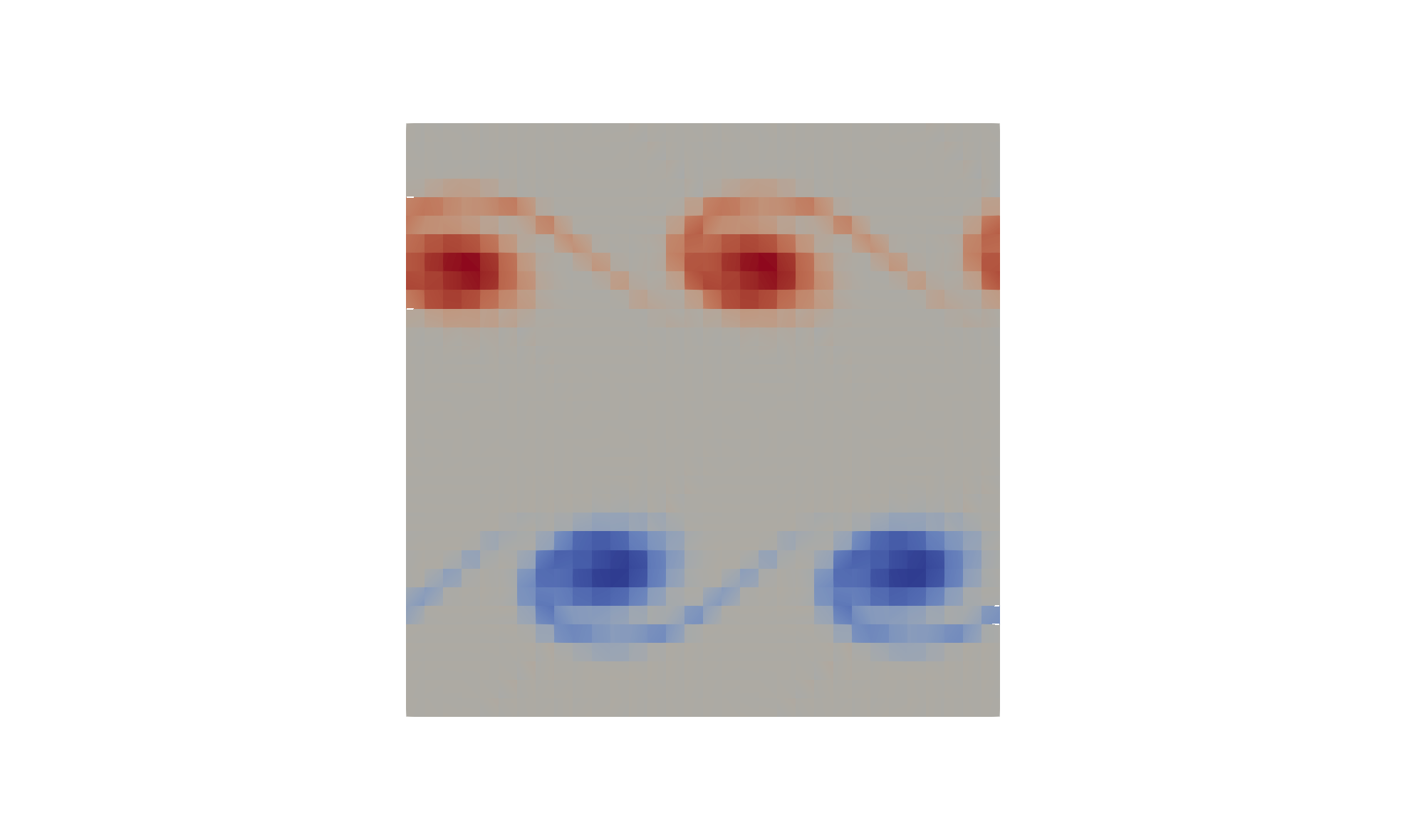}
  \includegraphics[trim=20cm 8cm 20cm 8cm,clip=true,width=0.29\textwidth]{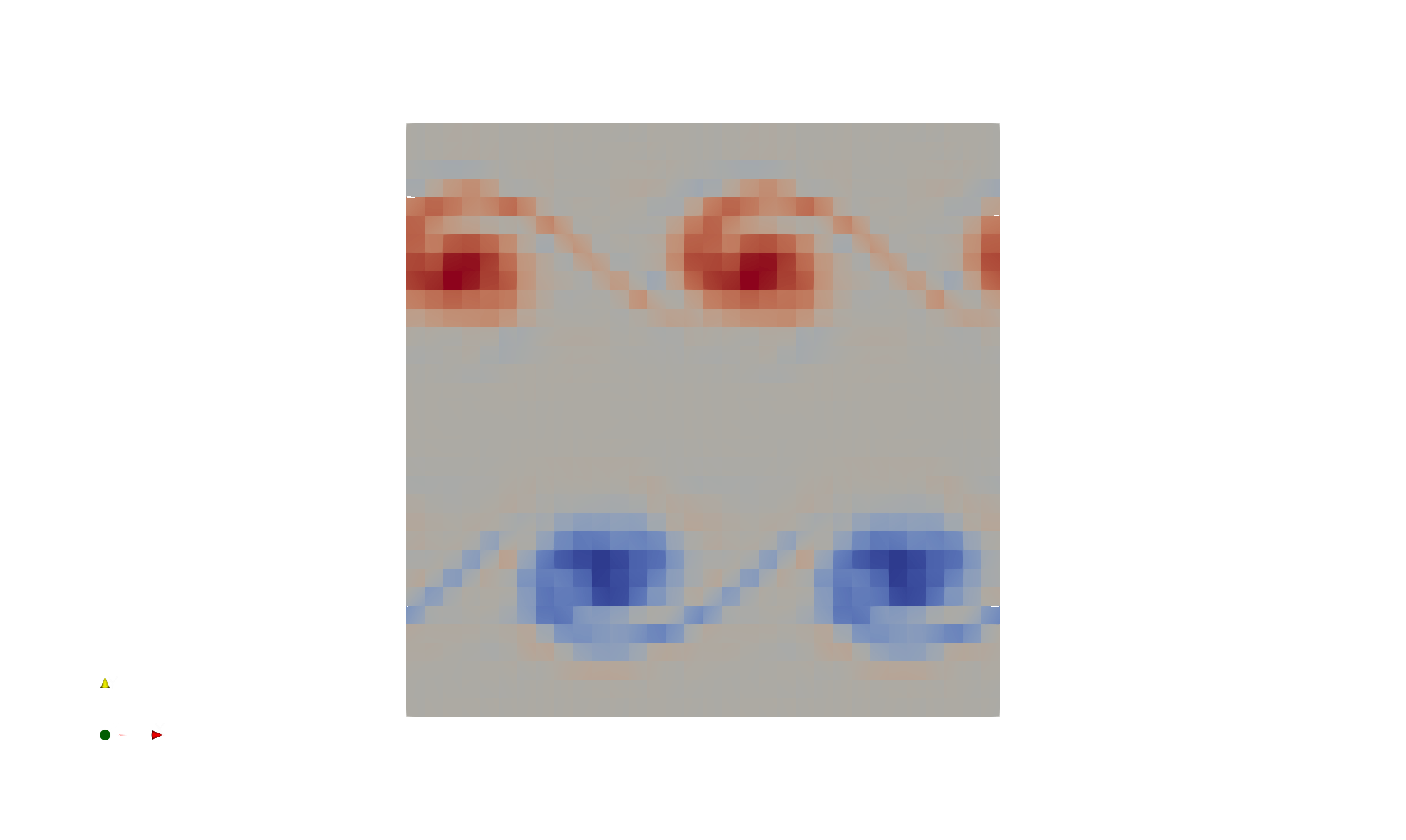}
  \includegraphics[trim=20cm 8cm 20cm 8cm,clip=true,width=0.29\textwidth]{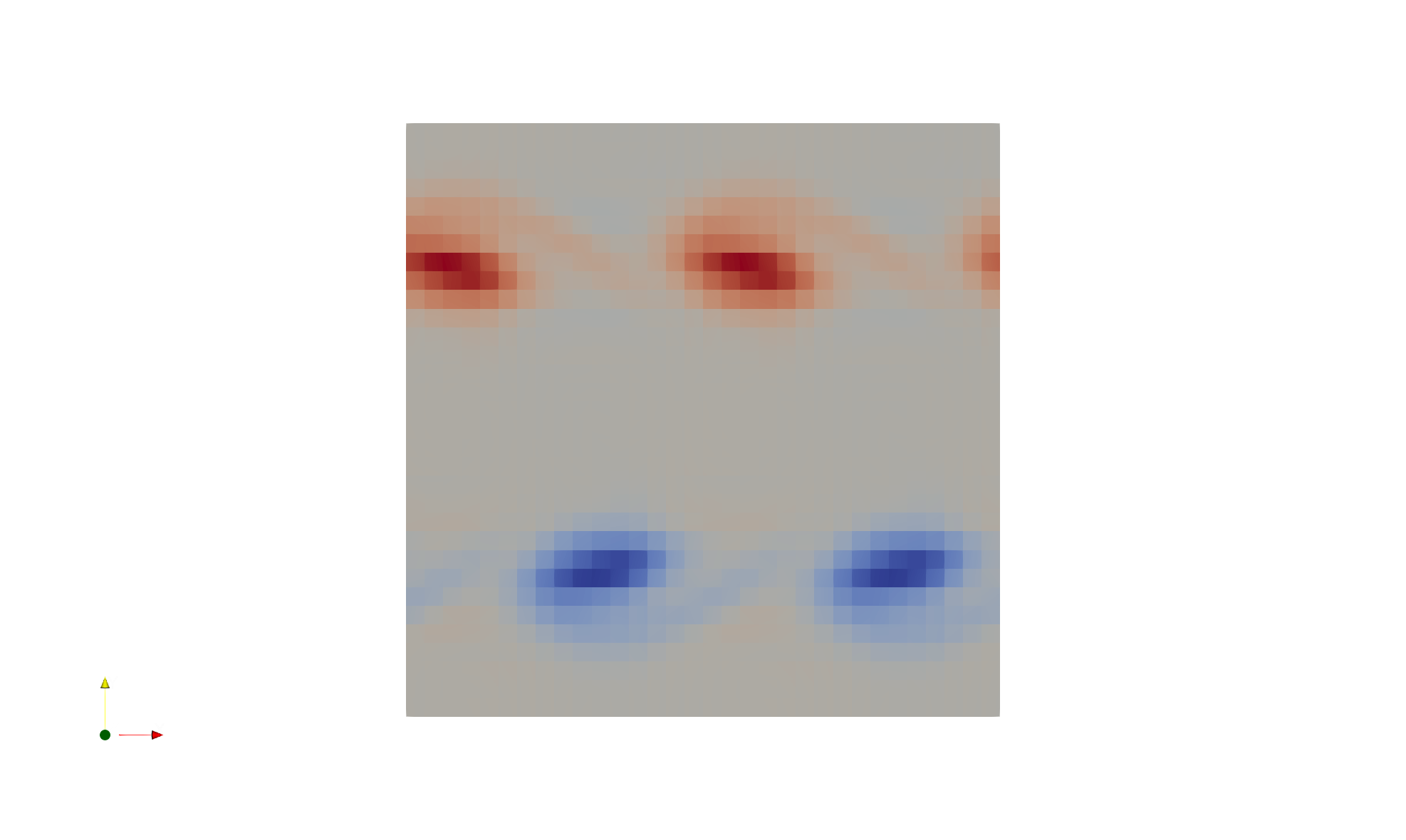}
  \\
  \includegraphics[trim=20cm 8cm 20cm 8cm,clip=true,width=0.29\textwidth]{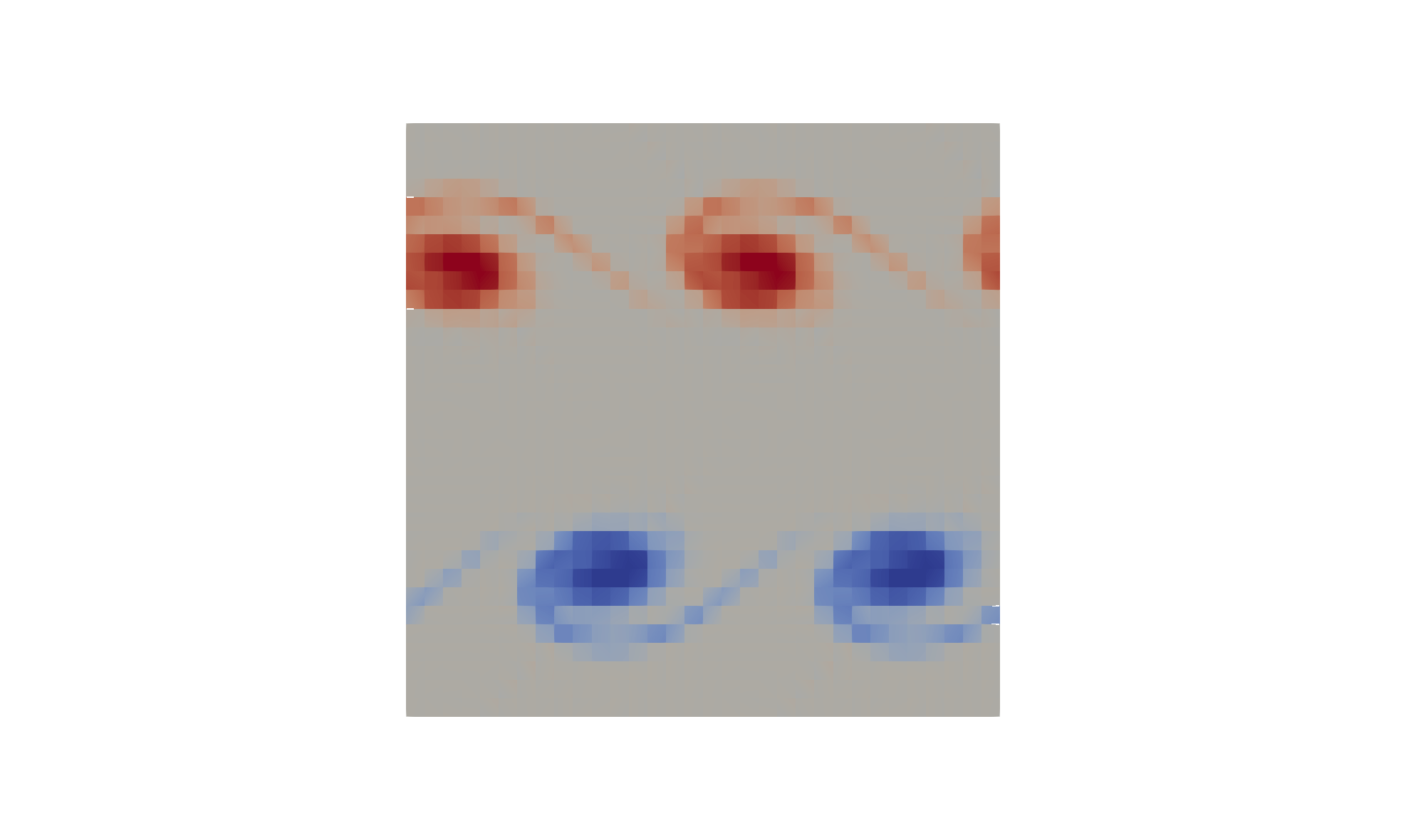}
  \includegraphics[trim=20cm 8cm 20cm 8cm,clip=true,width=0.29\textwidth]{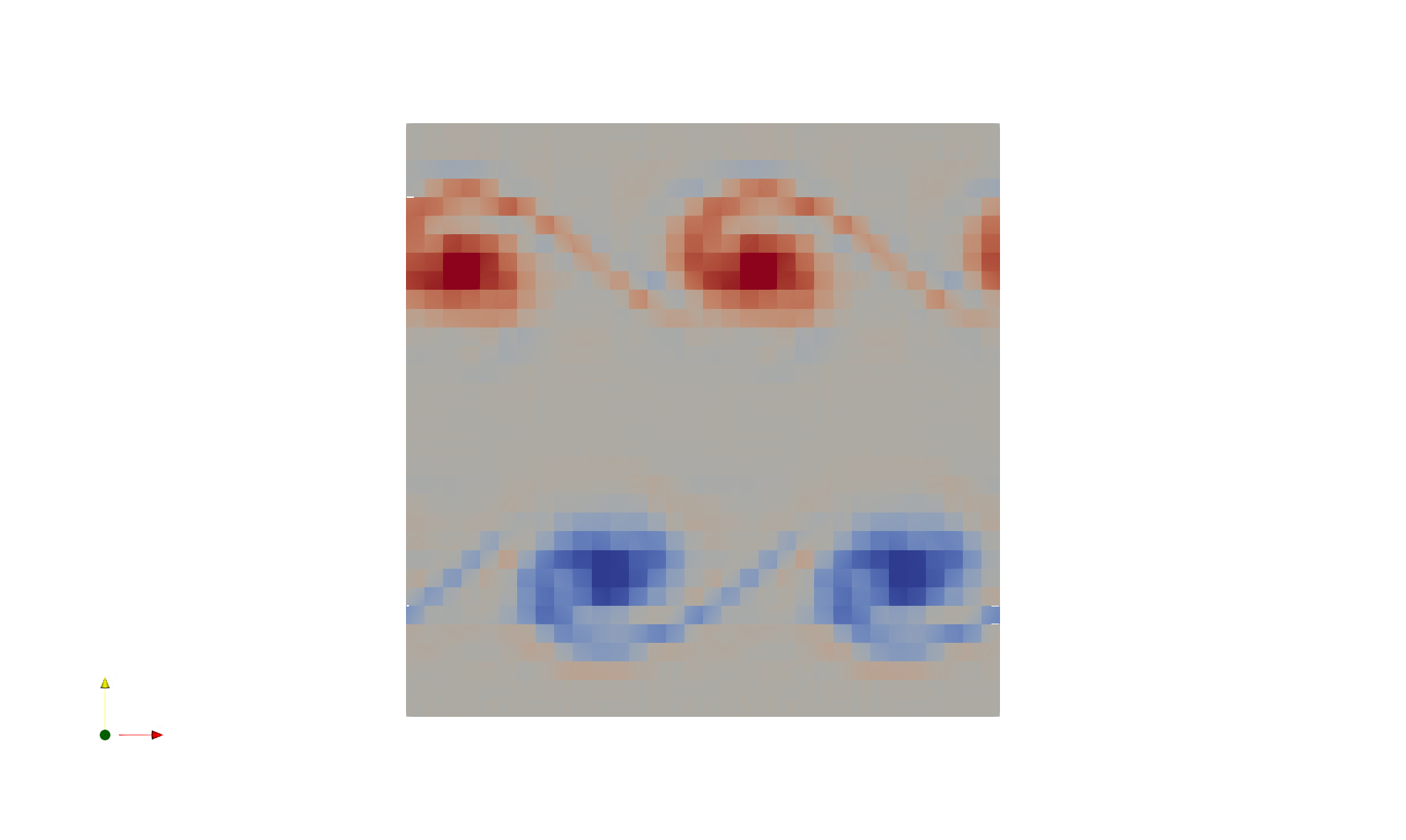}
  \includegraphics[trim=20cm 8cm 20cm 8cm,clip=true,width=0.29\textwidth]{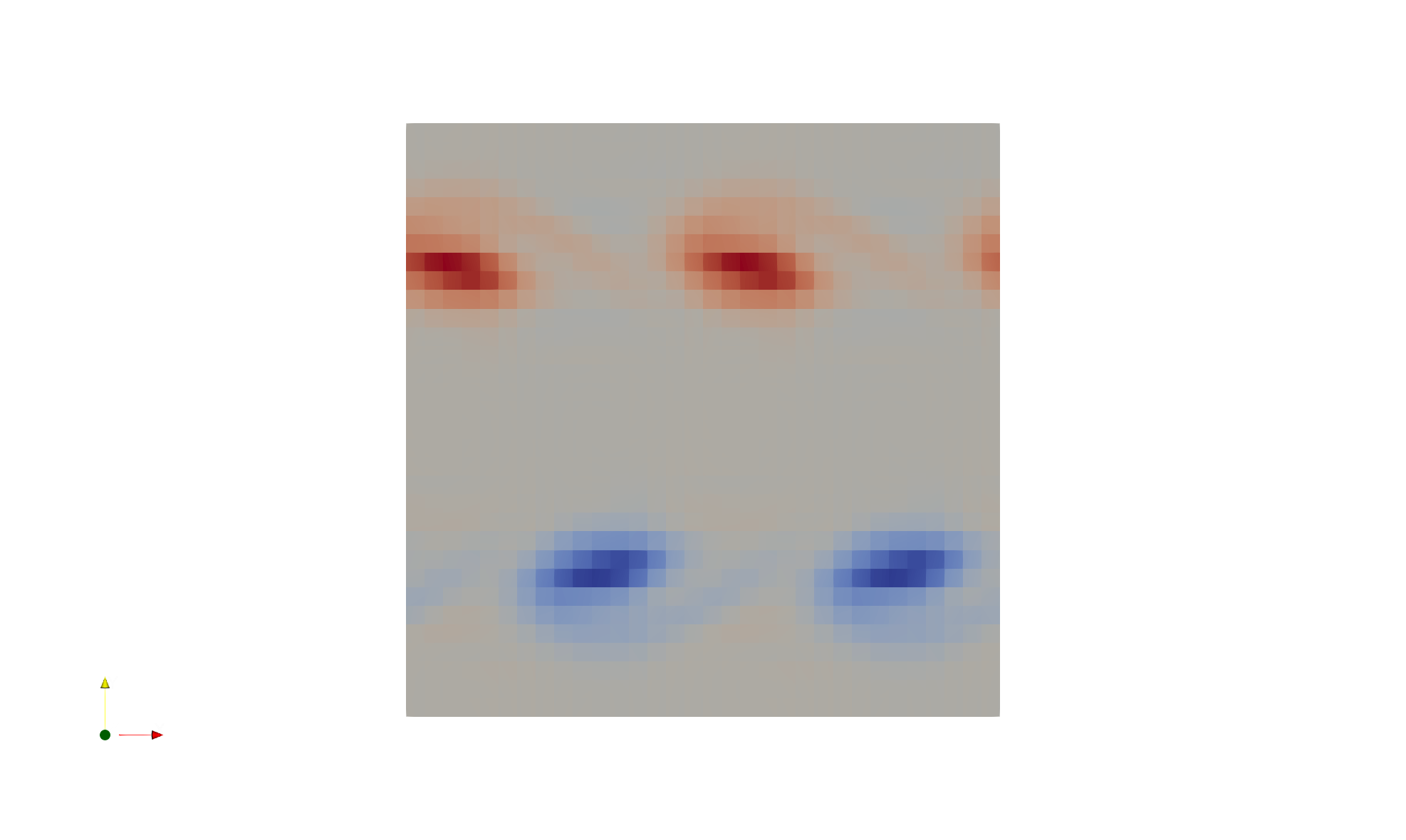}
  \caption{2D Kelvin--Helmholtz instability: snapshots of the vorticity fields at $t=3$ for the projected solution (left), augmented solution (middle), and low-order solution (right).
  From top to bottom, results are shown for $Re=100$, $Re=150$, $Re=200$, $Re=250$, and $Re=300$. Our approach successfully predicts for $Re=150$ and $Re=250$, which were not included in the training data.
  }
  \figlab{cns2d-khi-rexxx2-vort-from-t2-to-t3}
\end{figure}

\subsubsection{
Handling missing data}

The proposed NODE-DG framework is capable of handling missing data.
To demonstrate this capability, 
we consider a subset of $m$ consecutive instances. For example, when $m=11$, 
instead of using the full set of snapshots, 
$\LRc{\LRp{\Pmat^L u^H}_{j}\big\vert j=0,1,2,\cdots, 11}$, 
we select only a portion of them, such as 
$\LRc{\LRp{\Pmat^L u^H}_{j}\big\vert j=0,1,2,5,7,8,11}$. 
In this scenario we assume that the snapshots corresponding to $j=3,4,6,9,10$ are missing. 
We denote the augmented solution with this nonuniform data by $\acute{ \ub }^L$.
To account for the missing data, we adjust the loss function in \eqnref{loss-cns} as follows: 
\begin{align}
\eqnlab{loss-cns-part}
  \mc{L}(\theta) =  \frac{1}{n m}\sum_{i=1}^n \sum_{j\in idx} 
  \sum_{k=1}^\Nfield 
  \norm{\LRp{\Pmat^L u^H_k}_{j}^{(i)} - \LRp{\acute{u}^L_k}_{j}^{(i)}  }^2  
\end{align}
where $idx = \LRc{1,2,5,7,8,11}$. 
Using the same dataset across Reynolds numbers $Re=\{100,200,300\}$, 
we train the neural network source term  following the same procedure described in Section \secref{reynoldsnumber} with the exception of the modified loss term in \eqnref{loss-cns-part}.
\footnote{
We used $k_w=3$ and $m=11$. 
}

Figure \figref{cns2d-khi-rexxx3-err-re150} 
and Figure \figref{cns2d-khi-rexxx3-err-re250} 
illustrate the relative error histories for (a) $\rho$, (b) $\rho {\boldsymbol\varphi}$, and (c) $\rho E$. The plots compare the low-order solution ($\ub^L$), the augmented solution with uniform data ($\hat{\ub}^L$), and the augmented solution with nonuniform data ($\acute{ \ub }^L$) at $Re=150$ and $Re=250$, respectively. 
In general, the relative error of $\acute{ \ub }^L$ closely matches that of $\hat{\ub}^L$, indicating that the NODE-DG framework effectively improves the accuracy of the low-order approximation even with nonuniform data.
Similar behavior was observed for the cases with $Re=\LRc{100,200,300}$, and thus they are omitted here for brevity.

 

\begin{figure}[h!t!b!]
  \centering
  \subfigure[$\rho$]{
    \figlab{cns2d-khi-re150-err-r}
      \includegraphics[trim=2cm 0cm 0cm 0cm,clip=true,width=0.285\textwidth]{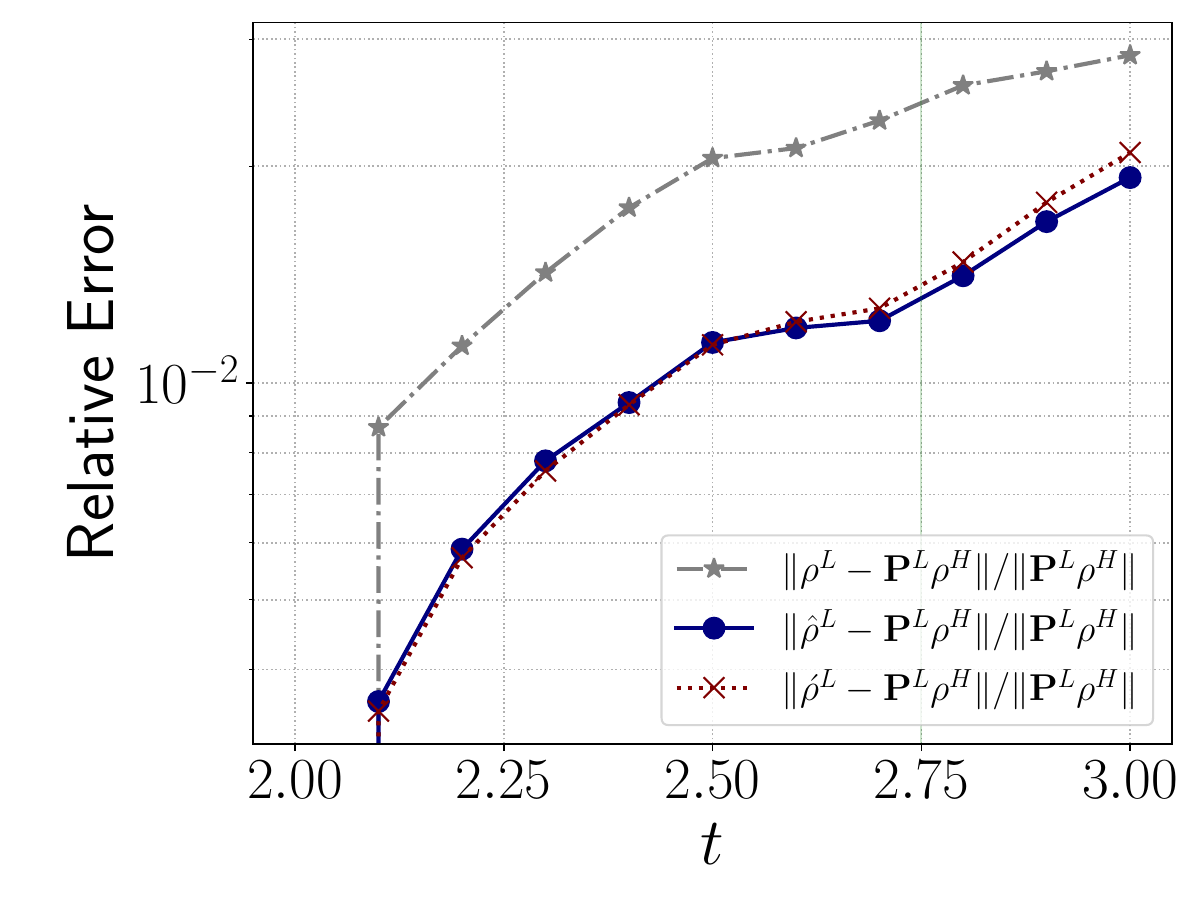}
  }
  \subfigure[$\rho {\boldsymbol\varphi}$]{
    \figlab{cns2d-khi-re150-err-rvel}
      \includegraphics[trim=0.5cm 0cm 0cm 0cm,clip=true,width=0.31\textwidth]{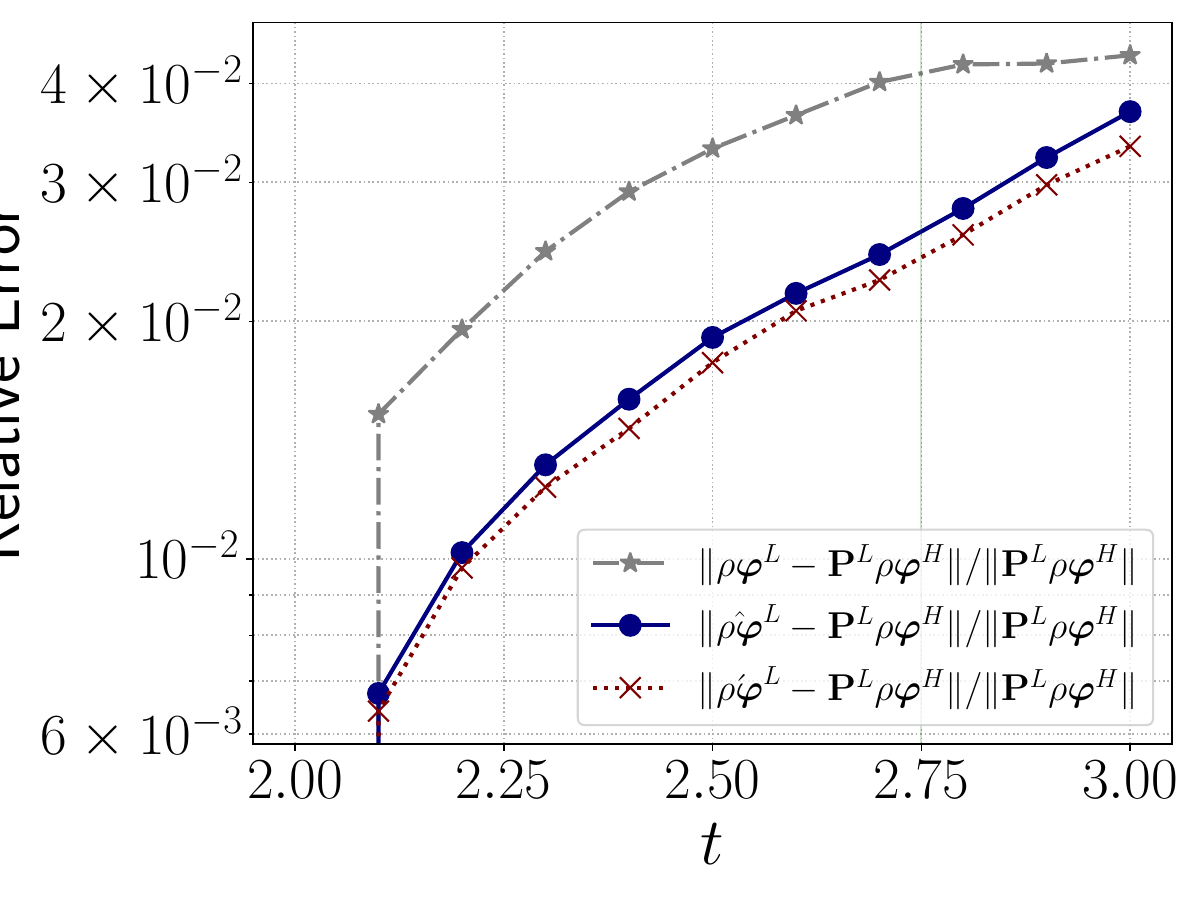}
  }
  \subfigure[$\rho E$]{
    \figlab{cns2d-khi-re150-err-rE}
      \includegraphics[trim=0.5cm 0cm 0cm 0cm,clip=true,width=0.31\textwidth]{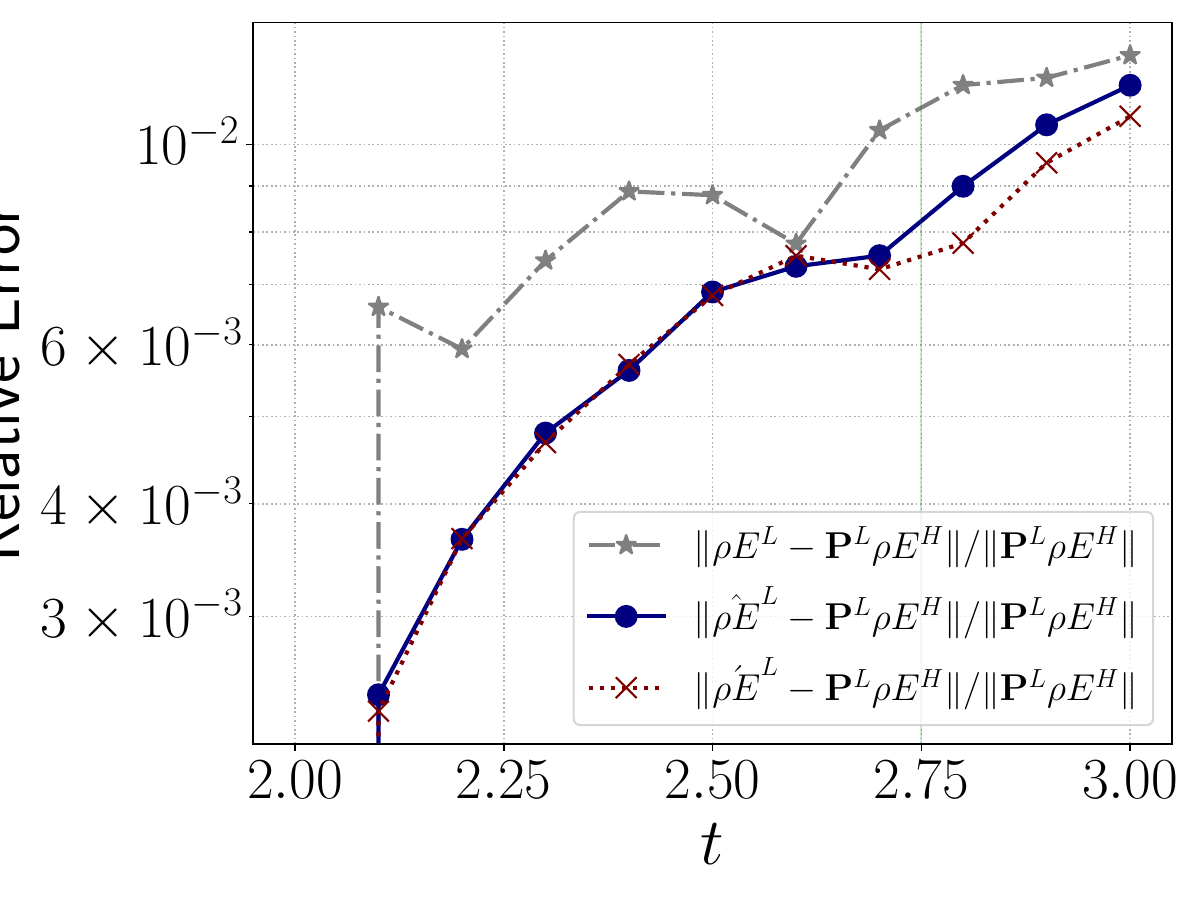}
  }
  \caption{2D Kelvin--Helmholtz instability: relative error histories of (a) $\rho$, (b) $\rho {\boldsymbol\varphi}$, and (c) $\rho E$ at $Re=150$ for low-order solution ($\ub^L$), augmented solution with uniform data ($\hat{\ub}^L$), and augmented solution with nonuniform data ($\acute{ \ub }^L$).
  }
  \figlab{cns2d-khi-rexxx3-err-re150}
\end{figure}  

\begin{figure}[h!t!b!]
  \centering
  \subfigure[$\rho$]{
    \figlab{cns2d-khi-re250-err-r}
      \includegraphics[trim=0.5cm 0cm 0cm 0cm,clip=true,width=0.31\textwidth]{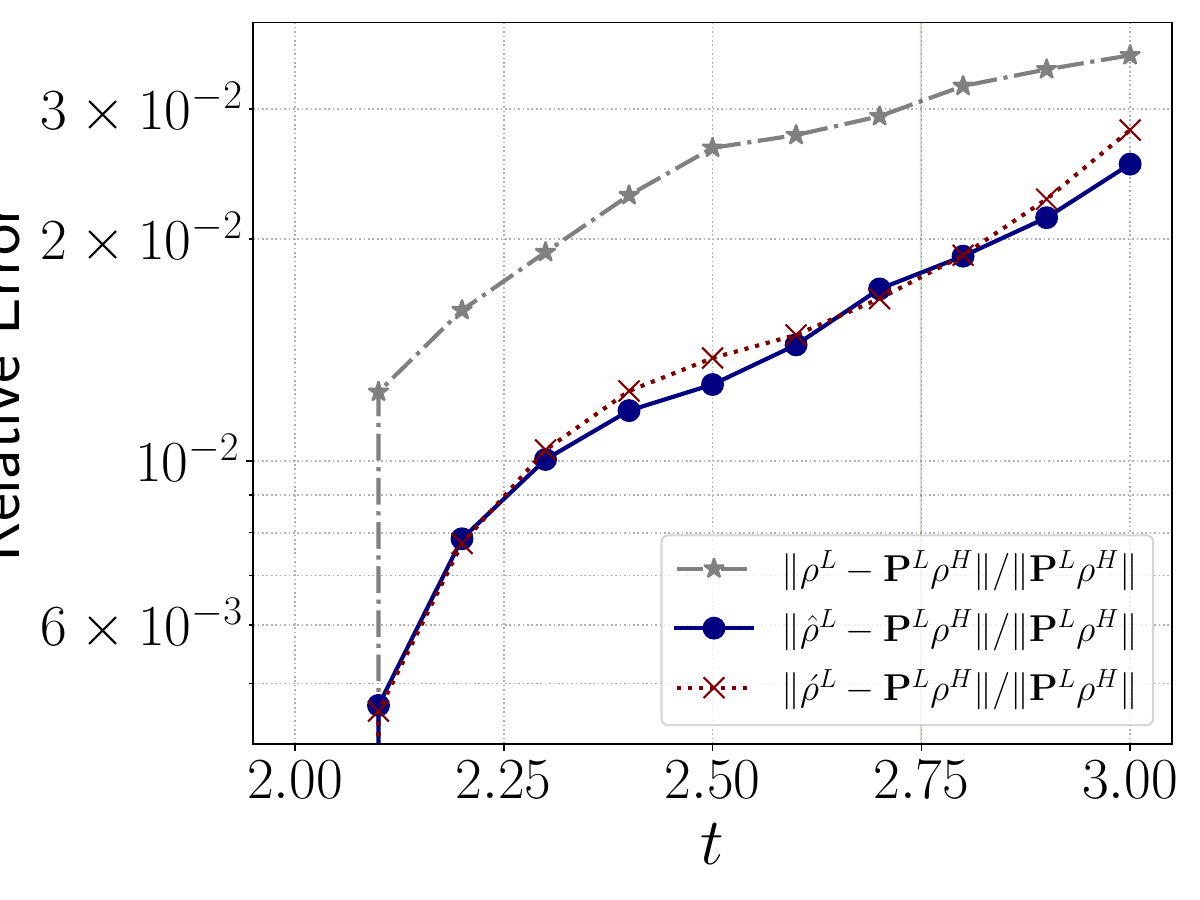}
  }
  \subfigure[$\rho {\boldsymbol\varphi}$]{
    \figlab{cns2d-khi-re250-err-rvel}
      \includegraphics[trim=0.5cm 0cm 0cm 0cm,clip=true,width=0.31\textwidth]{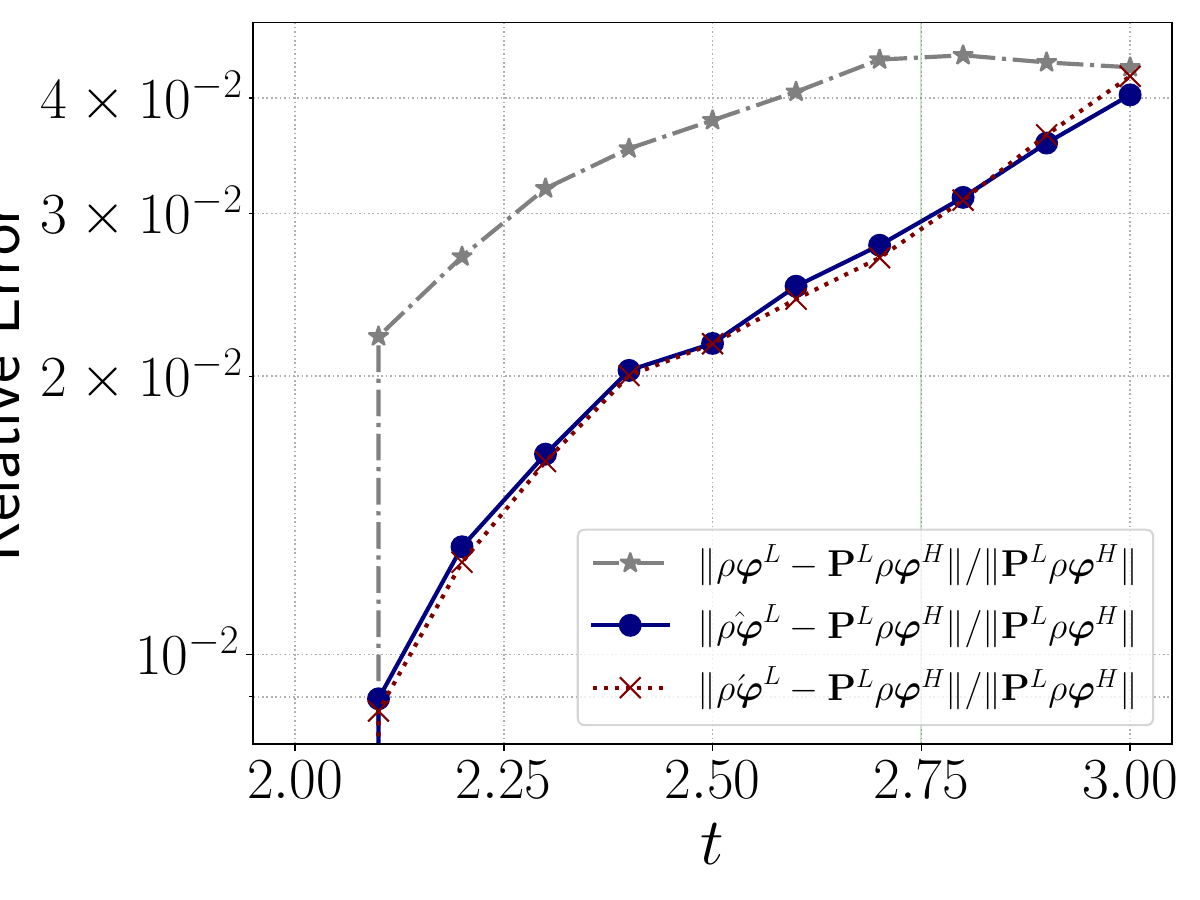}
  }
  \subfigure[$\rho E$]{
    \figlab{cns2d-khi-re250-err-rE}
      \includegraphics[trim=0.5cm 0cm 0cm 0cm,clip=true,width=0.31\textwidth]{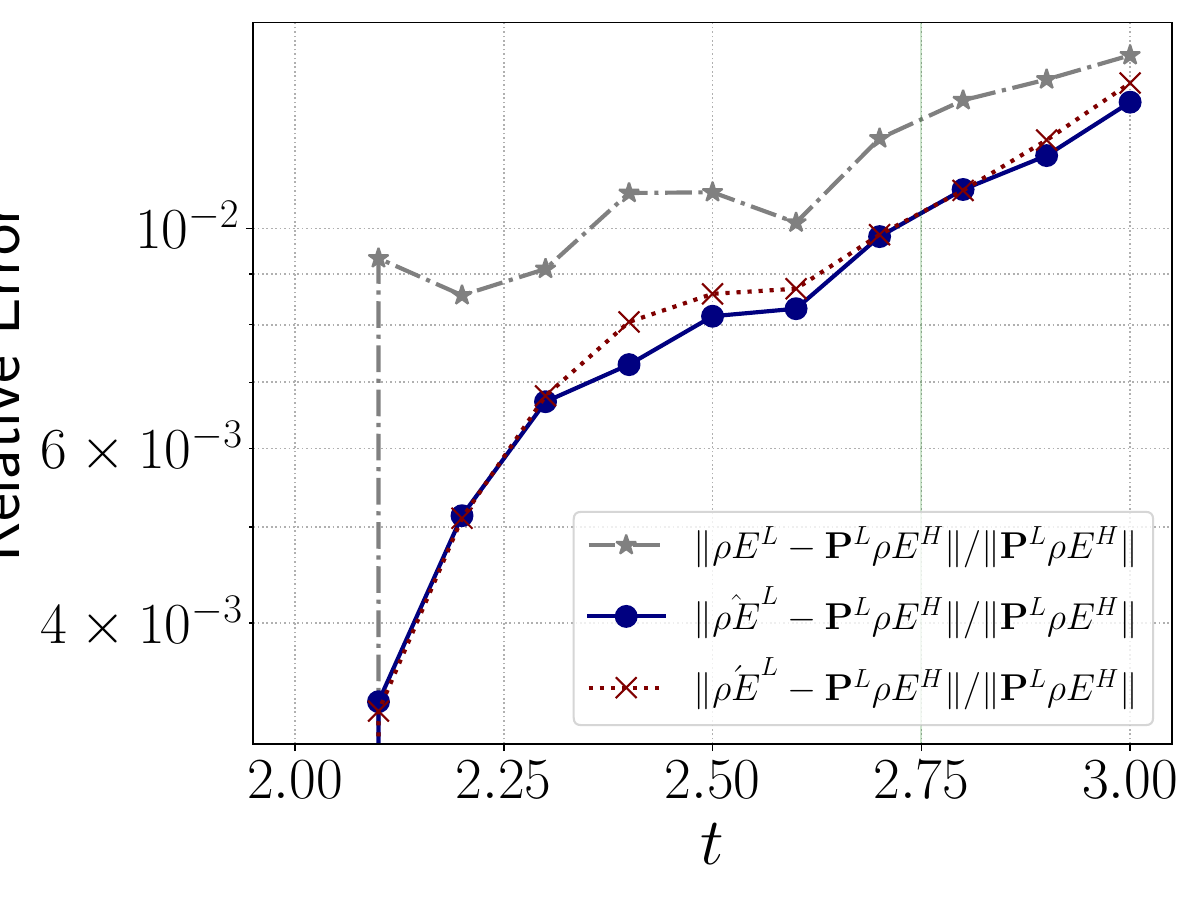}
  }
  \caption{2D Kelvin--Helmholtz instability: relative error histories of (a) $\rho$, (b) $\rho {\boldsymbol\varphi}$, and (c) $\rho E$ at $Re=250$ for low-order solution ($\ub^L$), augmented solution with uniform data ($\hat{\ub}^L$), and augmented solution with nonuniform data ($\acute{ \ub }^L$).
  }
  \figlab{cns2d-khi-rexxx3-err-re250}
\end{figure}

\subsubsection{
Mass-conservation property}

To verify the mass conservation property of source terms, we train the neural network source terms over $t\in[2,2.7]$ with $Re=[100,200,300]$, and then perform predictions for $t\in[2,10]$ with $Re=[150,250]$. Figure \figref{cns2d-khi-mc-re150-masshistory} shows the time evolution of the total mass and its relative loss with $Re=150$, which are defined by $mass(t)=(\rho(t),1)_\Omega$ and $massloss = |mass(t)-mass(0)|/mass(0)$. Since the results for $Re=250$ showed similar behavior, we report only the $Re=150$ case. The light green color represents the training region ($t\in[2,2.7]$). The na\"{\i}ve neural network source term approach does not preserve the total mass, whereas the mass conserving source term in \eqnref{mcnnsource} and low-order approximation $\ub^L$ do preserve the total mass. The $L_2$ regularization parameter $\alpha$ does not affect the total mass conservation. With the mass conserving source term, the observed mass loss is on the order of $10^{-7}$, which is comparable to the machine epsilon in single precision.

\begin{figure}[h!t!b!]
  \centering
  \subfigure[Total mass]{
    \figlab{cns2d-khi-re150-masshistory}
      \includegraphics[trim=0.8cm 0cm 0cm 0cm,clip=true,width=0.45\textwidth]{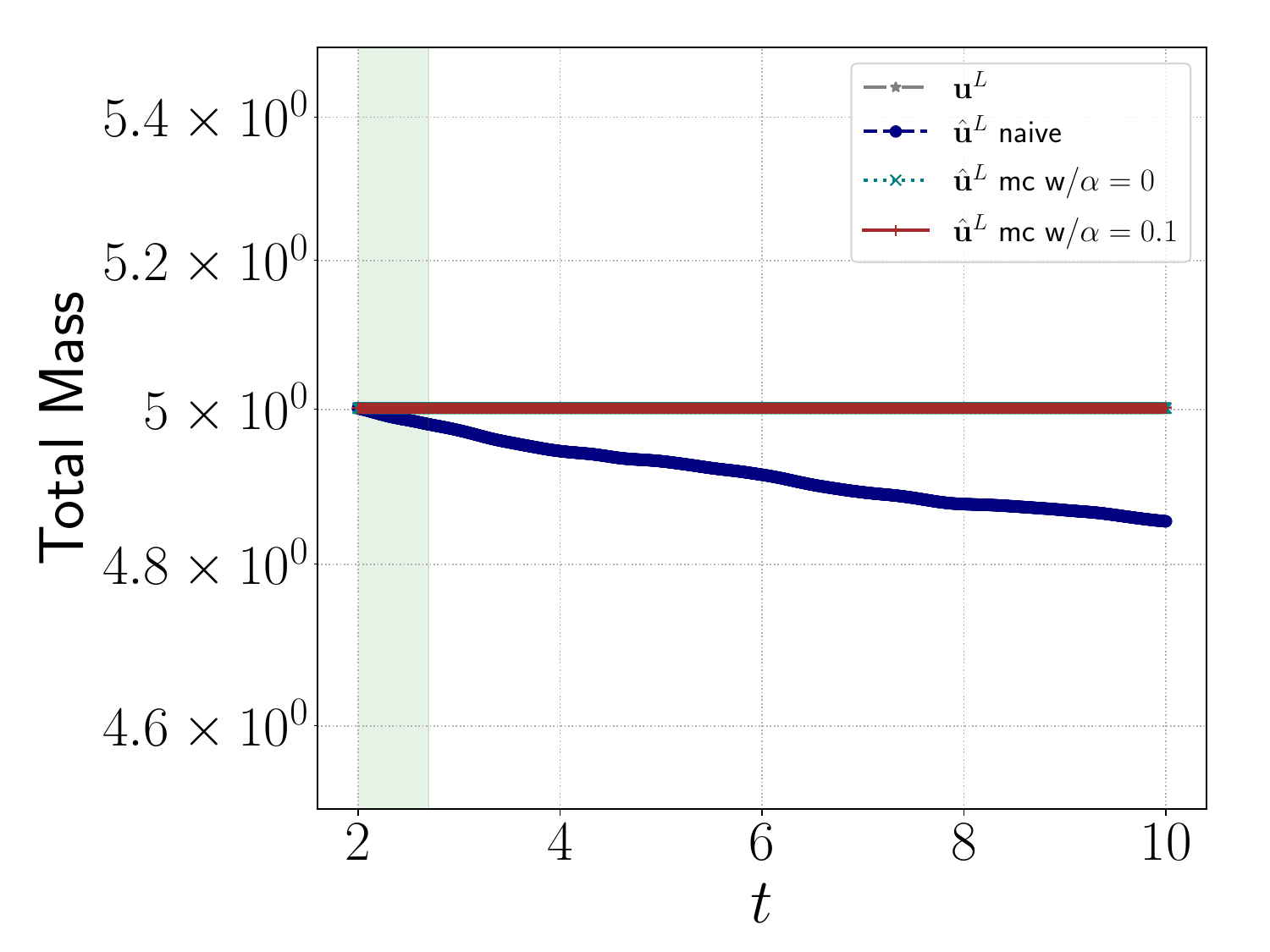}
  }
  \subfigure[Mass loss]{
    \figlab{cns2d-khi-re150-massloss}
      \includegraphics[trim=1.5cm 0cm 0cm 0cm,clip=true,width=0.45\textwidth]{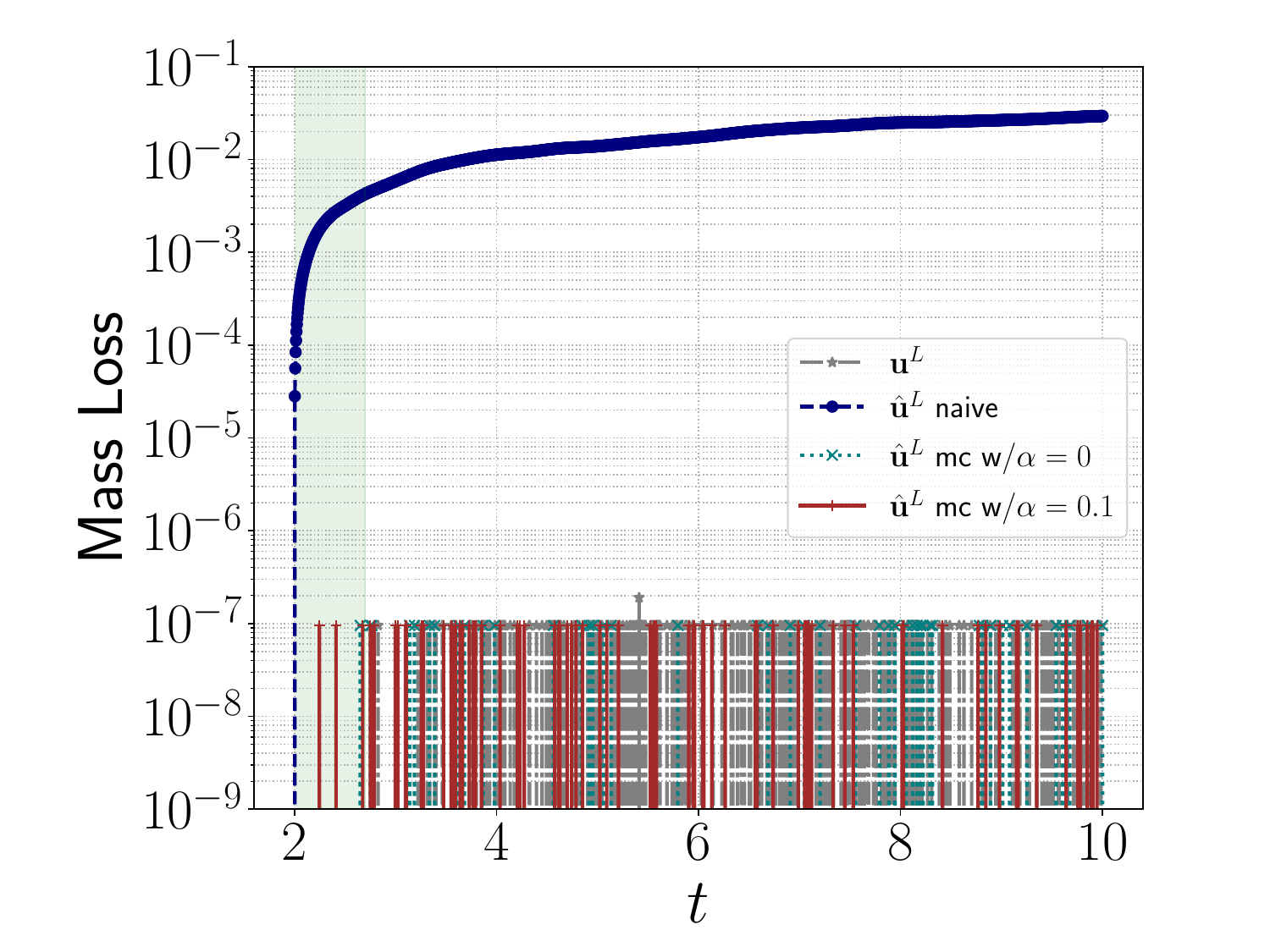}
  }
  \caption{2D Kelvin--Helmholtz instability: (a) total mass histories and (b) mass loss. Na\"{\i}ve augmented solution ($\hat{\ub} \texttt{ na\"{\i}ve}$) fails to preserve total mass, whereas mass-conserving augmented solution ($\hat{\ub} \texttt{ mc}$) and low-order solution  ($\ub^L$) conserves total mass.
  }
  \figlab{cns2d-khi-mc-re150-masshistory}
\end{figure}  

To assess the impact of the mass conserving source on accuracy, we plot the relative errors of density, momentum, and total energy in Figure \figref{cns2d-khi-mc-re150-errhistory} and Figure \figref{cns2d-khi-mc-re250-errhistory}. The grey dash-dot line indicates the relative error of the low-order approximation, while the blue dashed line corresponds to the augmented solution with the na\"{\i}ve source. The green dotted line and the brown solid line represent the augmented solutions with the mass conserving source for $\alpha=0$ and $\alpha=0.1$, respectively. 
In general, the na\"{\i}ve augmented solution starts with higher accuracy, but its error grows rapidly and eventually exceeds that of the low-order solution. In contrast, the mass conserving augmented solution with $\alpha=0$ begins less accurate, but its error increases more gradually, allowing it to maintain an accuracy advantage over a longer time span. When combined with $L_2$ regularization ($\alpha=0.1$), the mass conserving source shows behavior similar to the low-order solution, offering a limited accuracy gain while maintaining them over a longer period. 
For the density field in Figure \figref{cns2d-khi-mc-re150-errhistory}, the na\"{\i}ve augmented density $\hat{\rho}^L$ outperforms the low-order density $\rho^L$ up to $t=4$. The mass conserving augmented density with $\alpha=0$ maintains higher accuracy until $t=5$, while the one with $\alpha=0.1$ remains more accurate than the low-order density until $t=6.4$.
Similarly, in Figure \figref{cns2d-khi-mc-re250-errhistory}, the na\"{\i}ve augmented density $\hat{\rho}^L$ remains more accurate than the low-order density $\rho^L$ up to $t=3.6$, while the mass conserving augmented densities with $\alpha=0$ and regularization $\alpha=0.1$ extend this advantage until $t=4.3$ and $t=6.4$, respectively.
These results suggest a tradeoff between stability and accuracy.

\begin{figure}[h!t!b!]
  \centering
  \subfigure[$\rho$]{
    \figlab{cns2d-khi-mc-re150-err-r}
      \includegraphics[trim=2.8cm 0cm 1.0cm 0cm,clip=true,width=0.31\textwidth]{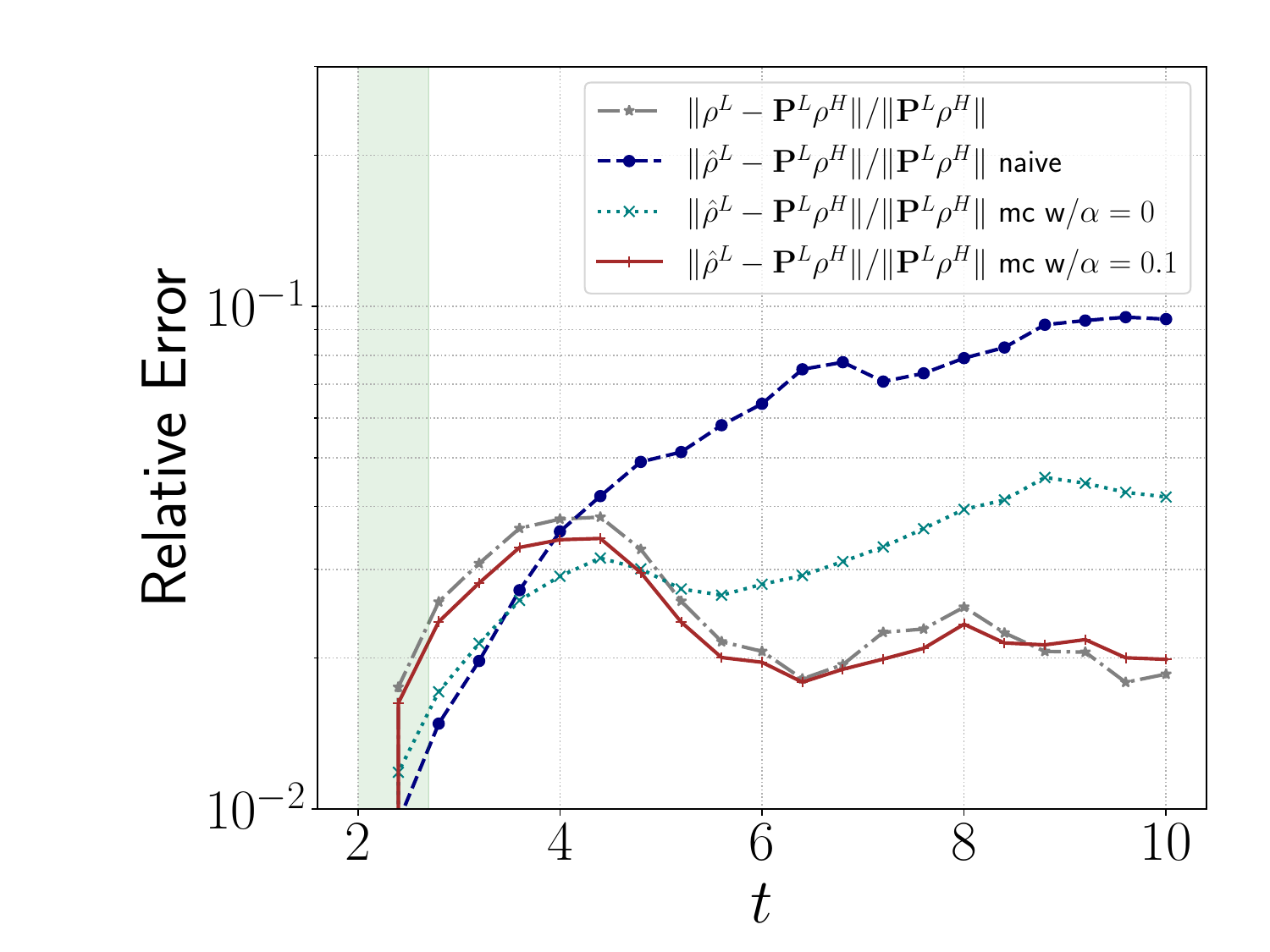}
  }
  \subfigure[$\rho {\boldsymbol\varphi}$]{
    \figlab{cns2d-khi-mc-re150-err-rvel}
      \includegraphics[trim=2.8cm 0cm 1.0cm 0cm,clip=true,width=0.31\textwidth]{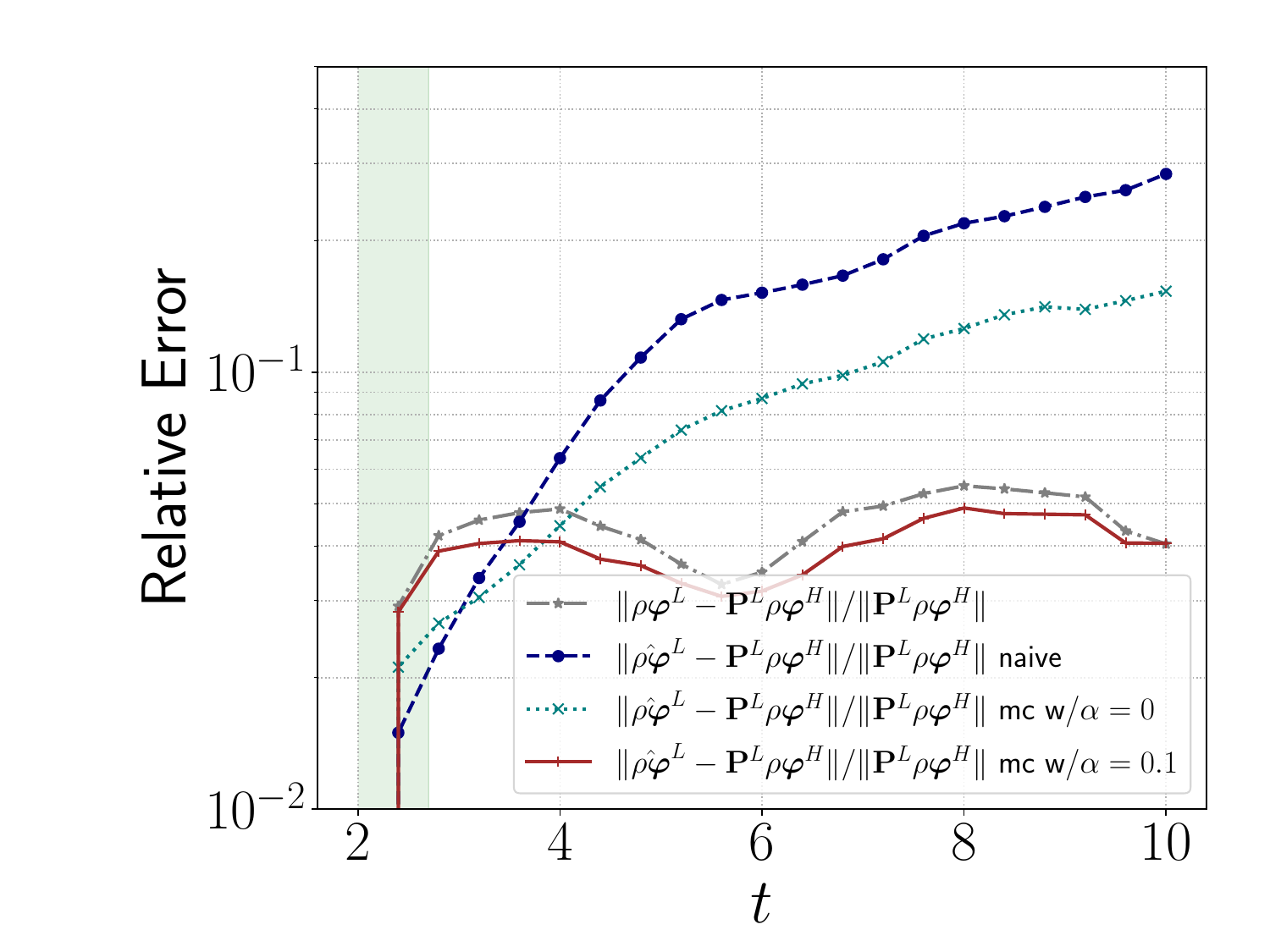}
  }
  \subfigure[$\rho E$]{
    \figlab{cns2d-khi-mc-re150-err-rE}
      \includegraphics[trim=2.8cm 0cm 1.0cm 0cm,clip=true,width=0.31\textwidth]{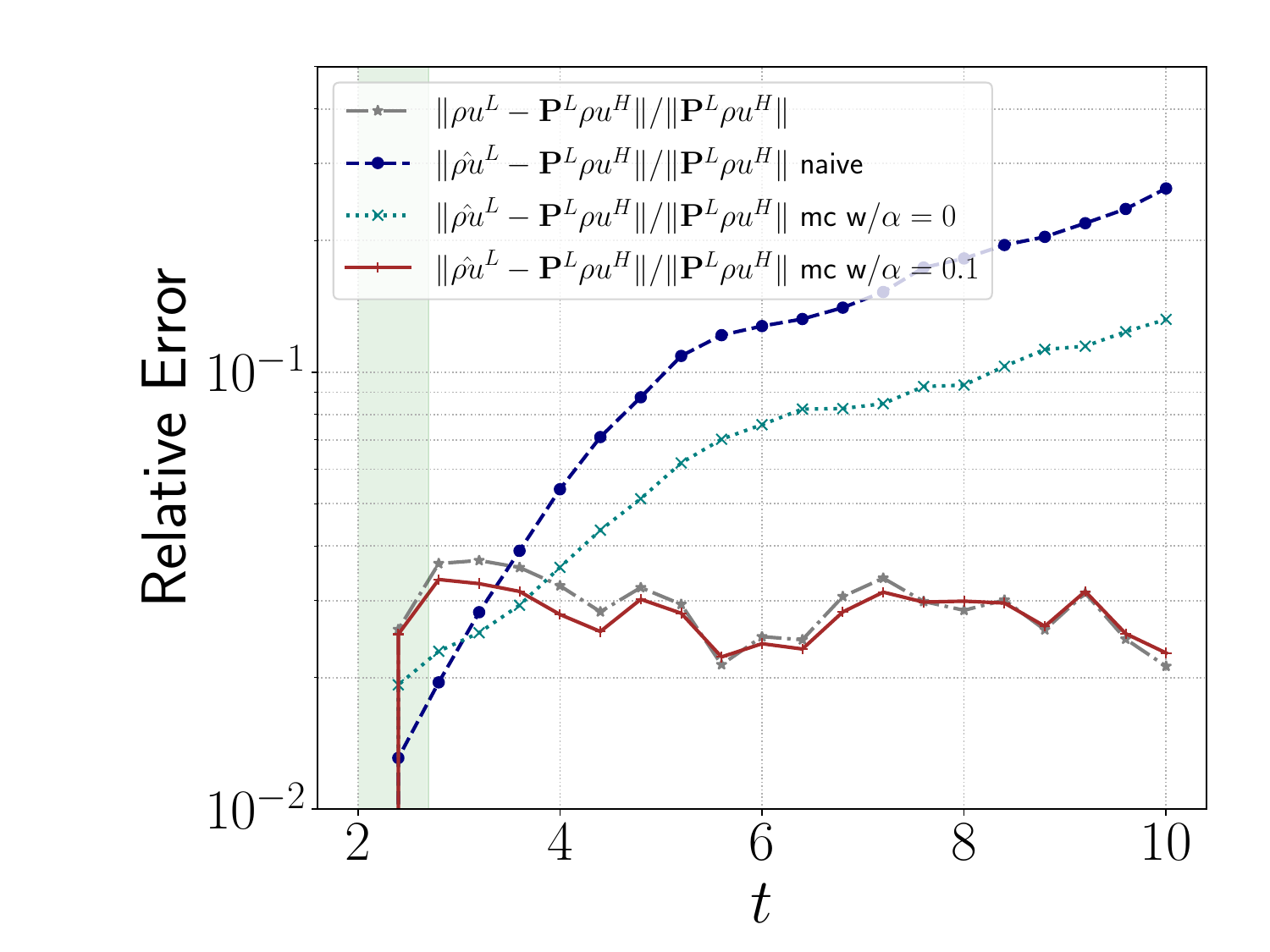}
  }
  \caption{2D Kelvin--Helmholtz instability: relative error histories of (a) $\rho$, (b) $\rho {\boldsymbol\varphi}$, and (c) $\rho E$ at $Re=150$ for low-order solution ($\ub^L$), na\"{\i}ve augmented solution ($\hat{\ub}^L\texttt{ na\"{\i}ve}$), and mass-conserving augmented solution ($\hat{\ub}^L\texttt{ mc}$).
  }
  \figlab{cns2d-khi-mc-re150-errhistory}
\end{figure}

\begin{figure}[h!t!b!]
  \centering
  \subfigure[$\rho$]{
    \figlab{cns2d-khi-mc-re250-err-r}
      \includegraphics[trim=2.8cm 0cm 1.0cm 0cm,clip=true,width=0.31\textwidth]{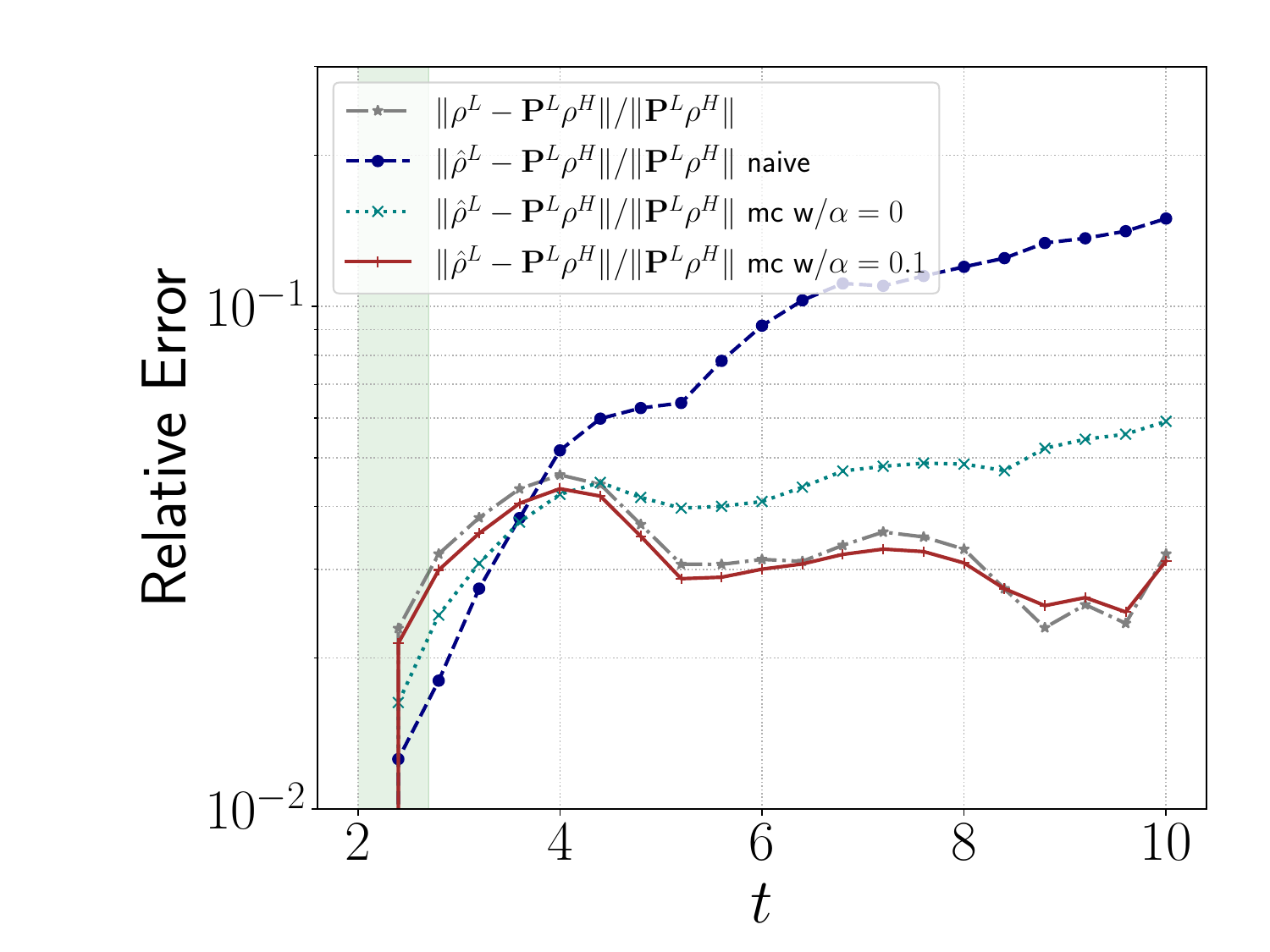}
  }
  \subfigure[$\rho {\boldsymbol\varphi}$]{
    \figlab{cns2d-khi-mc-re250-err-rvel}
      \includegraphics[trim=2.8cm 0cm 1.0cm 0cm,clip=true,width=0.31\textwidth]{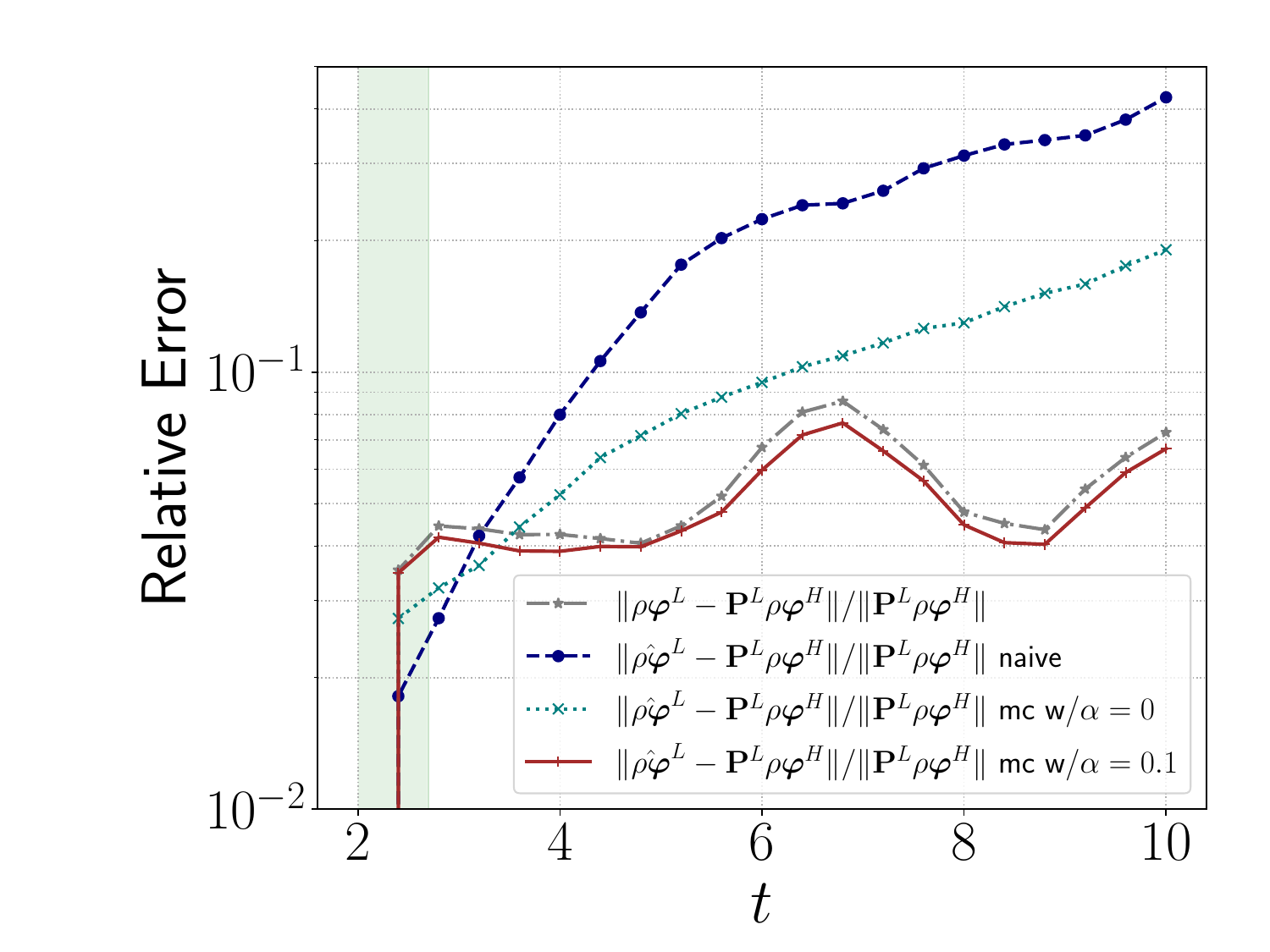}
  }
  \subfigure[$\rho E$]{
    \figlab{cns2d-khi-mc-re250-err-rE}
      \includegraphics[trim=2.8cm 0cm 1.0cm 0cm,clip=true,width=0.31\textwidth]{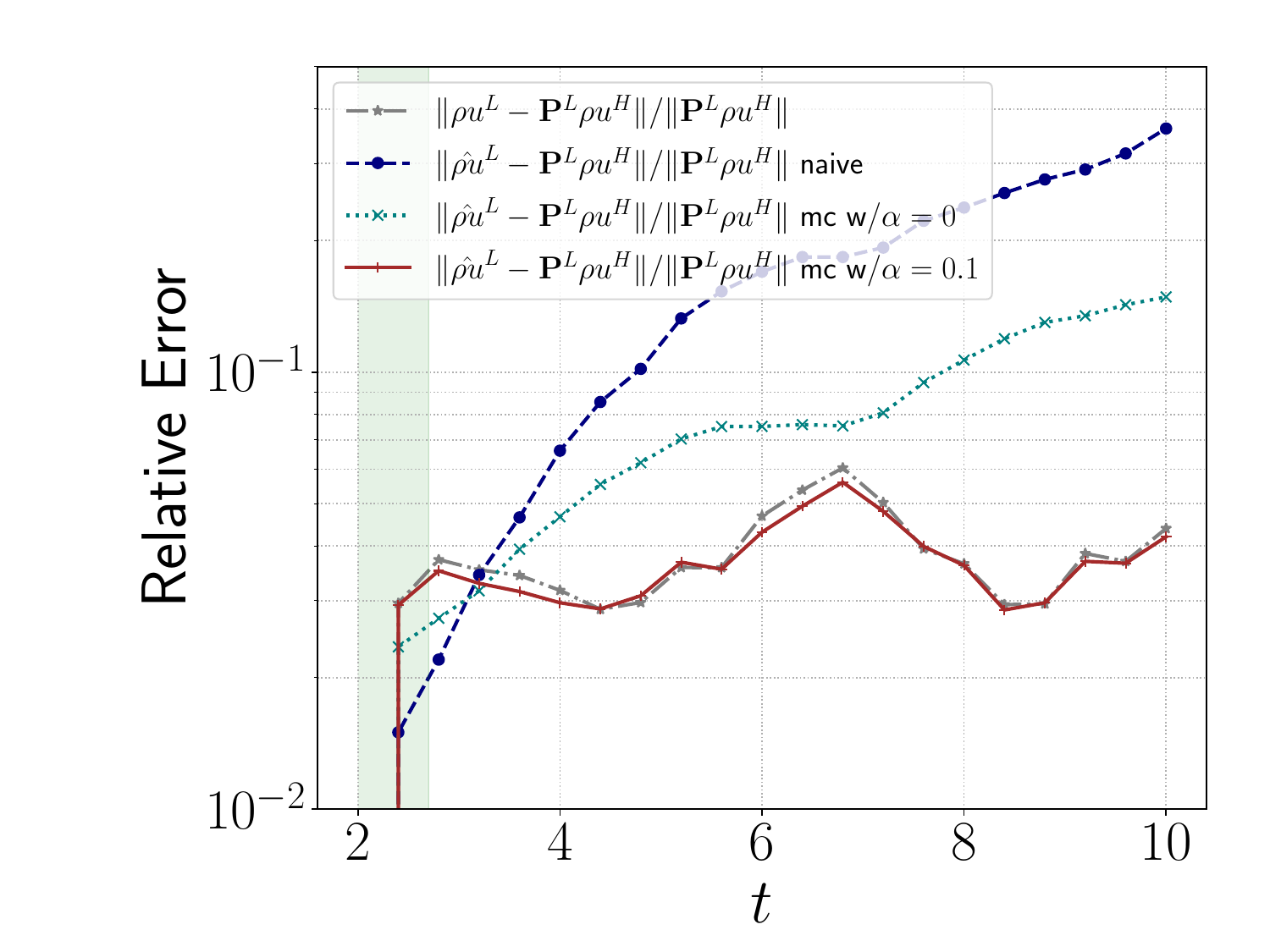}
  }
  \caption{2D Kelvin--Helmholtz instability: relative error histories of (a) $\rho$, (b) $\rho {\boldsymbol\varphi}$, and (c) $\rho E$ at $Re=250$ for low-order solution ($\ub^L$), na\"{\i}ve augmented solution ($\hat{\ub}^L\texttt{ na\"{\i}ve}$), and mass-conserving augmented solution ($\hat{\ub}^L\texttt{ mc}$).
  }
  \figlab{cns2d-khi-mc-re250-errhistory}
\end{figure}

\subsection{Three-dimensional Taylor--Green vortex}
\seclab{sec-tgv}

The Taylor--Green vortex \cite{taylor1937mechanism} is an exact solution to the two-dimensional, incompressible Navier--Stokes equations and is widely used to validate both the incompressible Navier--Stokes model and the compressible Navier--Stokes model at a low Mach number ($M_a$) regime. 
In two dimensions, the vortex remains stable and retains its shape but gradually loses energy because of viscous dissipation. 
In three dimensions, 
the vortex can stretch and twist across all spatial dimensions, facilitating energy transfer from larger to smaller scales. This energy cascade enables the transition from laminar flow to turbulence. 
At the smallest scales of turbulence, kinetic energy is dissipated into thermal energy because of viscosity.

\begin{figure}[h!t!b!]
  \centering
  \subfigure[$t=0$]{
    \figlab{cns3d-tgv-ss-t0}
      \includegraphics[trim=8cm 3cm 8cm 3cm,clip=true,width=0.31\textwidth]{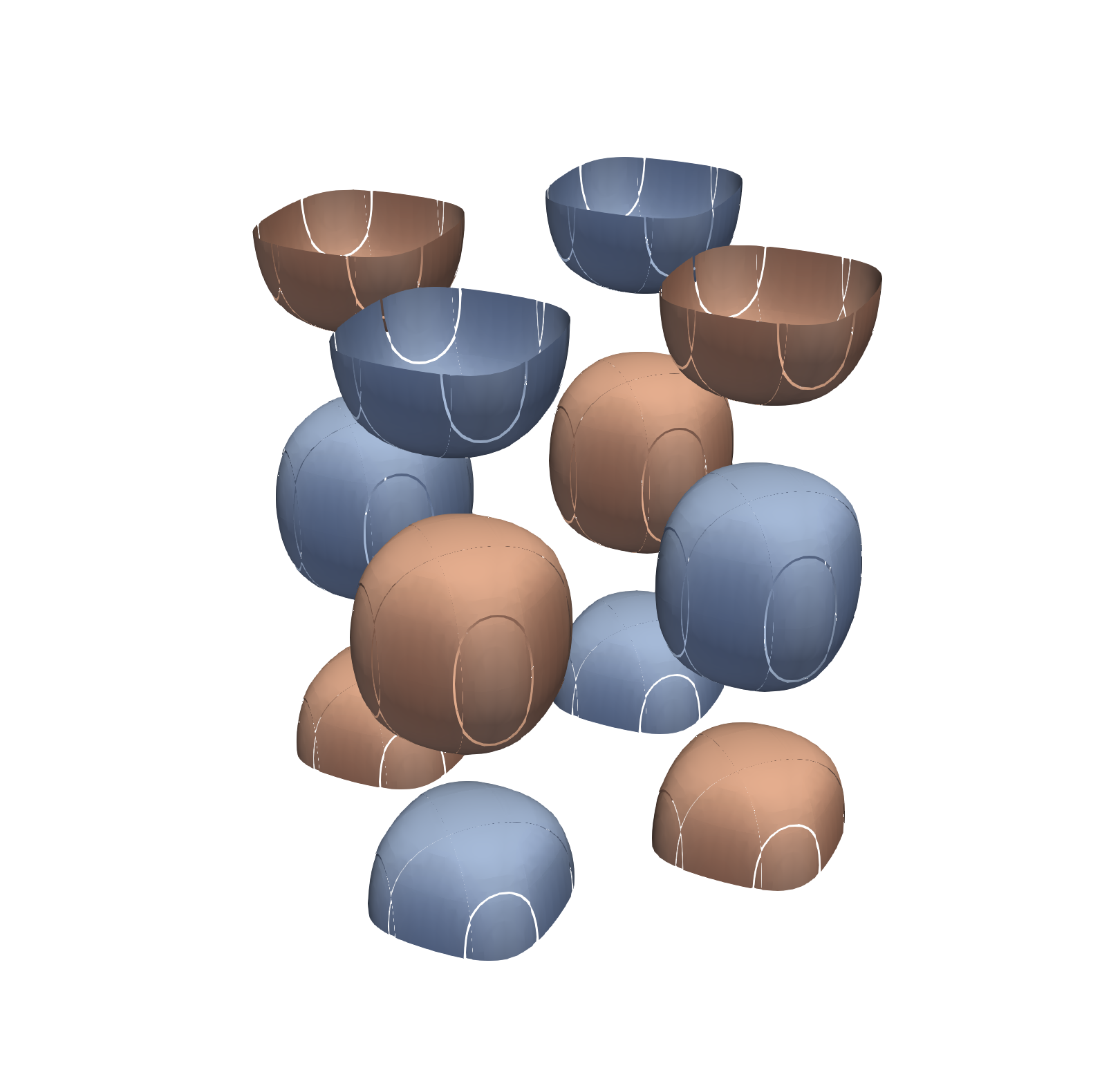}
  }
  \subfigure[$t=10$]{
    \figlab{cns3d-tgv-ss-t10}
      \includegraphics[trim=8cm 3cm 8cm 3cm,clip=true,width=0.31\textwidth]{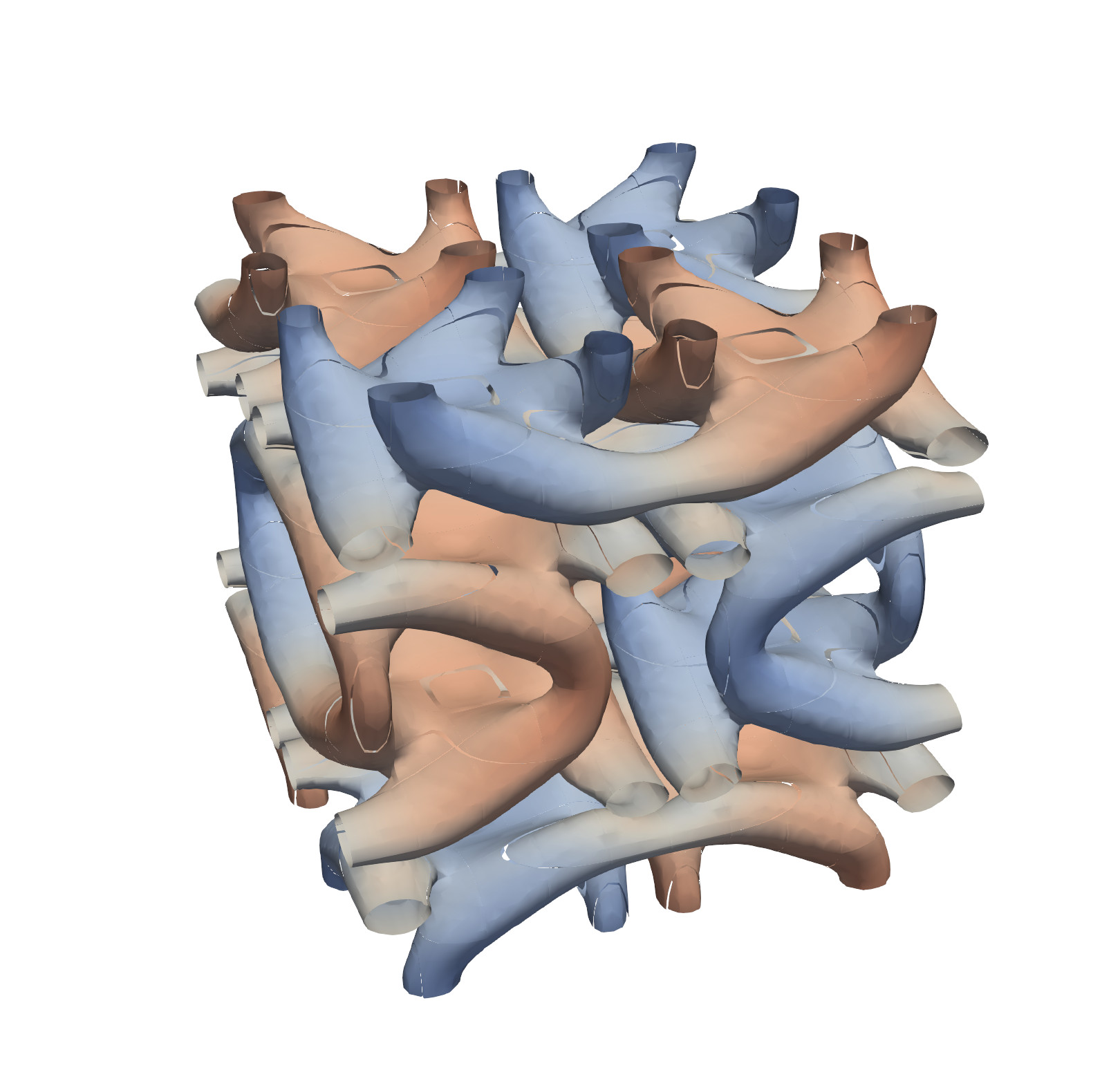}
  }
  \subfigure[$t=20$]{
    \figlab{cns3d-tgv-ss-t20}
      \includegraphics[trim=8cm 3cm 8cm 3cm,clip=true,width=0.31\textwidth]{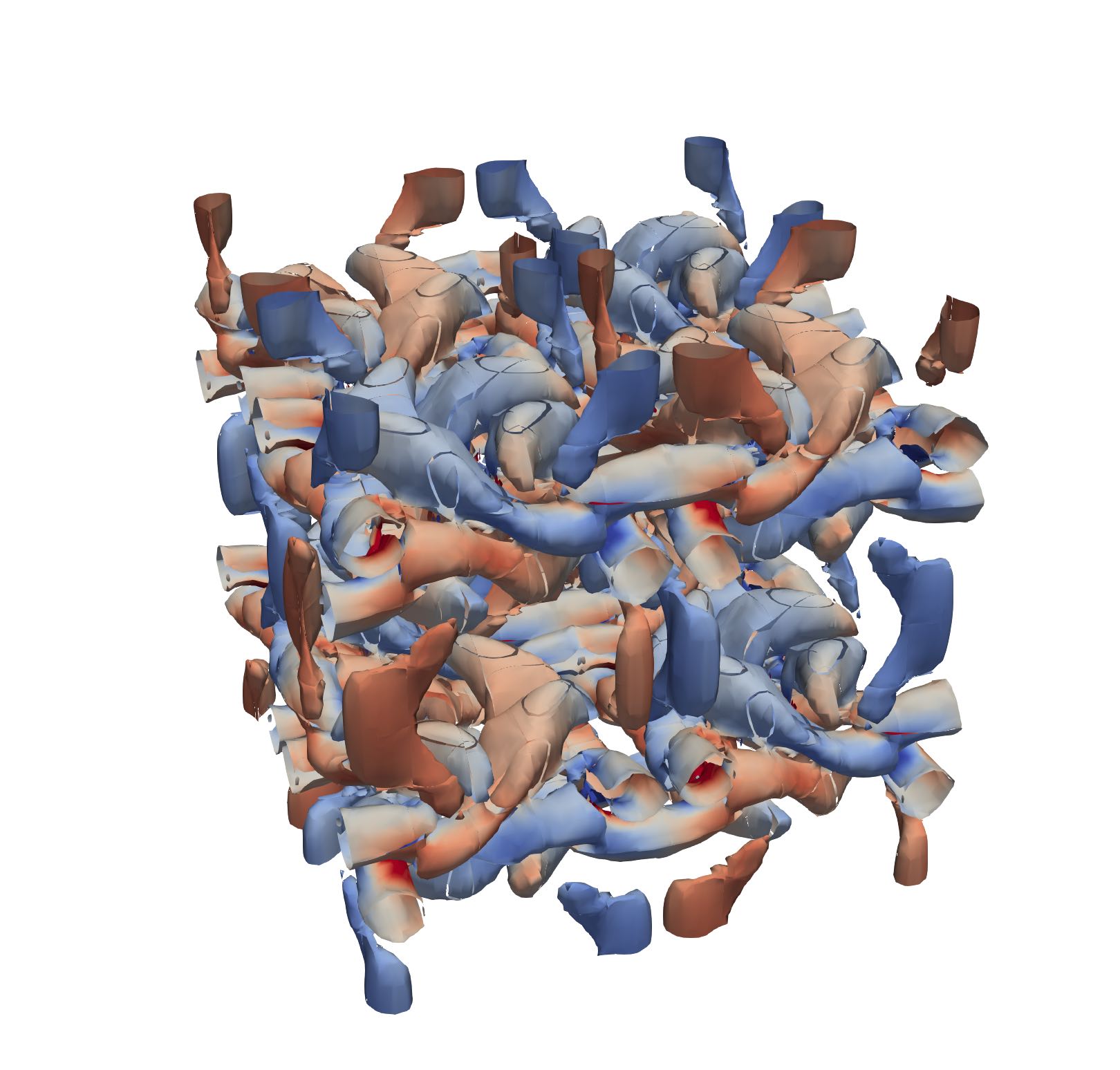}
  }
  \caption{3D Taylor--Green vortex: Q-criterion isosurfaces for $Re=1600$ with $10^3$ elements and $N=8$ at (a) $t=0$, (b) $t=10$, and (c) $t=20$. 
  The isosurfaces, contoured at a value of 0.02, are colored according to the z-component of the vorticity, ranging from $-1$ to $1$. 
  }
  \figlab{cns3d-tgv-ss}
\end{figure}

We examine the evolution of three-dimensional vortices within the domain $\Omega = [-1,1]^3$, 
where the initial field is given by 
\begin{align*}
\rho &= 1,\\
u &=   M_a~\sin\LRp{\pi x} \cos\LRp{\pi y} \cos\LRp{\pi z},\\
v &= - M_a~\cos\LRp{\pi x} \sin\LRp{\pi y} \cos\LRp{\pi z},\\
w & = 0,\\
\pres &= \frac{1}{\gamma} + \frac{M_a^2}{16} \LRp{ \cos\LRp{2\pi x} + \cos\LRp{2\pi y} } \LRp{ \cos\LRp{2\pi z} + 2}. 
\end{align*}
We set $\gamma=1.4$ and $Pr=0.72$ with $M_a=0.1$ so that the flow is in the incompressible flow regime. The initial temperature is calculated by using $T=\gamma p \rho^{-1}$ 
based on the normalized equation of state. 
Periodic boundary conditions are imposed on all the boundaries. We integrate the high-order DG model in   \eqnref{gov-ode-high} for the compressible Navier--Stokes equations using a third-order ERK with a time step of $\dtt=10^{-3}$ over a mesh of $10\times10\times10$ elements ($\Nel=10^3$) and the eighth-order polynomial ($N=8$).
%
%
Figure \figref{cns3d-tgv-ss} shows the evolution of three-dimensional vortices at $t=\LRc{0,10,20}$ for $Re=1600$.
In order to visualize the coherent vortical structures in the flow field, Q-criterion isosurfaces, contoured at a value of $0.02$, are colored based on the z-component of vorticity, ranging from $-1$ to $1$. 
\footnote{
Q-criterion is defined by 
$ Q=\half\LRp{\norm{\vort} - \norm{\strain}},$
where $\vort_{ij}=\frac{1}{2}\LRp{\dd{\varphi_i}{x_j}-\dd{\varphi_j}{x_i}}$ is vorticity and $\strain_{ij}=\frac{1}{2}\LRp{\dd{\varphi_i}{x_j}+\dd{\varphi_j}{x_i}}$ is strain rate. 
A positive Q value means the relative dominance of the rotational component over the stretching component in the velocity gradient $\Grad \varphi$. The smooth initial vortices undergo stretching, twisting, and splitting due to the high Reynolds number, resulting in turbulent motion.
} 
\begin{figure}[h!t!b!]
  \centering
  \subfigure[$Re=200$]{
    \figlab{cns3d-tgv-ss-t10-Re20}
      \includegraphics[trim=8cm 3cm 8cm 3cm,clip=true,width=0.31\textwidth]{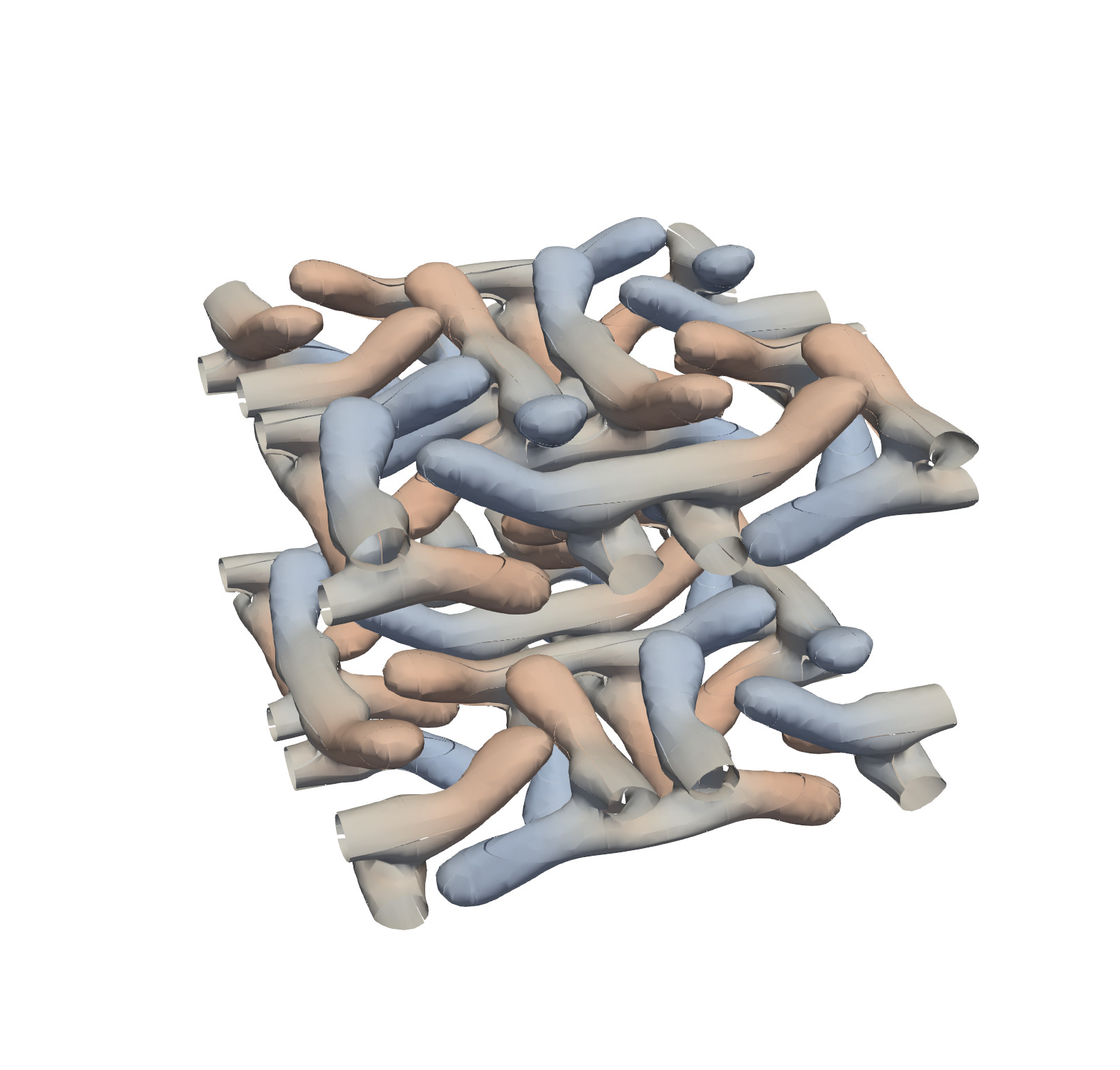}
  }
  \subfigure[$Re=400$]{
    \figlab{cns3d-tgv-ss-t10-Re400}
      \includegraphics[trim=8cm 3cm 8cm 3cm,clip=true,width=0.31\textwidth]{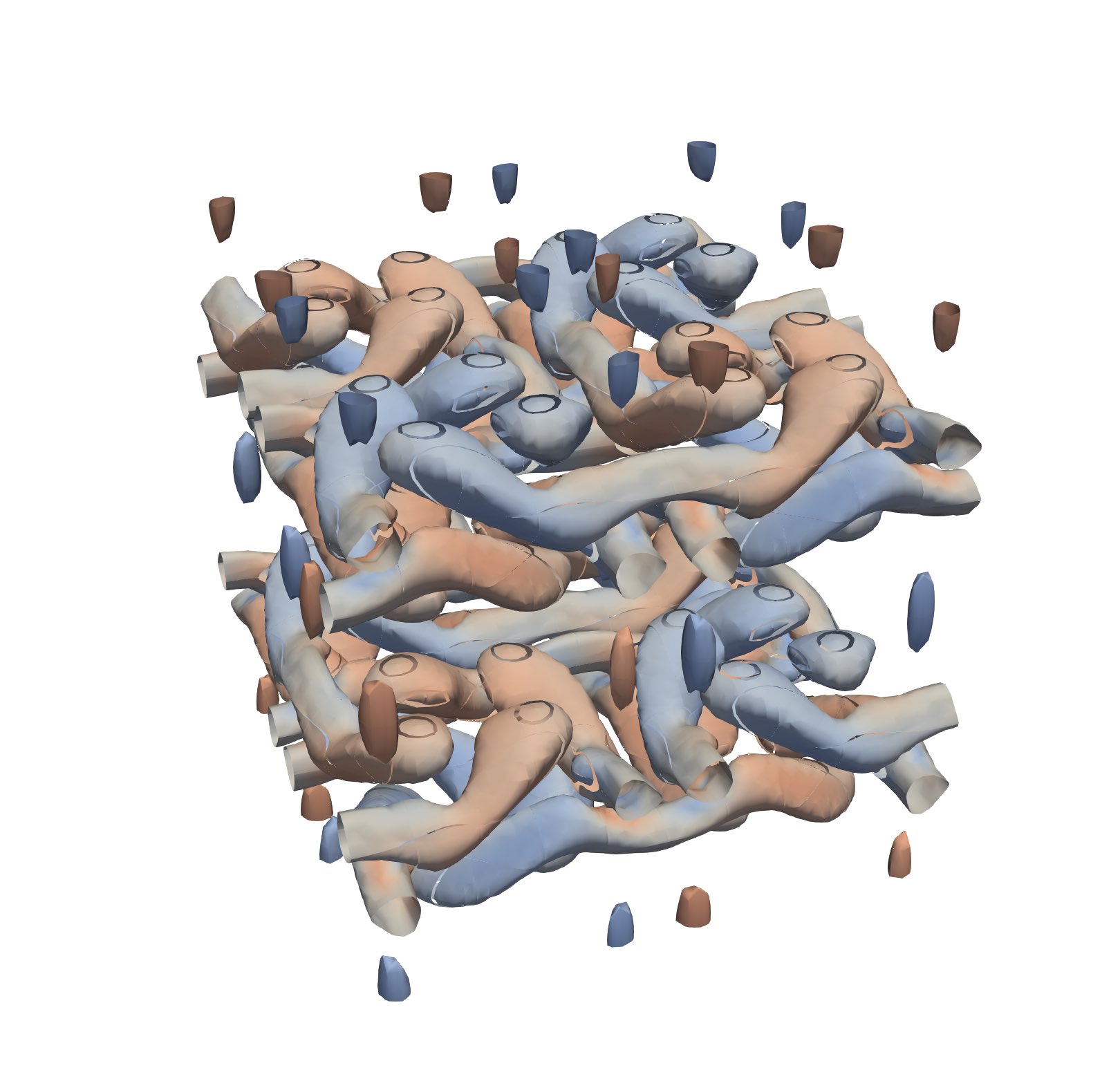}
  }
  \subfigure[$Re=800$]{
    \figlab{cns3d-tgv-ss-t10-Re800}
      \includegraphics[trim=8cm 3cm 8cm 3cm,clip=true,width=0.31\textwidth]{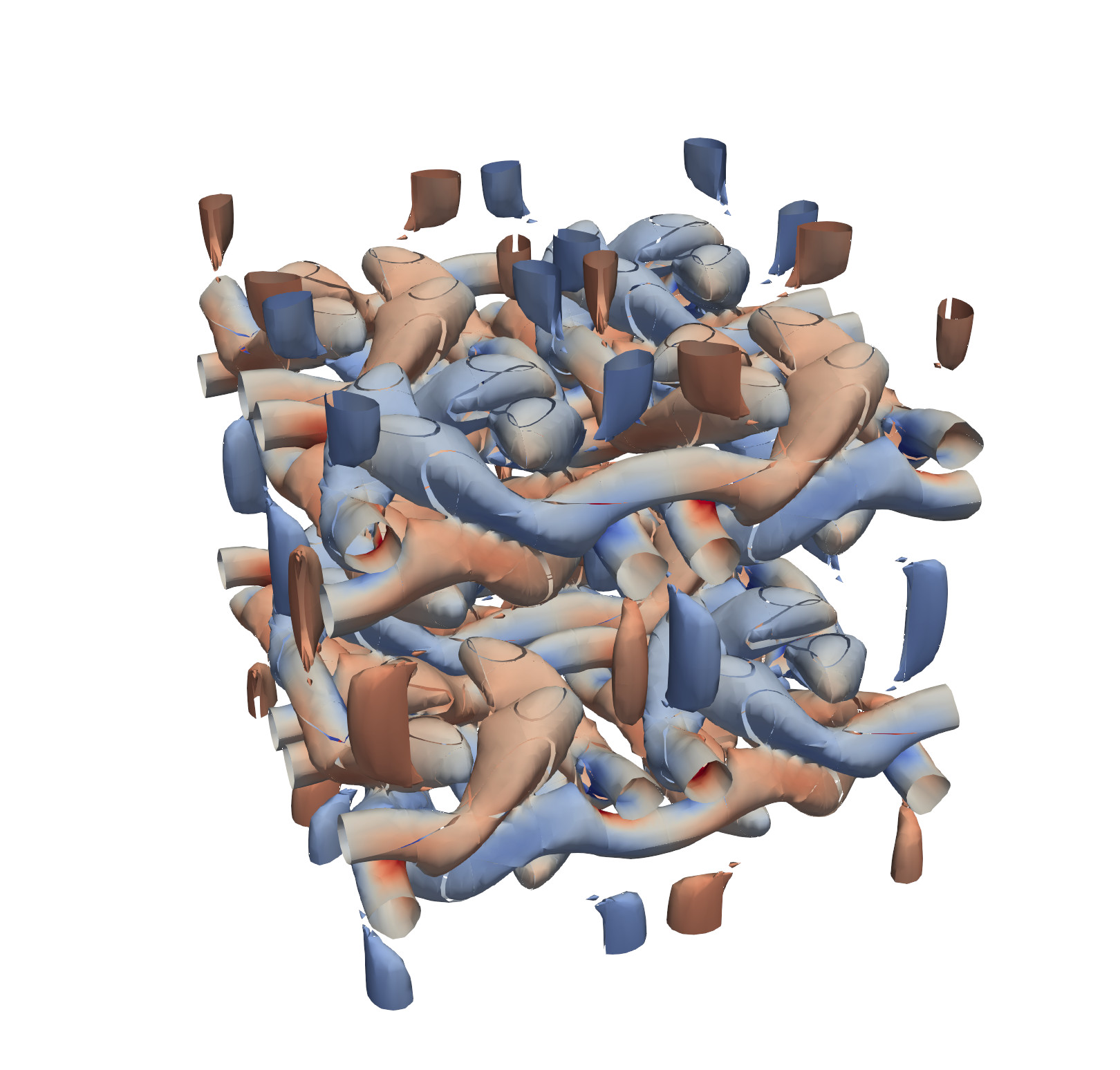}
  }
  \caption{3D Taylor--Green vortex: Q-criterion isosurfaces for (a) $Re=200$, (b) $Re=400$, and (c) $Re=800$ with $N=8$ and $10\times10\times10$ elements at $t=20$. 
  }
  \figlab{cns3d-tgv-Re}
\end{figure}
Figure \figref{cns3d-tgv-Re} shows that small-scale vortices emerge and become more prominent with increasing Reynolds number from $Re=200$ to $Re=800$. 
%


Following the approach of Lara and Ferrer~\cite{de2022accelerating,de2023accelerating}, 
we examine laminar, transitional, and turbulent regimes. 
For this purpose we create 12 datasets with Reynolds numbers $Re\in\LRc{100,200, 400,1600}$ and the starting times $t_1 \in \LRc{0,7,14}$.
\footnote{
A detailed analysis of the flow behavior at different Reynolds numbers can be found in Brachet et al.
~\cite{Brachet_Meiron_Orszag_Nickel_Morf_Frisch_1983}. 
}
 The high-order DG model is integrated by using the third-order ERK scheme and the time step size of $10^{-3}$ over one time unit, $t\in [t_1,t_1+1]$, on a mesh consisting of the sixth-order polynomial ($N=6$) and $20\times20\times20$ ($\Nel=8000$) elements. 
 We use the projection matrix $\Pmat^L$, defined in \eqnref{l2proj-high2low-op}, to project the high-order ($N=6$) approximations onto the low-order ($N=1$) approximations. 
 The time series of $\projL\ub^H$ is then split into training data for $t=[t_1,t_1+\frac{3}{4}]$ and 
test data for $t=[t_1+\frac{3}{4},t_1+1]$. 
A total of 12 neural network source terms, each corresponding to a specific dataset, are trained  
using the AdaBelief~\cite{zhuang2020adabelief} optimizer with a learning rate of $10^{-3}$ over $1,000$ epochs. 
Each neural network architecture is defined as $\LRc{n_0,64,32,64,n_{D}}$, where $n_0=40\times k_w^3$ and $n_D=40$. 
\footnote{
In a hexahedral element, the eight nodal points (located at vertices) are used for the first-order DG methods.
}
For training the neural network source term, two batch instances of $\projL \ub^H$ are randomly selected from the training data. The augmented system \eqnref{gov-ode-system} is then integrated for $m$ time steps by using the fifth-order ERK method~\cite{tsitouras2011runge} with a time step of $\dt=10^{-2}$, on a mesh consisting of $20\times20\times20$ elements with the first-order ($N=1$) polynomial. 
%
The simulations using the same third-order ERK method in different regimes and $Re$ numbers are discussed below.

\subsubsection{Results at various times for $Re=\LRc{100,200,400,1600}$}

 Figure \figref{cns3d-tgv-Re100-errhistory} shows the relative error histories for density, momentum, and total energy at $Re=100$ with $k_w=5$ and $m=11$.
 Across all variables (density, momentum, and total energy), the augmented solution (blue solid line) demonstrates smaller relative error compared with its low-order counterpart (gray dash-dot line).
 In the incompressible regime ($M < 0.3$), fluctuations in density and total energy are much smaller than those in momentum. Consequently, the momentum error dominates and is about two orders of magnitude larger than the errors in density and total energy. Additionally, as the flow evolves over time, the momentum error increases, with errors at $t=8$ and $t=15$ being greater than those at $t=1$. 
 This becomes evident when examining the snapshots of Q-criterion isosurfaces at $t=\LRc{1,8,15}$. Initially, the flow begins with smoothed profiles, resulting in a small difference between the low-order and augmented solutions, as shown in Figure \figref{cns3d-tgv-Re100-ss-qv-t1}. As flow evolves, however, the augmented solution demonstrates significantly improved accuracy compared with the low-order solution. At $t=8$ and $t=15$, the momentum error in the augmented solutions is half that of the low-order solution, as illustrated in Figure \figref{cns3d-tgv-Re100-p1-errhistory-rvel}. We see the remarkable improvements in the snapshots at $t=8$ (in Figure \figref{cns3d-tgv-Re100-ss-qv-t8}) and $t=15$ (in Figure \figref{cns3d-tgv-Re100-ss-qv-t15}). The augmented solution successfully recovers the vortical structures, while the low-order solution rapidly dissipates the vortices because of the numerical dissipation associated with the low-order solver in viscous-dominated flows.

\begin{figure}[h!t!b!]
  \centering
  \subfigure[Density ($\rho$)]{
    \figlab{cns3d-tgv-Re100-p1-errhistory-r}
      \includegraphics[trim=0.3cm 0.5cm 0.2cm 0.2cm,clip=true,width=0.31\textwidth]{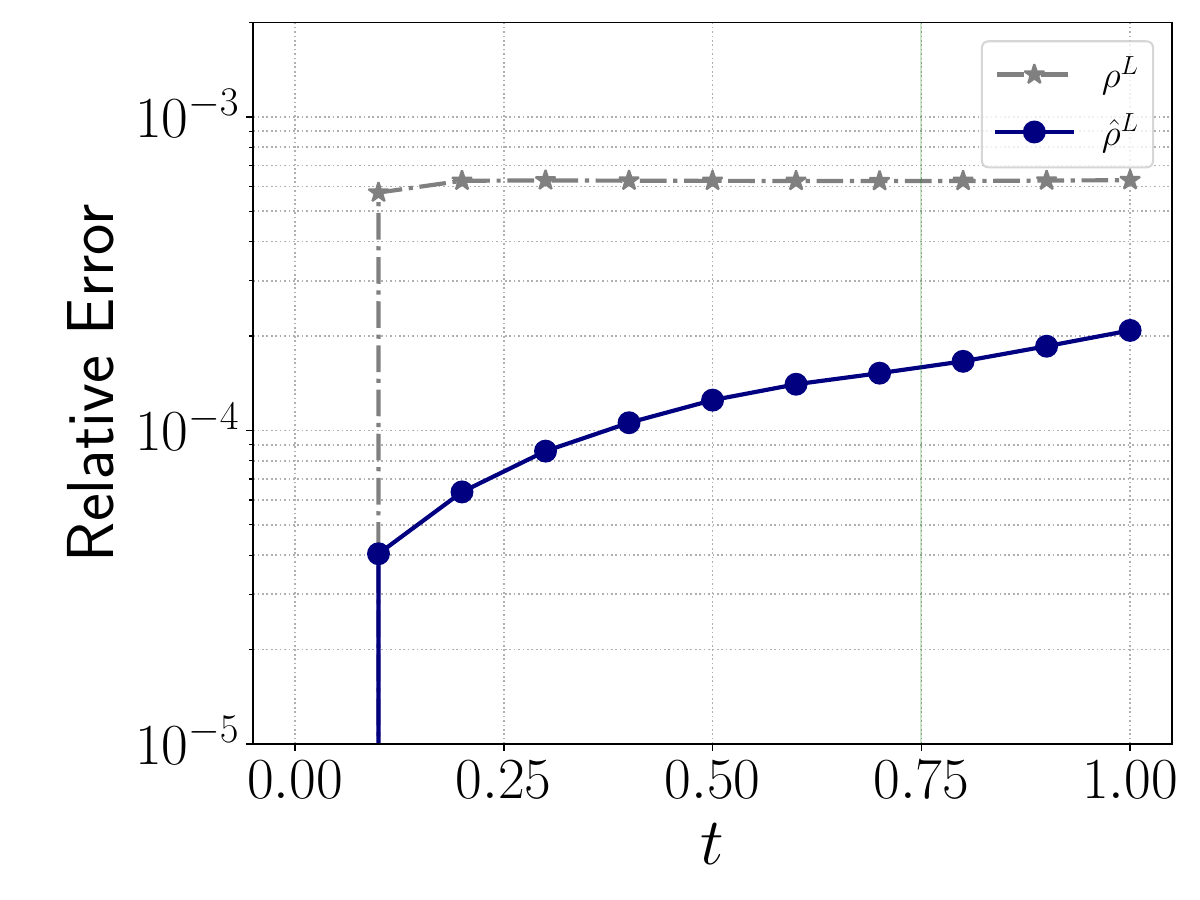}
    \includegraphics[trim=0.3cm 0.5cm 0.2cm 0.2cm,clip=true,width=0.31\textwidth]{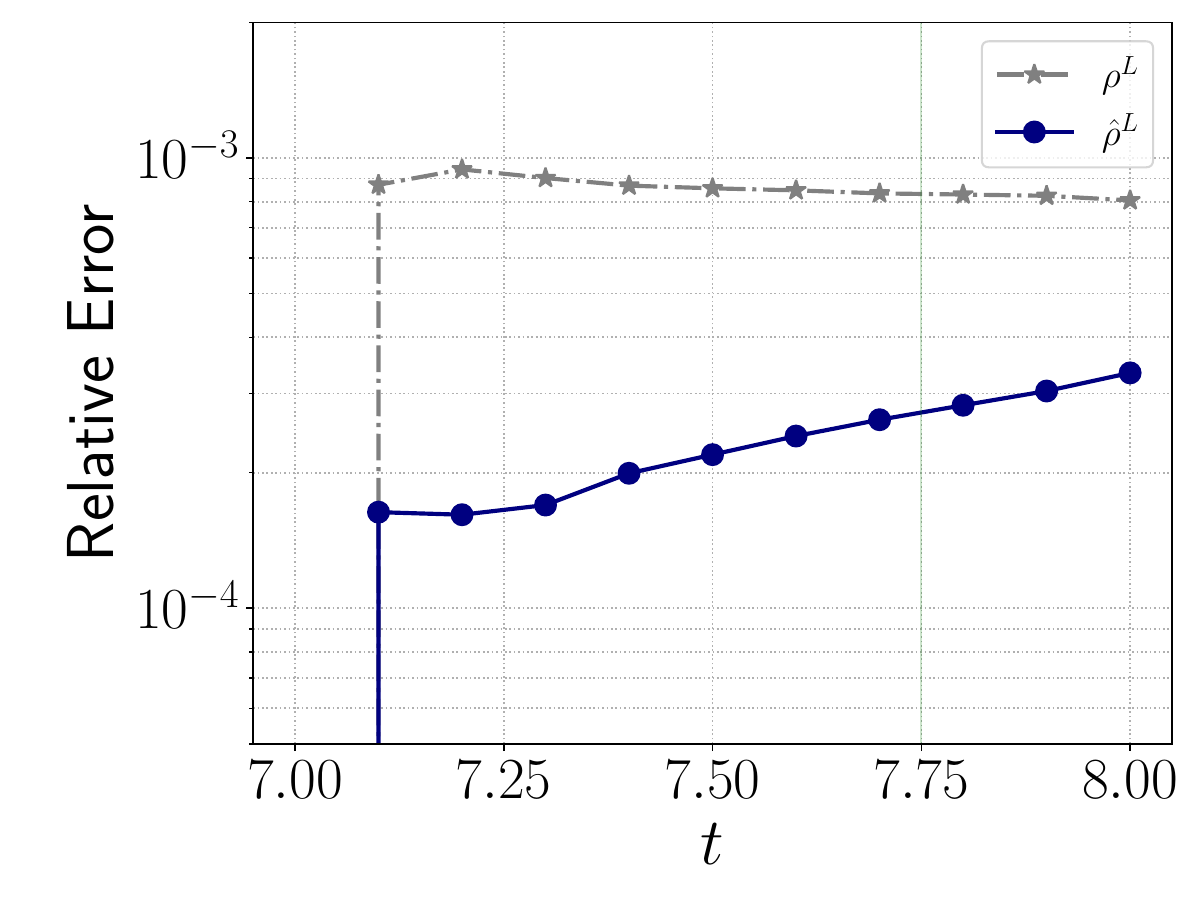}
    \includegraphics[trim=0.3cm 0.5cm 0.2cm 0.2cm,clip=true,width=0.31\textwidth]{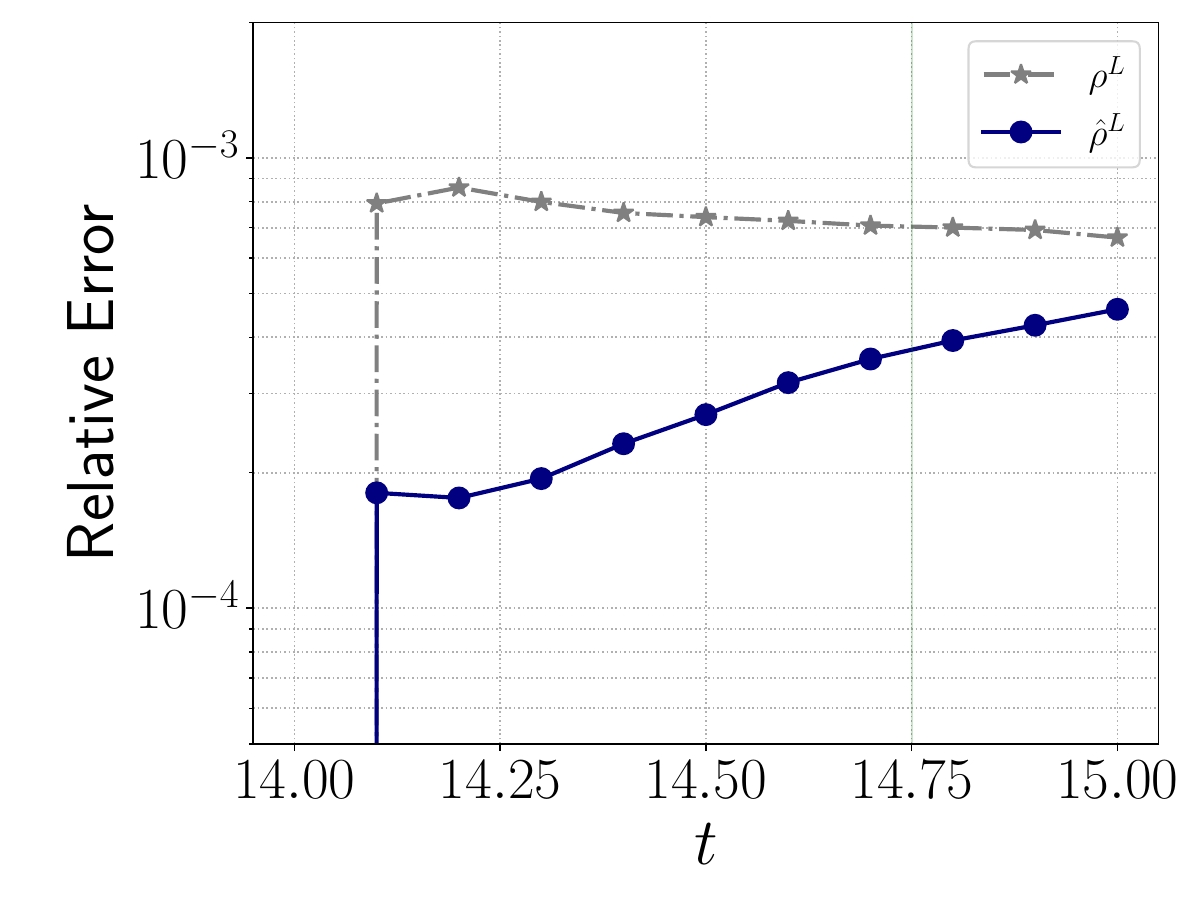}
  }
  \subfigure[Momentum ($\rho \boldsymbol{\varphi}$)]{
    \figlab{cns3d-tgv-Re100-p1-errhistory-rvel}
      \includegraphics[trim=0.3cm 0.5cm 0.2cm  0.2cm,clip=true,width=0.31\textwidth]{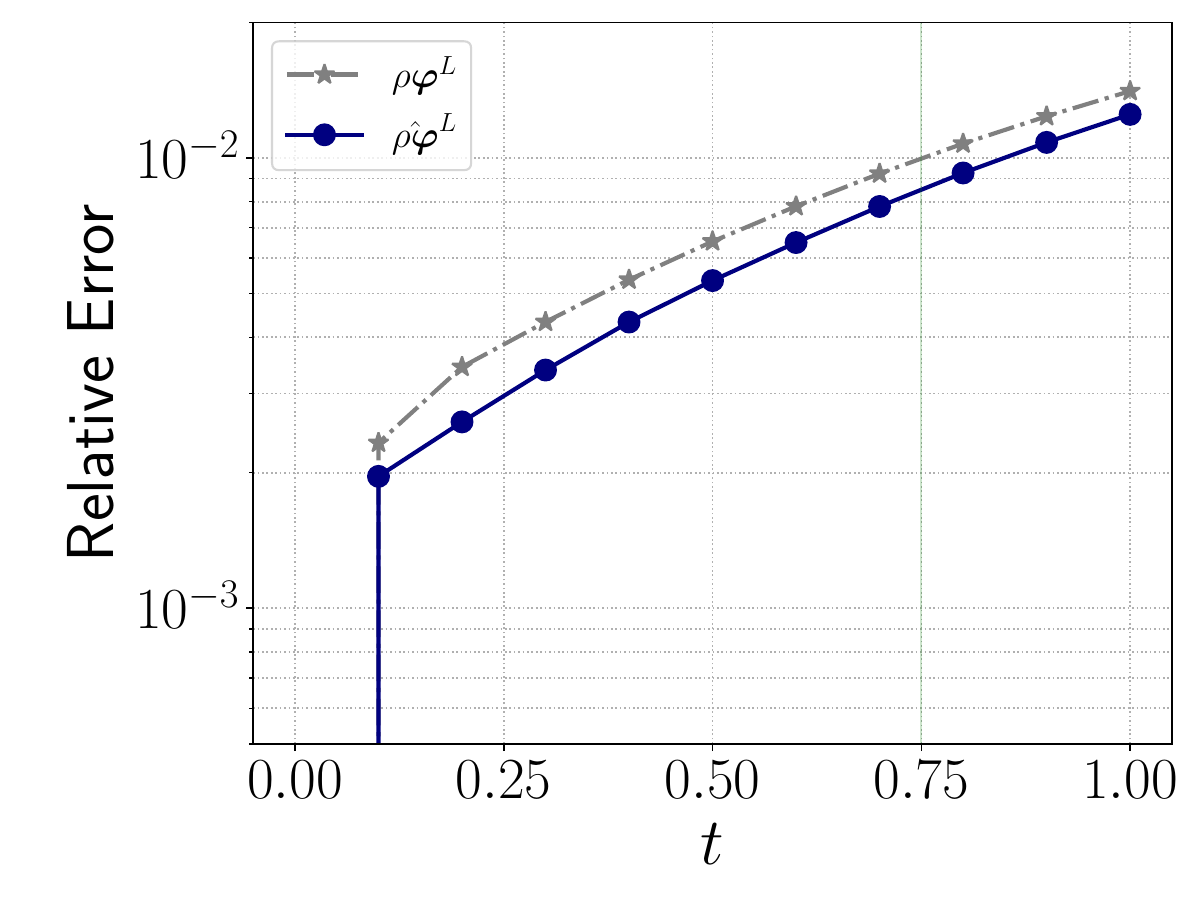}
      \includegraphics[trim=0.3cm 0.5cm 0.2cm  0.2cm,clip=true,width=0.31\textwidth]{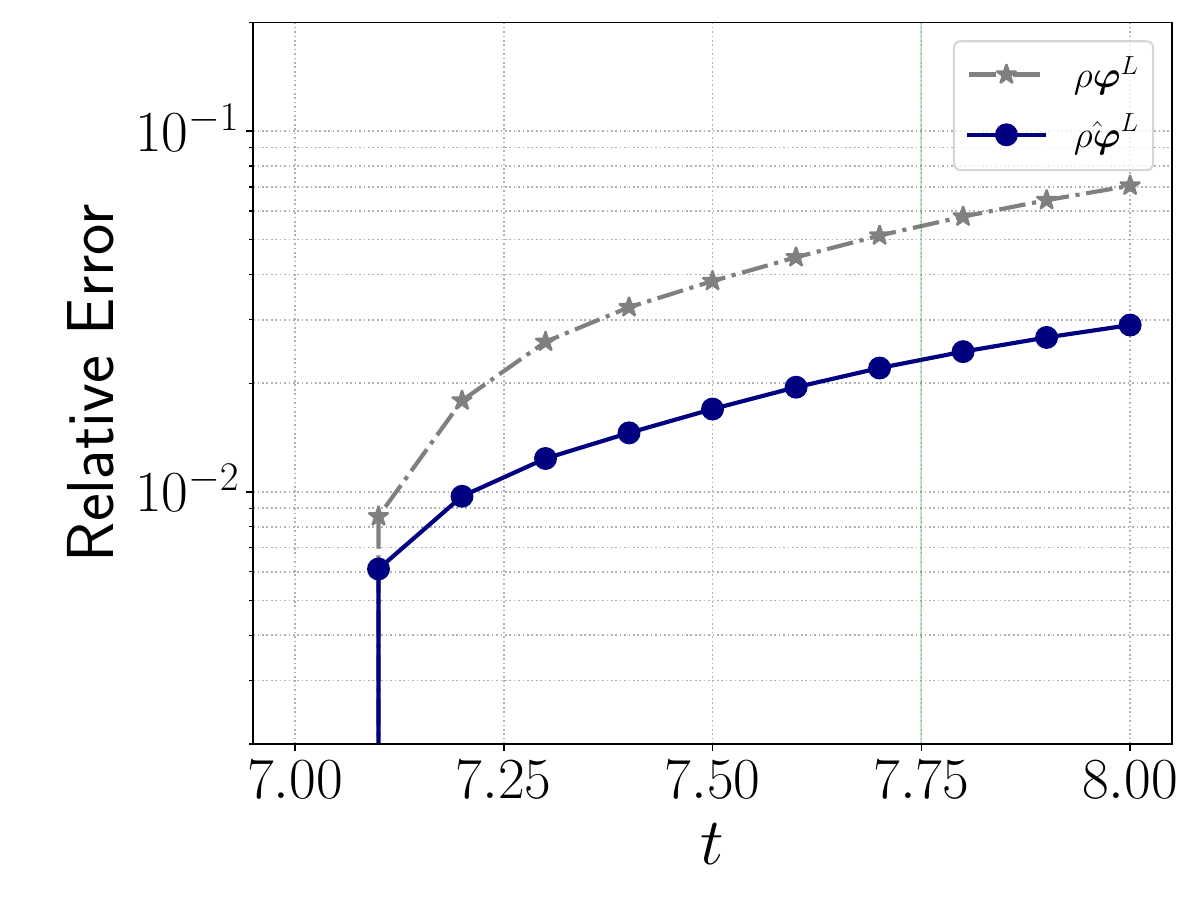}
      \includegraphics[trim=0.3cm 0.5cm 0.2cm  0.2cm,clip=true,width=0.31\textwidth]{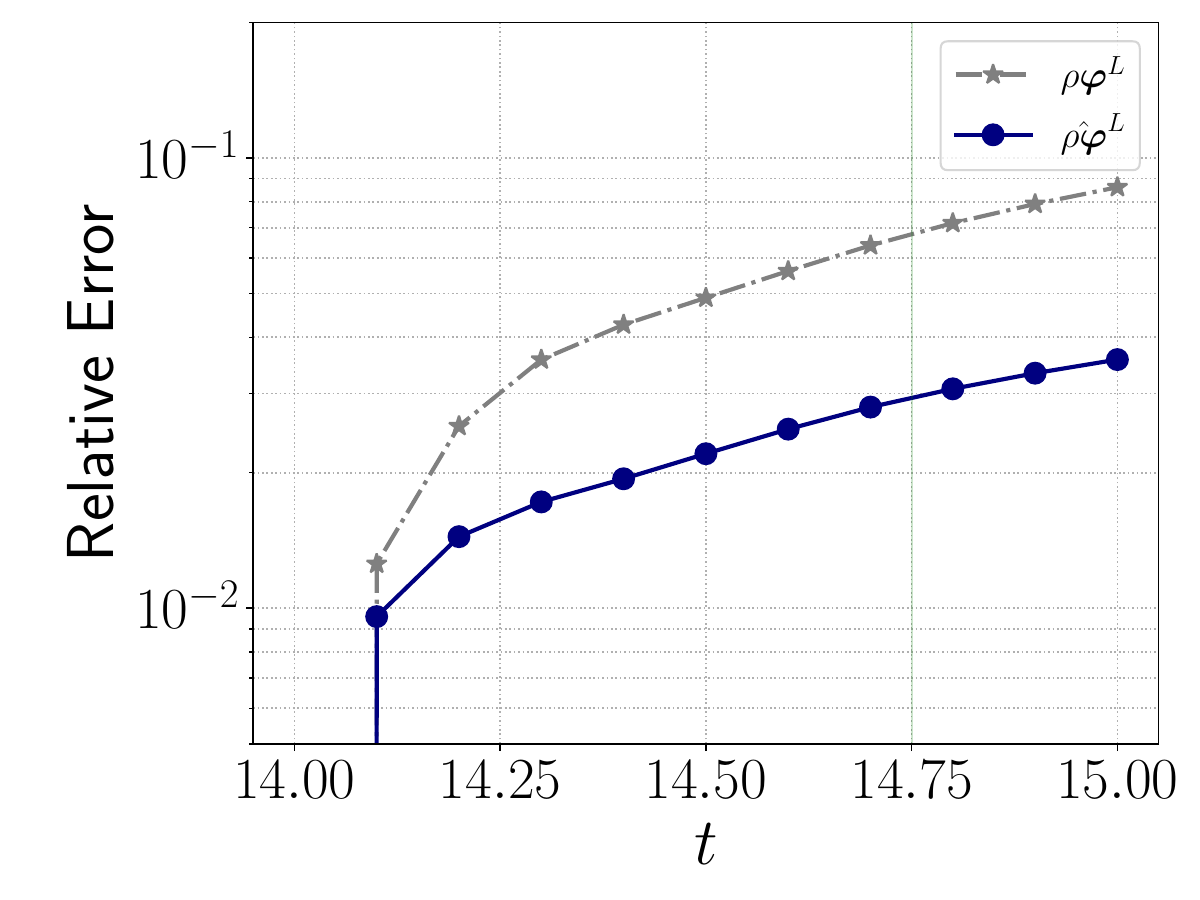}
  }
  \subfigure[Total energy ($\rho E$)]{
    \figlab{cns3d-tgv-Re100-p1-errhistory-rE}
      \includegraphics[trim=0.3cm 0.5cm 0.2cm 0.2cm,clip=true,width=0.31\textwidth]{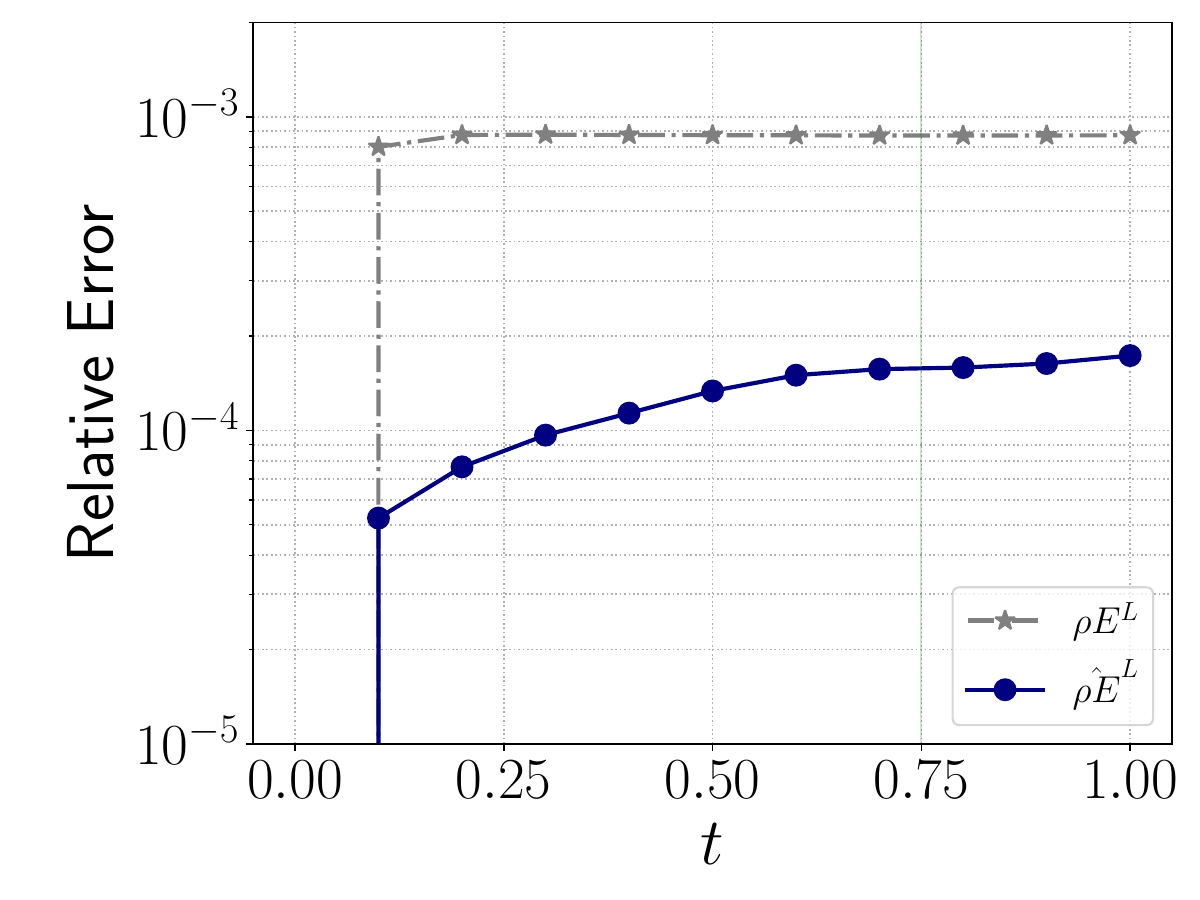}
      \includegraphics[trim=0.3cm 0.5cm 0.2cm 0.2cm,clip=true,width=0.31\textwidth]{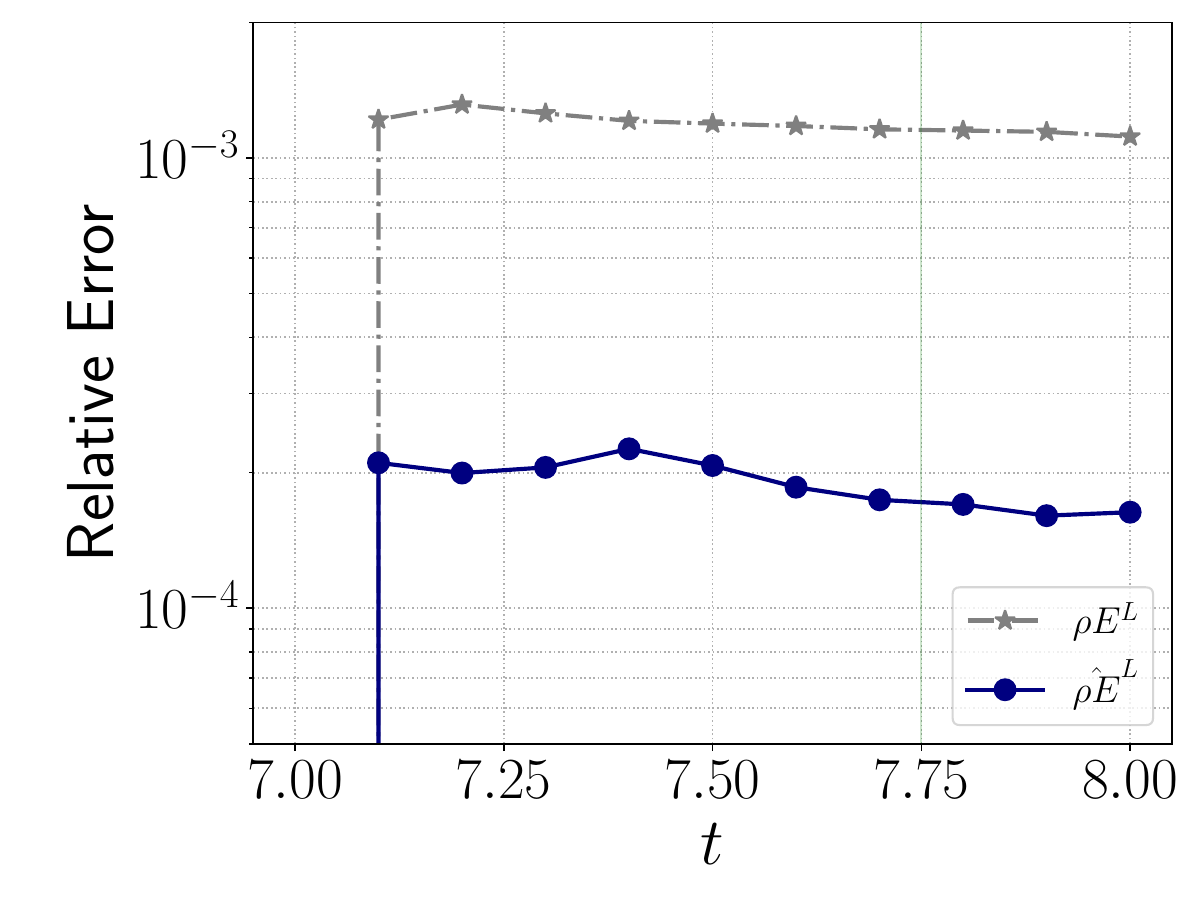}
      \includegraphics[trim=0.3cm 0.5cm 0.2cm 0.2cm,clip=true,width=0.31\textwidth]{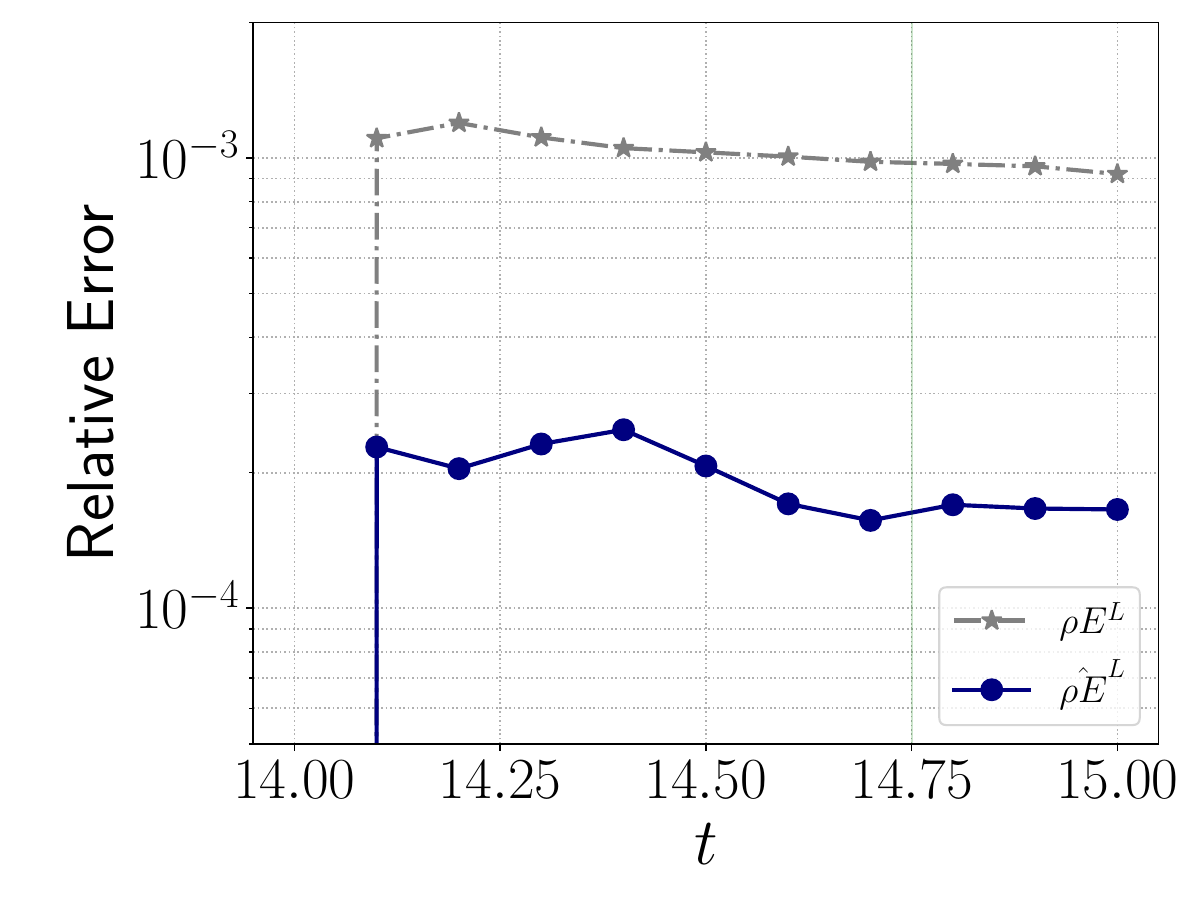}  
  }
  \caption{3D Taylor--Green vortex: relative error histories of (a) $\rho$, (b) $\rho\boldsymbol{\varphi}$, and (c) $\rho E$ at $Re=100$ 
    for low-order solution ($\ub^L$) and augmented solution ($\hat{\ub}^L$). 
  }
  \figlab{cns3d-tgv-Re100-errhistory}
\end{figure}

\begin{figure}[h!t!b!]
  \centering
  \subfigure[Projected]{
    \figlab{cns3d-tgv-Re100-p1-t1-qv-Gu-ss}
      \includegraphics[trim=8cm 3cm 8cm 3cm,clip=true,width=0.31\textwidth]{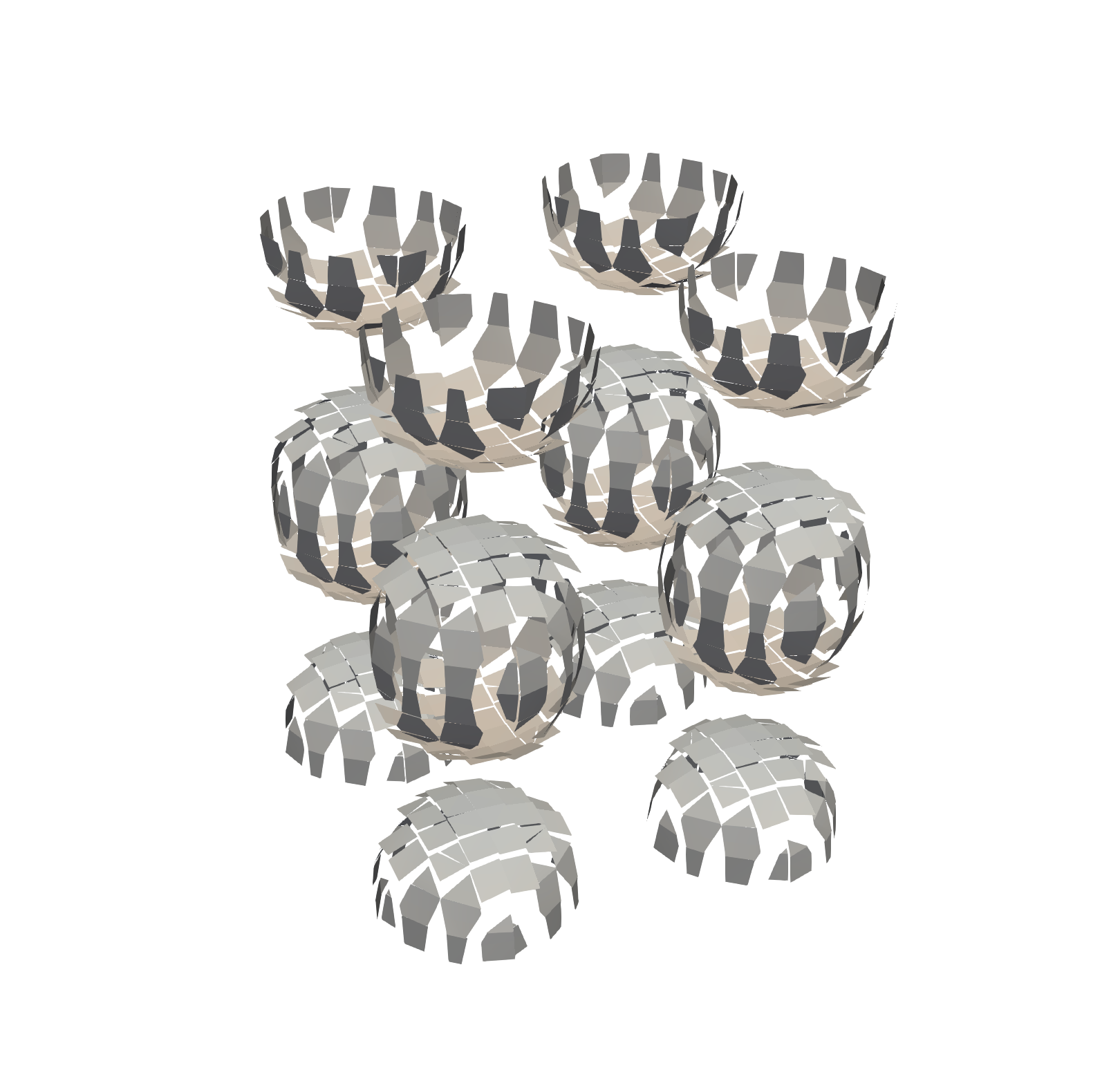}
  }
  \subfigure[Augmented]{
    \figlab{cns3d-tgv-Re100-p1-t1-qv-uh-ss}
      \includegraphics[trim=8cm 3cm 8cm 3cm,clip=true,width=0.31\textwidth]{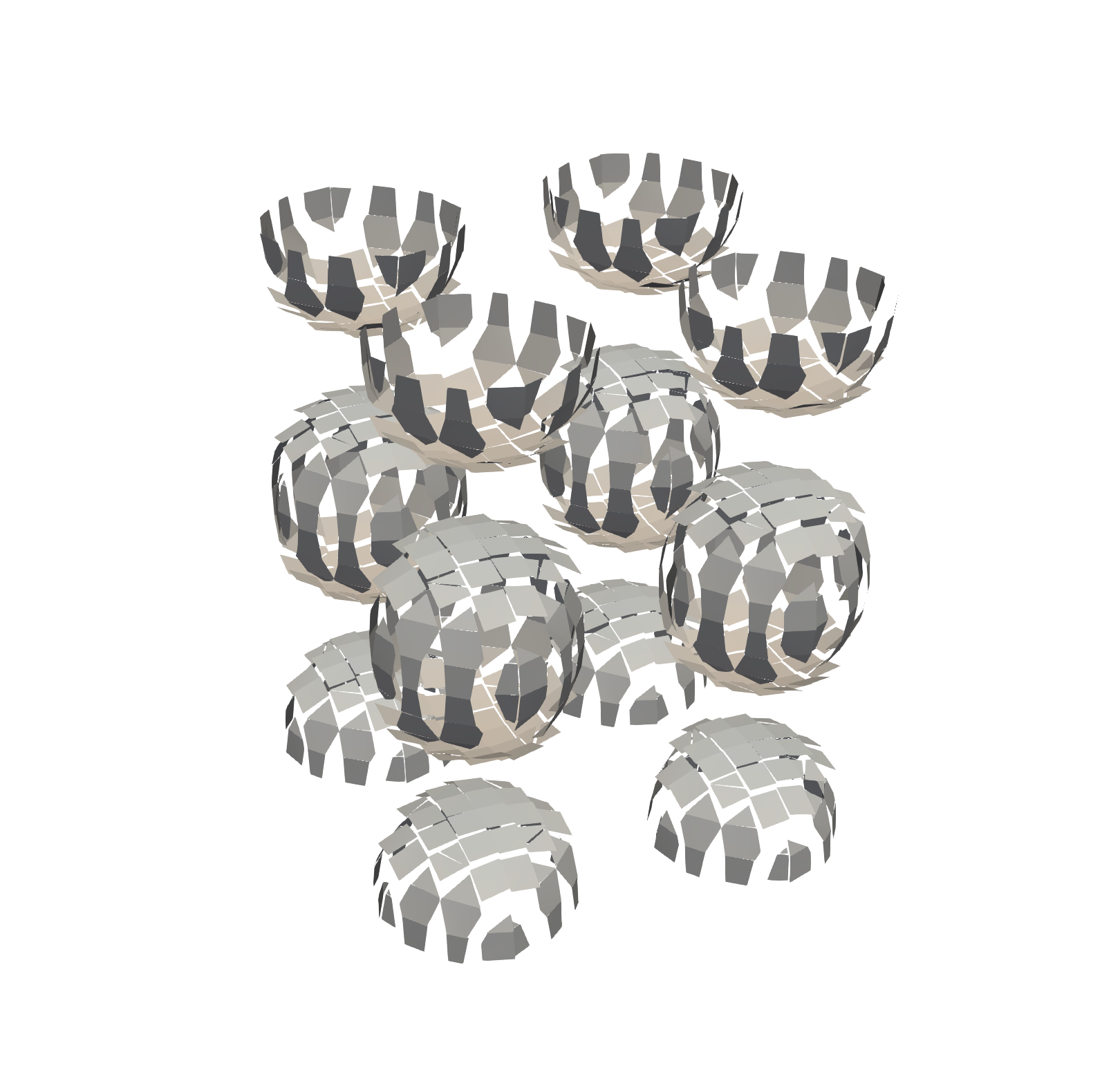}
  }
  \subfigure[Low-order]{
    \figlab{cns3d-tgv-Re100-p1-t1-qv-uL-ss}
      \includegraphics[trim=8cm 3cm 8cm 3cm,clip=true,width=0.31\textwidth]{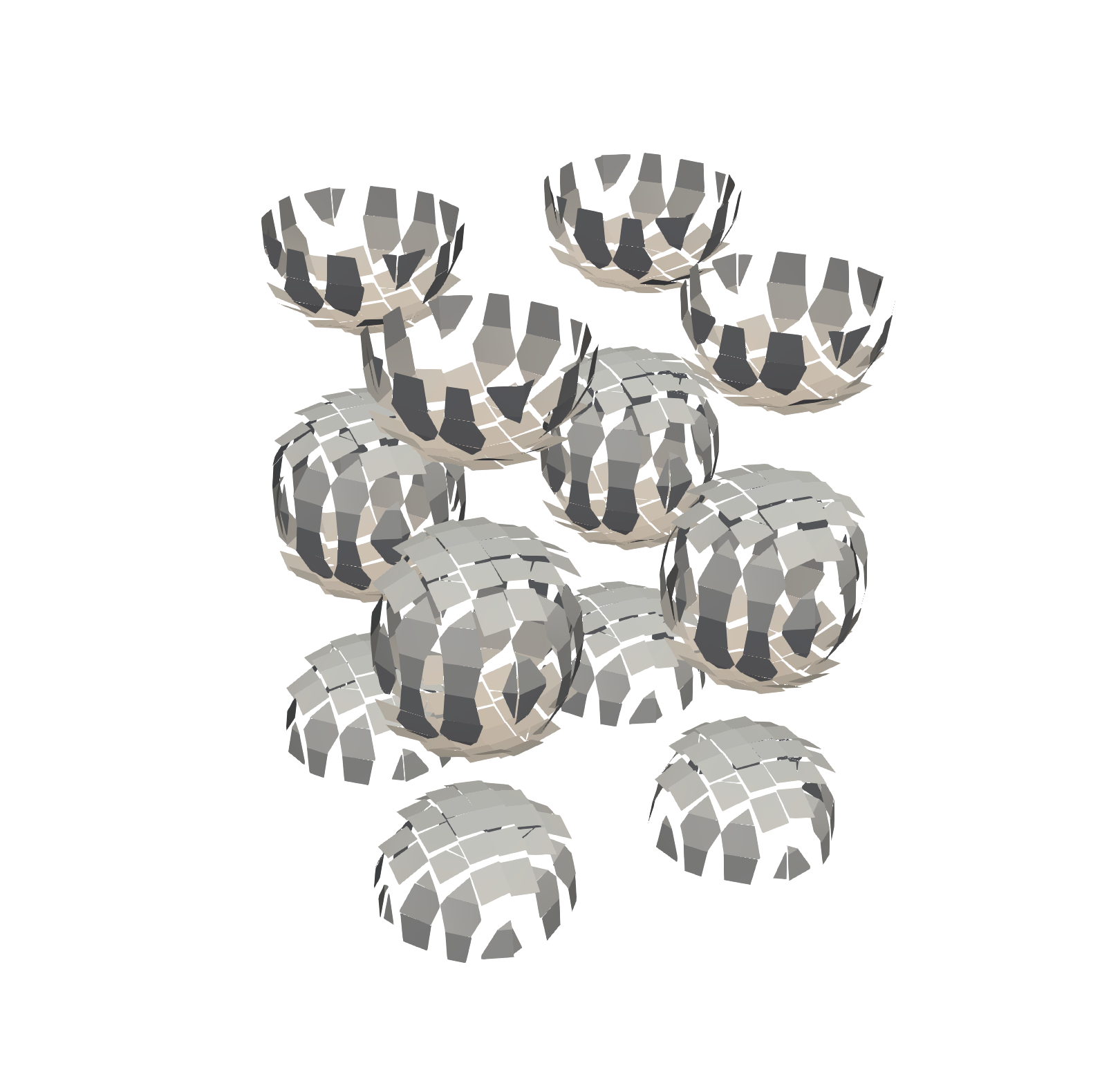}
  }
  \caption{3D Taylor--Green vortex: Q-criterion isosurfaces at $Re=100$ and $t=1$ are shown for (a) the projected solution ($\projL \ub^H$), (b) the augmented solution ($\hat{\ub}^L$), and (c) the low-order solution ($\ub^L$).
  The isosurfaces 
  are colored based on the z-component of the velocity, ranging from $-0.1$ to $0.1$. 
  }
  \figlab{cns3d-tgv-Re100-ss-qv-t1}
\end{figure}

\begin{figure}[h!t!b!]
  \centering
  \subfigure[Filtered]{
    \figlab{cns3d-tgv-Re100-p1-t8-qv-Gu-ss}
      \includegraphics[trim=8cm 3cm 8cm 3cm,clip=true,width=0.31\textwidth]{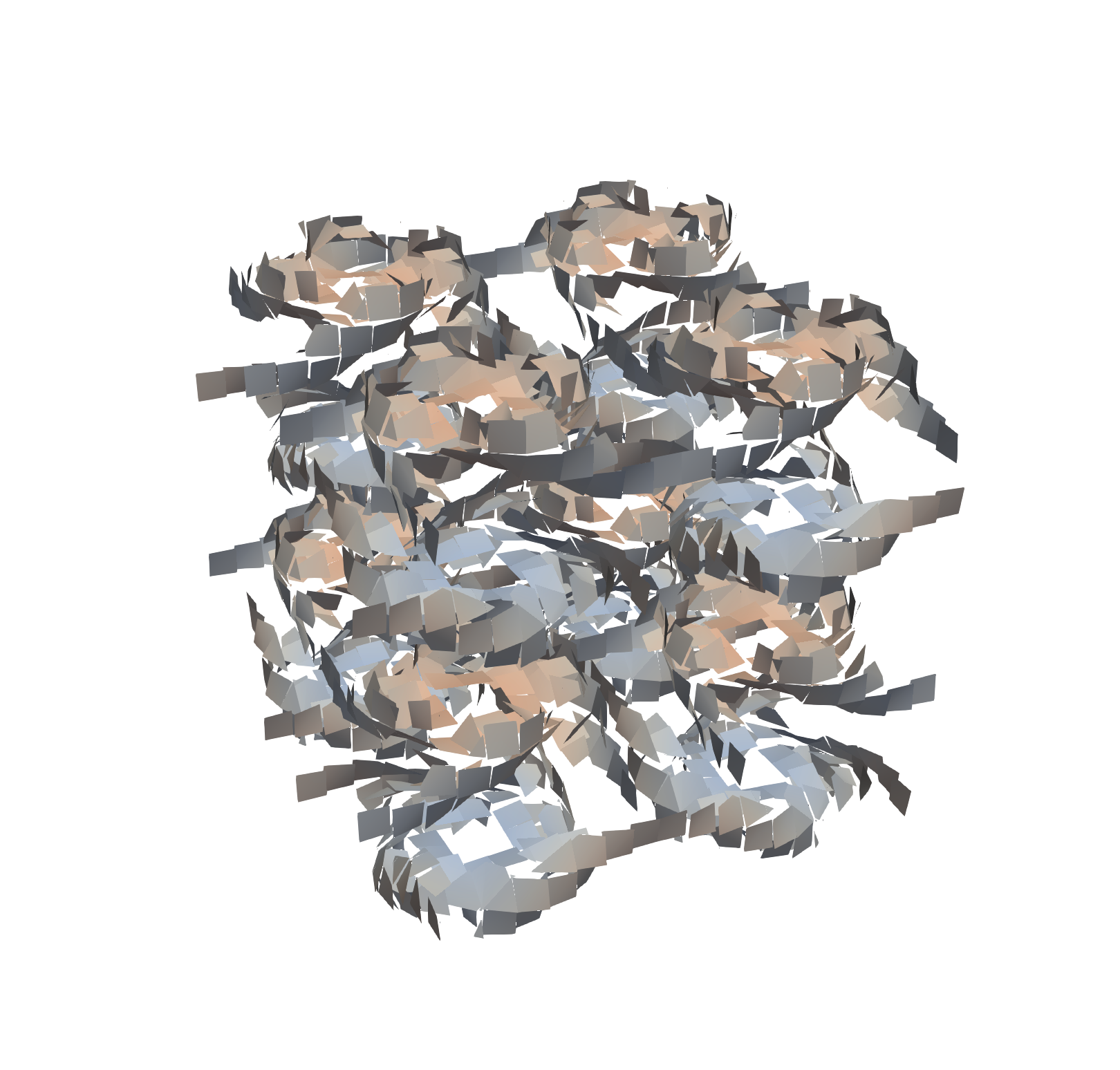}
  }
  \subfigure[Augmented]{
    \figlab{cns3d-tgv-Re100-p1-t8-qv-uh-ss}
      \includegraphics[trim=8cm 3cm 8cm 3cm,clip=true,width=0.31\textwidth]{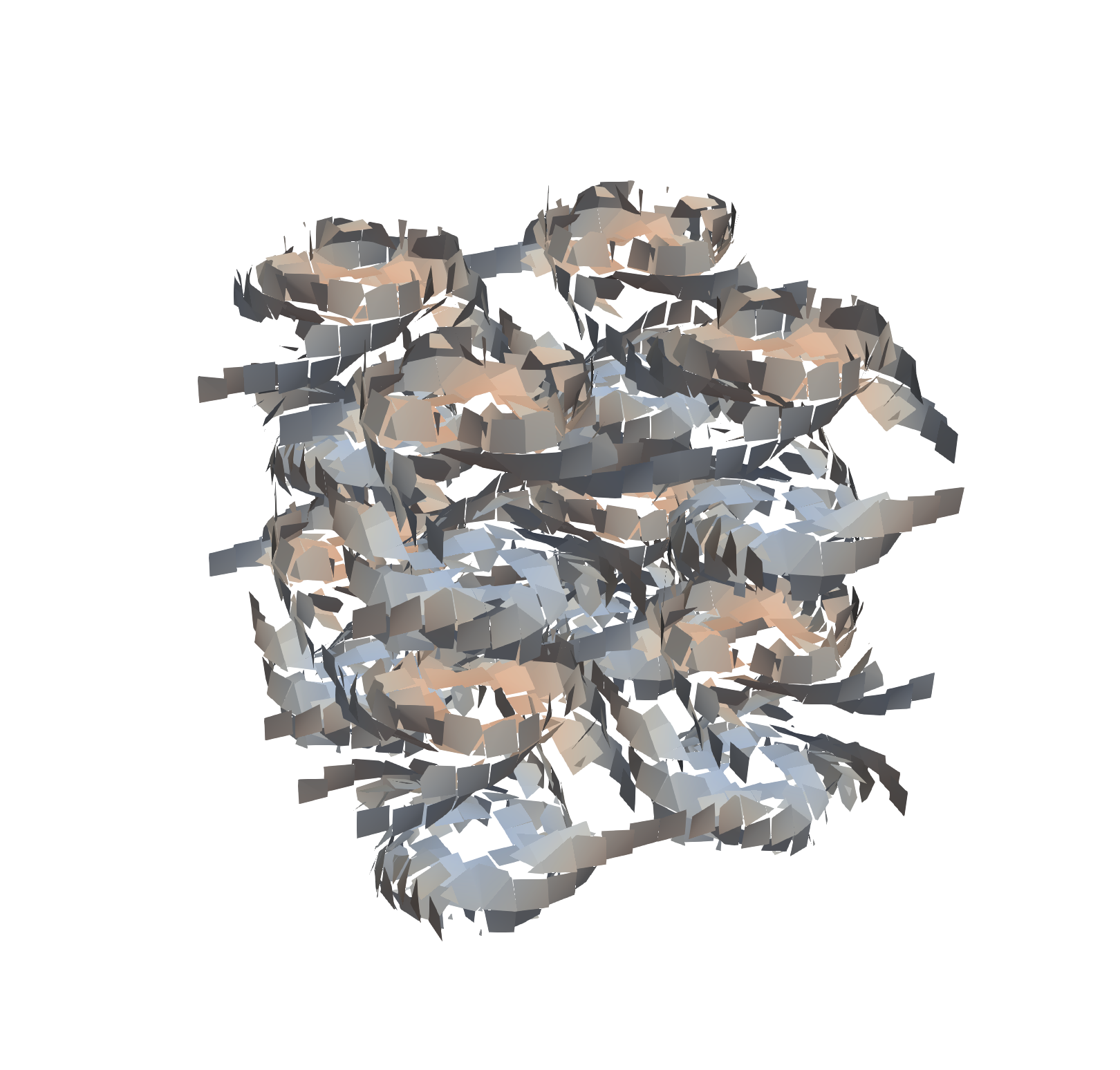}
  }
  \subfigure[Low-order]{
    \figlab{cns3d-tgv-Re100-p1-t8-qv-uL-ss}
      \includegraphics[trim=8cm 3cm 8cm 3cm,clip=true,width=0.31\textwidth]{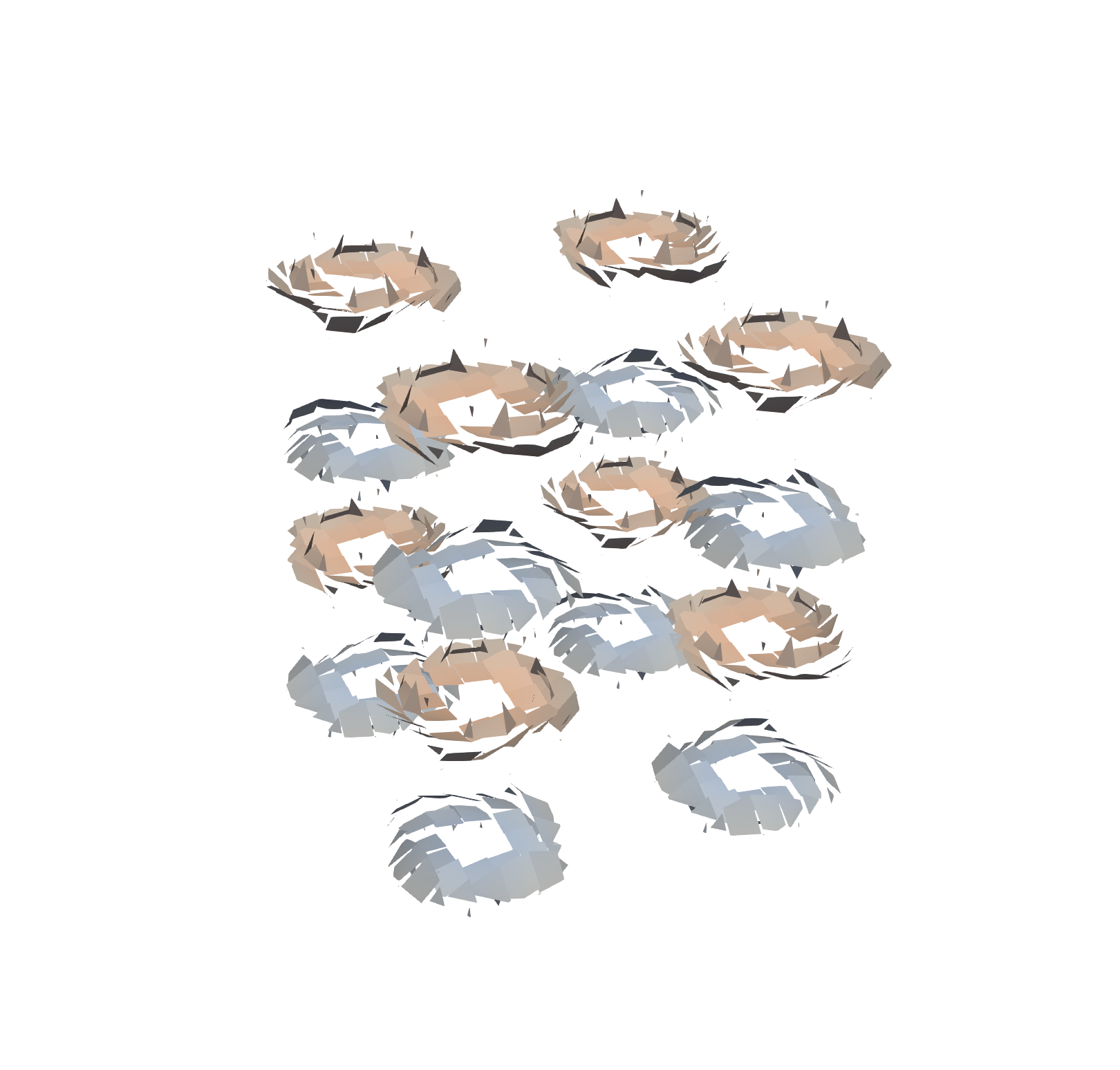}
  }
  \caption{3D Taylor--Green vortex: Q-criterion isosurfaces at $Re=100$ and $t=8$ are shown for (a) the projected solution ($\projL \ub^H$), (b) the augmented solution ($\hat{\ub}^L$), and (c) the low-order solution ($\ub^L$).
  The isosurfaces 
  are colored based on the z-component of the velocity, ranging from $-0.1$ to $0.1$. 
  }
  \figlab{cns3d-tgv-Re100-ss-qv-t8}
\end{figure}

\begin{figure}[h!t!b!]
  \centering
  \subfigure[Filtered]{
    \figlab{cns3d-tgv-Re100-p1-t15-qv-Gu-ss}
      \includegraphics[trim=8cm 3cm 8cm 3cm,clip=true,width=0.31\textwidth]{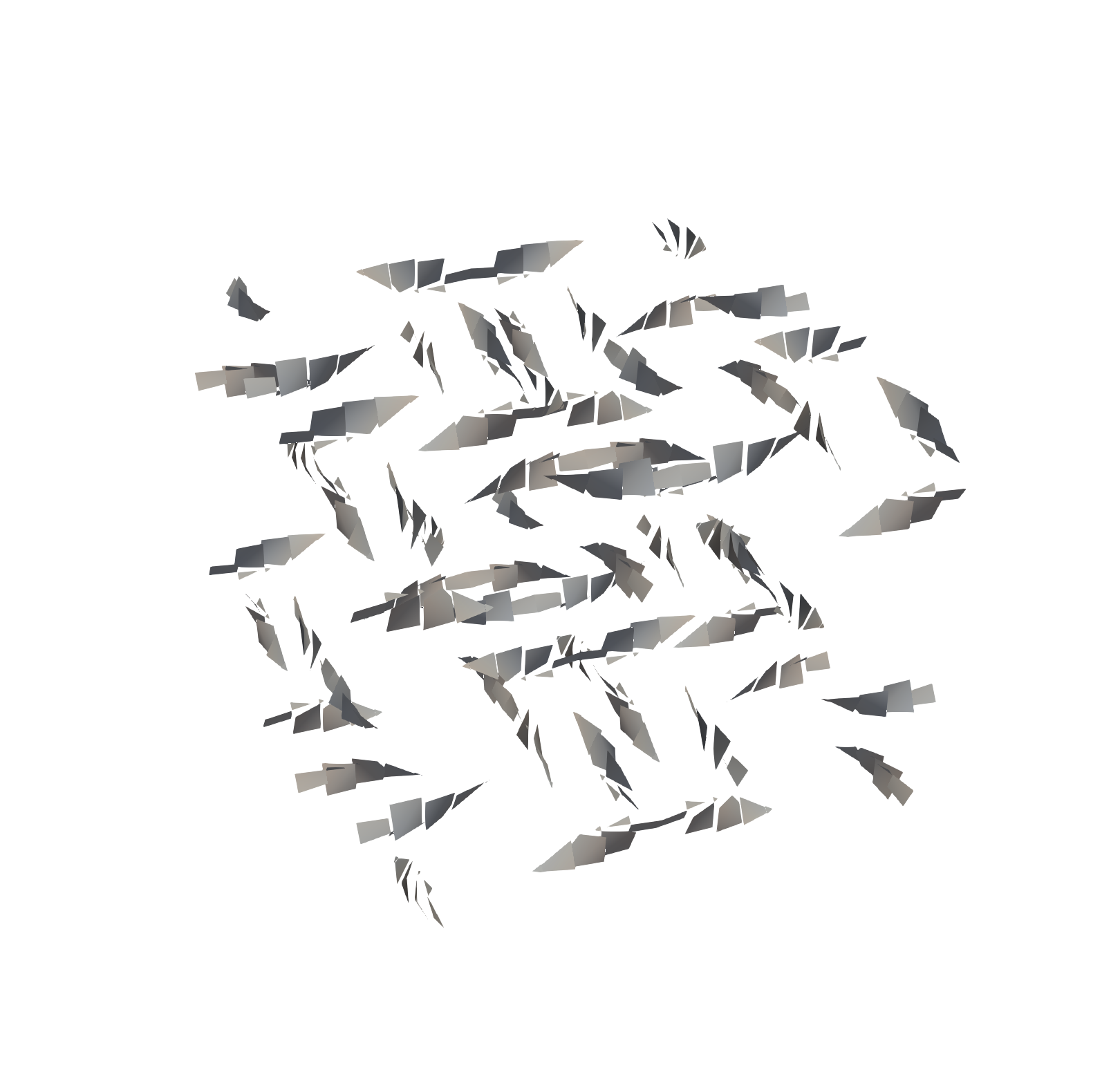}
  }
  \subfigure[Augmented]{
    \figlab{cns3d-tgv-Re100-p1-t15-qv-uh-ss}
      \includegraphics[trim=8cm 3cm 8cm 3cm,clip=true,width=0.31\textwidth]{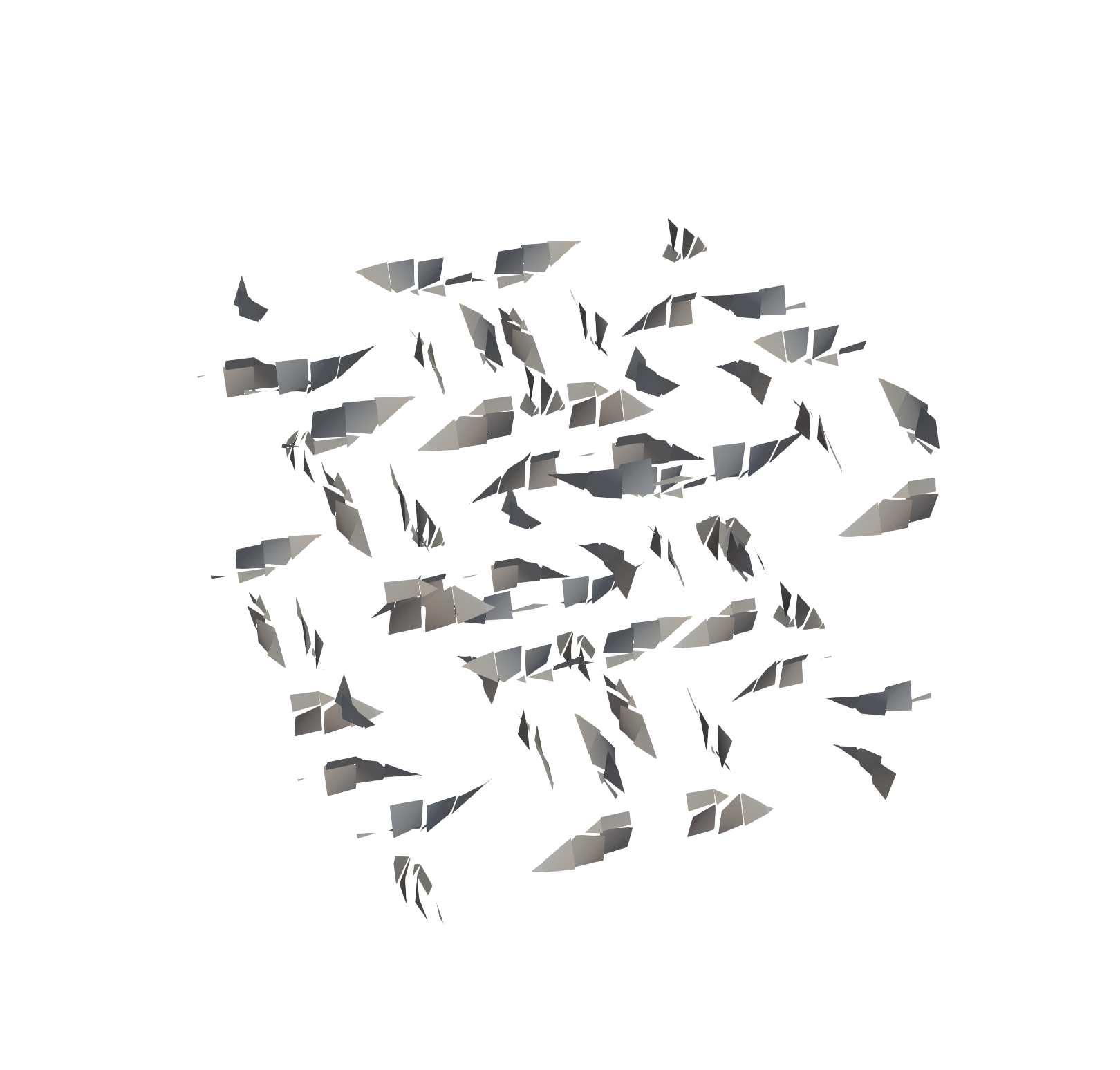}
  }
  \subfigure[Low-order]{
    \figlab{cns3d-tgv-Re100-p1-t15-qv-uL-ss}
      \includegraphics[trim=8cm 3cm 8cm 3cm,clip=true,width=0.31\textwidth]{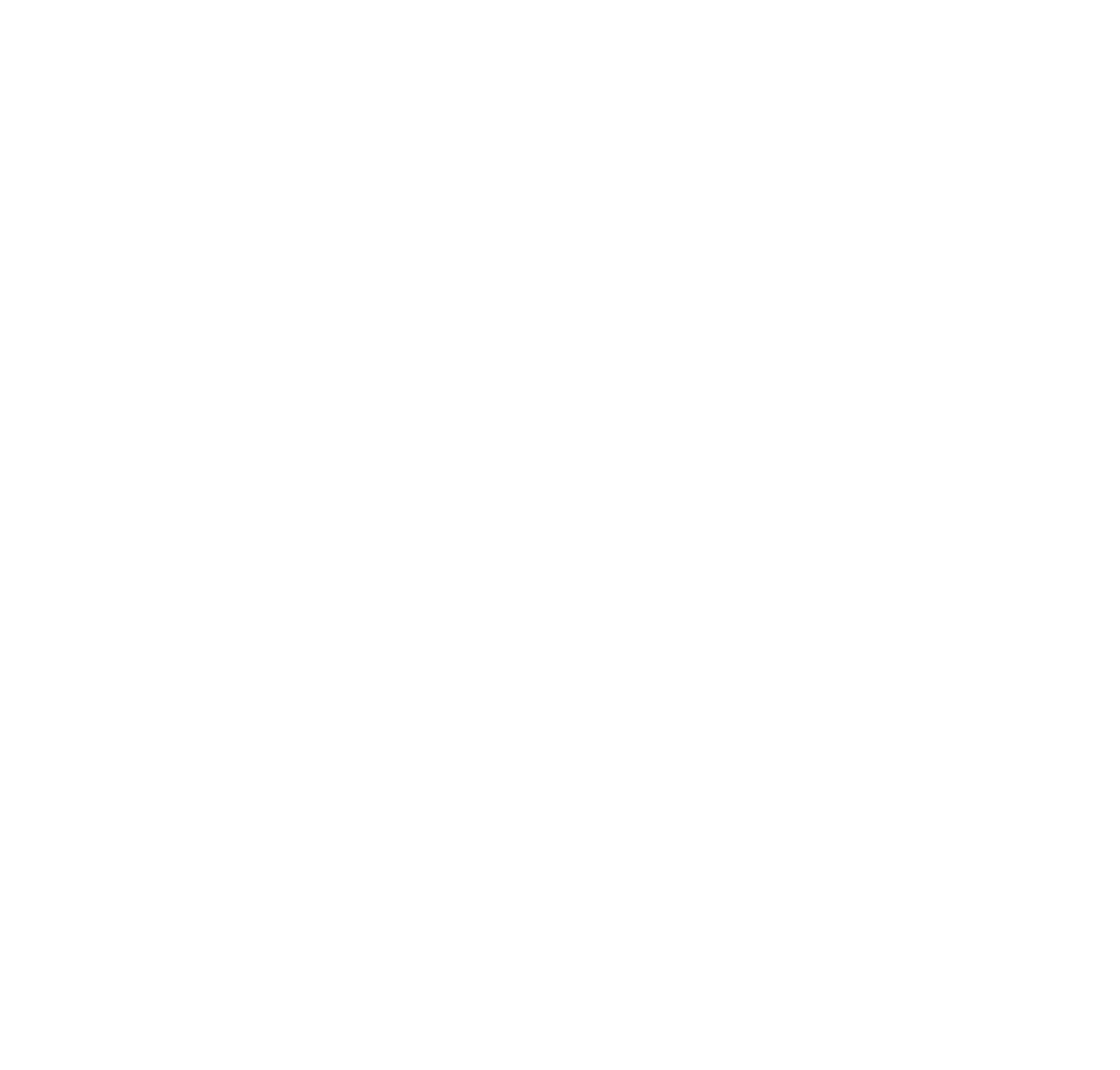}
  }
  \caption{3D Taylor--Green vortex: Q-criterion isosurfaces at $Re=100$ and $t=15$ are shown for (a) the projected solution ($\projL \ub^H$), (b) the augmented solution ($\hat{\ub}^L$), and (c) the low-order solution ($\ub^L$).
  The isosurfaces 
  are colored based on the z-component of the velocity, ranging from $-0.1$ to $0.1$. 
  }
  \figlab{cns3d-tgv-Re100-ss-qv-t15}
\end{figure}

%
%

\begin{figure}[h!t!b!]
  \centering
  \subfigure[Density ($\rho$)]{
    \figlab{cns3d-tgv-Re200-p1-errhistory-r}
      \includegraphics[trim=0.3cm 0.5cm 0.2cm 0.2cm,clip=true,width=0.31\textwidth]{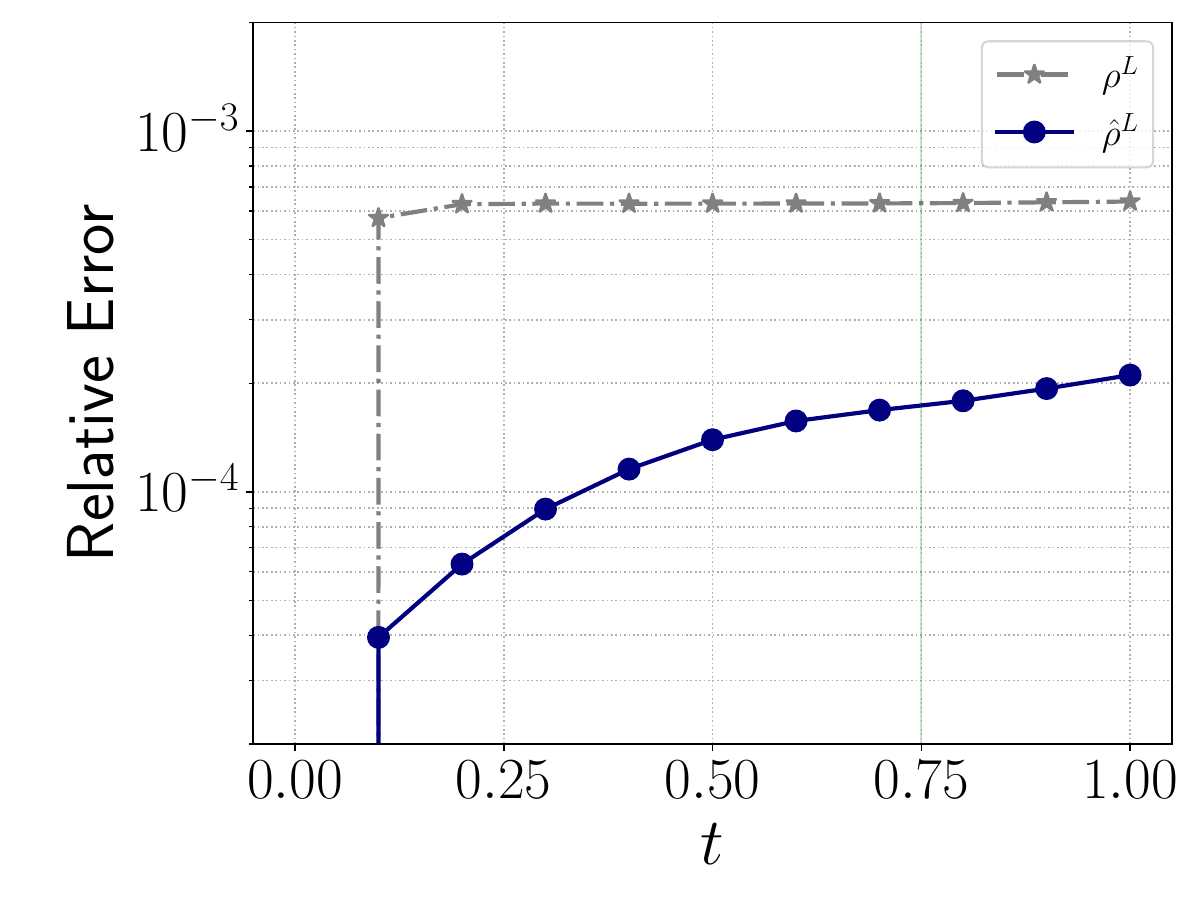}
    \includegraphics[trim=0.3cm 0.5cm 0.2cm 0.2cm,clip=true,width=0.31\textwidth]{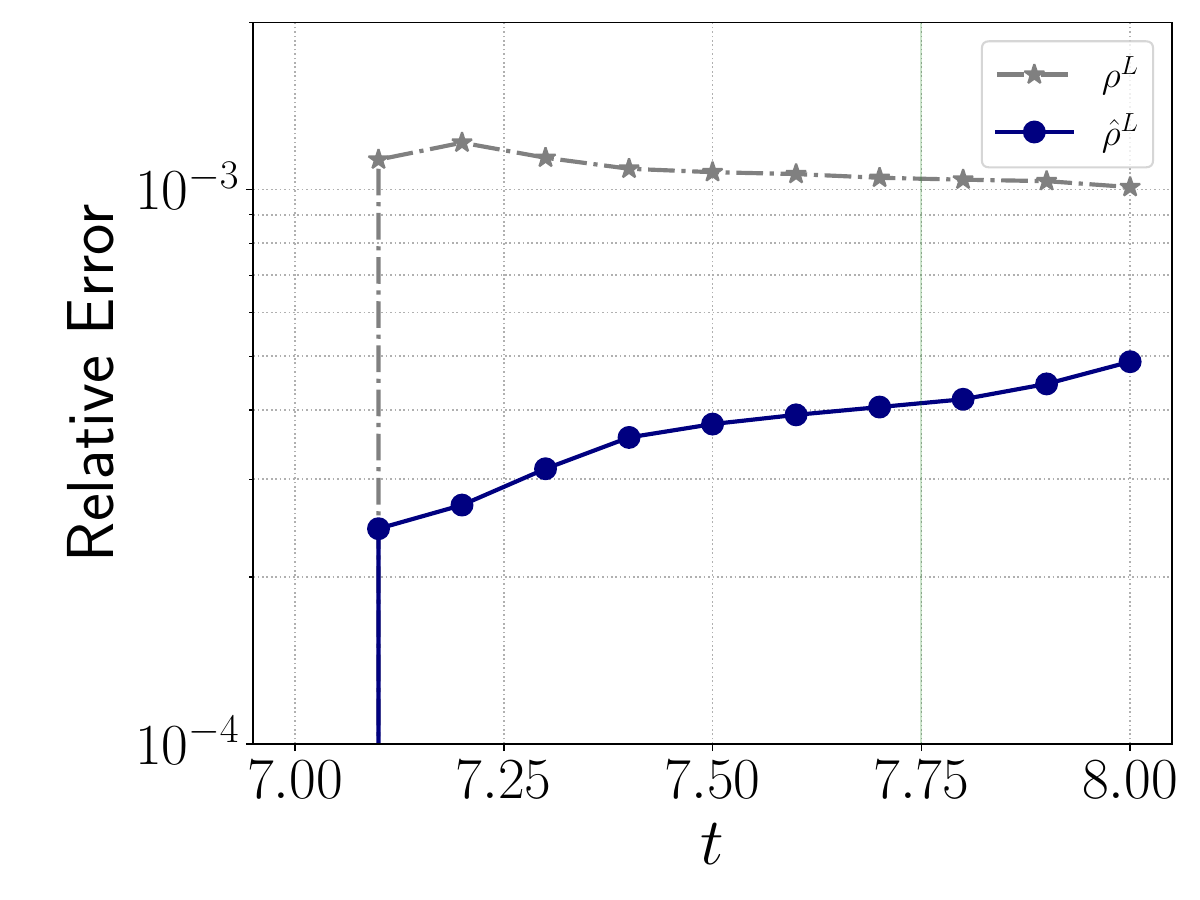}
    \includegraphics[trim=0.3cm 0.5cm 0.2cm 0.2cm,clip=true,width=0.31\textwidth]{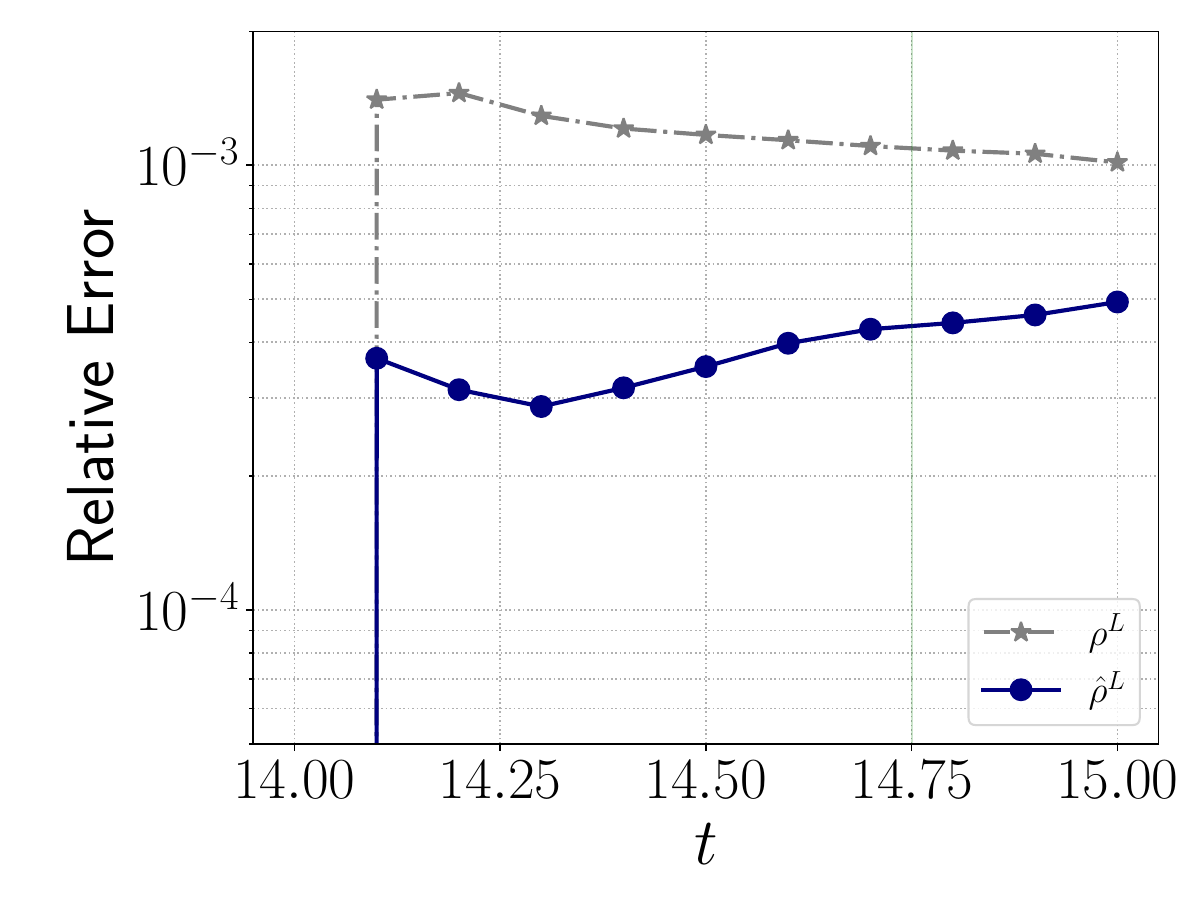}
  }
  \subfigure[Momentum ($\rho \boldsymbol{\varphi}$)]{
    \figlab{cns3d-tgv-Re200-p1-errhistory-rvel}
      \includegraphics[trim=0.3cm 0.5cm 0.2cm  0.2cm,clip=true,width=0.31\textwidth]{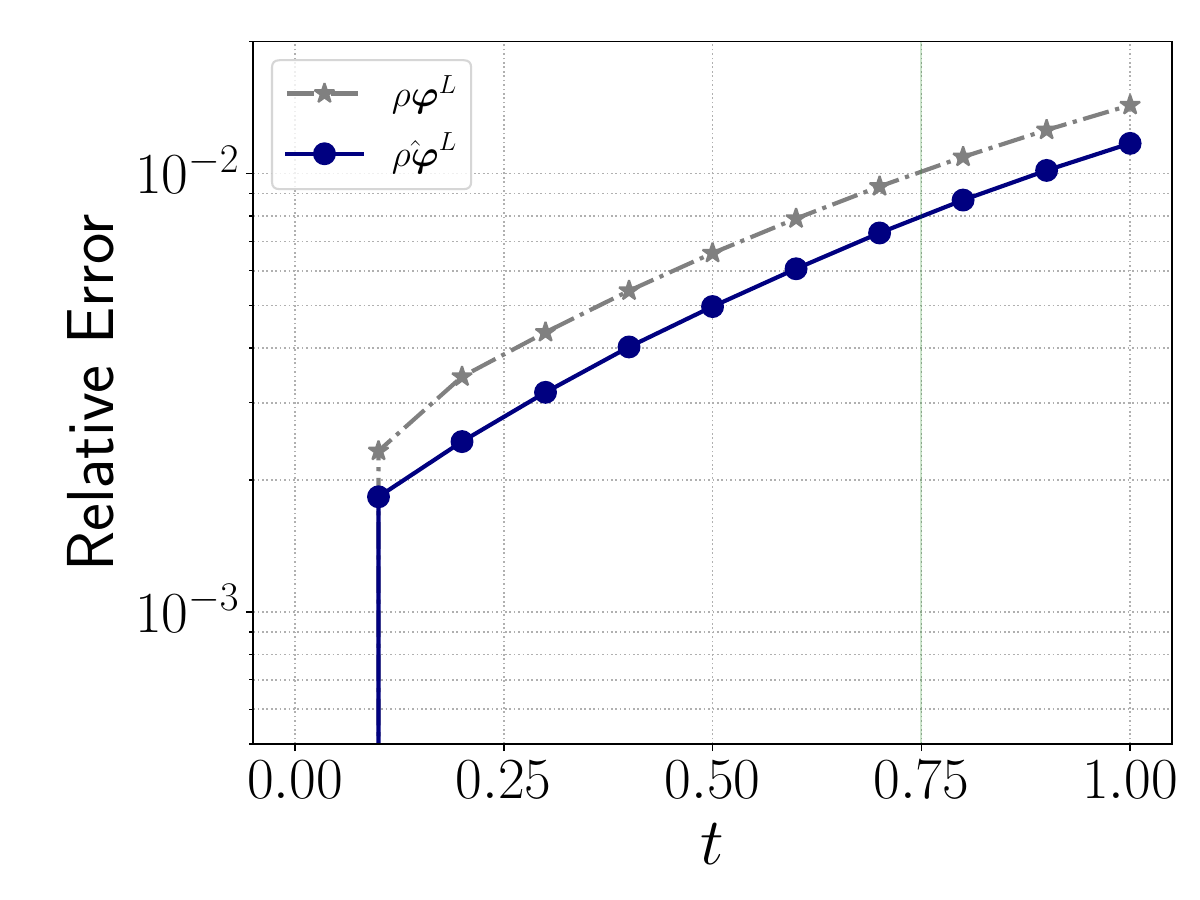}
      \includegraphics[trim=0.3cm 0.5cm 0.2cm  0.2cm,clip=true,width=0.31\textwidth]{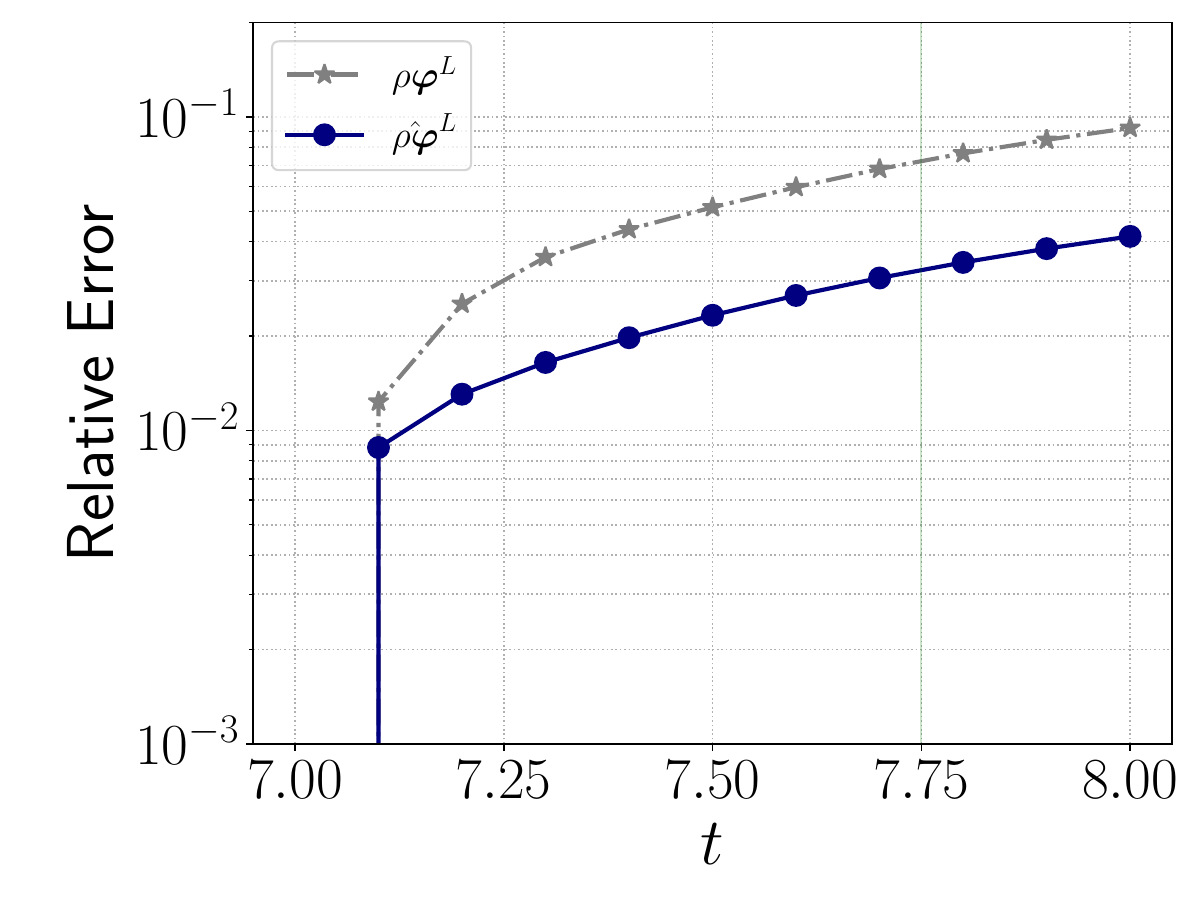}
      \includegraphics[trim=0.3cm 0.5cm 0.2cm  0.2cm,clip=true,width=0.31\textwidth]{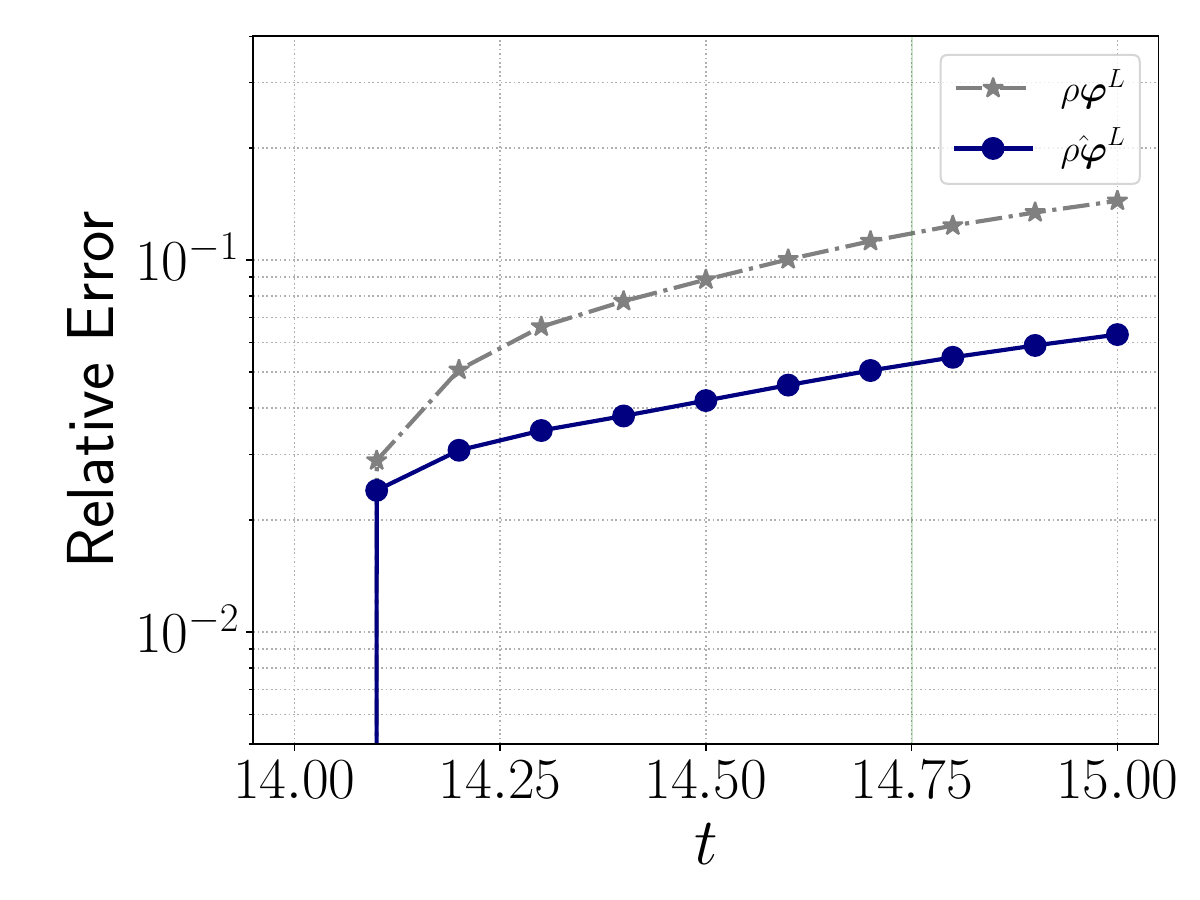}
  }
  \subfigure[Total energy ($\rho E$)]{
    \figlab{cns3d-tgv-Re200-p1-errhistory-rE}
      \includegraphics[trim=0.3cm 0.5cm 0.2cm 0.2cm,clip=true,width=0.31\textwidth]{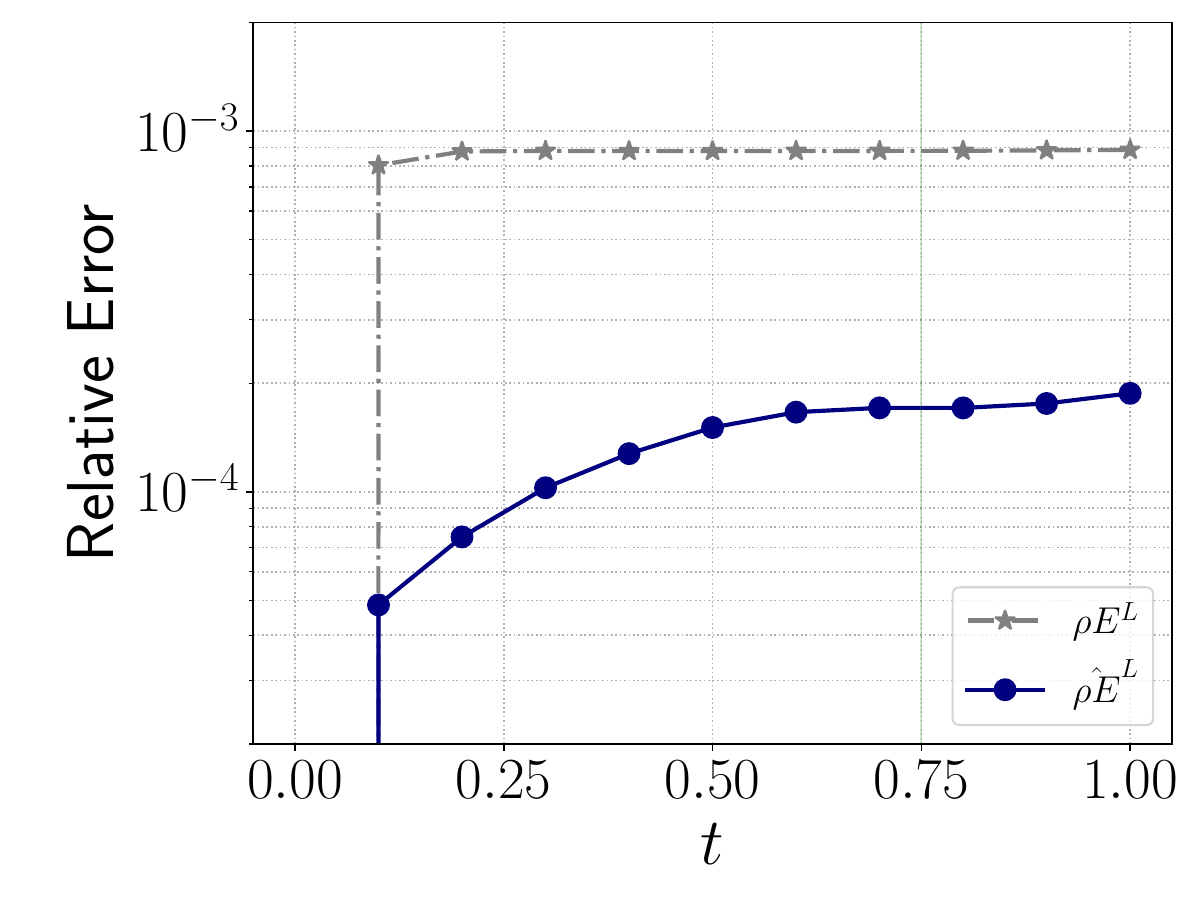}
      \includegraphics[trim=0.3cm 0.5cm 0.2cm 0.2cm,clip=true,width=0.31\textwidth]{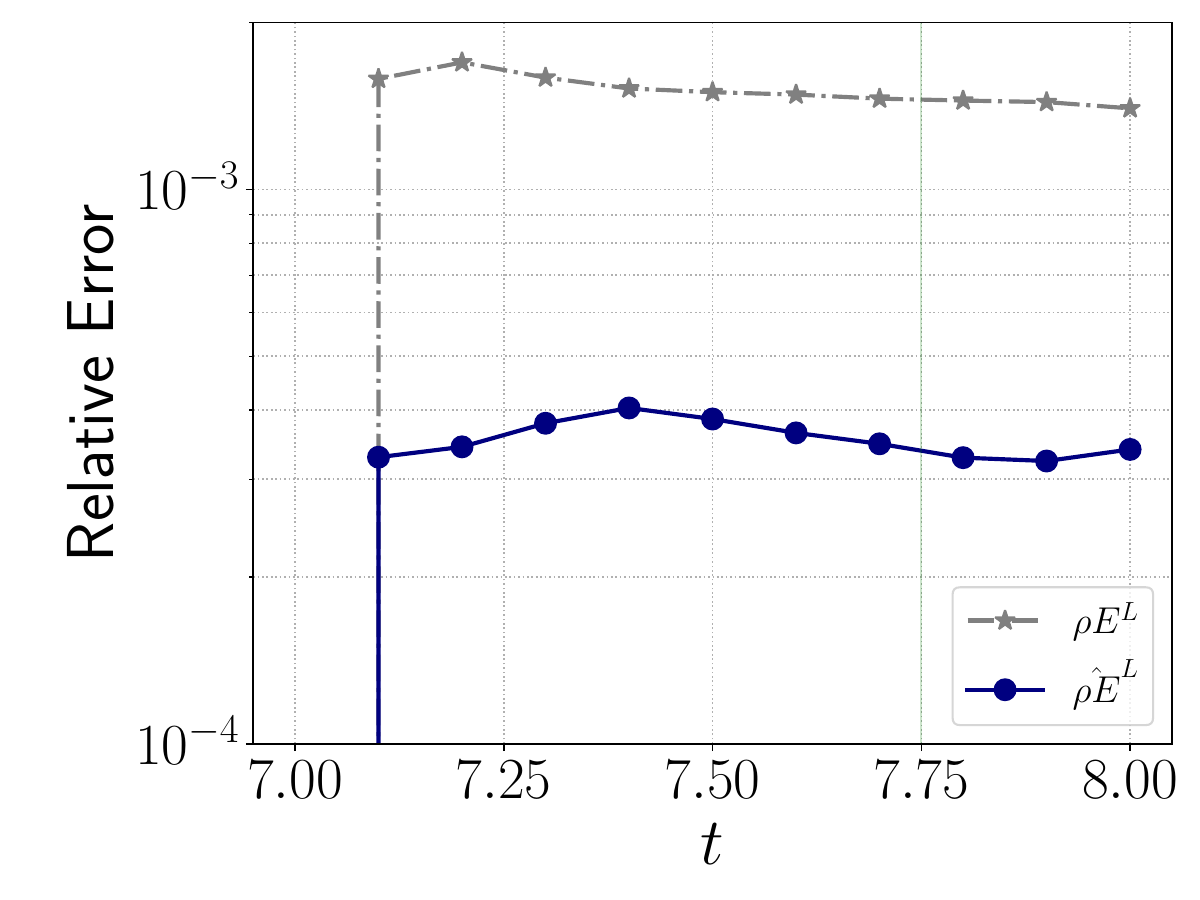}
      \includegraphics[trim=0.3cm 0.5cm 0.2cm 0.2cm,clip=true,width=0.31\textwidth]{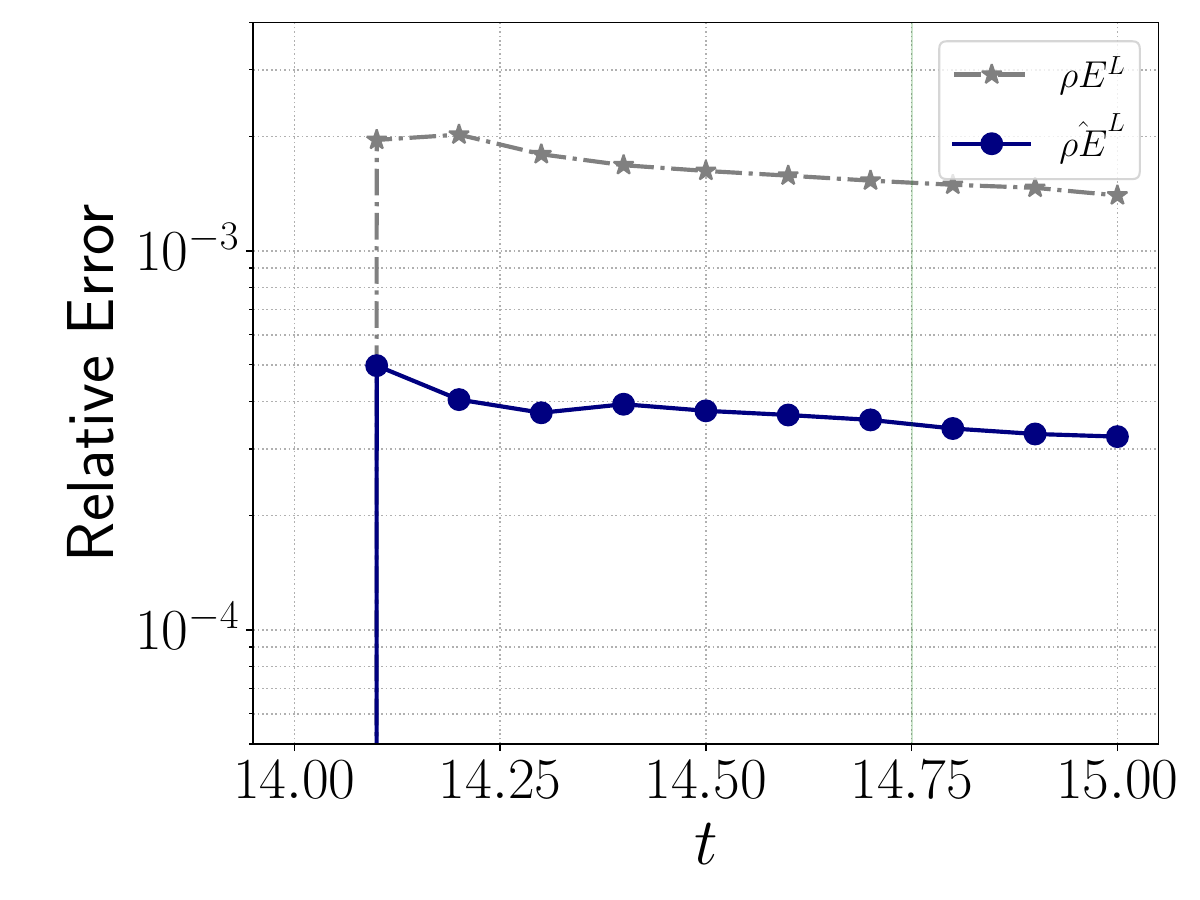}  
  }
  \caption{3D Taylor--Green vortex: relative error histories of (a) $\rho$, (b) $\rho\boldsymbol{\varphi}$, and (c) $\rho E$ at $Re=200$ 
    for low-order solution ($\ub^L$) and augmented solution ($\hat{\ub}^L$). 
  }
  \figlab{cns3d-tgv-Re200-errhistory}
\end{figure}

\begin{figure}[h!t!b!]
  \centering
  \subfigure[Density ($\rho$)]{
    \figlab{cns3d-tgv-Re400-p1-errhistory-r}
      \includegraphics[trim=0.3cm 0.5cm 0.2cm 0.2cm,clip=true,width=0.31\textwidth]{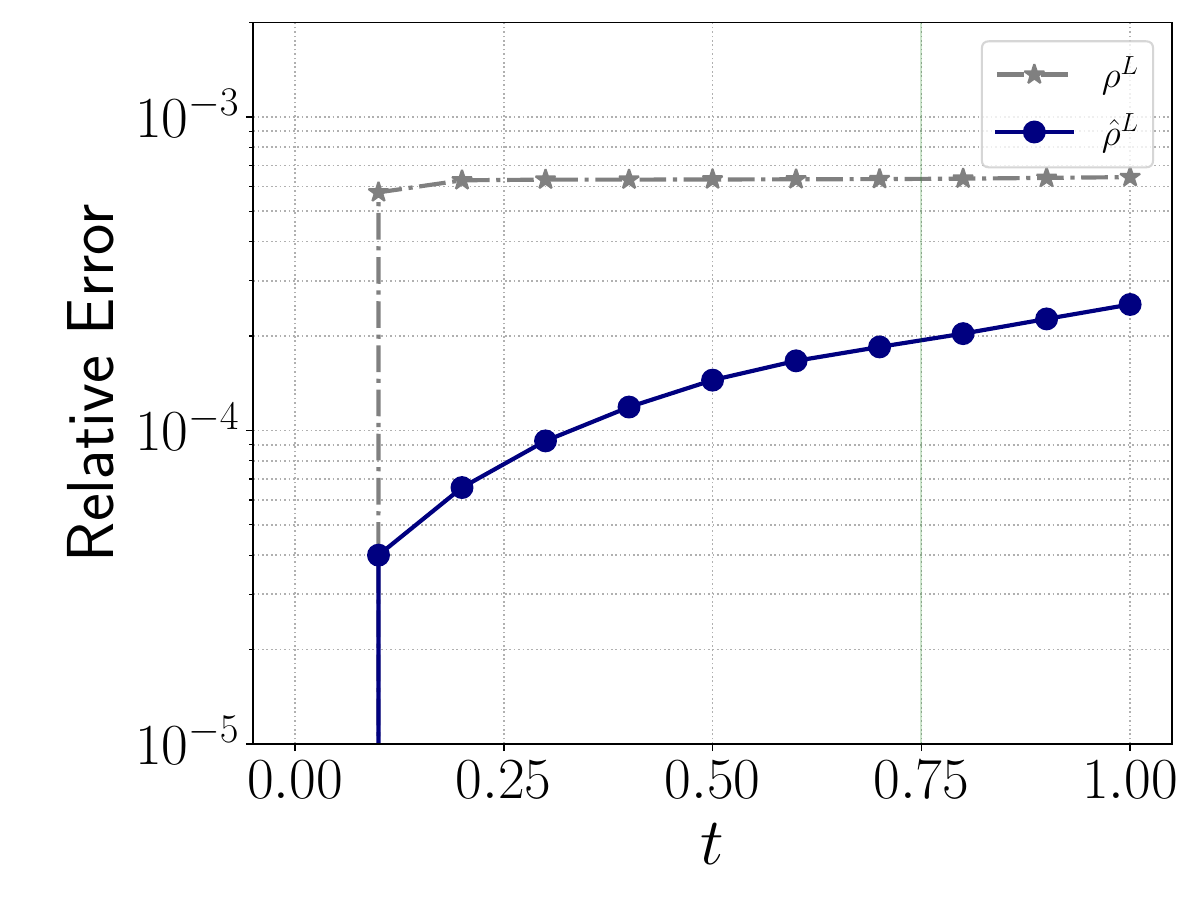}
    \includegraphics[trim=0.3cm 0.5cm 0.2cm 0.2cm,clip=true,width=0.31\textwidth]{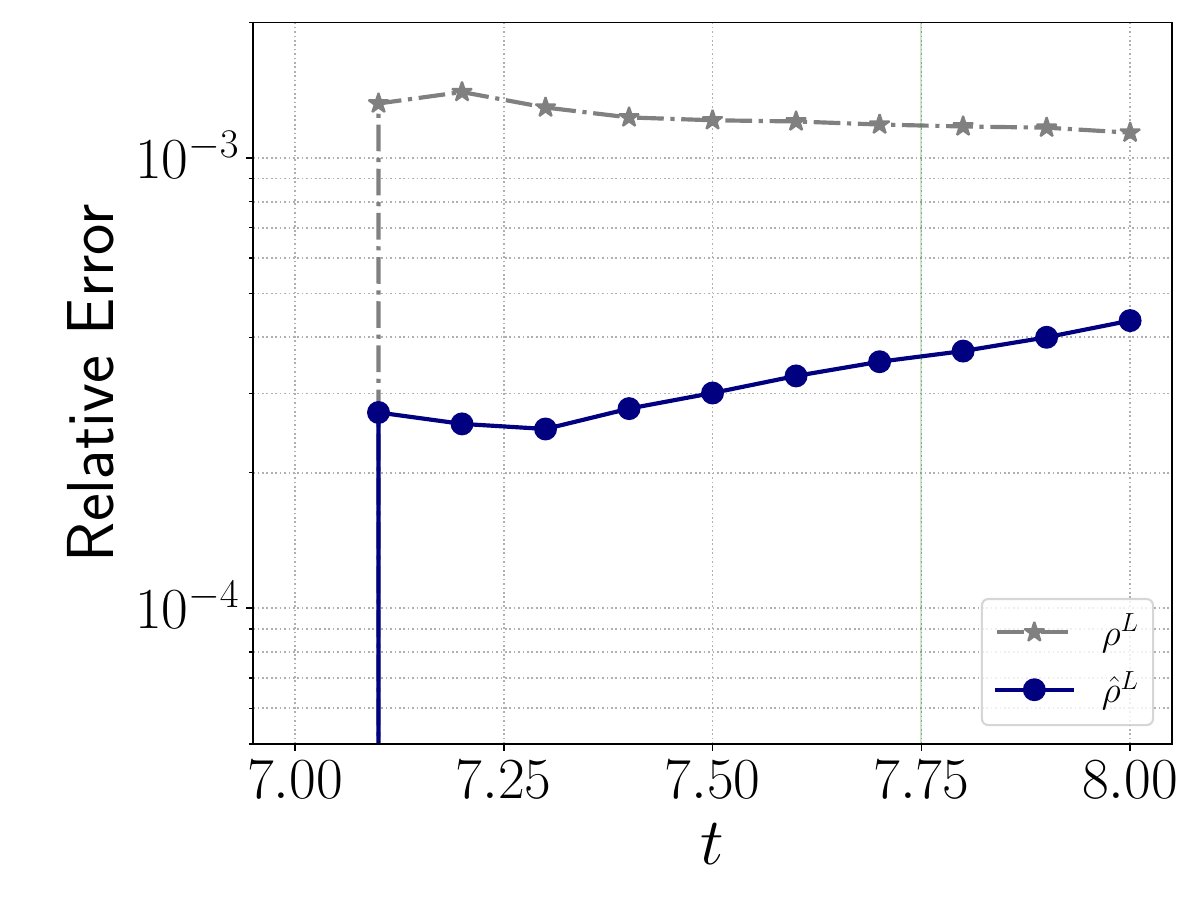}
    \includegraphics[trim=0.3cm 0.5cm 0.2cm 0.2cm,clip=true,width=0.31\textwidth]{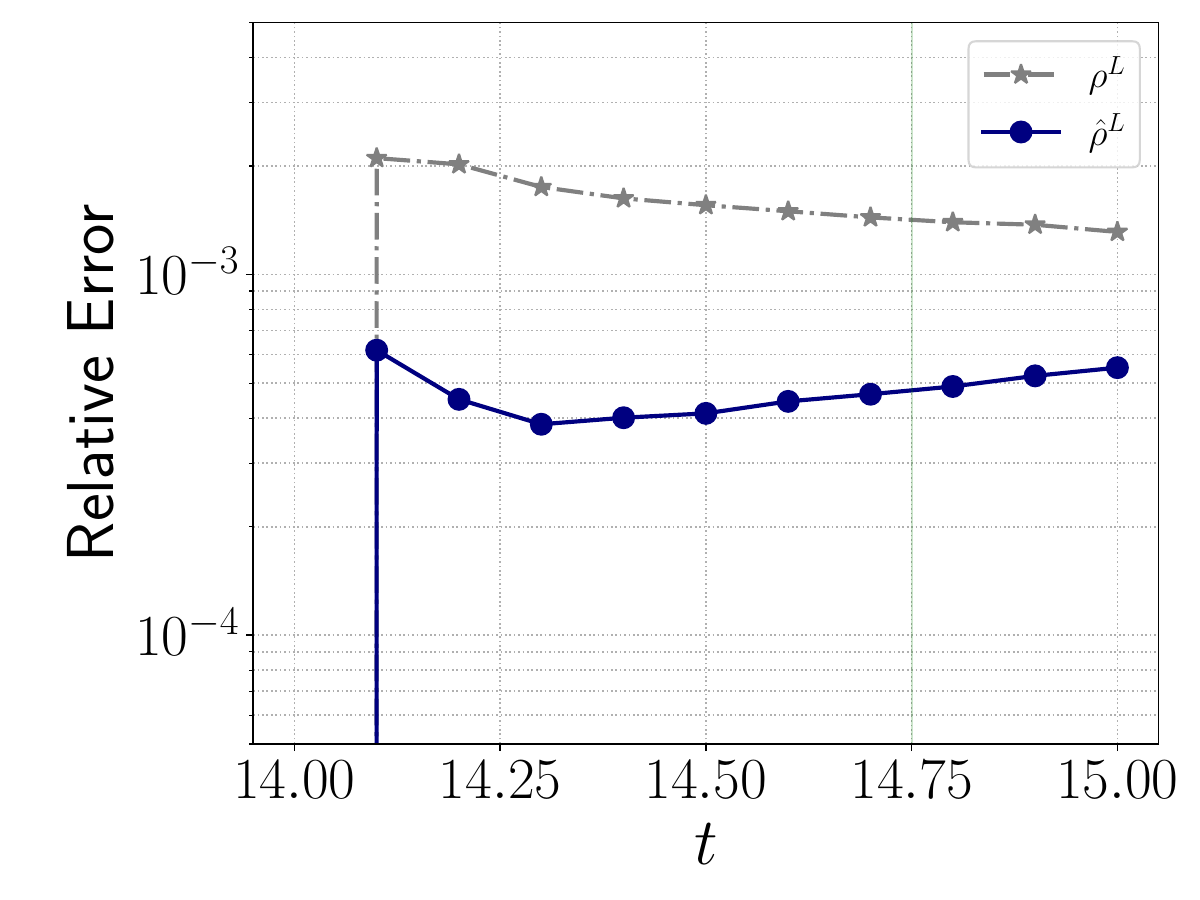}
  }
  \subfigure[Momentum ($\rho \boldsymbol{\varphi}$)]{
    \figlab{cns3d-tgv-Re400-p1-errhistory-rvel}
      \includegraphics[trim=0.3cm 0.5cm 0.2cm  0.2cm,clip=true,width=0.31\textwidth]{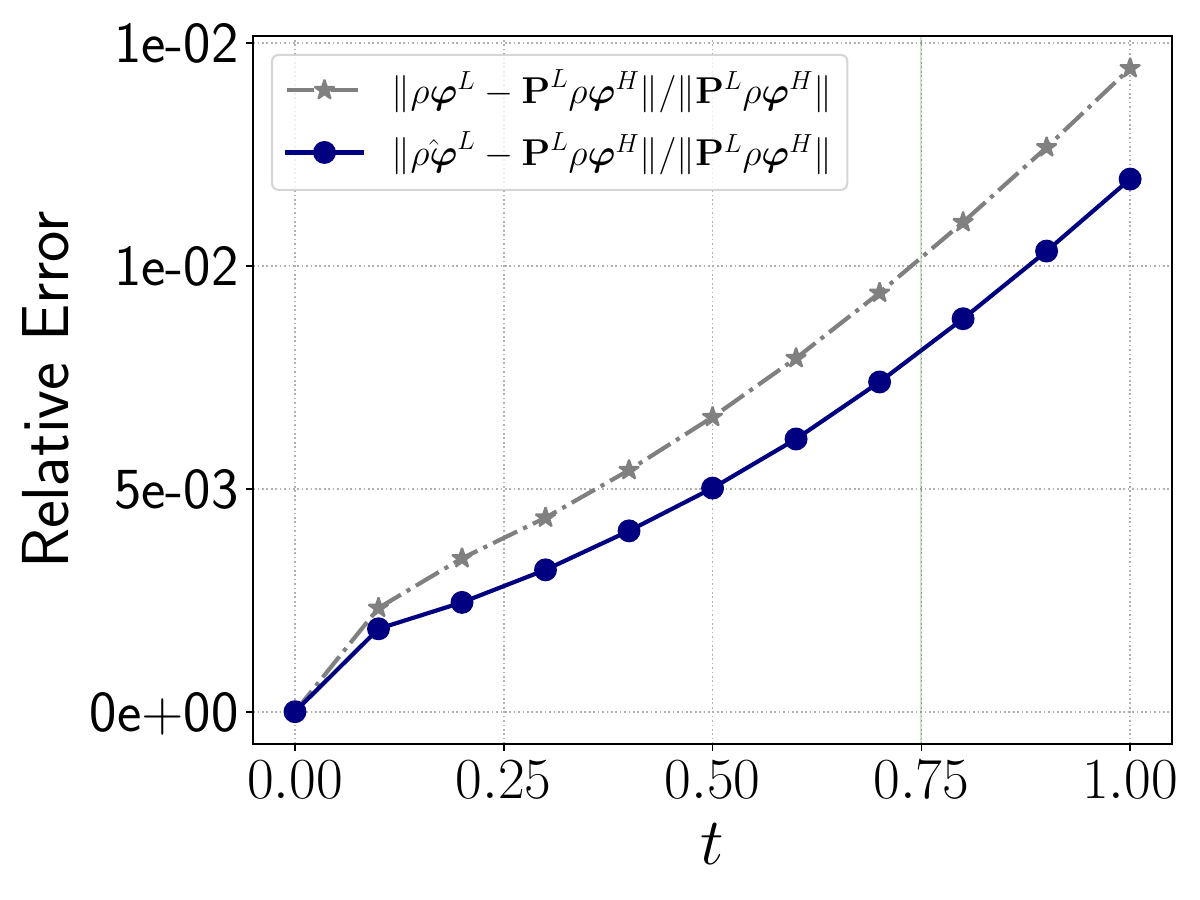}
      \includegraphics[trim=0.3cm 0.5cm 0.2cm  0.2cm,clip=true,width=0.31\textwidth]{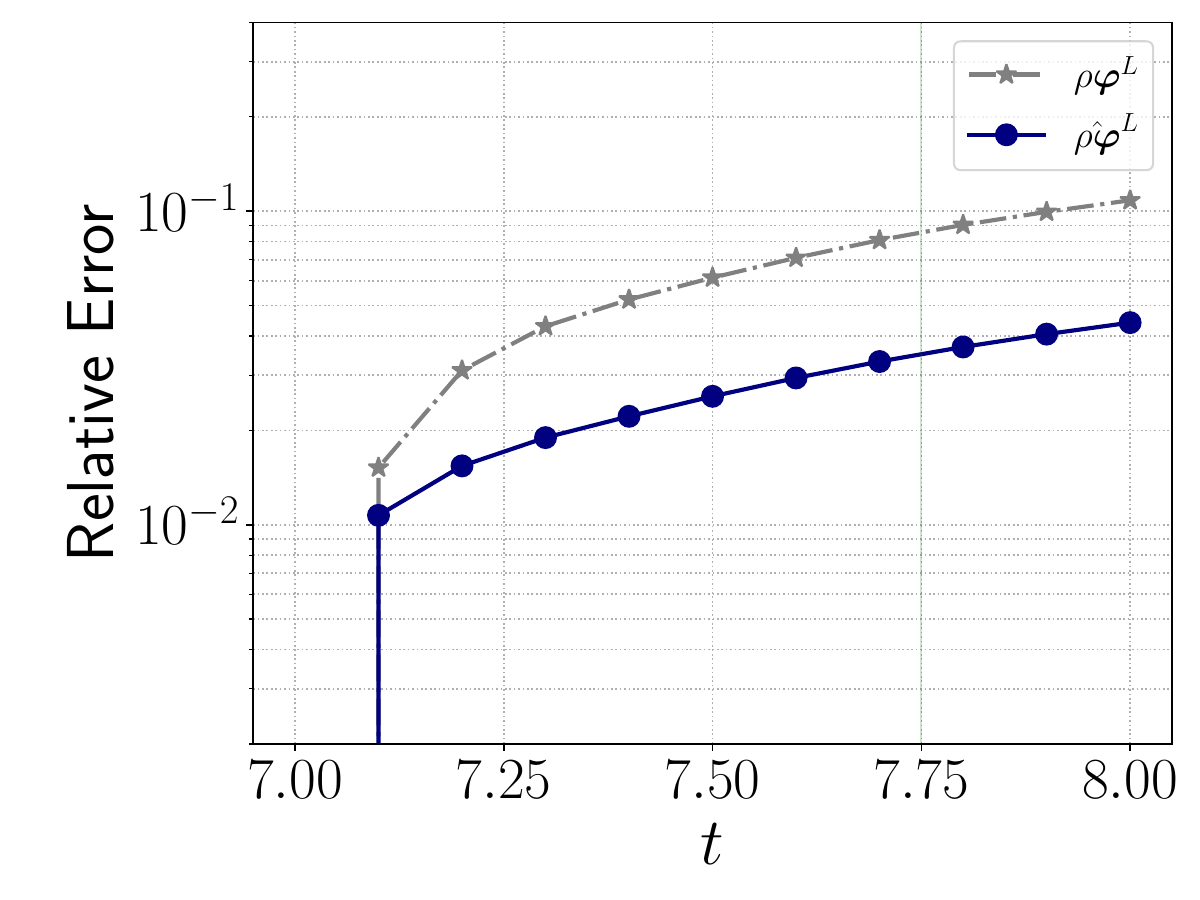}
      \includegraphics[trim=0.3cm 0.5cm 0.2cm  0.2cm,clip=true,width=0.31\textwidth]{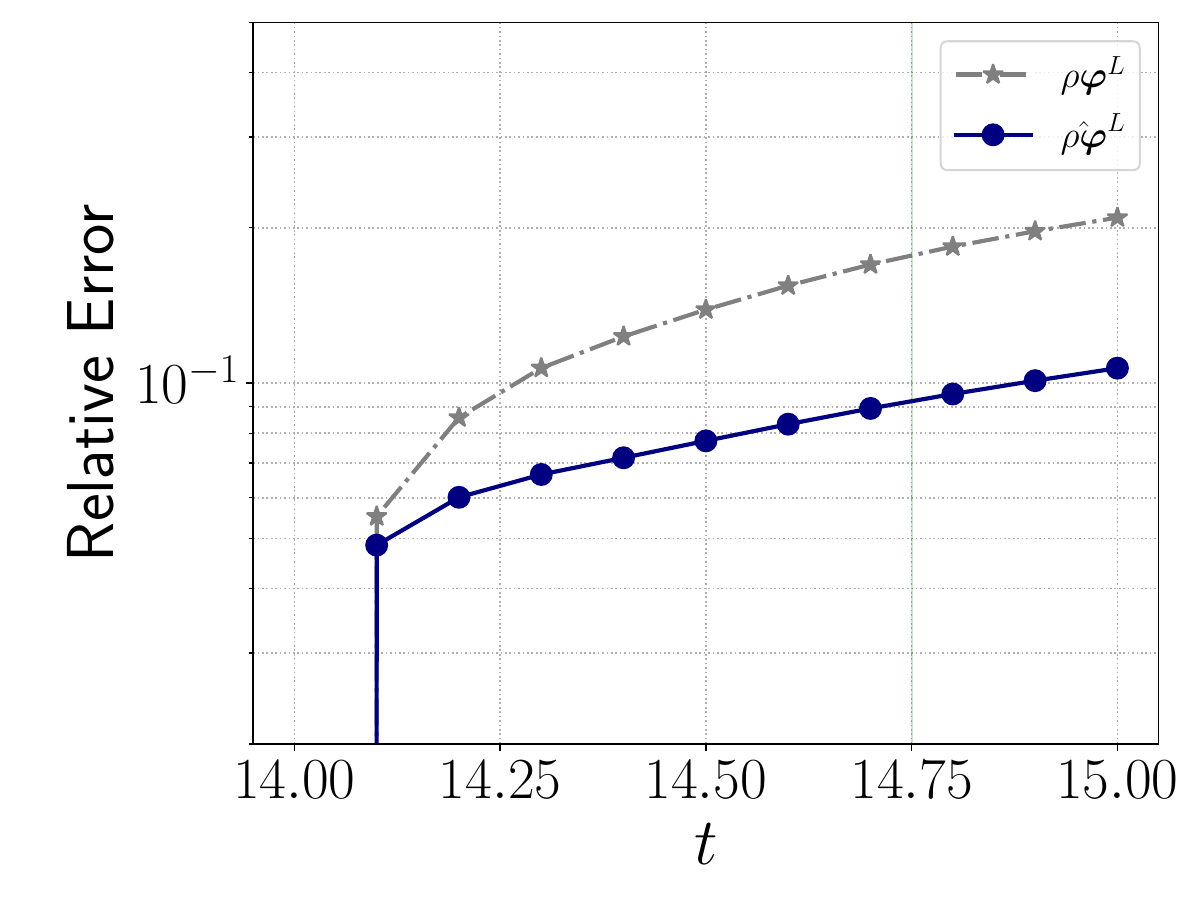}
  }
  \subfigure[Total energy ($\rho E$)]{
    \figlab{cns3d-tgv-Re400-p1-errhistory-rE}
      \includegraphics[trim=0.3cm 0.5cm 0.2cm 0.2cm,clip=true,width=0.31\textwidth]{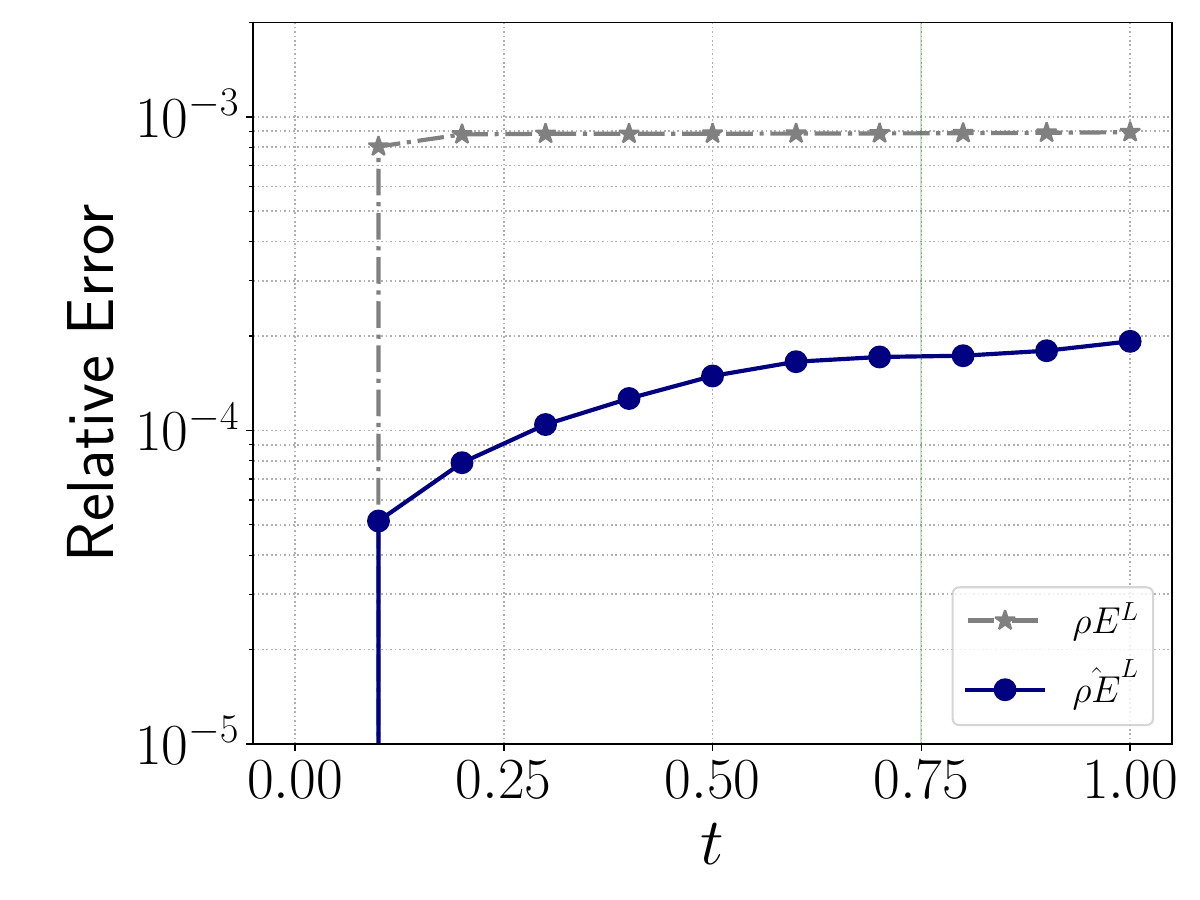}
      \includegraphics[trim=0.3cm 0.5cm 0.2cm 0.2cm,clip=true,width=0.31\textwidth]{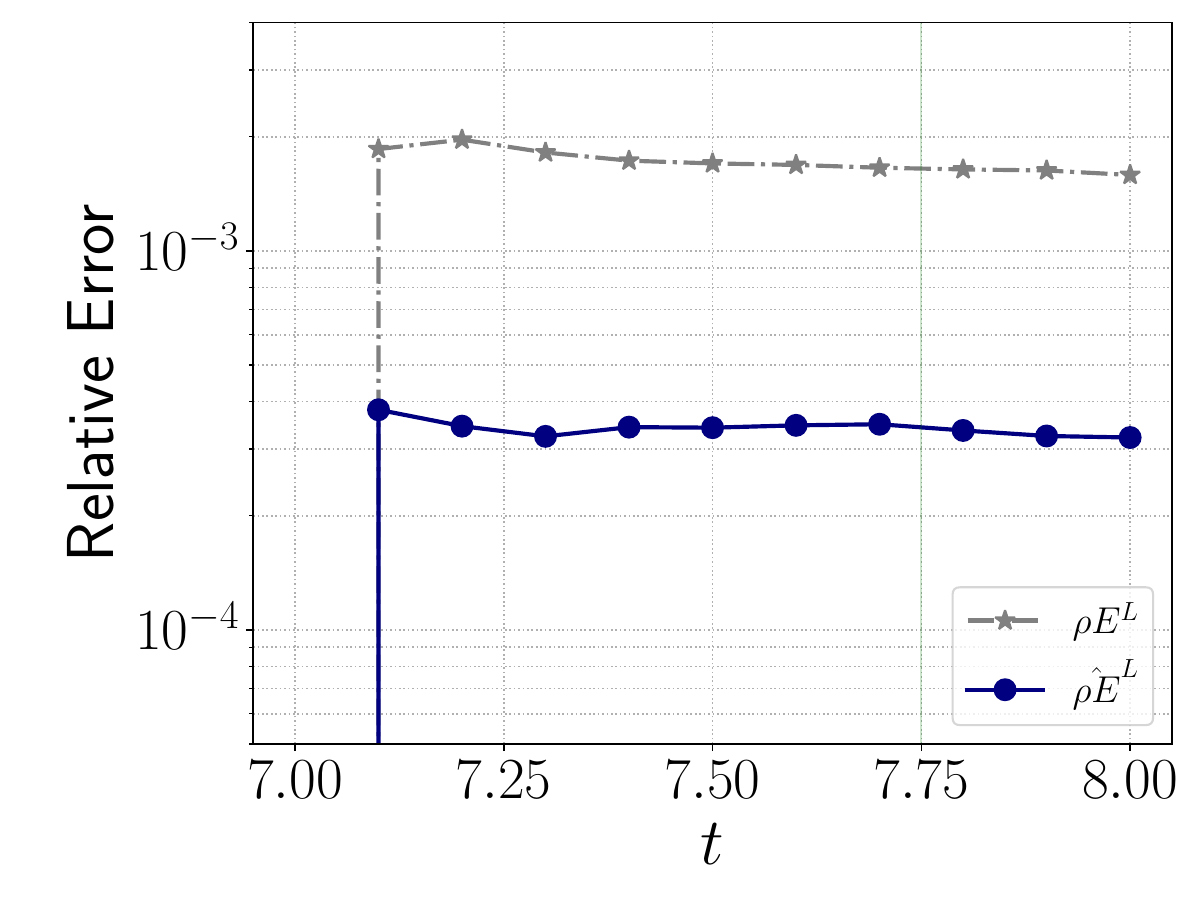}
      \includegraphics[trim=0.3cm 0.5cm 0.2cm 0.2cm,clip=true,width=0.31\textwidth]{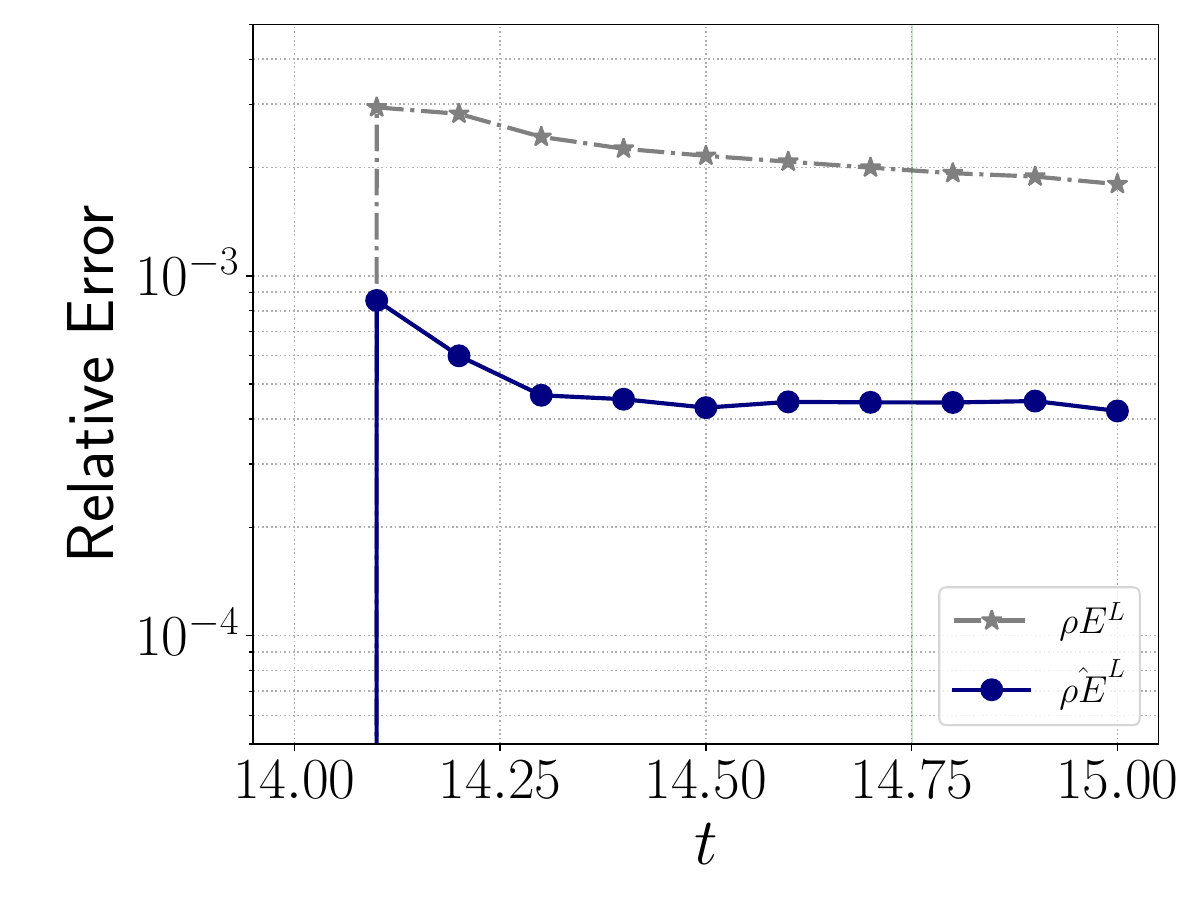}  
  }
  \caption{3D Taylor--Green vortex: relative error histories of (a) $\rho$, (b) $\rho\boldsymbol{\varphi}$, and (c) $\rho E$ at $Re=400$ 
    for low-order solution ($\ub^L$) and augmented solution ($\hat{\ub}^L$). 
  }
  \figlab{cns3d-tgv-Re400-errhistory}
\end{figure}

\begin{figure}[h!t!b!]
  \centering
  \subfigure[Density ($\rho$)]{
    \figlab{cns3d-tgv-Re1600-p1-errhistory-r}
      \includegraphics[trim=0.3cm 0.5cm 0.2cm 0.2cm,clip=true,width=0.31\textwidth]{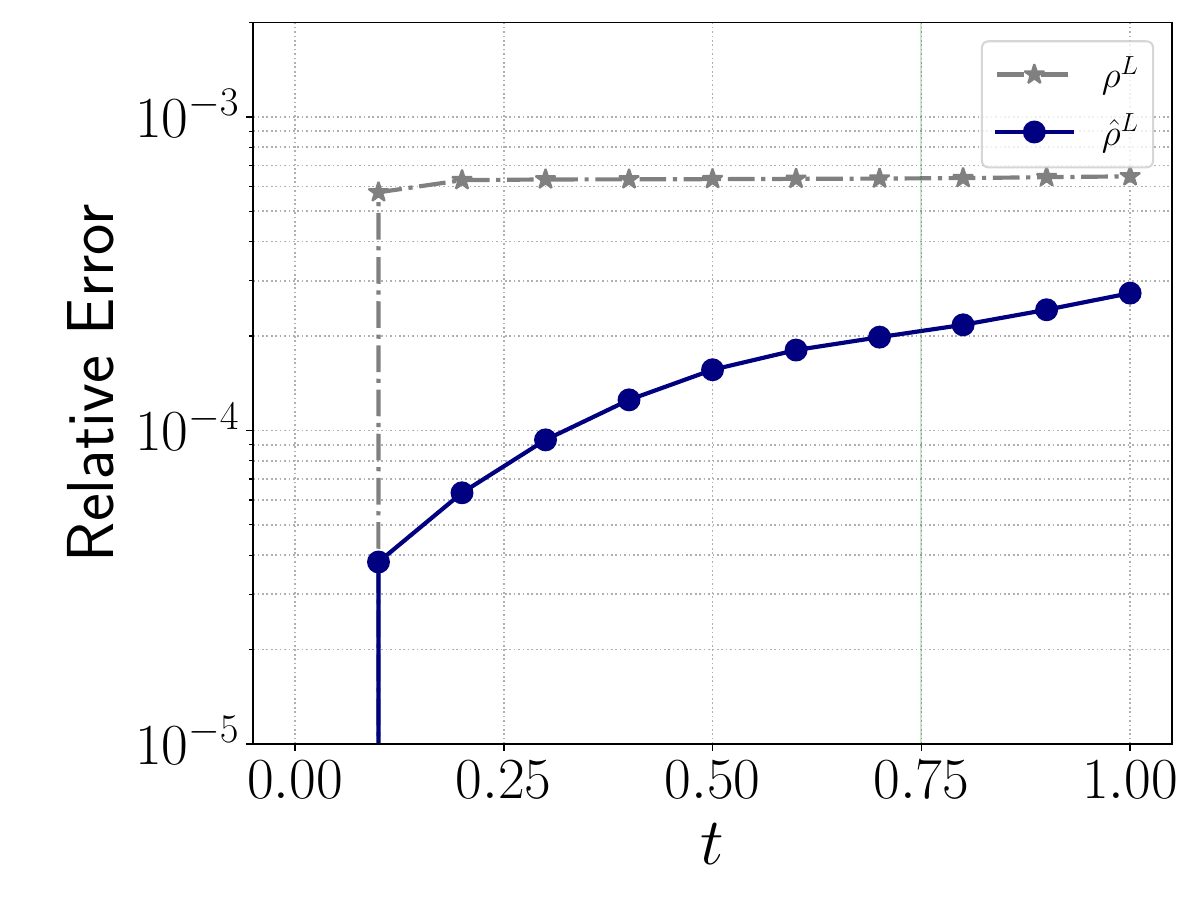}
    \includegraphics[trim=0.3cm 0.5cm 0.2cm 0.2cm,clip=true,width=0.31\textwidth]{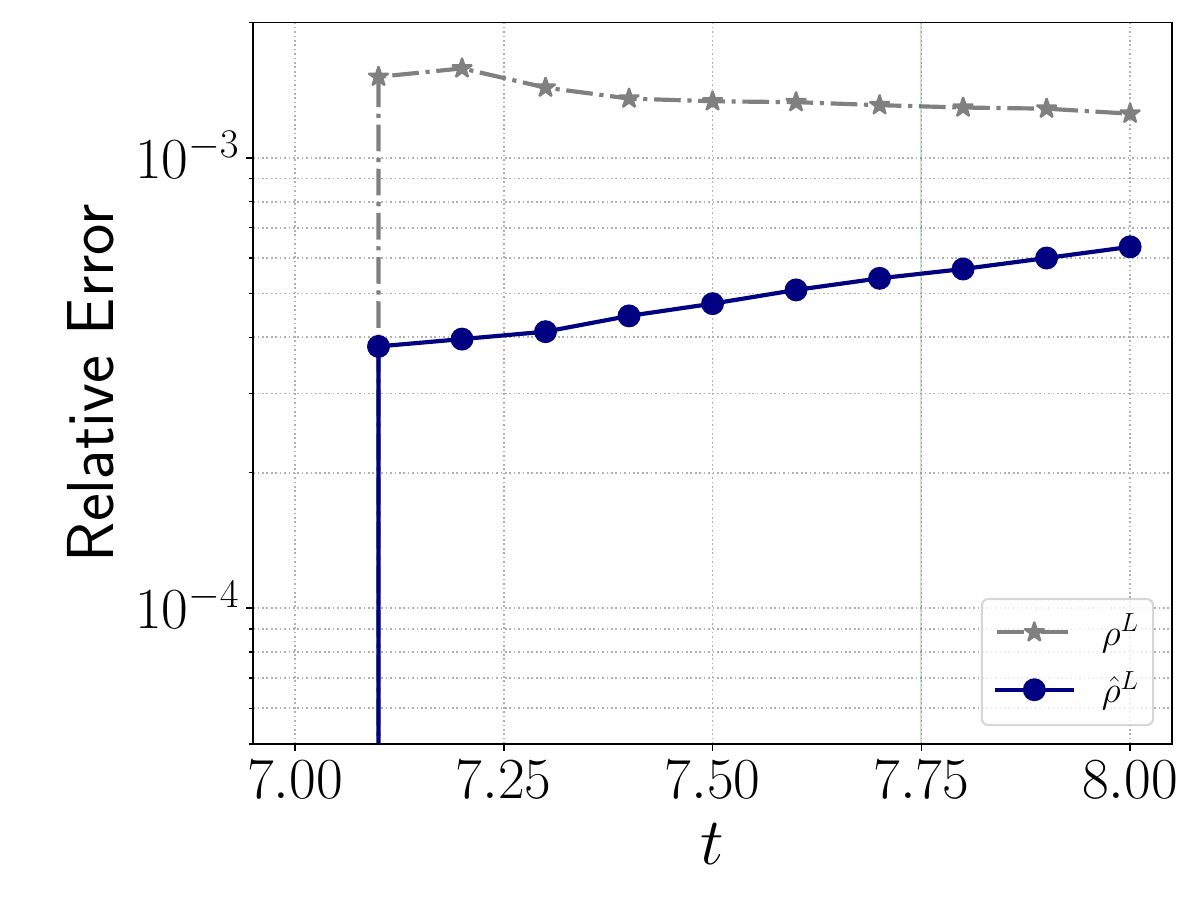}
    \includegraphics[trim=0.3cm 0.5cm 0.2cm 0.2cm,clip=true,width=0.31\textwidth]{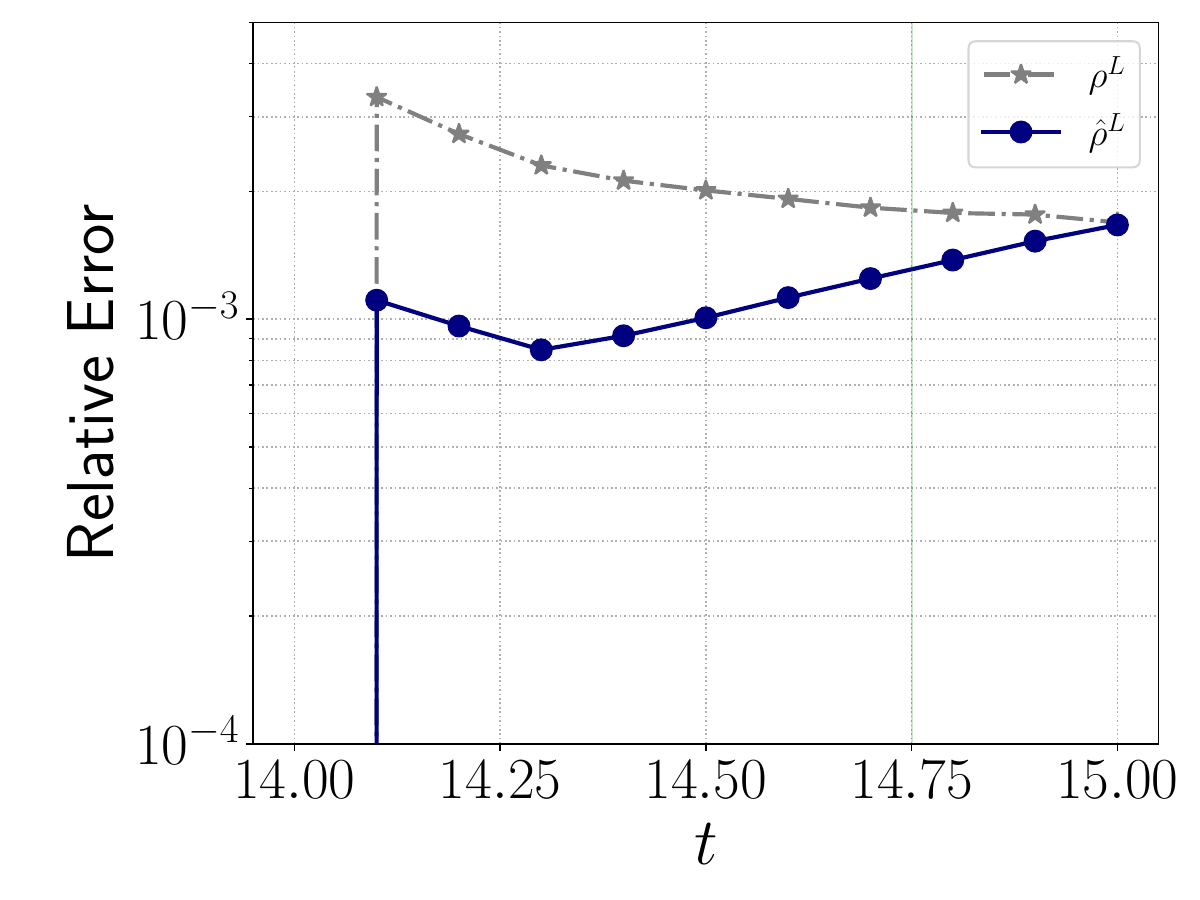}
  }
  \subfigure[Momentum ($\rho \boldsymbol{\varphi}$)]{
    \figlab{cns3d-tgv-Re1600-p1-errhistory-rvel}
      \includegraphics[trim=0.3cm 0.5cm 0.2cm  0.2cm,clip=true,width=0.31\textwidth]{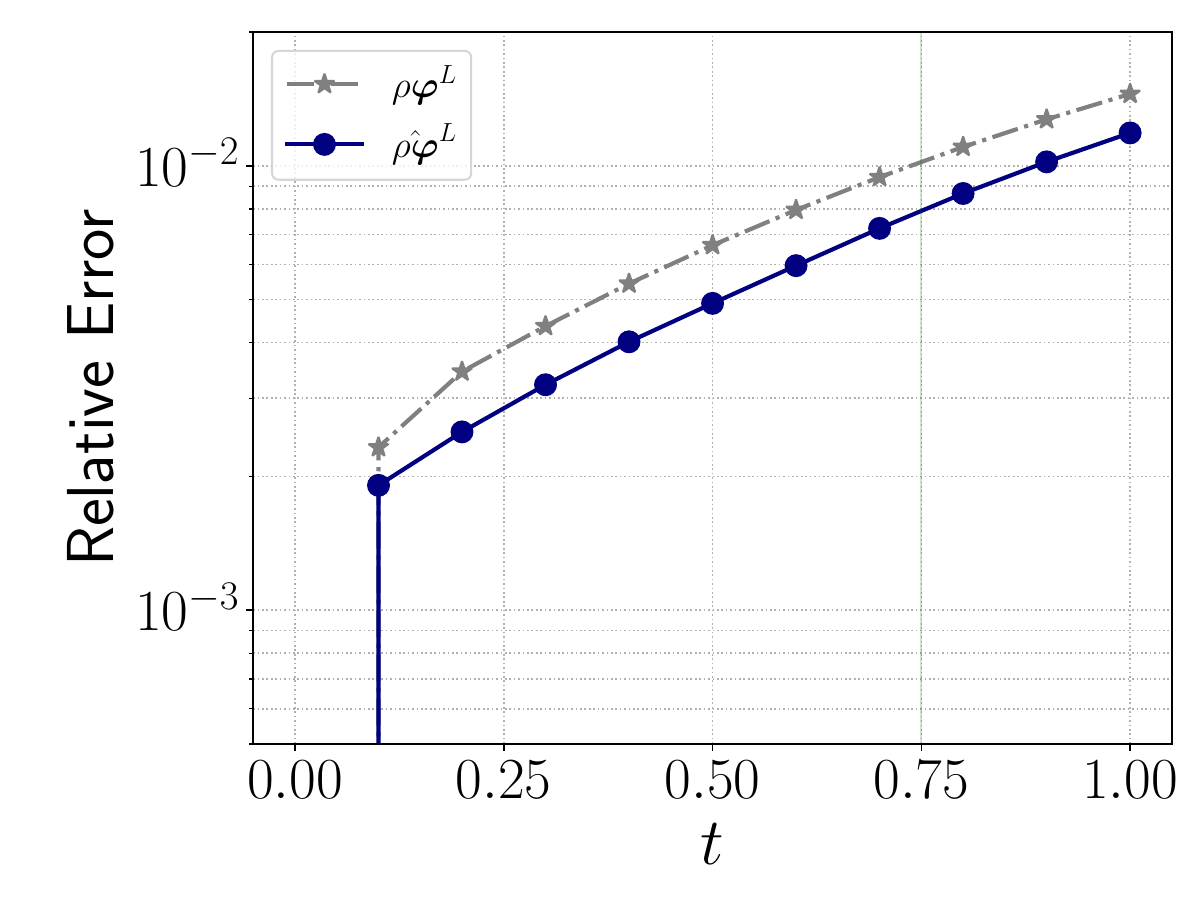}
      \includegraphics[trim=0.3cm 0.5cm 0.2cm  0.2cm,clip=true,width=0.31\textwidth]{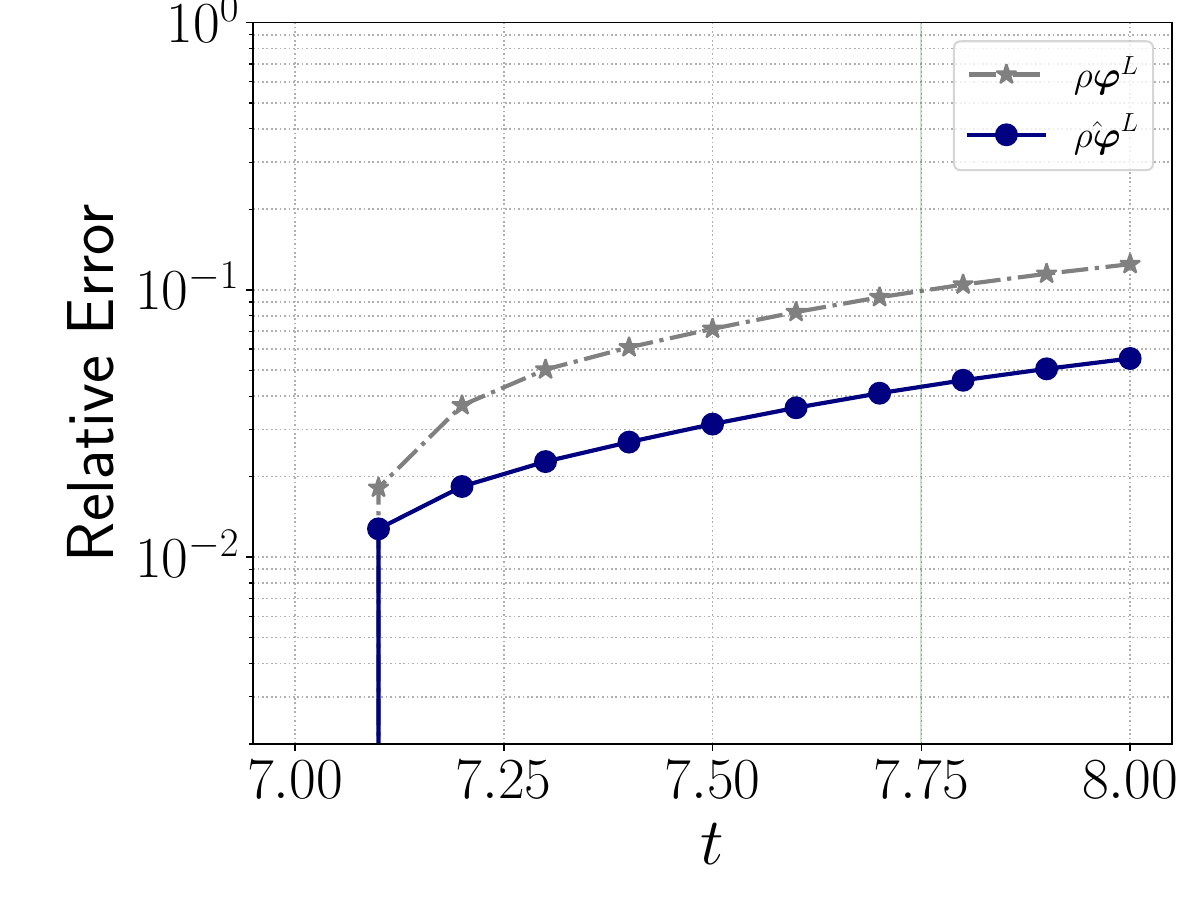}
      \includegraphics[trim=0.3cm 0.5cm 0.2cm  0.2cm,clip=true,width=0.31\textwidth]{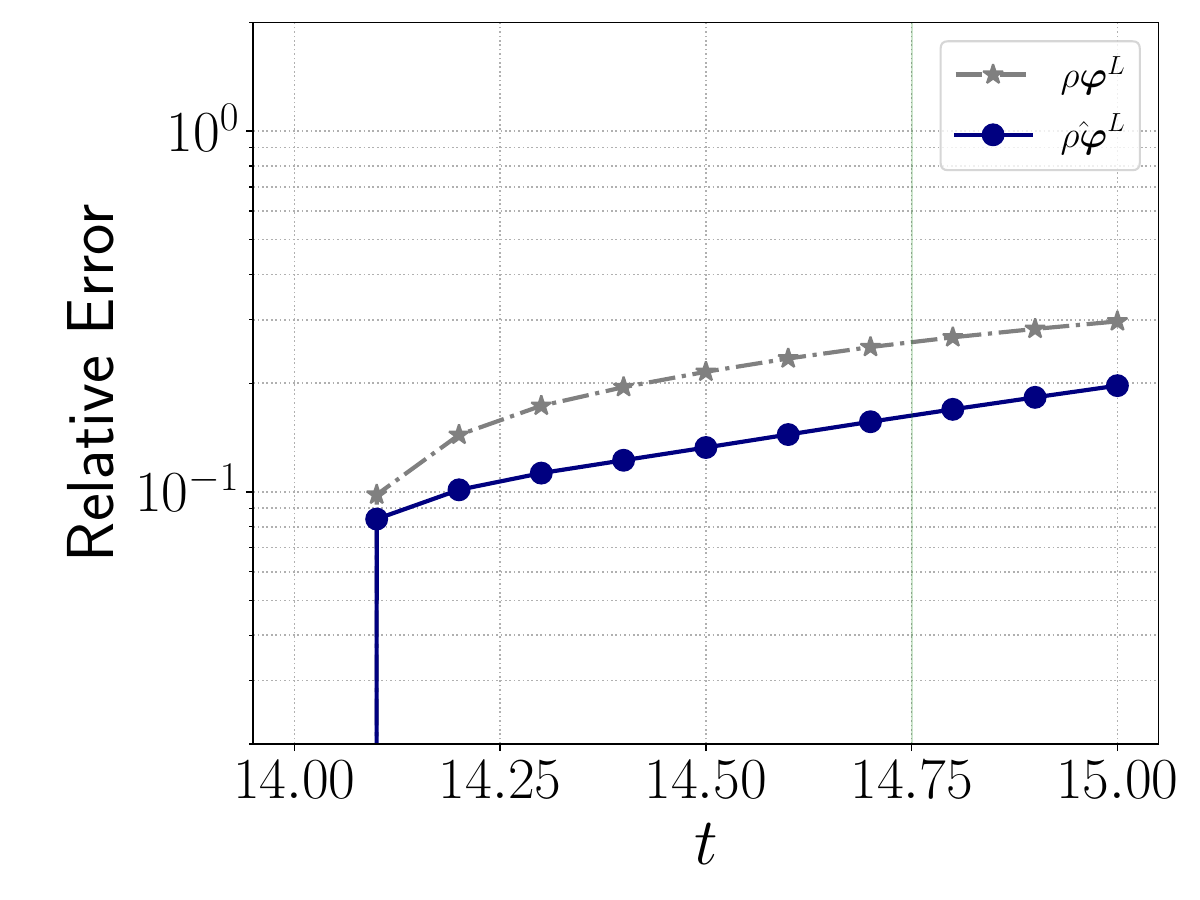}
  }
  \subfigure[Total energy ($\rho E$)]{
    \figlab{cns3d-tgv-Re1600-p1-errhistory-rE}
      \includegraphics[trim=0.3cm 0.5cm 0.2cm 0.2cm,clip=true,width=0.31\textwidth]{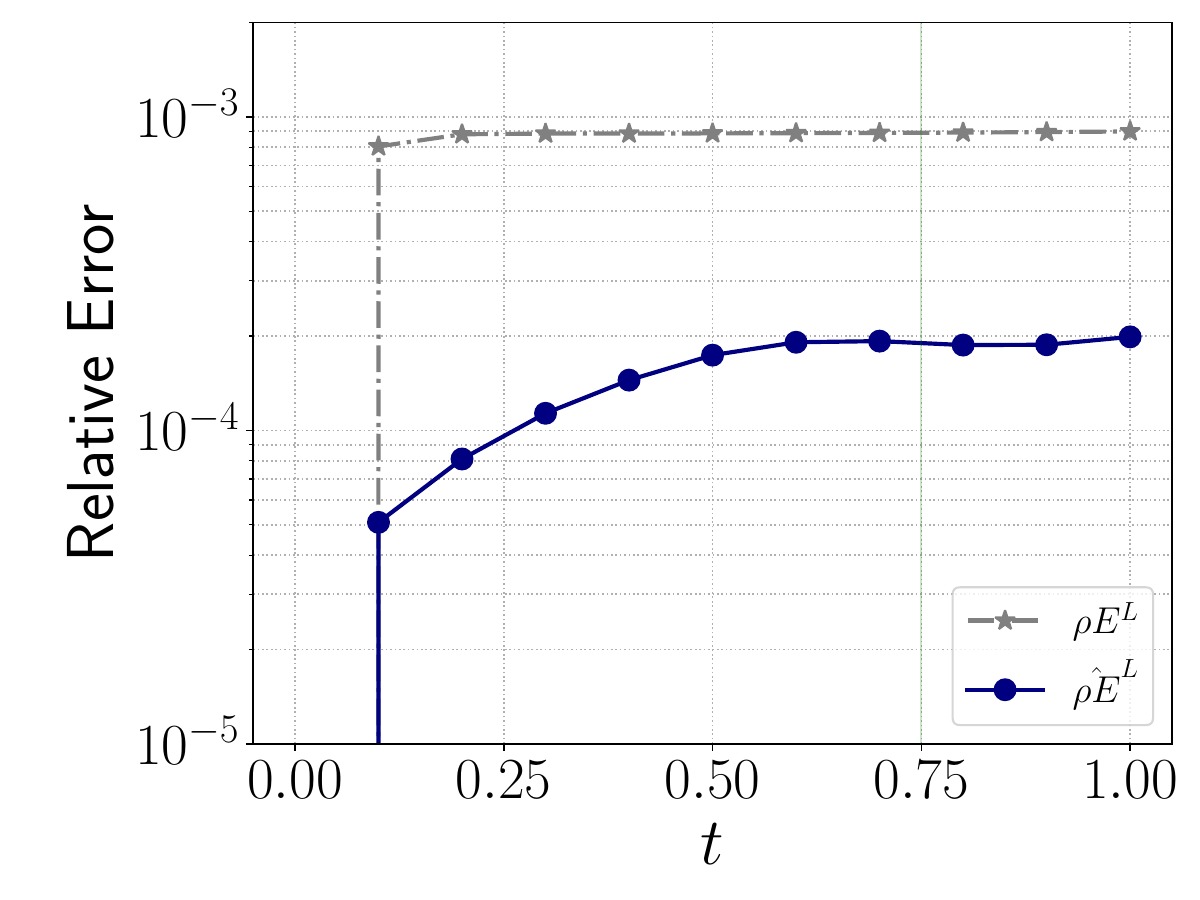}
      \includegraphics[trim=0.3cm 0.5cm 0.2cm 0.2cm,clip=true,width=0.31\textwidth]{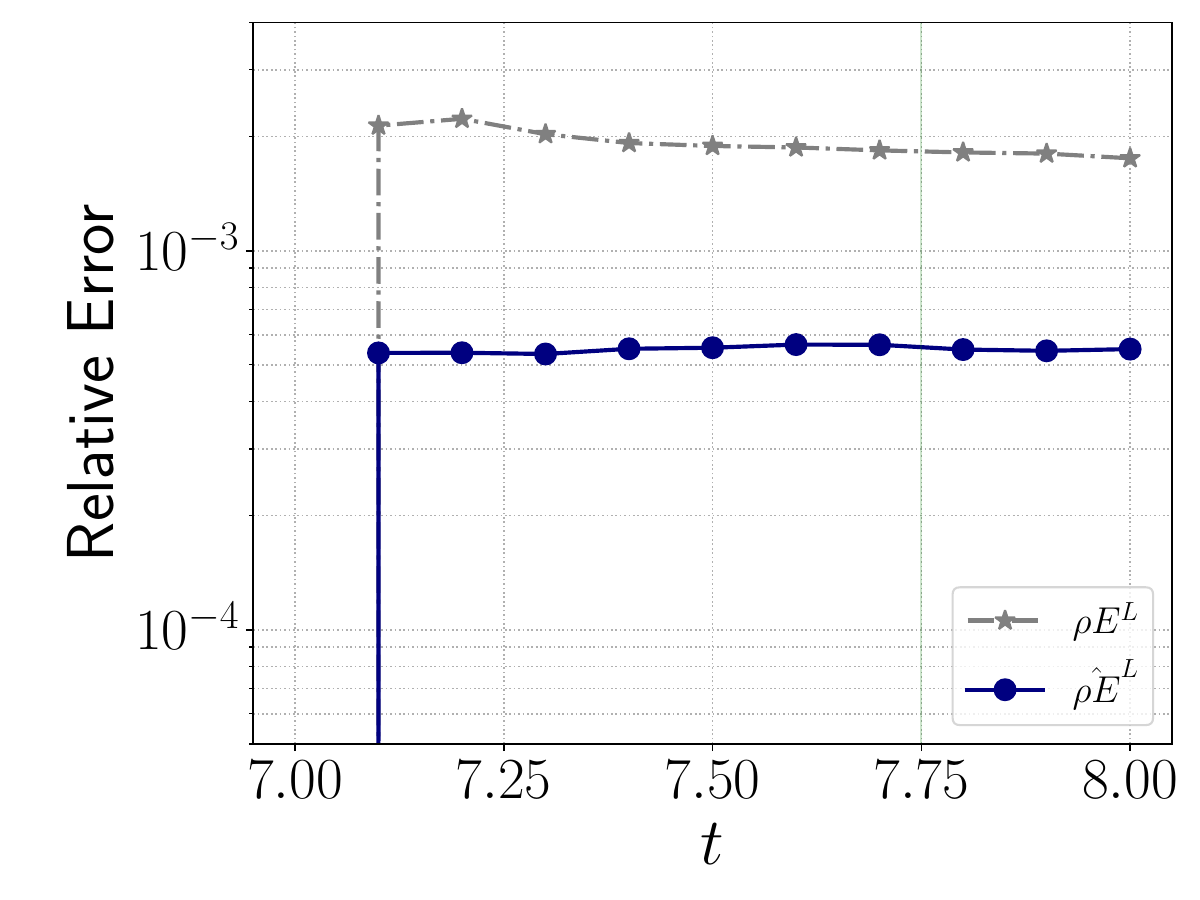}
      \includegraphics[trim=0.3cm 0.5cm 0.2cm 0.2cm,clip=true,width=0.31\textwidth]{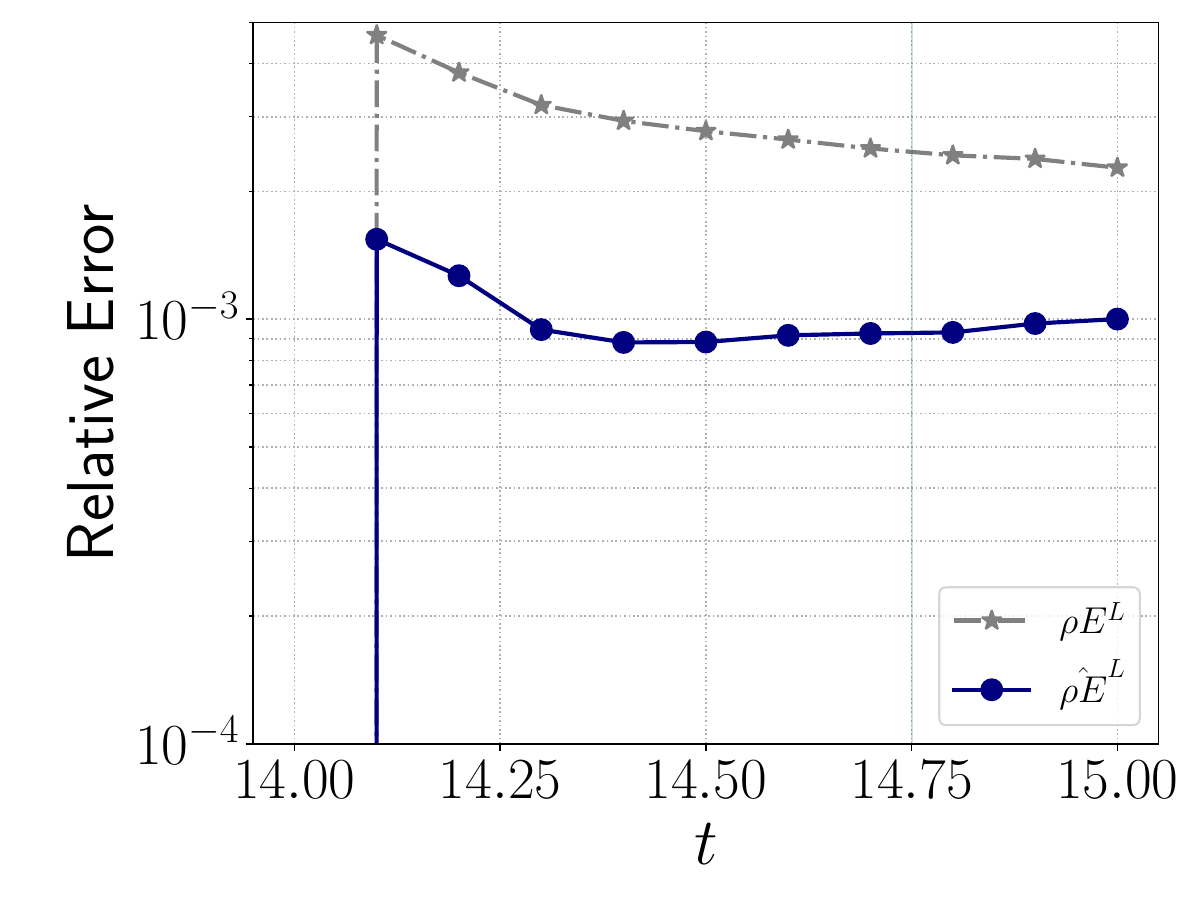}  
  }
  \caption{3D Taylor--Green vortex: relative error histories of (a) $\rho$, (b) $\rho\boldsymbol{\varphi}$, and (c) $\rho E$ at $Re=1600$
    for low-order solution ($\ub^L$) and augmented solution ($\hat{\ub}^L$). 
  }
  \figlab{cns3d-tgv-Re1600-errhistory}
\end{figure}

\begin{figure}[h!t!b!]
  \centering
  \subfigure[Projected $Q$ at $Re=200$]{
    \figlab{cns3d-tgv-Re200-p1-t8-qv-Gu-ss}
      \includegraphics[trim=8cm 5cm 8cm 3cm,clip=true,width=0.31\textwidth]{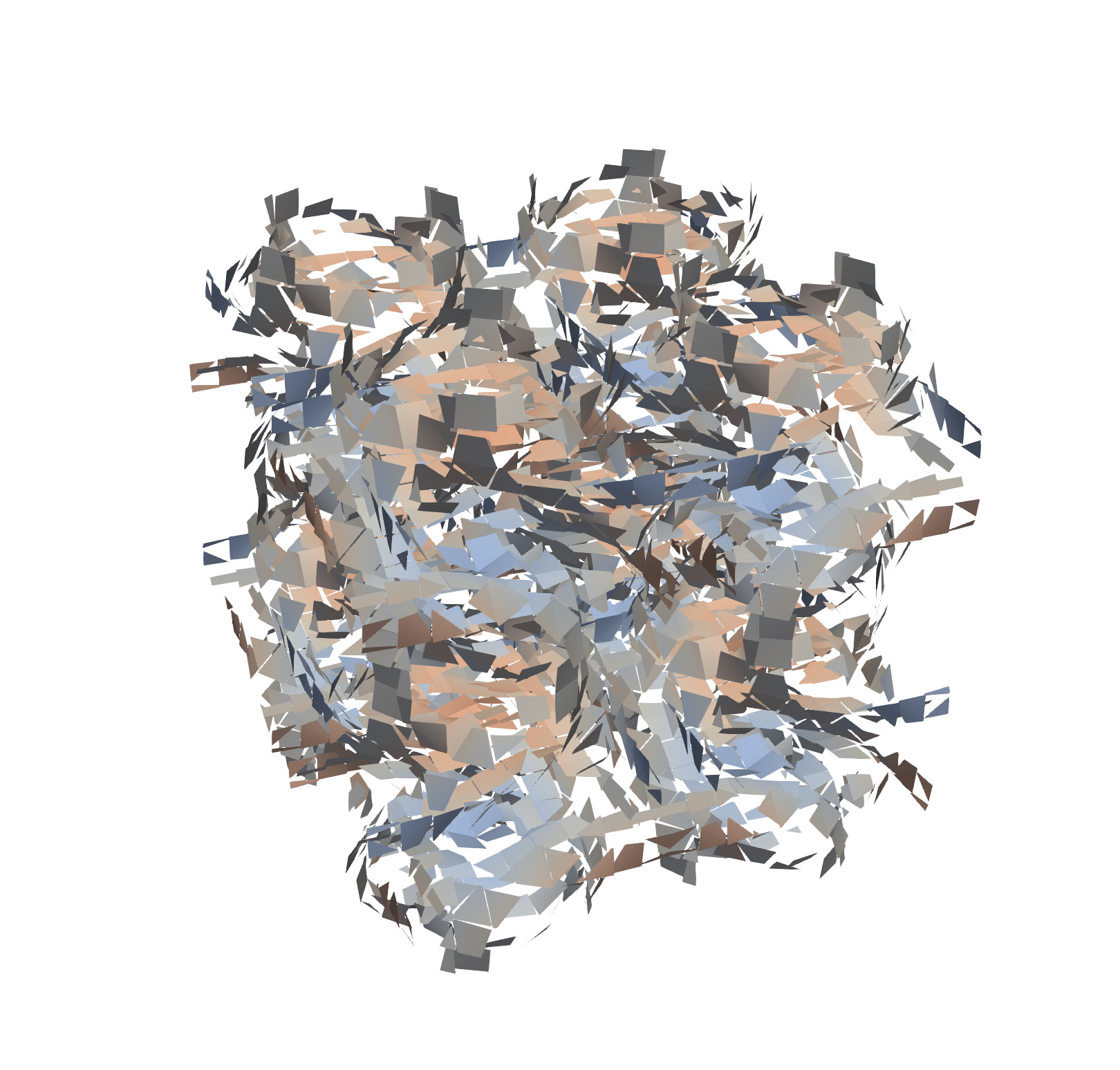}
  }
  \subfigure[Augmented $Q$ at $Re=200$]{
    \figlab{cns3d-tgv-Re200-p1-t8-qv-uh-ss}
      \includegraphics[trim=8cm 5cm 8cm 3cm,clip=true,width=0.31\textwidth]{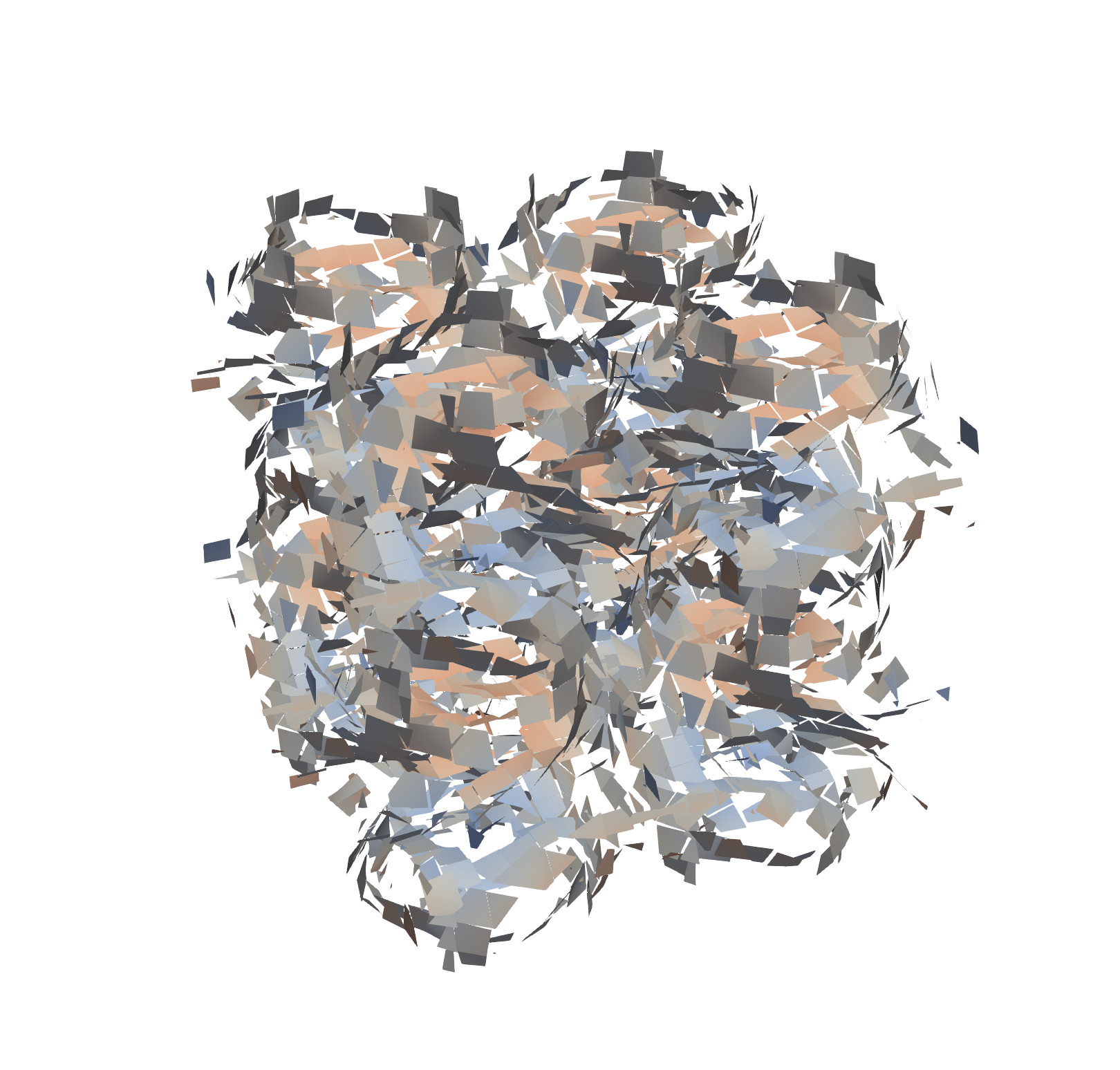}
  }
  \subfigure[Low-order $Q$ at $Re=200$]{
    \figlab{cns3d-tgv-Re200-p1-t8-qv-uL-ss}
      \includegraphics[trim=8cm 5cm 8cm 3cm,clip=true,width=0.31\textwidth]{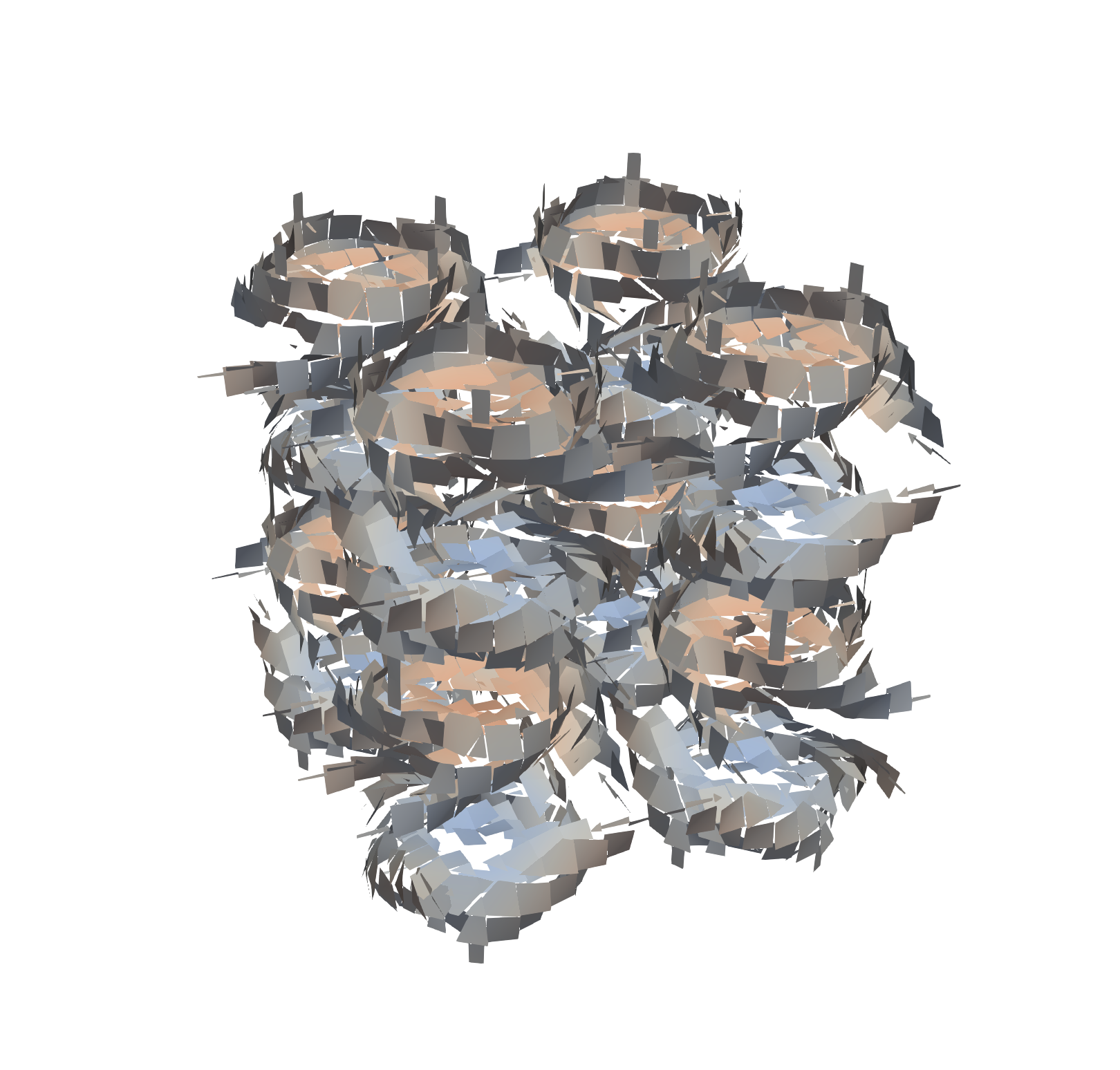}
  }
  \\ 
  \subfigure[Projected $Q$ at $Re=400$]{
    \figlab{cns3d-tgv-Re400-p1-t8-qv-Gu-ss}
      \includegraphics[trim=8cm 5cm 8cm 3cm,clip=true,width=0.31\textwidth]{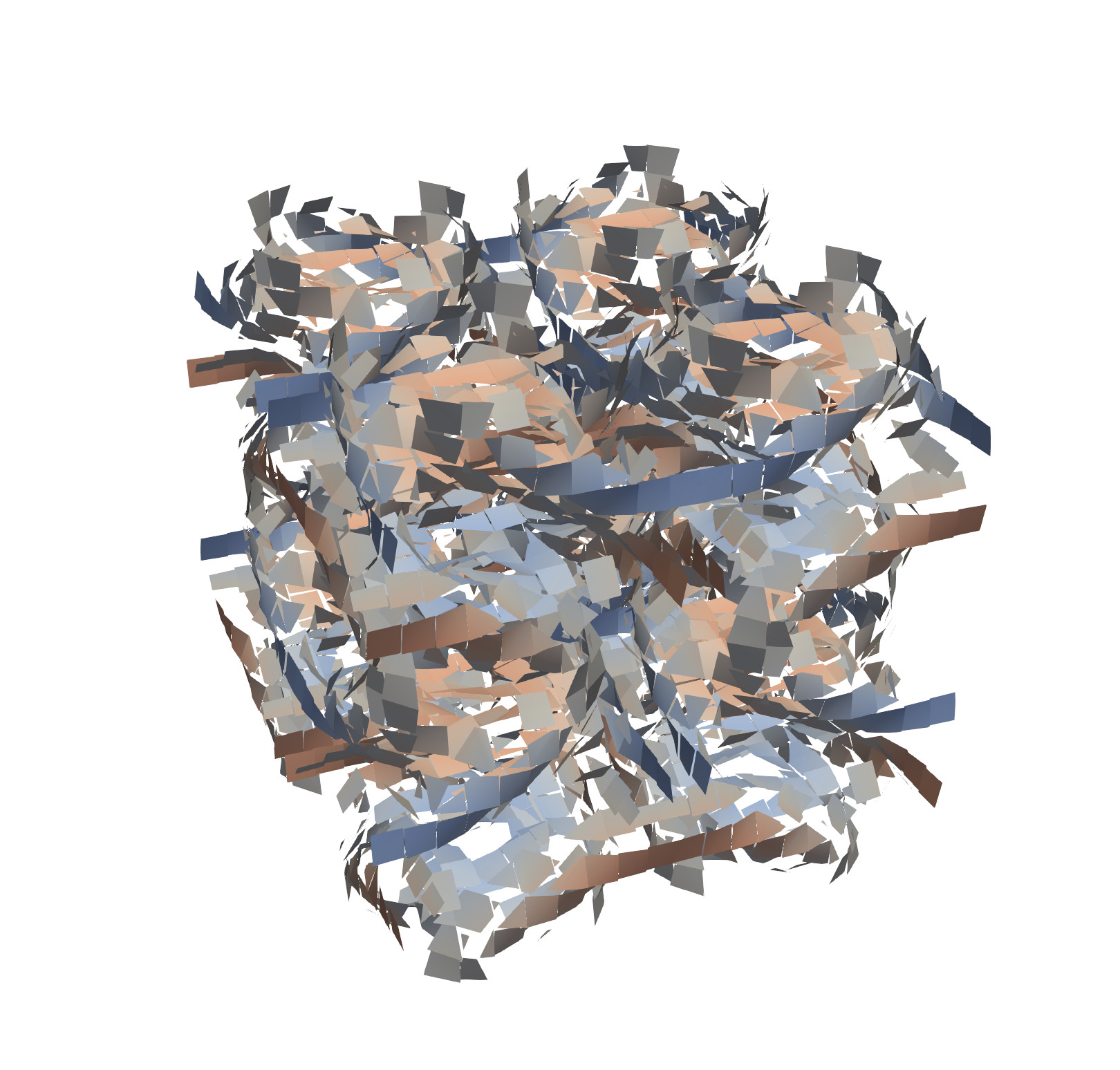}
  }
  \subfigure[Augmented $Q$ at $Re=400$]{
    \figlab{cns3d-tgv-Re400-p1-t8-qv-uh-ss}
      \includegraphics[trim=8cm 5cm 8cm 3cm,clip=true,width=0.31\textwidth]{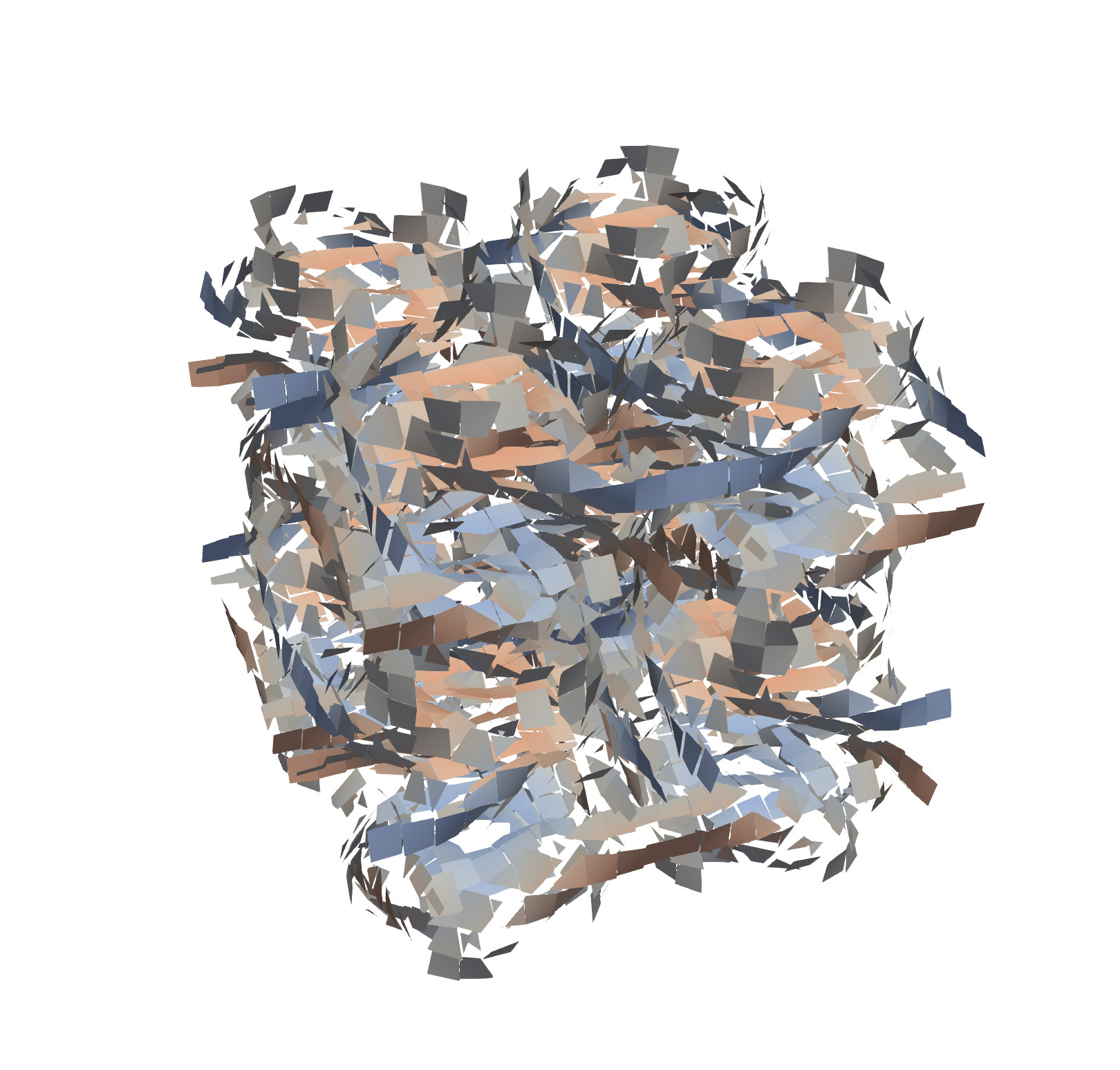}
  }
  \subfigure[Low-order $Q$ at $Re=400$]{
    \figlab{cns3d-tgv-Re400-p1-t8-qv-uL-ss}
      \includegraphics[trim=8cm 5cm 8cm 3cm,clip=true,width=0.31\textwidth]{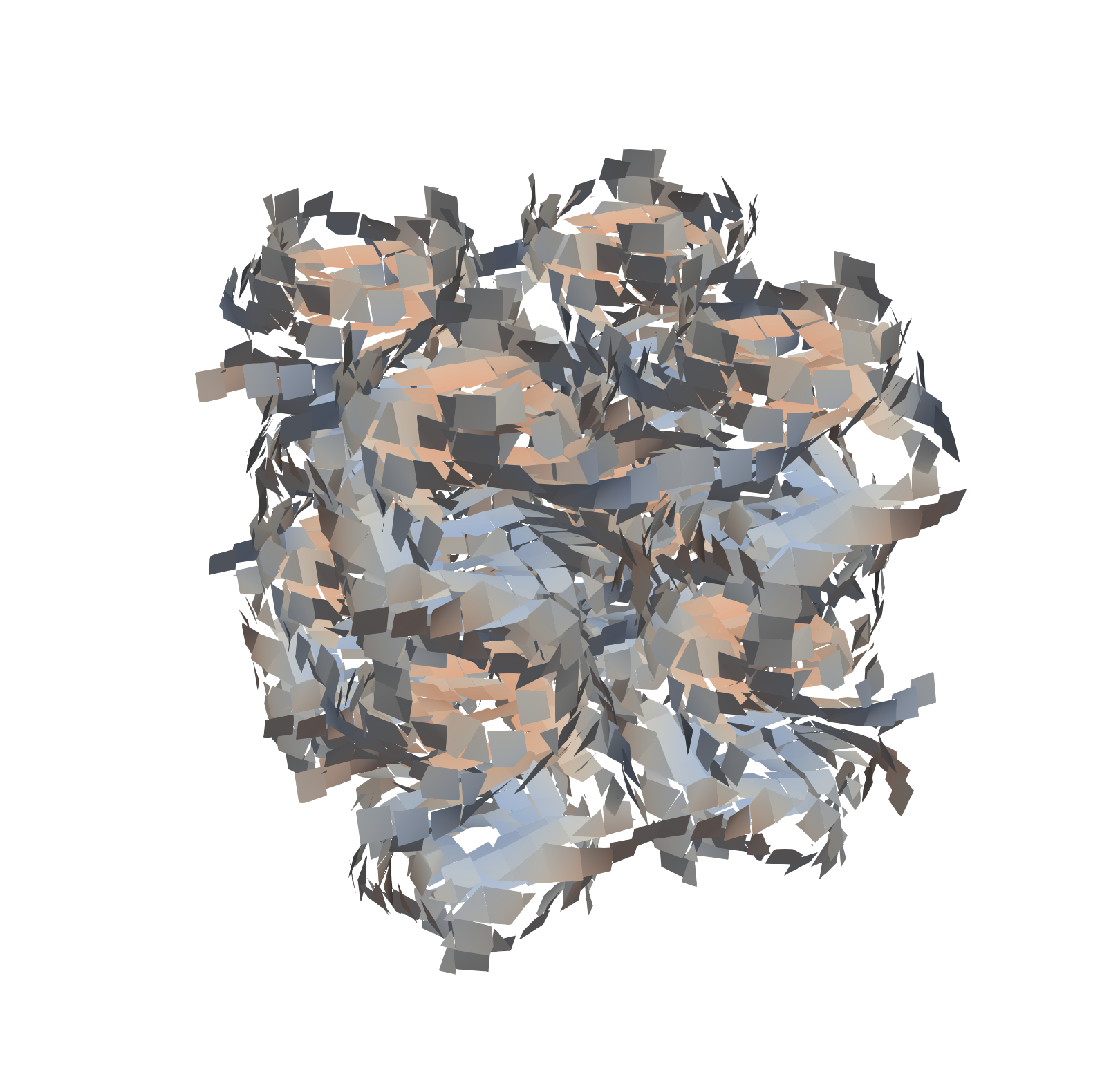}
  }
  \\
  \subfigure[Projected $Q$ at $Re=1600$]{
    \figlab{cns3d-tgv-Re1600-p1-t8-qv-Gu-ss}
      \includegraphics[trim=8cm 5cm 8cm 3cm,clip=true,width=0.31\textwidth]{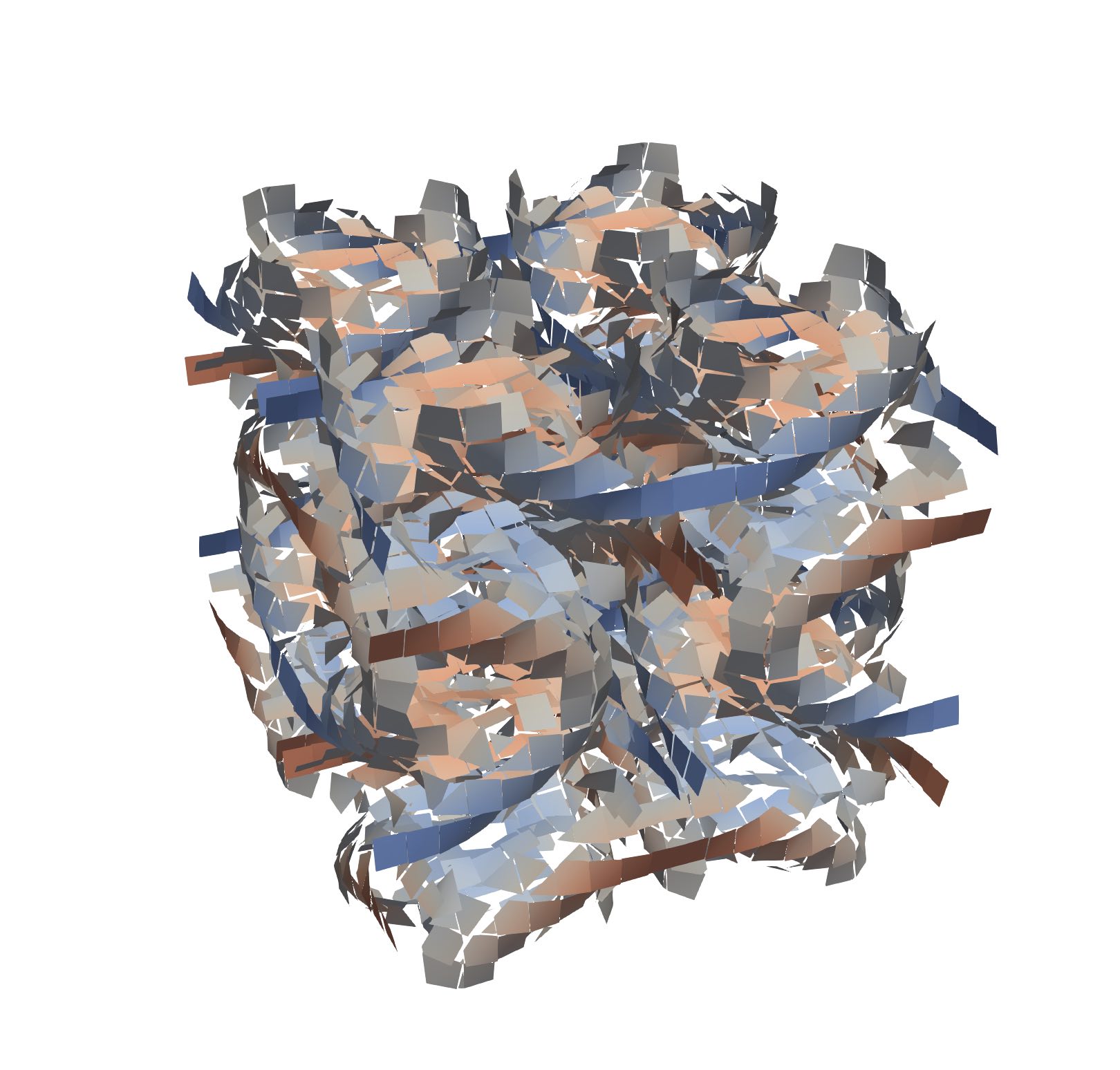}
  }
  \subfigure[Augmented $Q$ at $Re=1600$]{
    \figlab{cns3d-tgv-Re1600-p1-t8-qv-uh-ss}
      \includegraphics[trim=8cm 5cm 8cm 3cm,clip=true,width=0.31\textwidth]{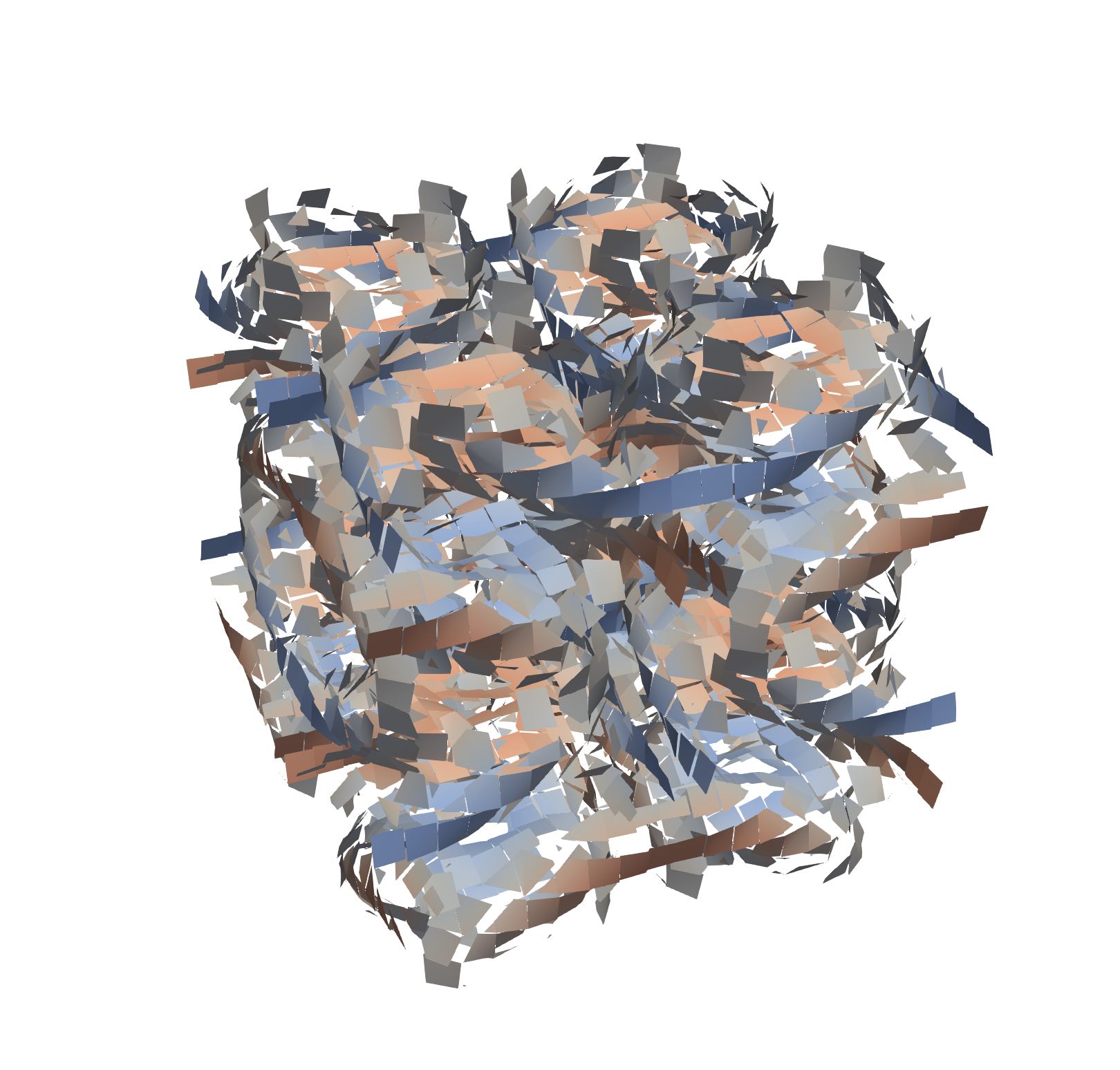}
  }
  \subfigure[Low-order $Q$ at $Re=1600$]{
    \figlab{cns3d-tgv-Re1600-p1-t8-qv-uL-ss}
      \includegraphics[trim=8cm 5cm 8cm 3cm,clip=true,width=0.31\textwidth]{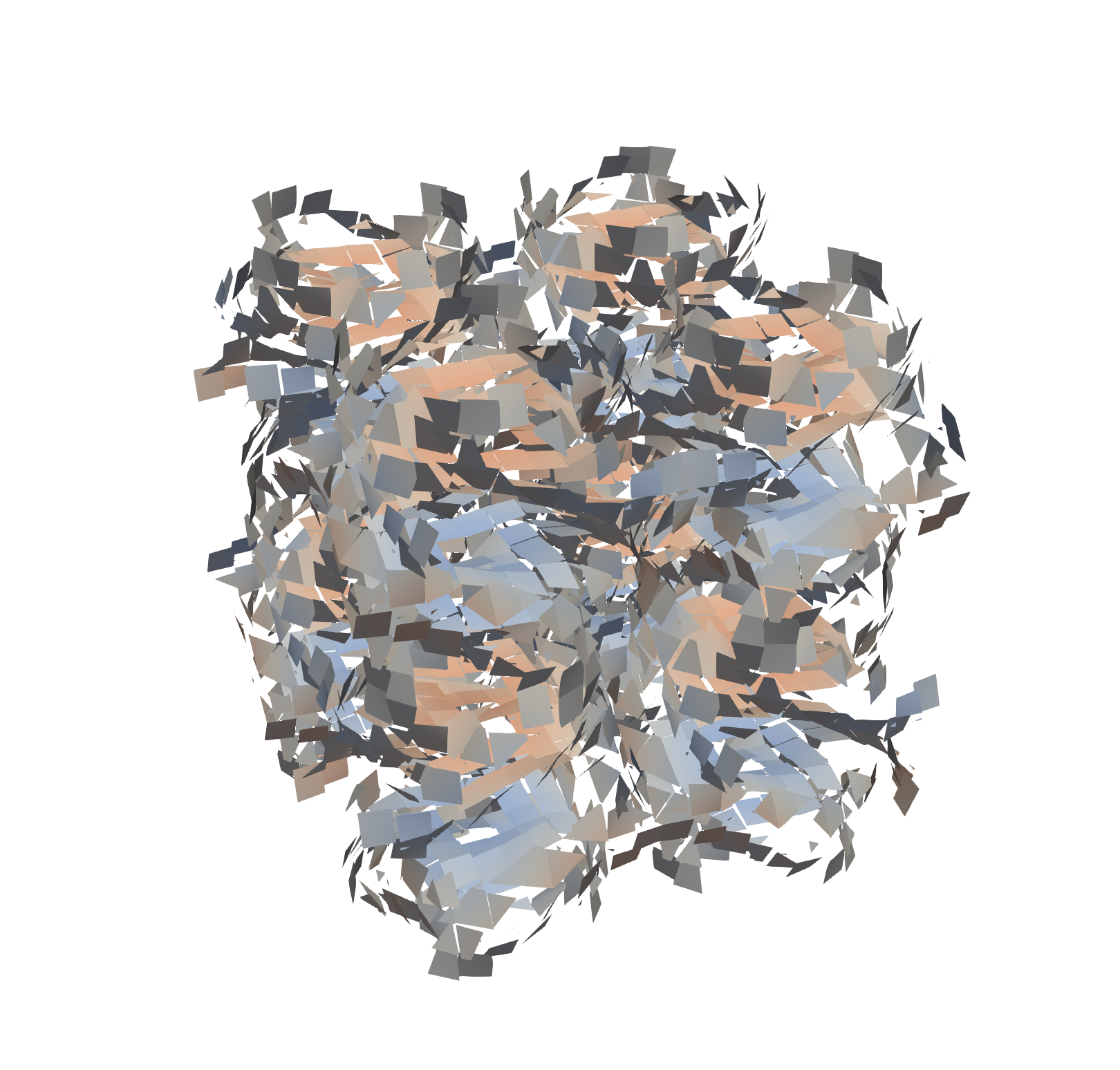}
  }
  \caption{3D Taylor--Green vortex:
  Q-criterion isosurfaces at $Re=\LRc{200,400,1600}$ and $t=8$ are shown for (a) the projected solution ($\projL \ub^H$), (b) the augmented solution ($\hat{\ub}^L$), and (c) the low-order solution ($\ub^L$).
  The isosurfaces 
  are colored based on the z-component of the velocity, ranging from $-0.1$ to $0.1$. 
  }
  \figlab{cns3d-tgv-ss-qv-t8}
\end{figure}

\begin{figure}[h!t!b!]
  \centering
  \subfigure[Projected $Q$ at $Re=200$]{
    \figlab{cns3d-tgv-Re200-p1-t15-qv-Gu-ss}
      \includegraphics[trim=8cm 5cm 8cm 3cm,clip=true,width=0.31\textwidth]{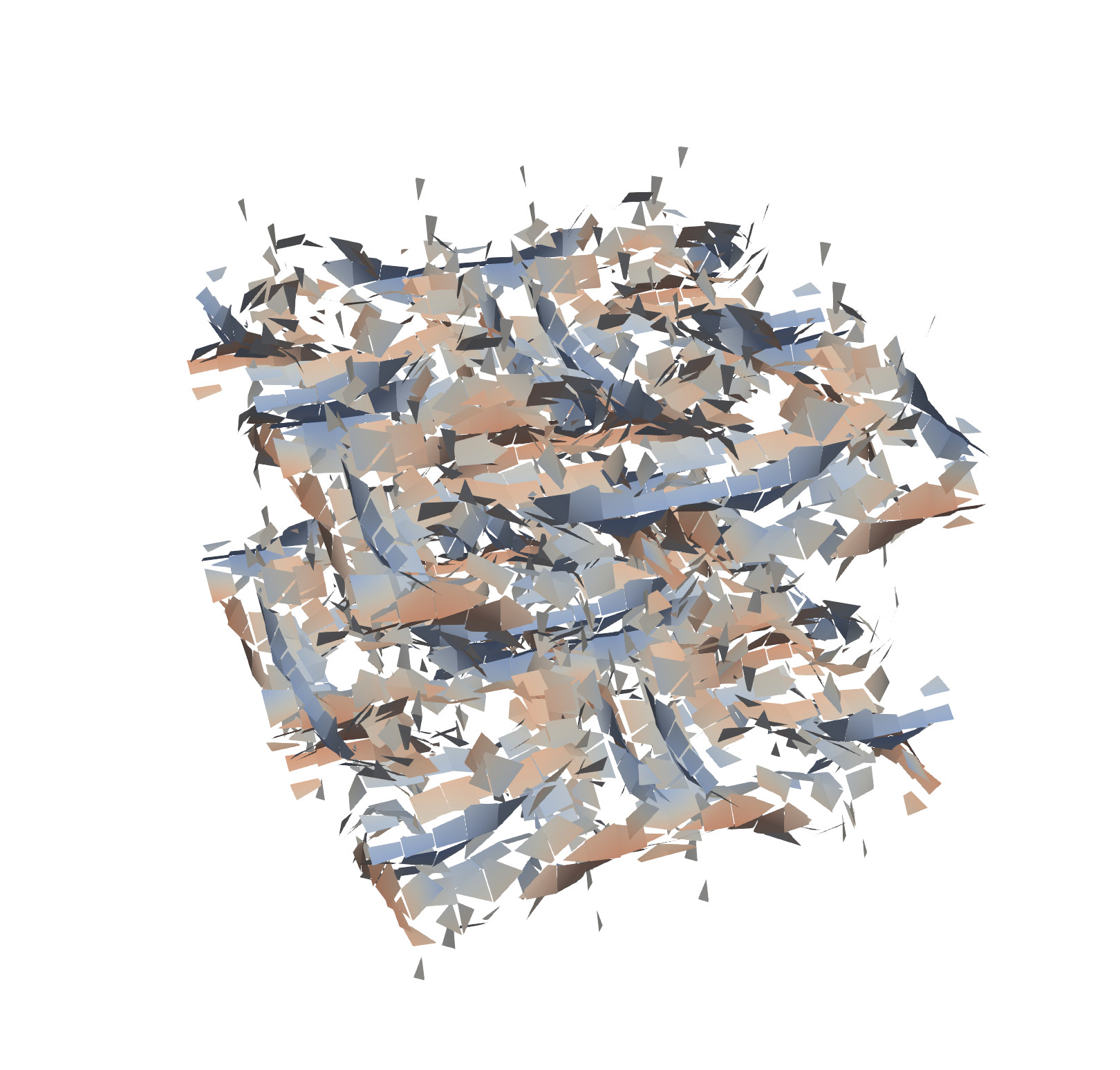}
  }
  \subfigure[Augmented $Q$ at $Re=200$]{
    \figlab{cns3d-tgv-Re200-p1-t15-qv-uh-ss}
      \includegraphics[trim=8cm 5cm 8cm 3cm,clip=true,width=0.31\textwidth]{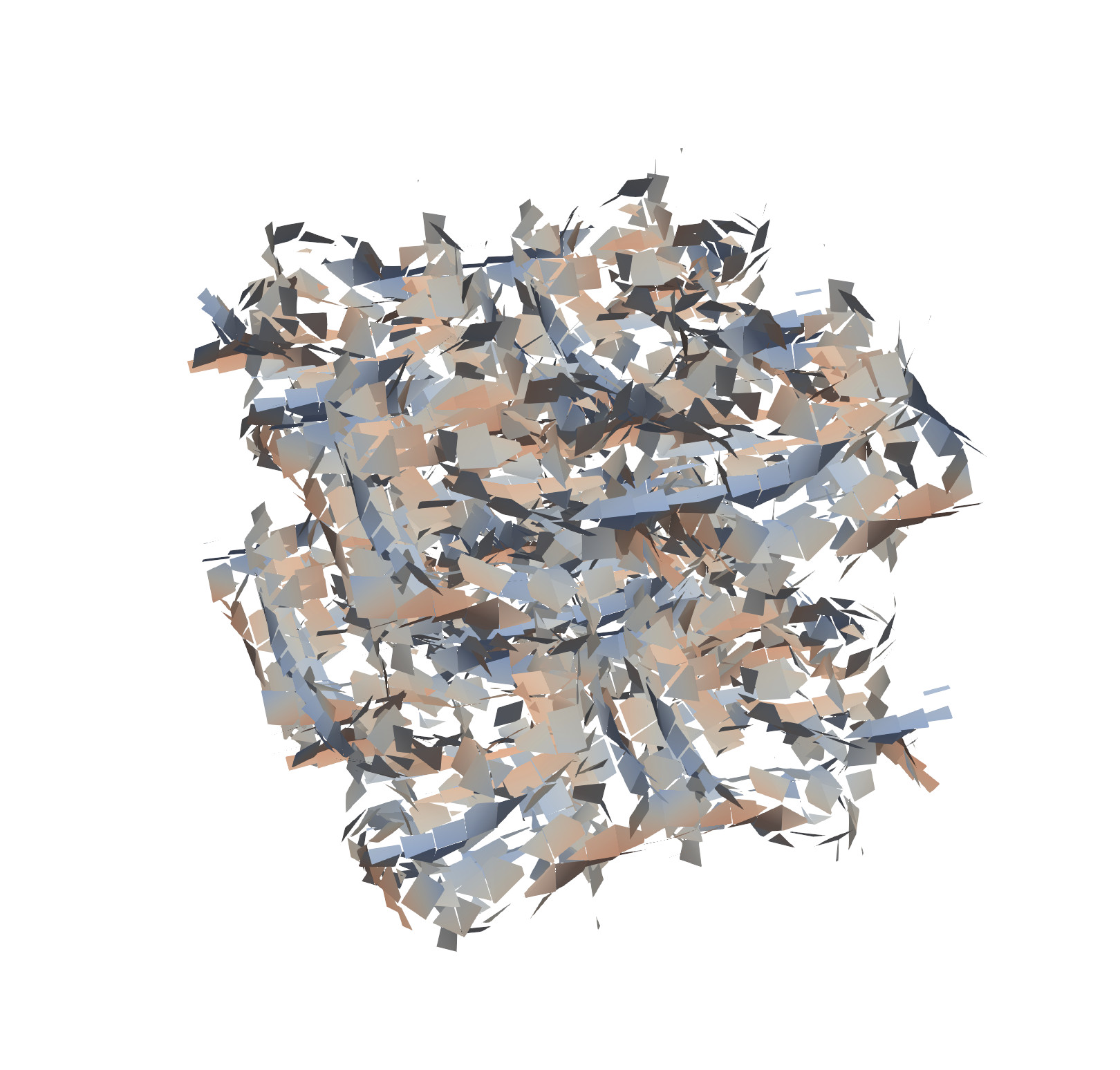}
  }
  \subfigure[Low-order $Q$ at $Re=200$]{
    \figlab{cns3d-tgv-Re200-p1-t15-qv-uL-ss}
      \includegraphics[trim=8cm 5cm 8cm 3cm,clip=true,width=0.31\textwidth]{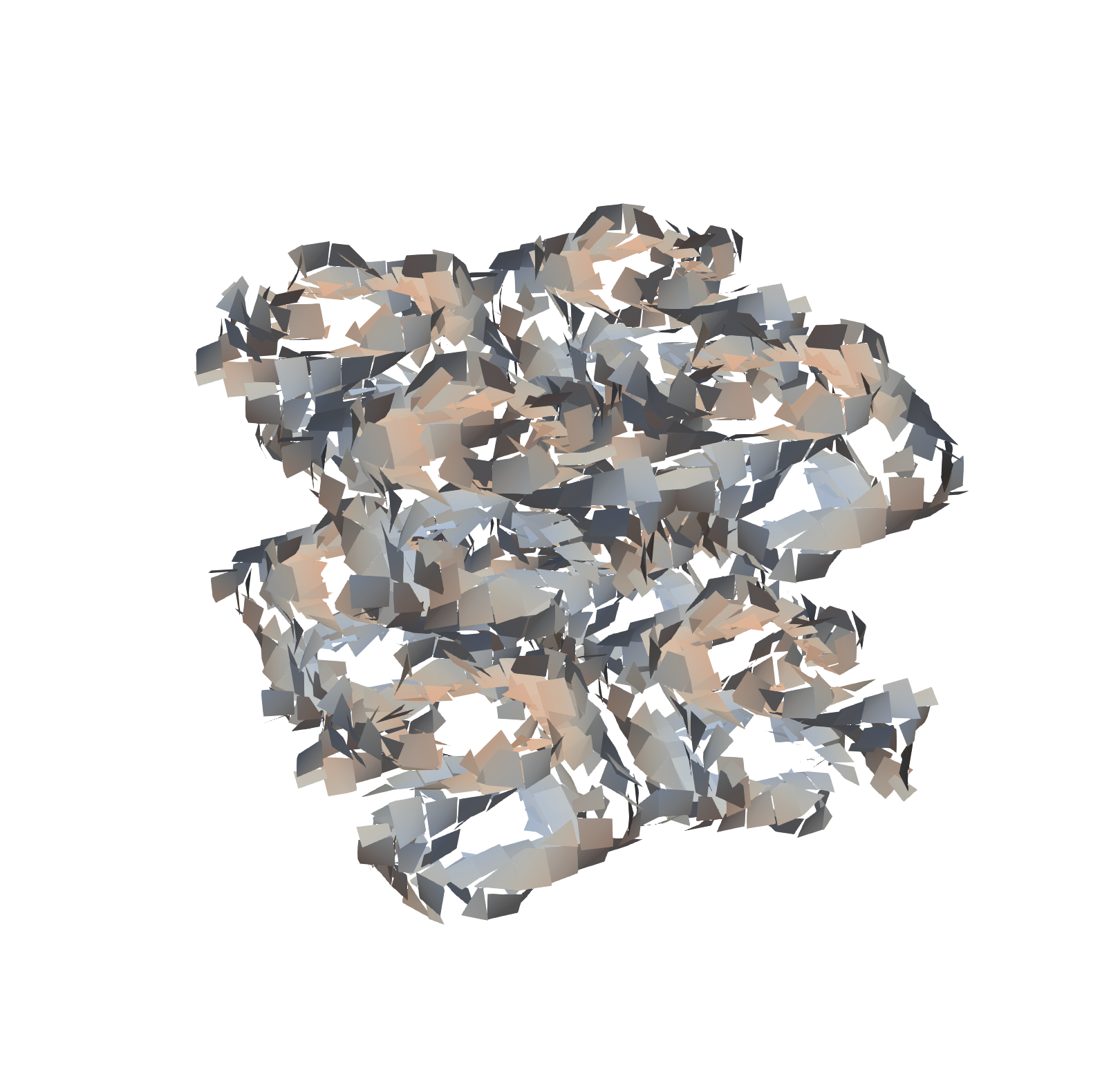}
  }
  \\ 
  \subfigure[Projected $Q$ at $Re=400$]{
    \figlab{cns3d-tgv-Re400-p1-t15-qv-Gu-ss}
      \includegraphics[trim=8cm 5cm 8cm 3cm,clip=true,width=0.31\textwidth]{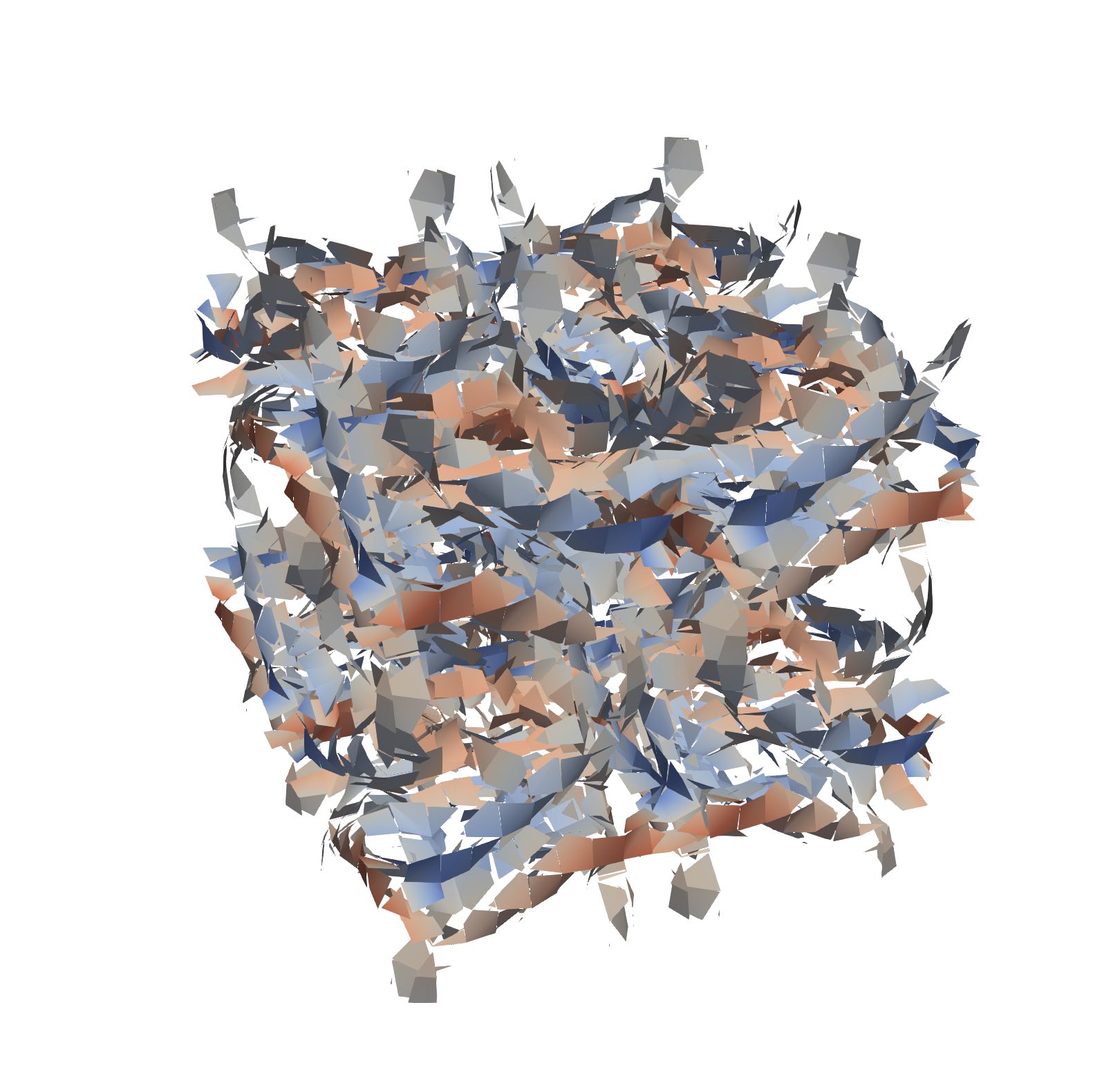}
  }
  \subfigure[Augmented $Q$ at $Re=400$]{
    \figlab{cns3d-tgv-Re400-p1-t15-qv-uh-ss}
      \includegraphics[trim=8cm 5cm 8cm 3cm,clip=true,width=0.31\textwidth]{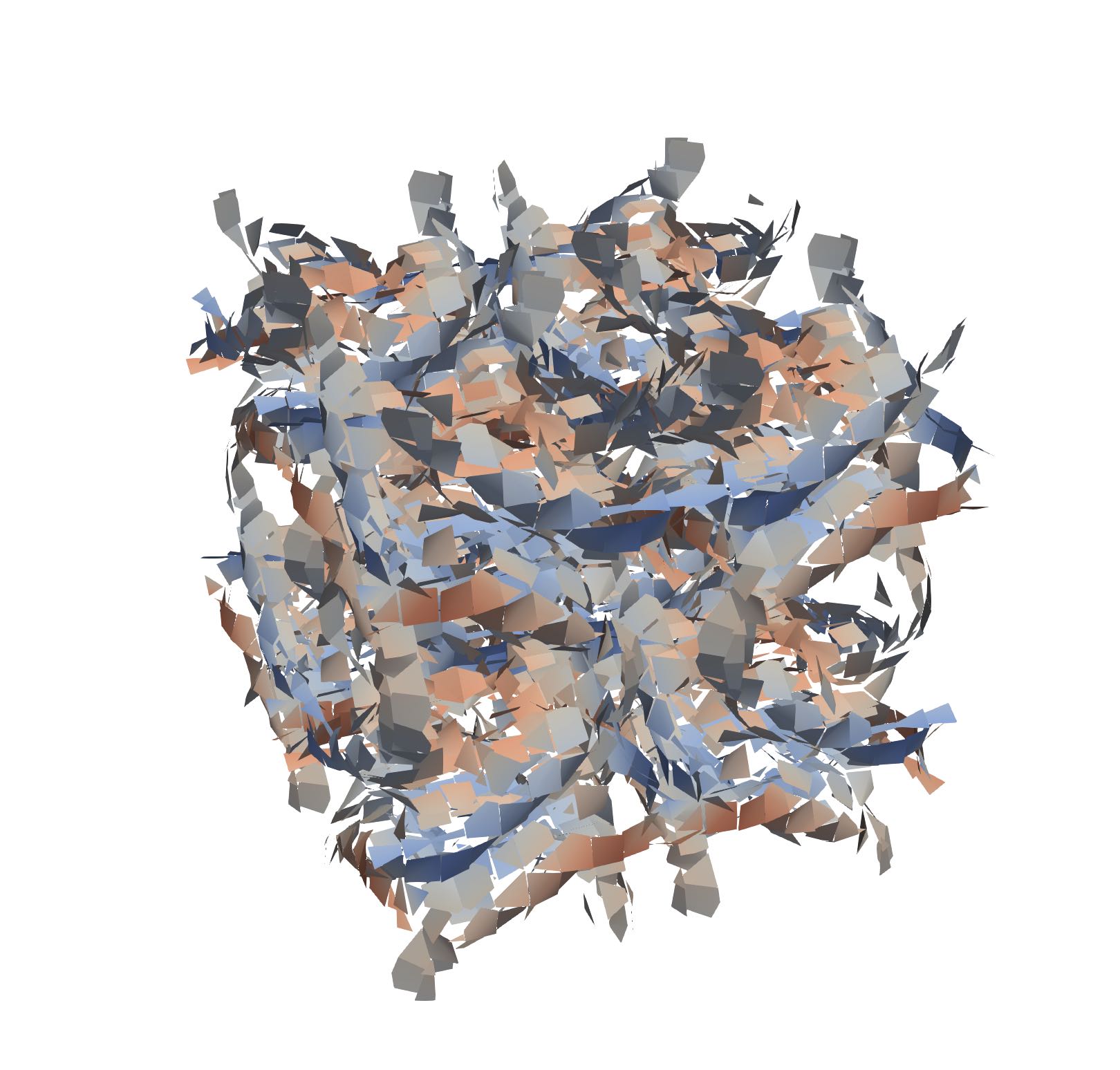}
  }
  \subfigure[Low-order $Q$ at $Re=400$]{
    \figlab{cns3d-tgv-Re400-p1-t15-qv-uL-ss}
      \includegraphics[trim=8cm 5cm 8cm 3cm,clip=true,width=0.31\textwidth]{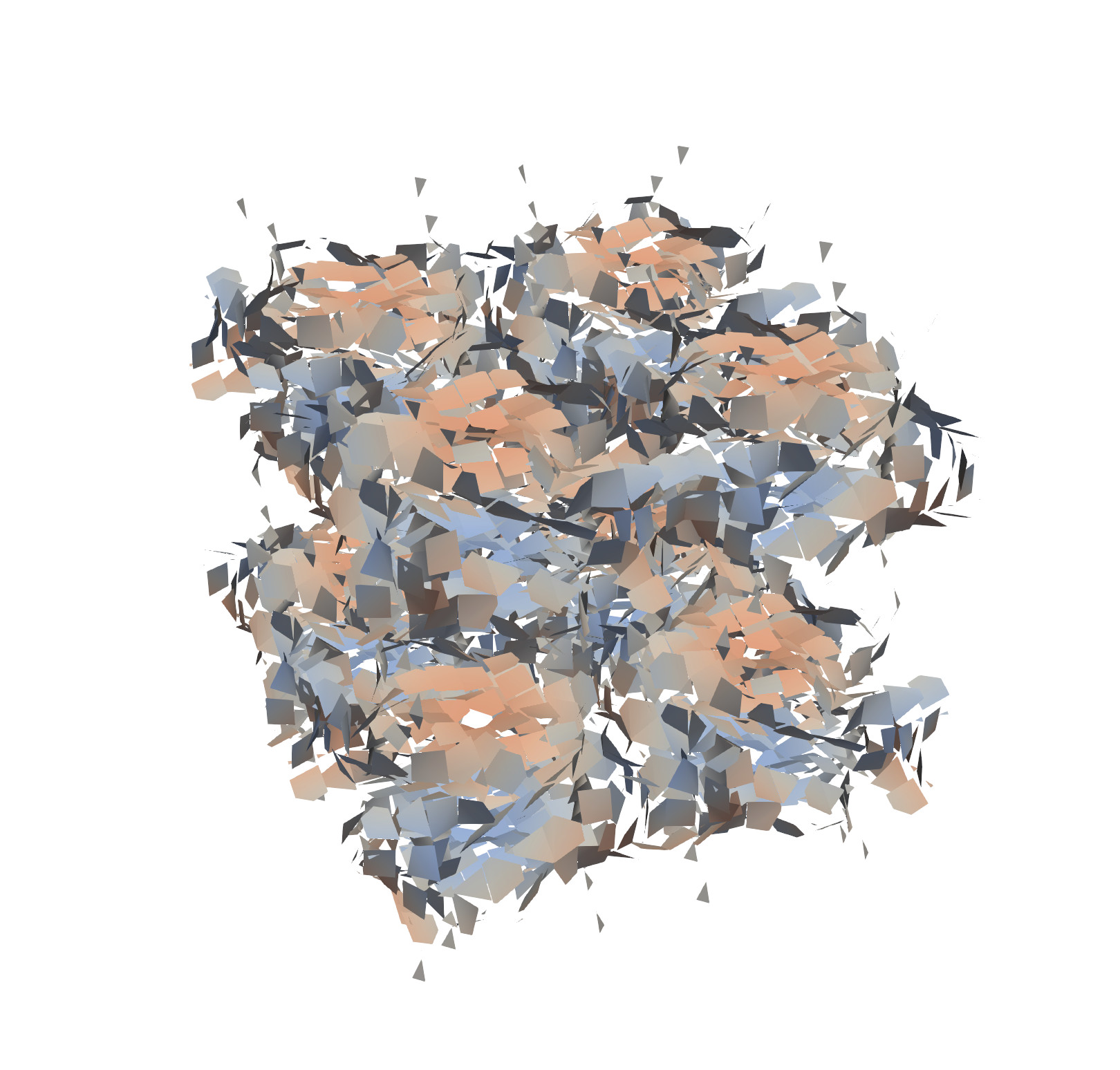}
  }
  \\
  \subfigure[Projected $Q$ at $Re=1600$]{
    \figlab{cns3d-tgv-Re1600-p1-t15-qv-Gu-ss}
      \includegraphics[trim=8cm 5cm 8cm 3cm,clip=true,width=0.31\textwidth]{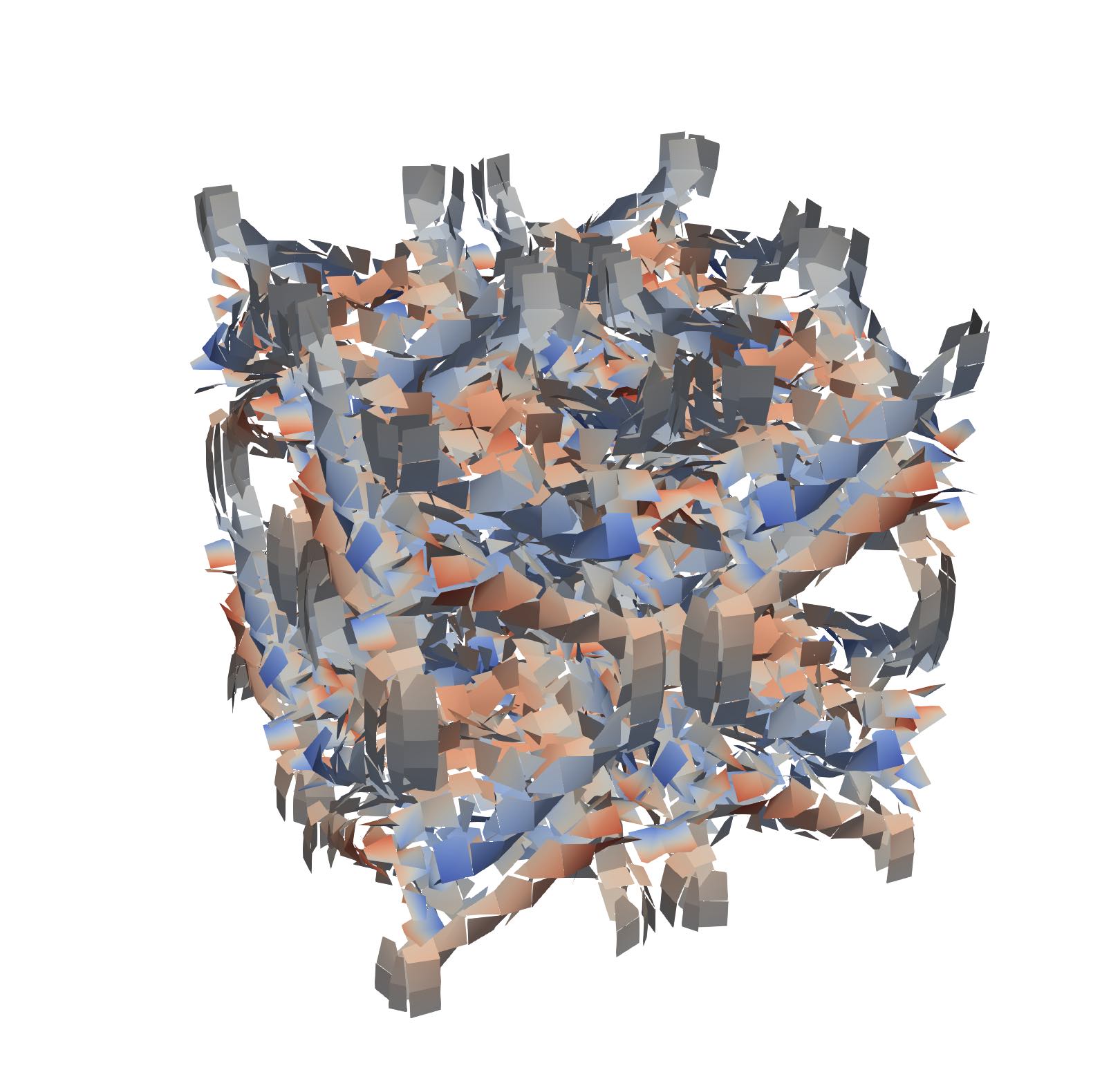}
  }
  \subfigure[Augmented $Q$ at $Re=1600$]{
    \figlab{cns3d-tgv-Re1600-p1-t15-qv-uh-ss}
      \includegraphics[trim=8cm 5cm 8cm 3cm,clip=true,width=0.31\textwidth]{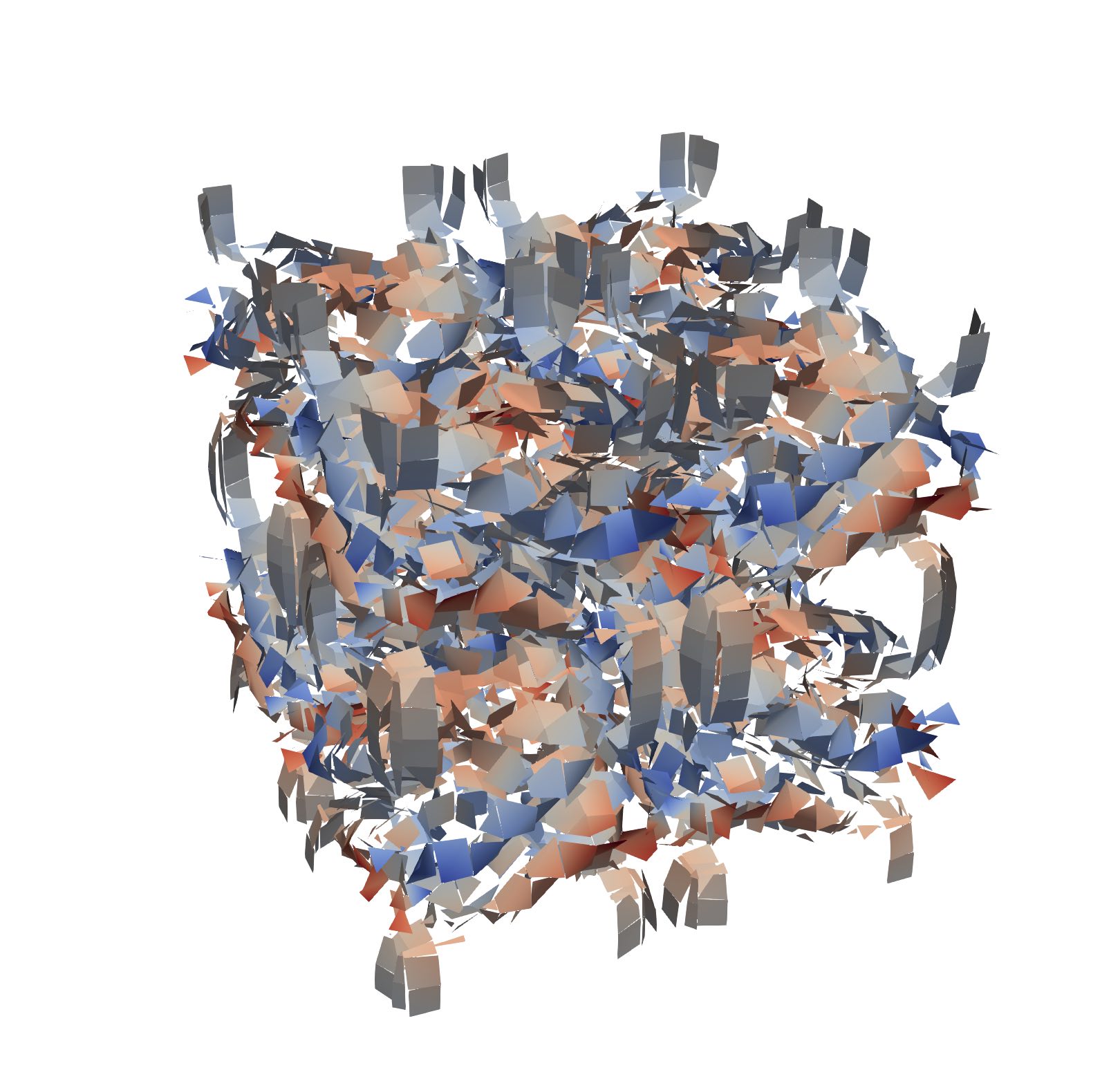}
  }
  \subfigure[Low-order at $Re=1600$]{
    \figlab{cns3d-tgv-Re1600-p1-t15-qv-uL-ss}
      \includegraphics[trim=8cm 5cm 8cm 3cm,clip=true,width=0.31\textwidth]{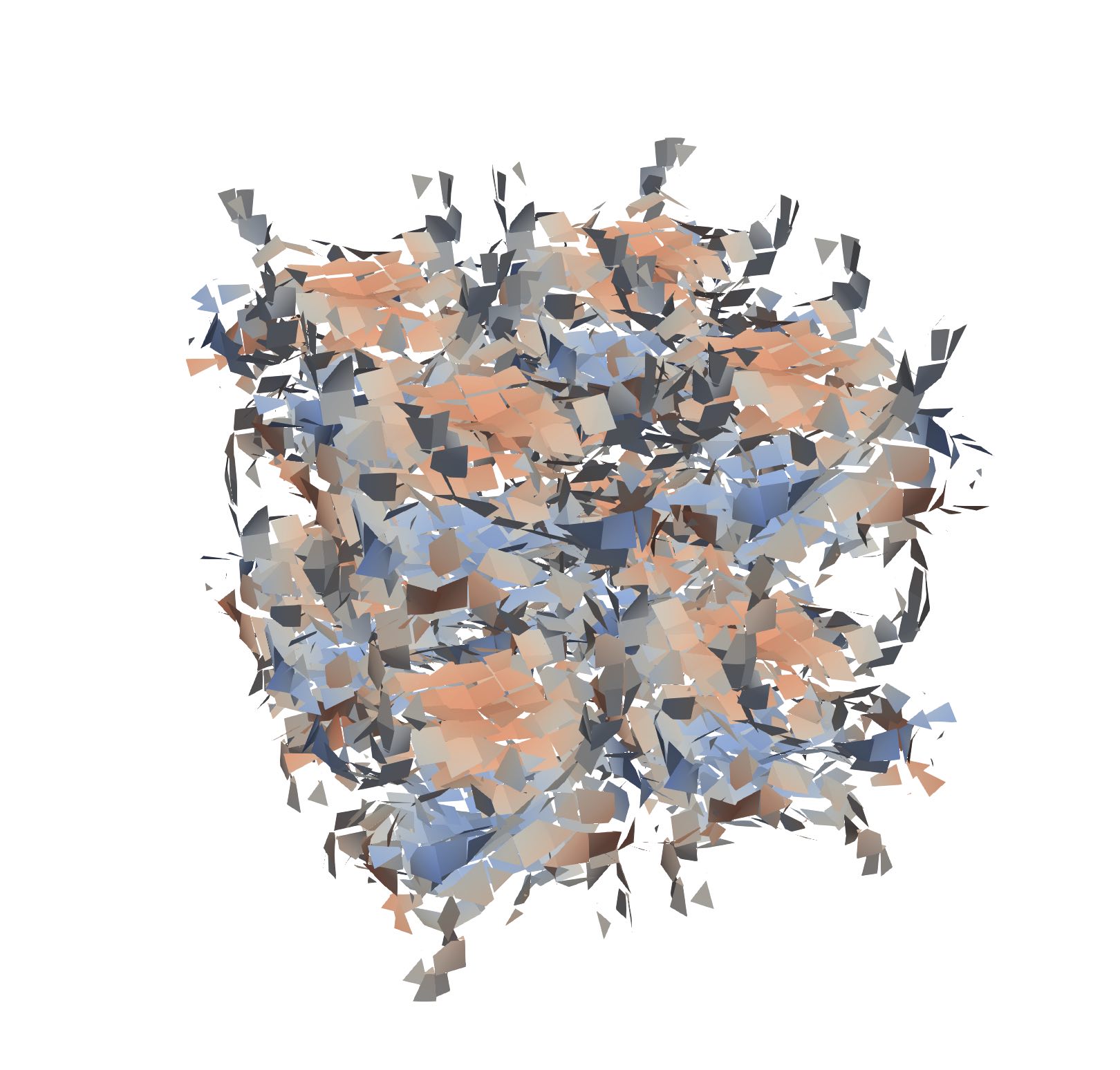}
  }
  \caption{3D Taylor--Green vortex: 
  Q-criterion isosurfaces at $Re=\LRc{200,400,1600}$ and $t=15$ are shown for (a) the projected solution ($\projL \ub^H$), (b) the augmented solution ($\hat{\ub}^L$), and (c) the low-order solution ($\ub^L$).
  The isosurfaces 
  are colored based on the z-component of the velocity, ranging from $-0.1$ to $0.1$. 
  }
  \figlab{cns3d-tgv-ss-qv-t15}
\end{figure}

Similar trends are observed in Figure \figref{cns3d-tgv-Re200-errhistory}, Figure \figref{cns3d-tgv-Re400-errhistory}, and Figure \figref{cns3d-tgv-Re1600-errhistory} for $Re=\LRc{200,400,1600}$. The accuracy of the augmented solutions is highly influenced by hyperparameters such as the kernel width ($k_w$) and the number of steps ($m$). Following a sensitivity analysis with respect to $k_w$ and $m$, we selected the optimal values for each Reynolds number: $k_w=1$ and $m=11$ for $Re=\LRc{200,400,1600}$.
In all these cases the momentum error dominates over the density and the total energy errors, similar to the case with $Re=100$. Additionally, the momentum error at $t=1$ is lower than the errors at $t=8$ and $t=15$.
Compared with the low-order solution, the augmented solution (blue solid line) exhibits consistently lower error overall. At $t=8$, the momentum errors in the augmented solutions are nearly half those of the low-order solutions for $Re=\LRc{200, 400, 1600}$. At $t=15$, the momentum errors of the augmented solution are $6.3 \times 10^{-2}$, $1.1 \times 10^{-1}$, and $2.0 \times 10^{-1}$ for $Re=200$, $Re=400$, and $Re=1600$, respectively, which are $2.3$, $2.0$, and $1.5$ times smaller than their low-order counterparts.
Snapshots of Q-criterion isosurfaces at $t=8$ and $t=15$ for the projected, augmented, and low-order solutions are shown in Figure \figref{cns3d-tgv-ss-qv-t8} and Figure \figref{cns3d-tgv-ss-qv-t15}.
In both figures, the augmented solutions demonstrate significant improvements as expected. 
These results demonstrate that the continuous neural network source term in \eqnref{gov-ode-system} successfully recovers misrepresented scales and enhances the accuracy of the low-order approximation.

\subsubsection{Sensitivity test for the kernel width $k_w$ and the number of steps $m$}

In this subsection we present a sensitivity analysis of the accuracy of the augmented solution with respect to the kernel width ($k_w \in \LRc{1,3,5}$) and the number of steps ($m\in\LRc{1,3,7,11}$) for $Re=100$ and $Re=1600$; see Figure \figref{cns3d-tgv-sensitivity-m}.
Generally, the relative error decreases with an increasing number of steps ($m$), although there are exceptions. For $Re=1600$ and $k_w=5$, the error for $m=7$ is smaller than for $m=11$. Even in such cases, however, $m=11$ (green solid line) outperforms $m=1$ (red dash-dot line) and $m=3$ (blue dash line) in accuracy.
The optimal configurations are $k_w=5$, $m=11$ for $Re=100$ and $k_w=1$, $m=11$ for $Re=1600$.
\footnote{
For $Re=100$, the relative error for $k_w=5$ and $m=11$ is comparable to that of $k_w=3$ and $m=11$. However, $k_w=5$ and $m=11$ provides better snapshot results than those of $k_w=3$ and $m=11$. 
}
Turbulent flows seem to achieve better performance with a single kernel width, while laminar flows tend to perform better with multiple kernel widths.

\begin{figure}[h!t!b!]
  \centering
  \subfigure[$Re=100$]{
    \figlab{cns3d-tgv-sensitivity-m-Re100}
      \includegraphics[trim=0.3cm 0.5cm 0.5cm 0.0cm,clip=true,width=0.31\textwidth]{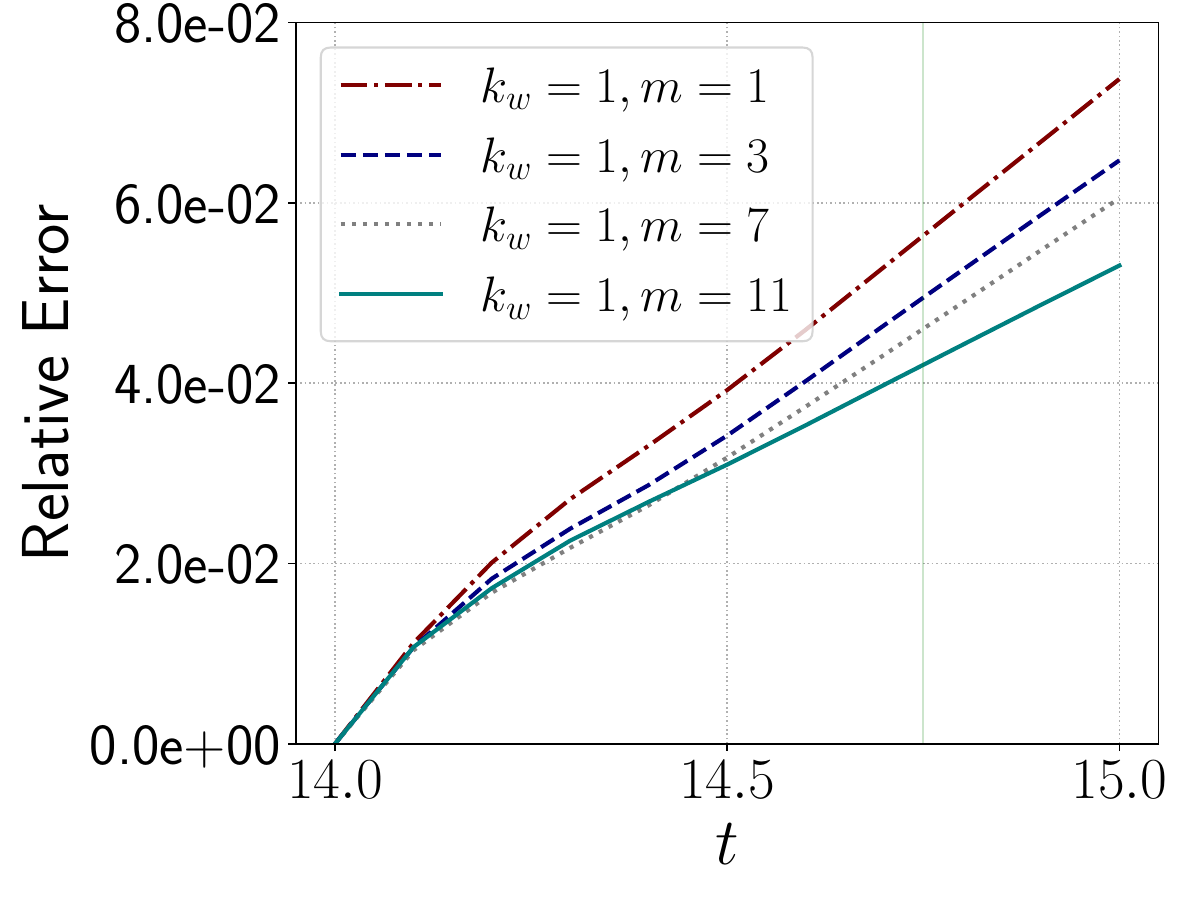}
    \includegraphics[trim=0.3cm     0.5cm 0.5cm 0.0cm,clip=true,width=0.31\textwidth]{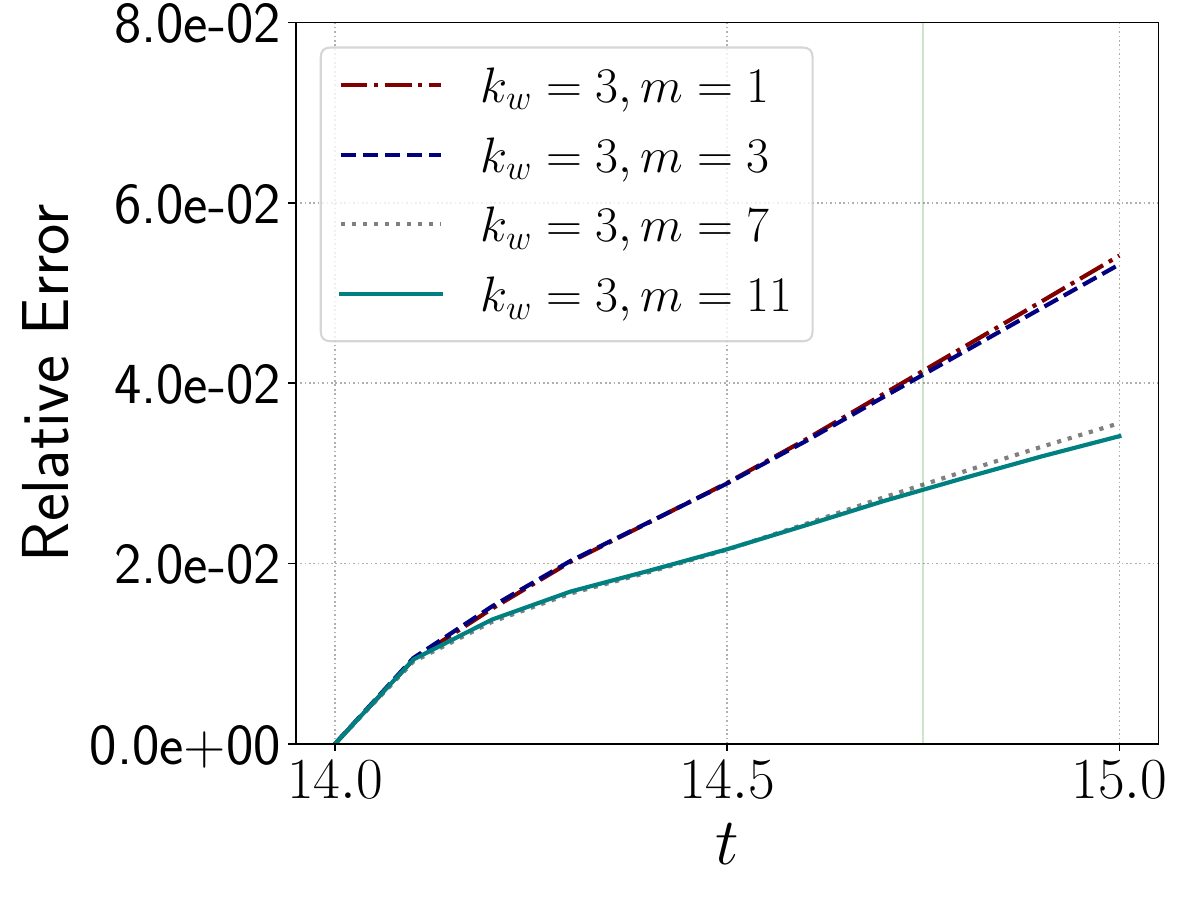}
    \includegraphics[trim=0.3cm     0.5cm 0.5cm 0.0cm,clip=true,width=0.31\textwidth]{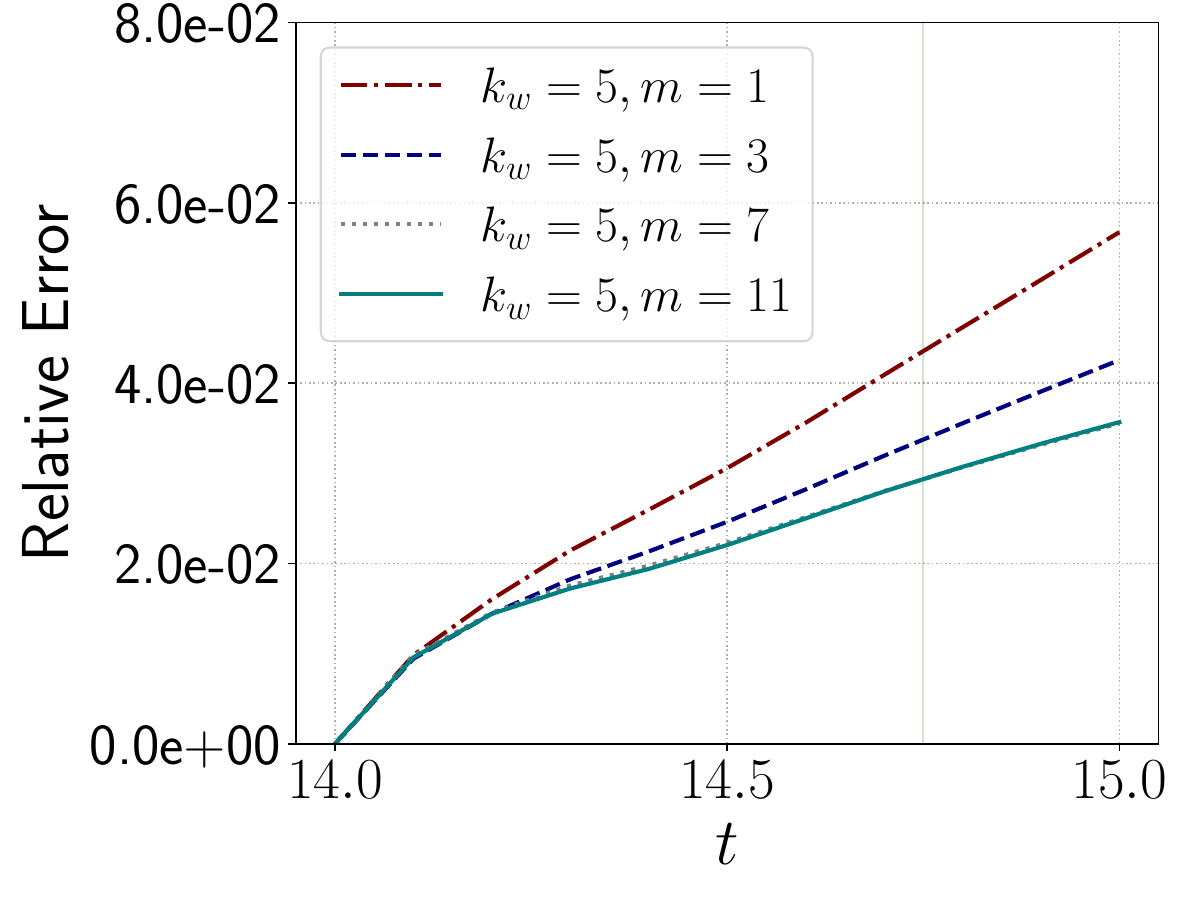}
  }
  \subfigure[$Re=1600$]{
    \figlab{cns3d-tgv-sensitivity-m-Re1600}
      \includegraphics[trim=0.3cm 0.5cm 0.5cm 0.0cm,clip=true,width=0.31\textwidth]{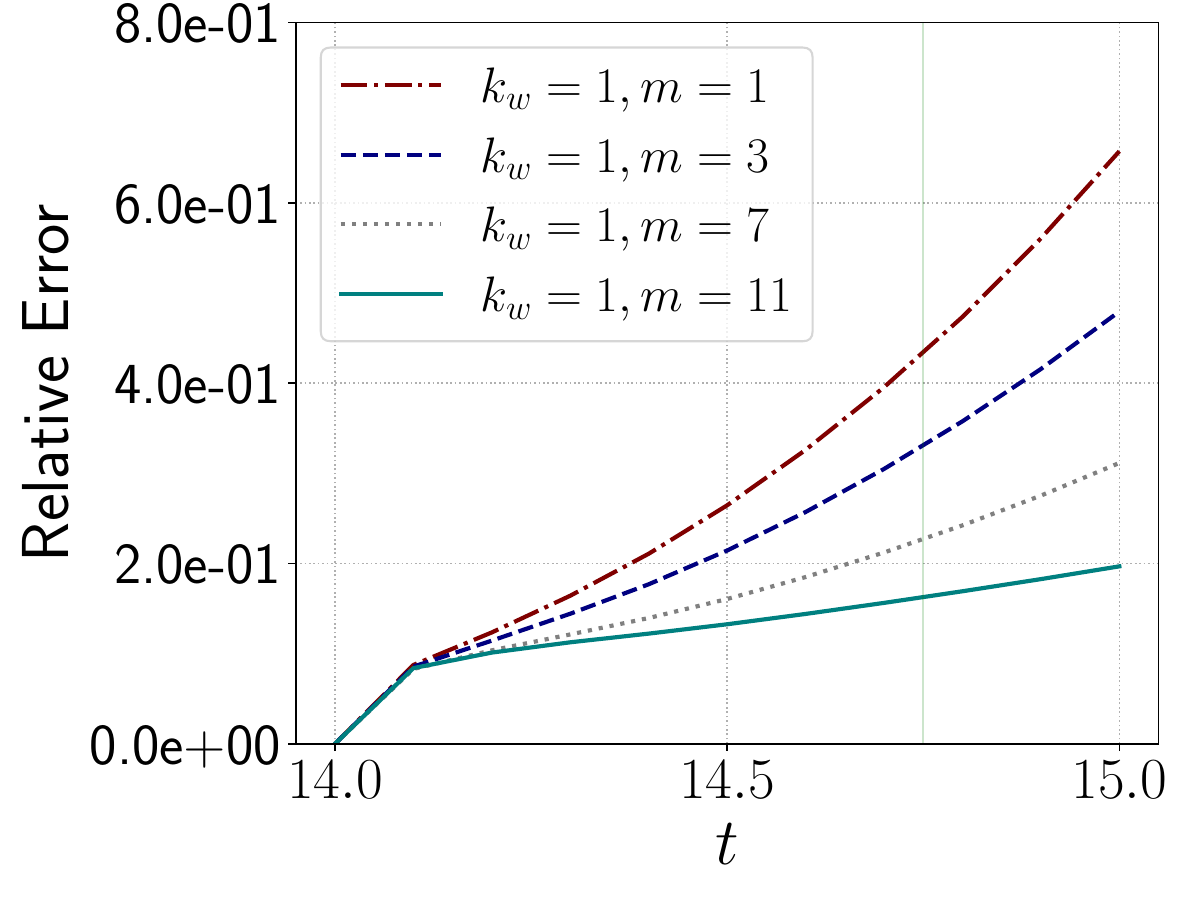}
    \includegraphics[trim=0.3cm 0.5cm 0.5cm 0.0cm,clip=true,width=0.31\textwidth]{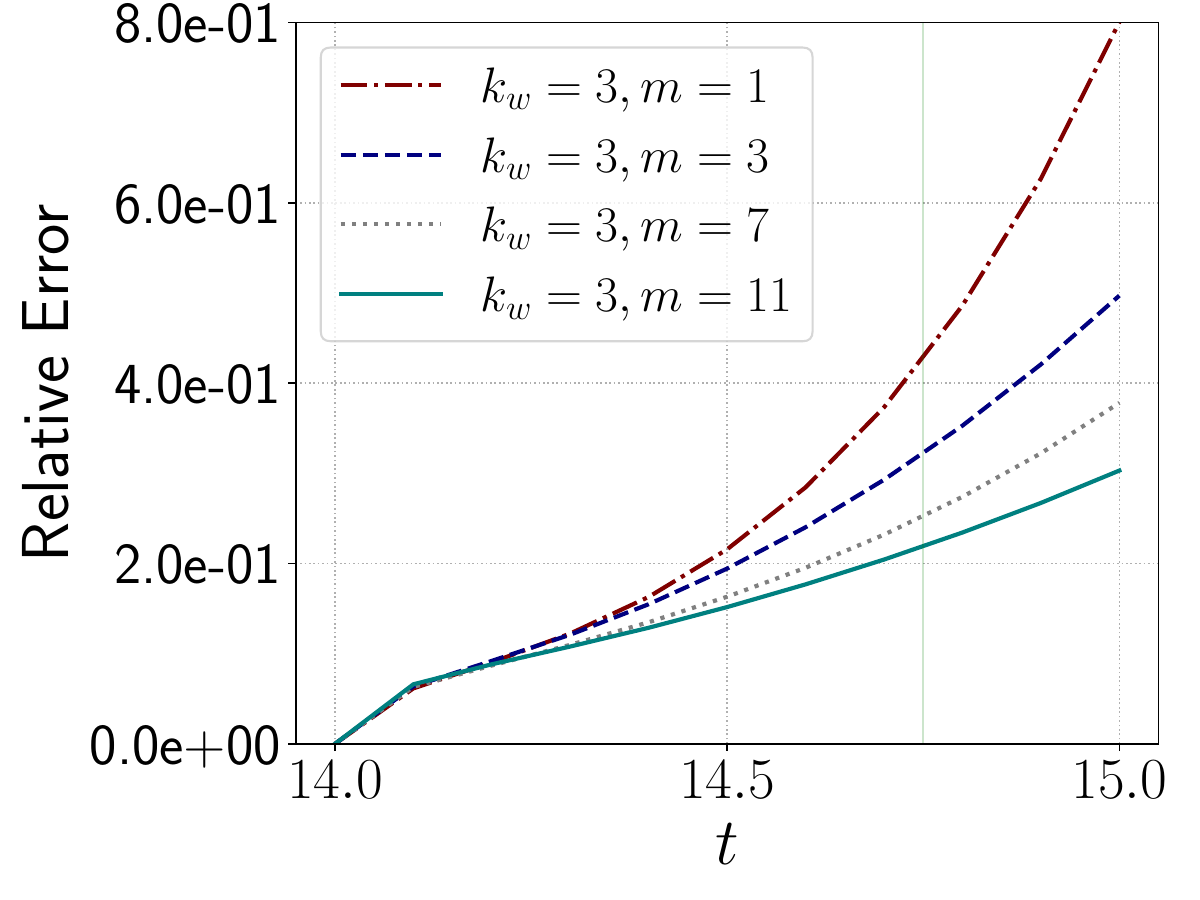}
    \includegraphics[trim=0.3cm 0.5cm 0.5cm 0.0cm,clip=true,width=0.31\textwidth]{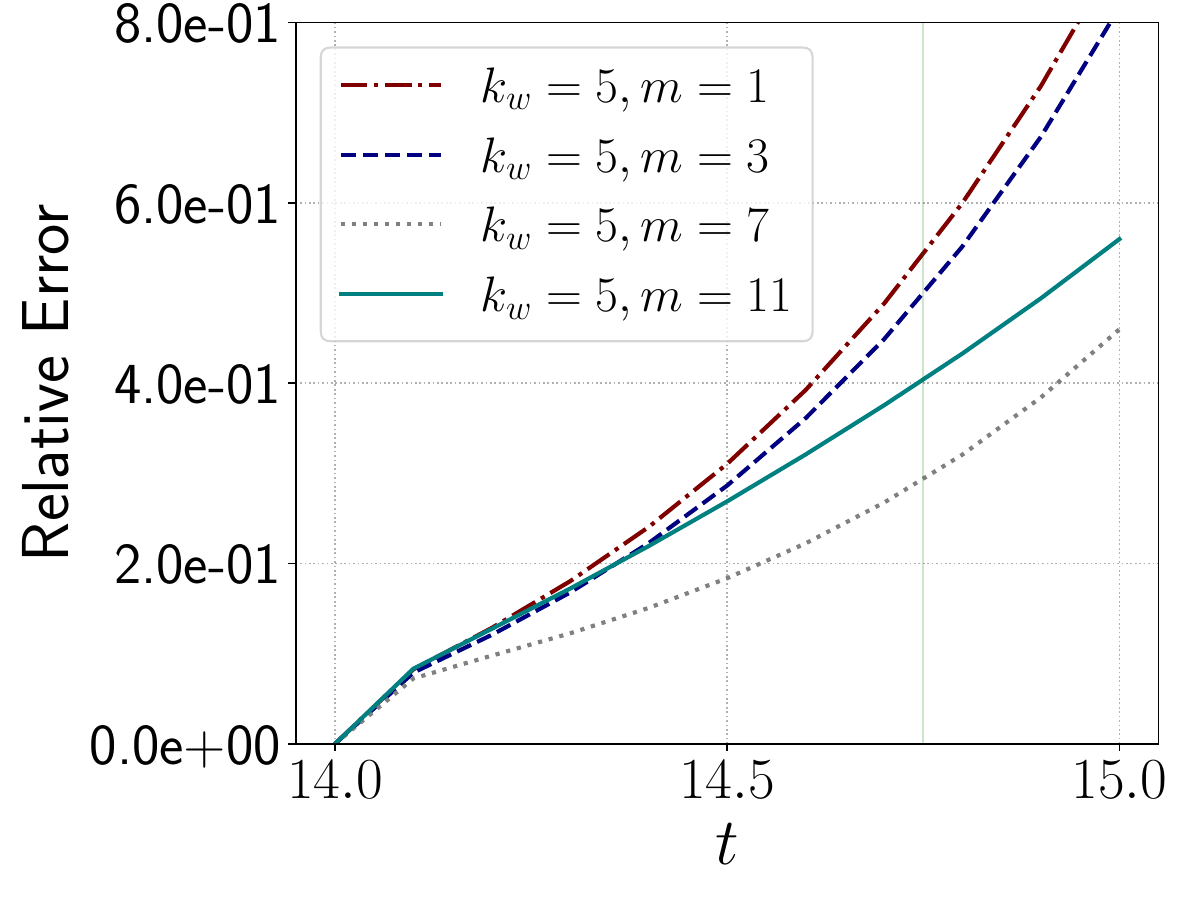}
  }
  \caption{3D Taylor--Green vortex: error histories of $\hat{\rho\boldsymbol{\varphi}}$ for $Re\in\LRc{100,1600}$ with respect to the number of steps $m$.
  }
  \figlab{cns3d-tgv-sensitivity-m}
\end{figure}

\subsubsection{Computational cost}
    
We train the neural network source term and measure the wall-clock times for predicting $\hat{\ub}^L$, $\ub^L$, and $\ub^H$ on ThetaGPU at the Argonne Leadership Computing Facility using a single NVIDIA DGX A100.
For the predictions we select the maximum stable time step size $\dtt$, where doubling the time step leads to a blowup. The stable time step sizes are $\dtt^H=10^{-3}$ for high-order solution and $\dtt^L=2\times10^{-2}$ for low-order and augmented solutions. 

We perform five prediction runs from $t=14$ to $t=15$ and report the average wall-clock times in Table \tabref{cns3d-walltime-re100} for $Re=100$ and Table \tabref{cns3d-walltime-re1600} for $Re=1600$. 
For brevity, we denote $\hat{\ub}^L$ with $k_w=1$, $\hat{\ub}^L$ with $k_w=3$, and $\hat{\ub}^L$ with $k_w=5$ by $\hat{\ub}^L(k_w=1)$, $\hat{\ub}^L(k_w=3)$, and $\hat{\ub}^L(k_w=5)$, respectively. 
We measure the wall-clock time twice due to the just-in-time (JIT) compilation in \texttt{JAX}. 
When a function is called for the first time, \texttt{JAX} compiles it and caches the resulting code, leading to significantly faster execution in subsequent runs. The JIT compilation time is calculated by subtracting the second wall-clock time from the first. 
The JIT compilation time for $\hat{\ub}^H$ is approximately $2.5$ times longer than for the other solutions. In contrast, $\ub^L$, $\hat{\ub}^L(k_w=1)$, $\hat{\ub}^L(k_w=3)$, and $\hat{\ub}^L(k_w=5)$ have similar compilation times. 
For the second run using cached code, 
the wall-clock time for $\hat{\ub}^L(k_w=1)$ is comparable to that of the low-order solution $\ub^L$. As the kernel width $k_w$ increases, however, the wall-clock time for $\hat{\ub}^L$ grows accordingly.  
The wall-clock time for $\hat{\ub}^L(k_w=3)$ is $14\%$ higher than that for $\hat{\ub}^L(k_w=1)$, while 
the wall-clock time for $\hat{\ub}^L(k_w=5)$ is $48\%$ higher. This behavior is expected because the input size of the neural network source term increases with larger kernel widths. 
The wall-clock time for the low-order solution $\ub^L$ is $620$ times shorter than that of the high-order solution $\ub^H$. The augmented solutions $\hat{\ub}^L$ with $k_w=1$, $k_w=3$, and $k_w=5$ are approximately $620$, $543$, and $425$ times faster than the high-order solution $\ub^H$, respectively. 
Compared with the high-order solution, 
the augmented solutions achieve a speedup of two orders of magnitude. This represents a substantial improvement over the one-dimensional cases reported in \cite{kang2023learning}, where augmenting the neural network source is $3$-fold and $19$-fold more efficient for the one-dimensional convection-diffusion model and the one-dimensional viscous Burgers' model, respectively.  

\begin{table}[t] 
    \caption{3D Taylor--Green vortex: wall-clock time for the predictions of $\ub^H$, $\ub^L$, and $\hat{\ub}^L$ from $t=14$ to $t=15$ at $Re=100$. }
    \tablab{cns3d-walltime-re100} 
    \begin{center} 
    \begin{tabular}{*{1}{c}|*{1}{c}|*{1}{c}|*{1}{c}} 
    \hline 
      & $\dt $  & JIT Compile wc~[$s$] & Simulation wc~[$s$] \tabularnewline 
    \hline\hline 
    $\ub^H$ &   $0.001$       &    47.0  & 217 \tabularnewline
    $\ub^L$ &   $0.02$        &    17.72  & 0.35 \tabularnewline
    $\hat{\ub}^L (k_w=1)$ & $0.02$    &    18.05  & 0.35  \tabularnewline
    $\hat{\ub}^L (k_w=3)$ & $0.02$    &    18.04  & 0.40  \tabularnewline
    $\hat{\ub}^L (k_w=5)$ & $0.02$    &    18.35  & 0.51  \tabularnewline
    \hline\hline 
    \end{tabular} 
    \end{center} 
  \end{table}

\begin{table}[t] 
    \caption{3D Taylor--Green vortex: wall-clock time for the predictions of $\ub^H$, $\ub^L$, and $\hat{\ub}^L$ from $t=14$ to $t=15$ at $Re=1600$. }
    \tablab{cns3d-walltime-re1600} 
    \begin{center} 
    \begin{tabular}{*{1}{c}|*{1}{c}|*{1}{c}|*{1}{c}} 
    \hline 
      & $\dt $  & JIT Compile wc~[$s$] & Simulation wc~[$s$] \tabularnewline 
    \hline\hline 
    $\ub^H$ &   $0.001 $       &    47.4 & 217 \tabularnewline
    $\ub^L$ &   $0.02$        &    17.83  & 0.33 \tabularnewline
    $\hat{\ub}^L (k_w=1)$ & $0.02$    &    18.56 & 0.36  \tabularnewline
    $\hat{\ub}^L (k_w=3)$ & $0.02$    &    18.11 & 0.39  \tabularnewline
    $\hat{\ub}^L (k_w=5)$ & $0.02$    &    18.53 & 0.49  \tabularnewline
    \hline\hline 
    \end{tabular} 
    \end{center} 
  \end{table}

\subsubsection{Comparison with the second- and the third-order DG approximations}

\begin{figure}[h!t!b!]
  \centering
  \subfigure[$\rho u$]{
    \figlab{cns3d-tgv-errhistory-high-ru}
      \includegraphics[trim=0.3cm 0.5cm 0.2cm 0.2cm,clip=true,width=0.31\textwidth]{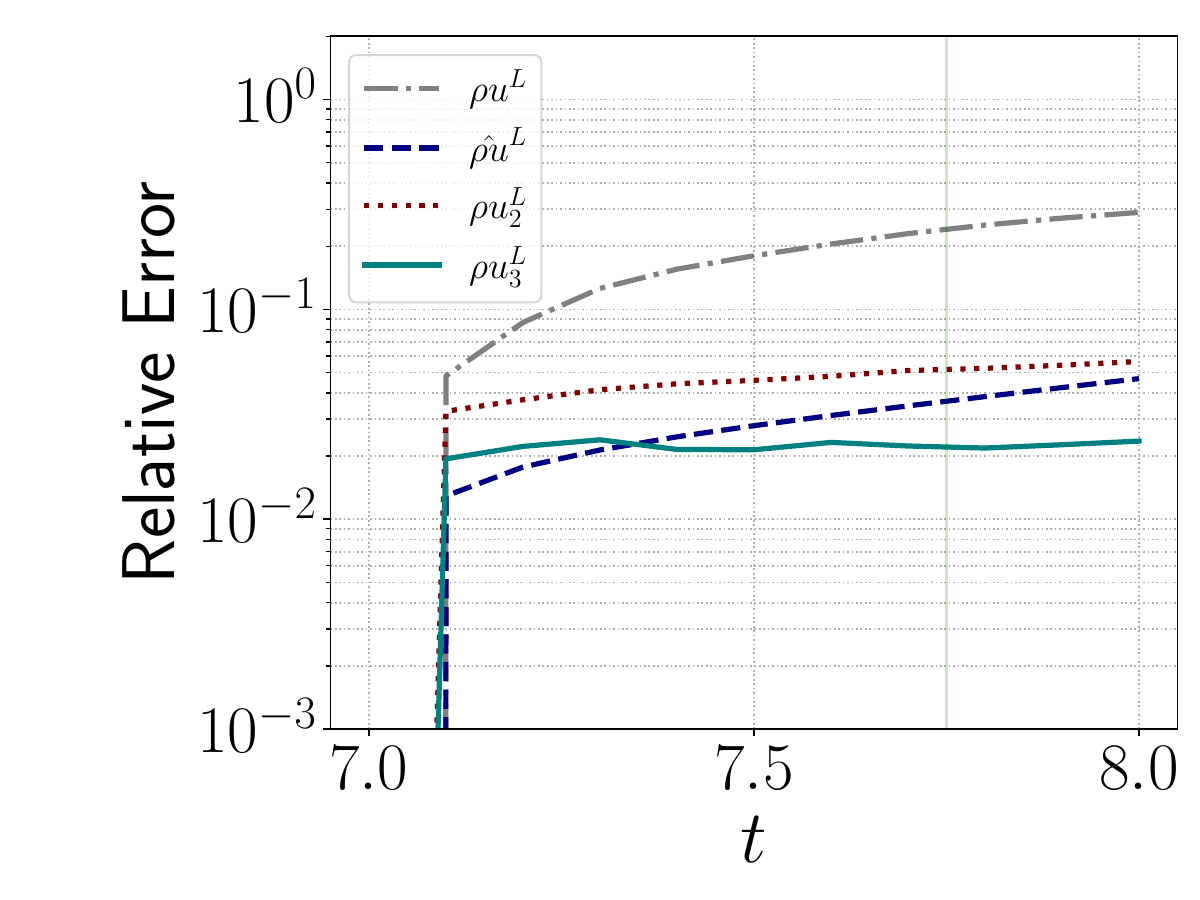}
  }
  \subfigure[$\rho v$]{
    \figlab{cns3d-tgv-errhistory-high-rv}
    \includegraphics[trim=0.3cm 0.5cm 0.2cm 0.2cm,clip=true,width=0.31\textwidth]{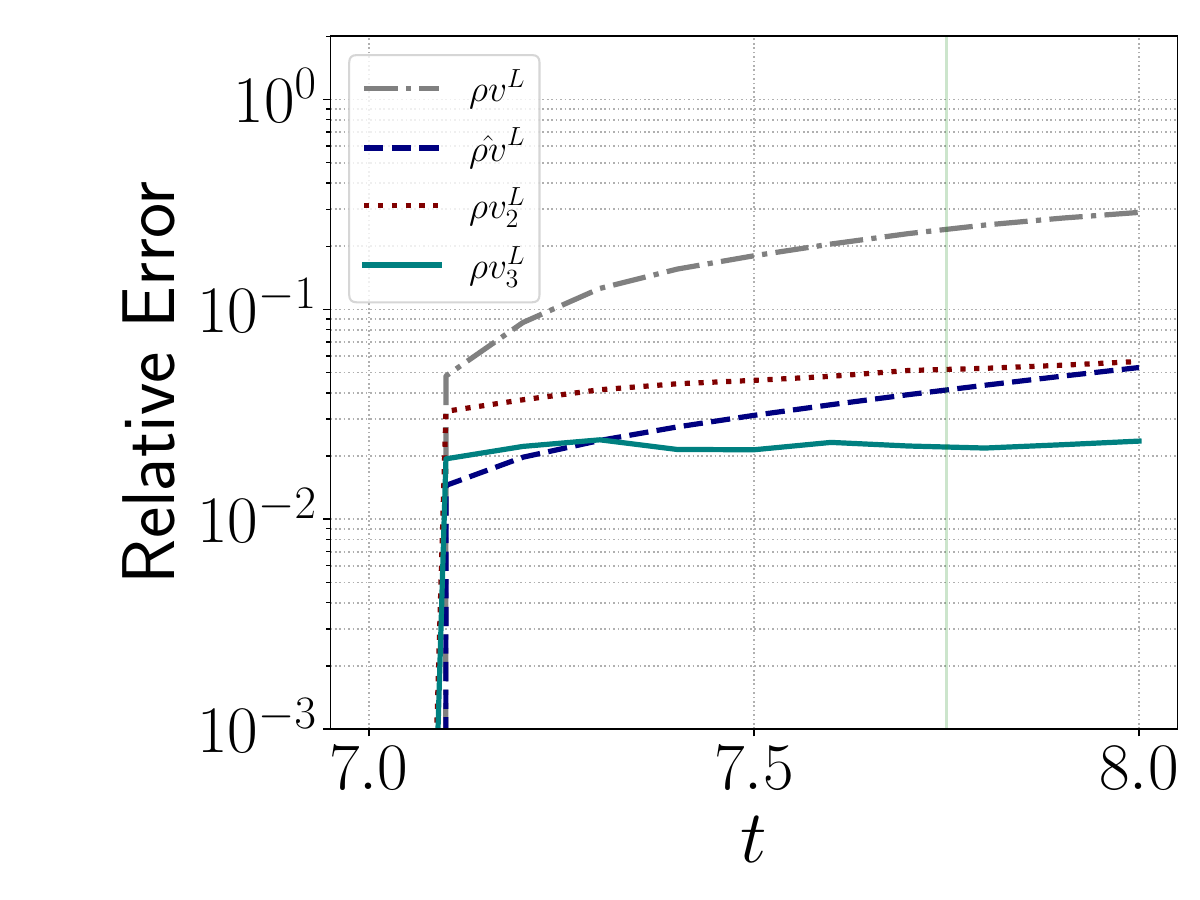}
  }
  \subfigure[$\rho w$]{
    \figlab{cns3d-tgv-errhistory-high-rw}
      \includegraphics[trim=0.3cm 0.5cm 0.2cm 0.2cm,clip=true,width=0.31\textwidth]{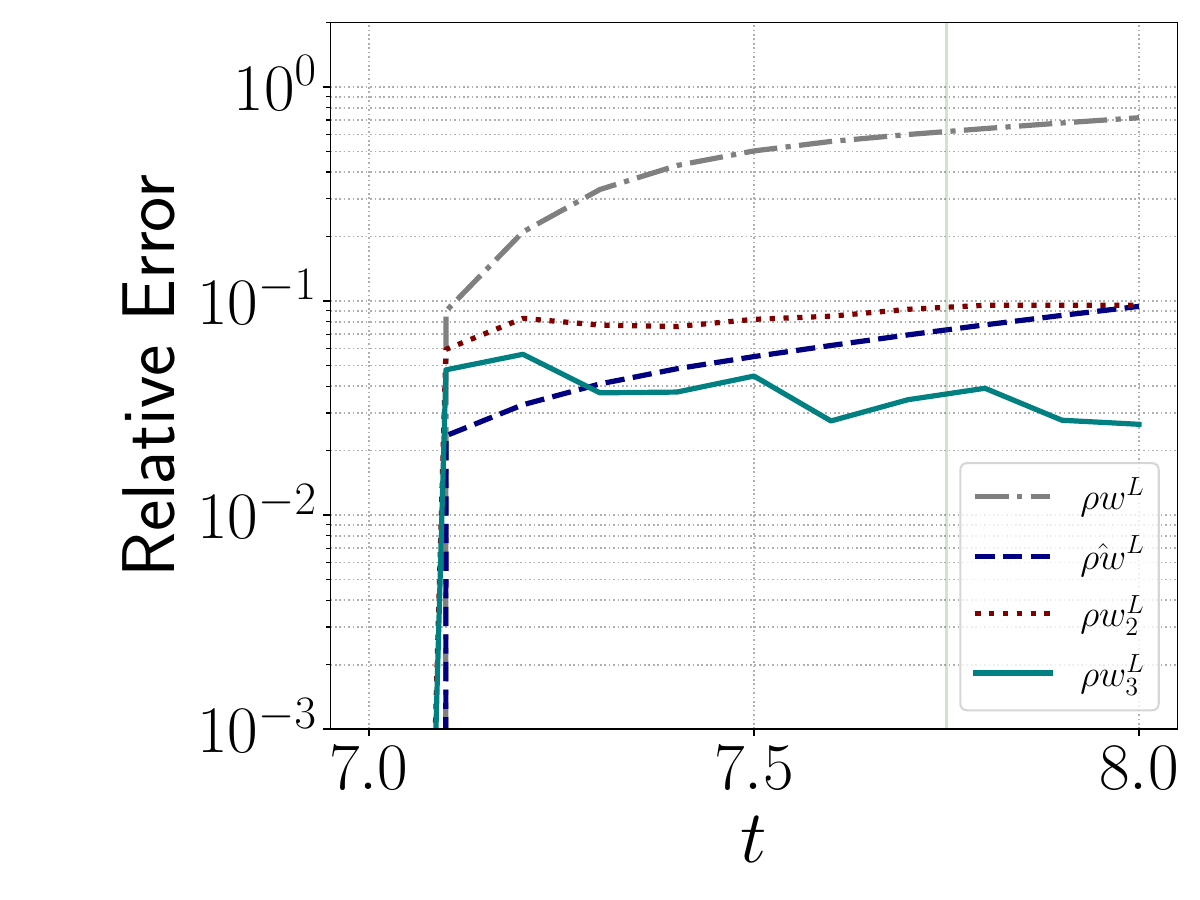}
  }
  \caption{3D Taylor--Green vortex: error histories of (a) $\rho u$, (b) $\rho v$, and (c) $\rho w$ at $Re=1600$ for $\ub^L$, $\hat{\ub}^L$, $\ub^L_2$, and $\ub^L_3$. 
  }
  \figlab{cns3d-tgv-errhistory-high-order}
\end{figure}

Up to this point we numerically showed that the augmented approximation has better accuracy than the low-order approximation. Now we examine how much our approach improves the accuracy of the low-order approximation by comparing the second- and the third-order DG approximations. 
We use a mesh with the eighth-order polynomial ($N=8$) and $8\times8\times8$ ($\Nel=512$) elements. The high-order model is integrated by using the third-order ERK scheme with a time step size of $10^{-3}$ over one time unit, $t\in [7,8]$, at $Re=1600$. The eighth-order solution is projected to the first-order space to create the projected data. Then we divide its time series into both the training data for $t=[7,7.75]$ and the test data for $t=[7.75,8]$.
The neural network source has the dimensions of $\LRc{40,128,128,128,40}$. We employ the AdaBelief \cite{zhuang2020adabelief} optimizer with a learning rate of $10^{-3}$ and train the network for $2,000$ epochs. 
We denote the second- and the third-order DG approximations by $\ub_2$ and $\ub_3$, respectively. We let $\ub^L_2$ and $\ub^L_3$ respectively be the projected solutions of $\ub_2$ and $\ub_3$ onto the first-order polynomial space on each element. 
To initialize $\ub_2$ and $\ub_3$, we interpolate the initial profile of $\ub^L$ to the second- and the third-order polynomial spaces. We then integrate $\ub^L$, $\hat{\ub}^L$, $\ub_2$, and $\ub_3$ with a time step size of $\dt=10^{-3}$ over the interval $t=[7,8]$.
We measure the relative $L^2$ errors for $\ub^L$, $\hat{\ub}^L$, $\ub_2^L$, and $\ub^L_3$, using $\Pmat^L\ub^H$ as the ground truth. 
Since the momentum error dominates over the errors in density and the total energy, we plot the relative errors for the momentum variables ($\rho u$, $\rho v$, and $\rho w$) in Figure \figref{cns3d-tgv-errhistory-high-order}. 
  For $t=[7,7.25]$, the augmented approximations of $\hat{\rho u}^L$, $\hat{\rho v}^L$, and $\hat{\rho w}^L$ demonstrate superior accuracy compared with the others. 
At $t=7.25$, the third-order approximations catch up with the accuracy levels of $\hat{\rho u}^L$, $\hat{\rho v}^L$, and $\hat{\rho w}^L$. 
After $t=7.25$, the relative errors of ${\rho u}^L_3$, ${\rho v}^L_3$, and ${\rho w}^L_3$ become smaller than those of $\hat{\rho u}^L$, $\hat{\rho v}^L$, and $\hat{\rho w}^L$.
For $t=[7.25,8]$, the augmented approximations of $\hat{\rho u}^L$, $\hat{\rho v}^L$, and $\hat{\rho w}^L$ continue to exhibit higher accuracy than the first- and the second-order approximation counterparts. 
At $t=8.0$, the relative errors of the augmented approximations become comparable to those of ${\rho u}^L_2$, ${\rho v}^L_2$, and ${\rho w}^L_2$. 

\subsubsection{Comparison with the discrete corrective forcing approach}

\begin{figure}[h!t!b!]
  \centering
  \subfigure[$Re=100$]{
    \figlab{cns3d-tgv-errhistory-lara-Re100}
    \includegraphics[trim=0.3cm 0.5cm 0.2cm 0.2cm,clip=true,width=0.31\textwidth]{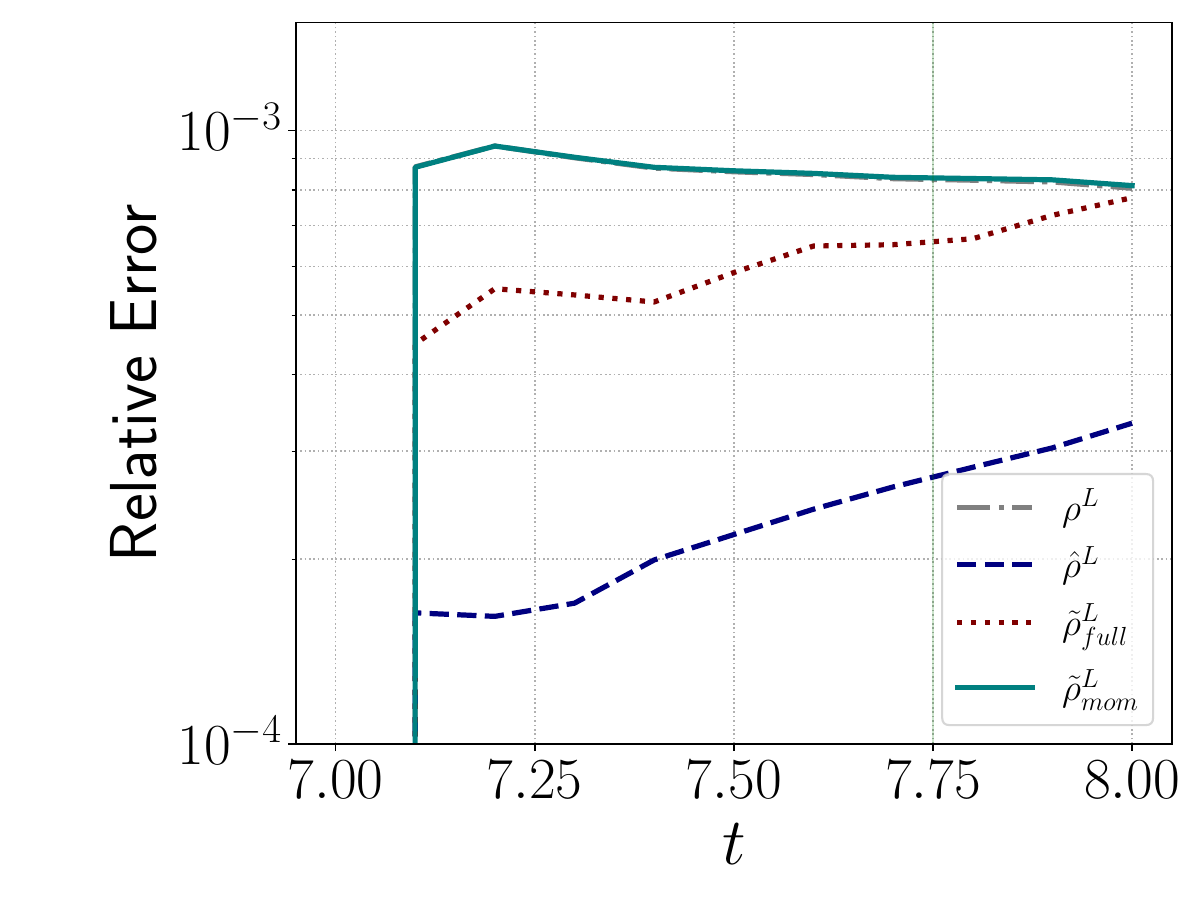}
    \includegraphics[trim=0.3cm 0.5cm 0.2cm 0.2cm,clip=true,width=0.31\textwidth]{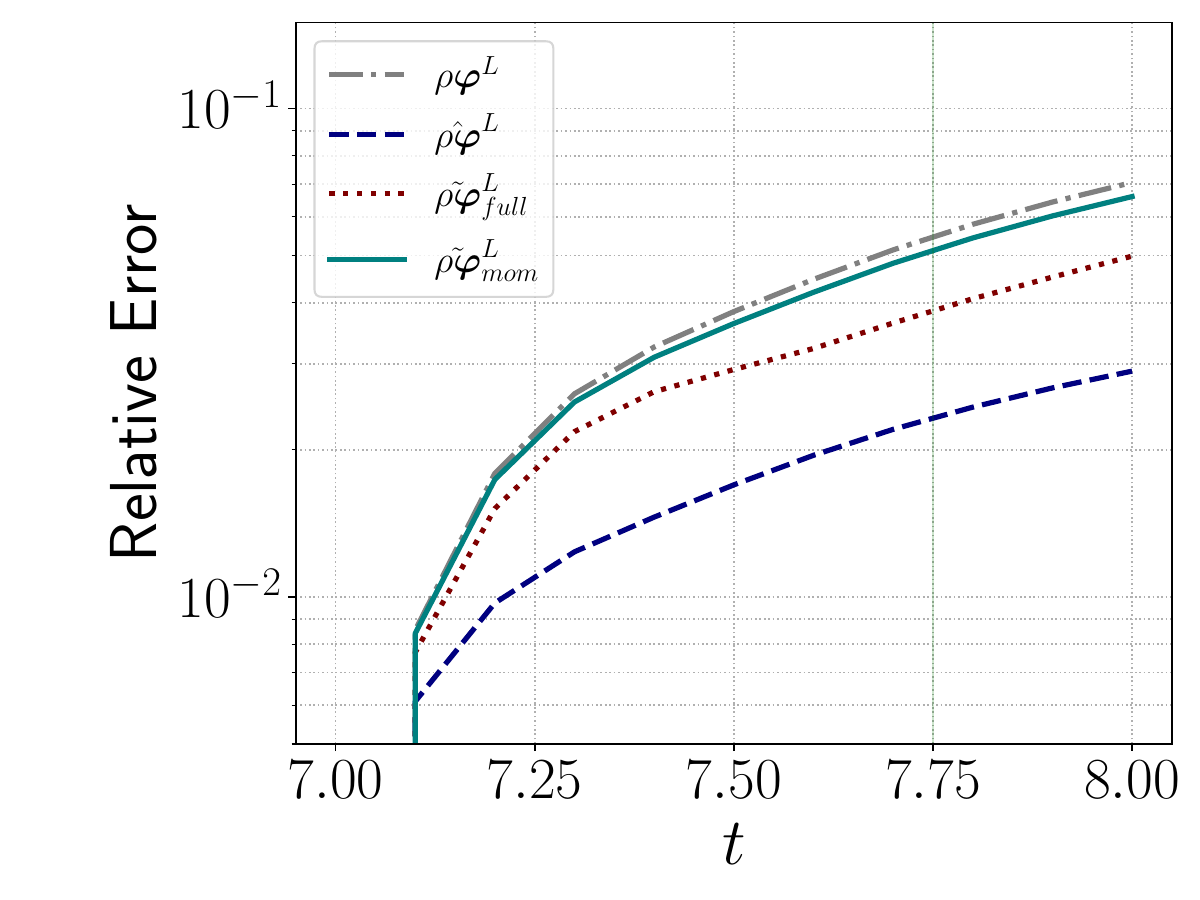}
    \includegraphics[trim=0.3cm 0.5cm 0.2cm 0.2cm,clip=true,width=0.31\textwidth]{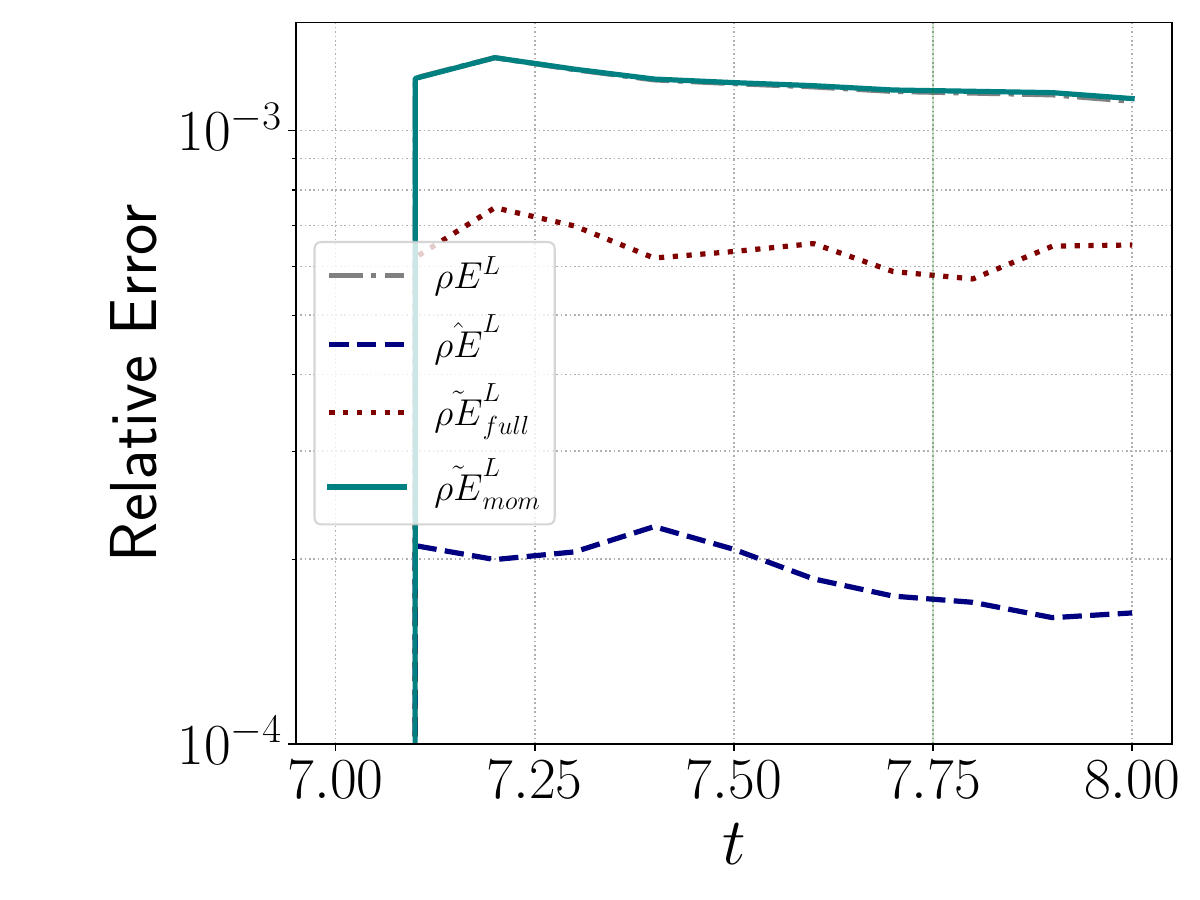}
  }
  \subfigure[$Re=1600$]{
    \figlab{cns3d-tgv-errhistory-lara-Re1600}
    \includegraphics[trim=0.3cm 0.5cm 0.2cm 0.2cm,clip=true,width=0.31\textwidth]{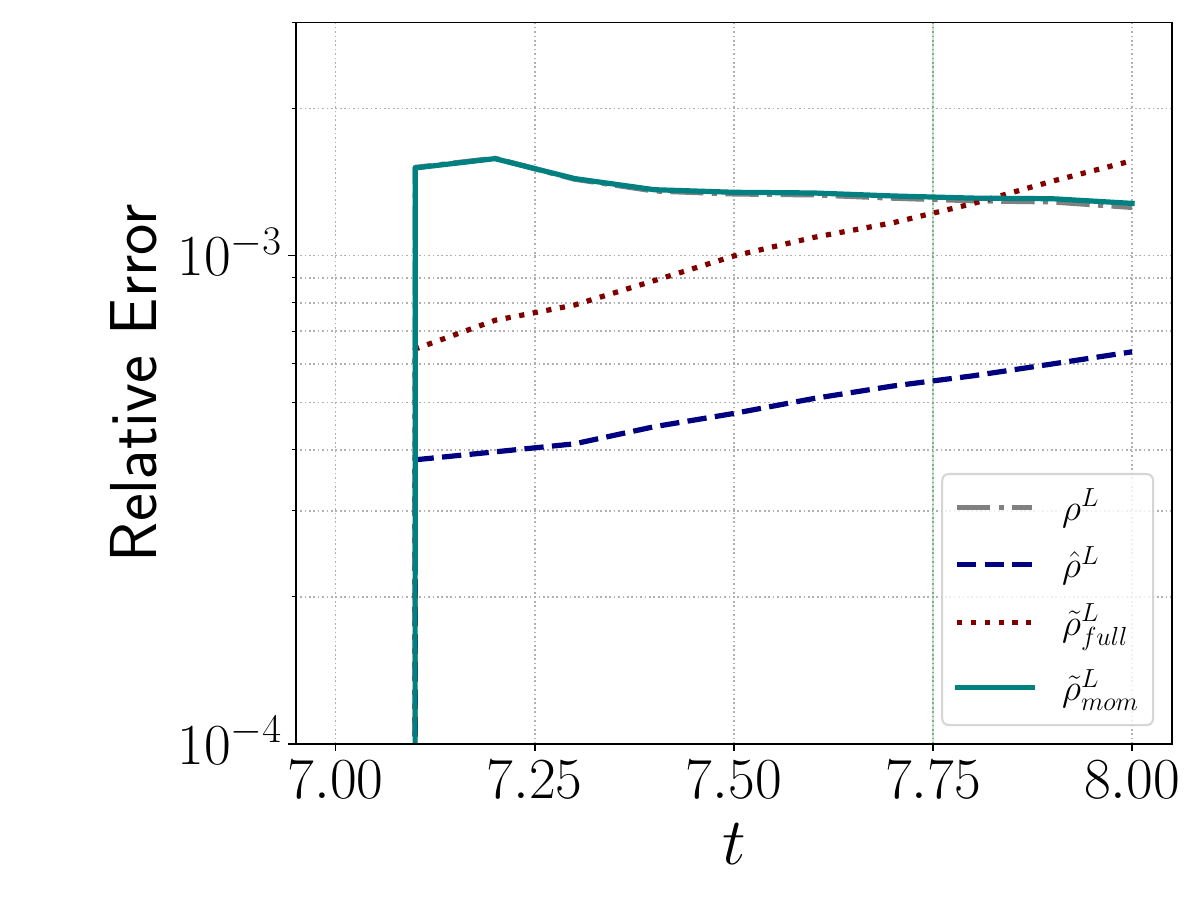}
    \includegraphics[trim=0.3cm 0.5cm 0.2cm 0.2cm,clip=true,width=0.31\textwidth]{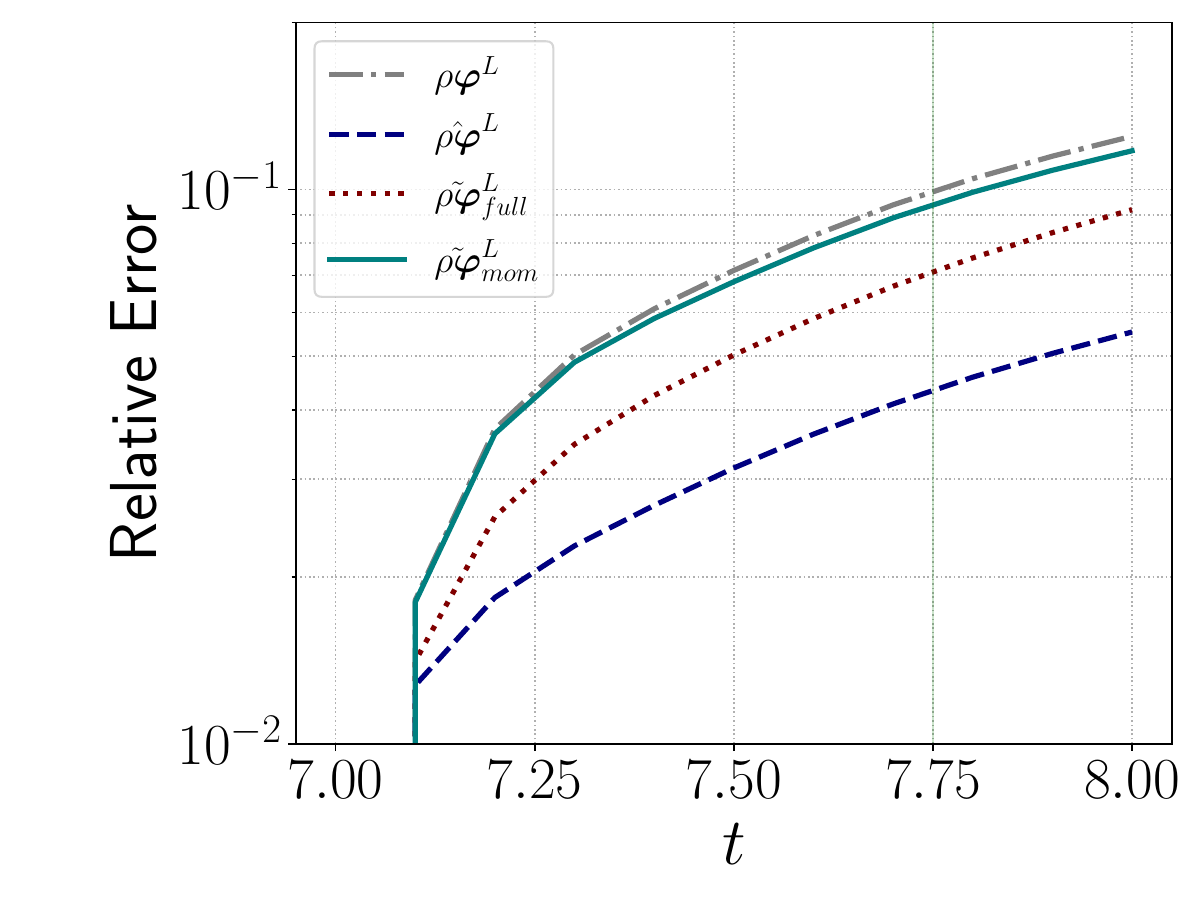}
    \includegraphics[trim=0.3cm 0.5cm 0.2cm 0.2cm,clip=true,width=0.31\textwidth]{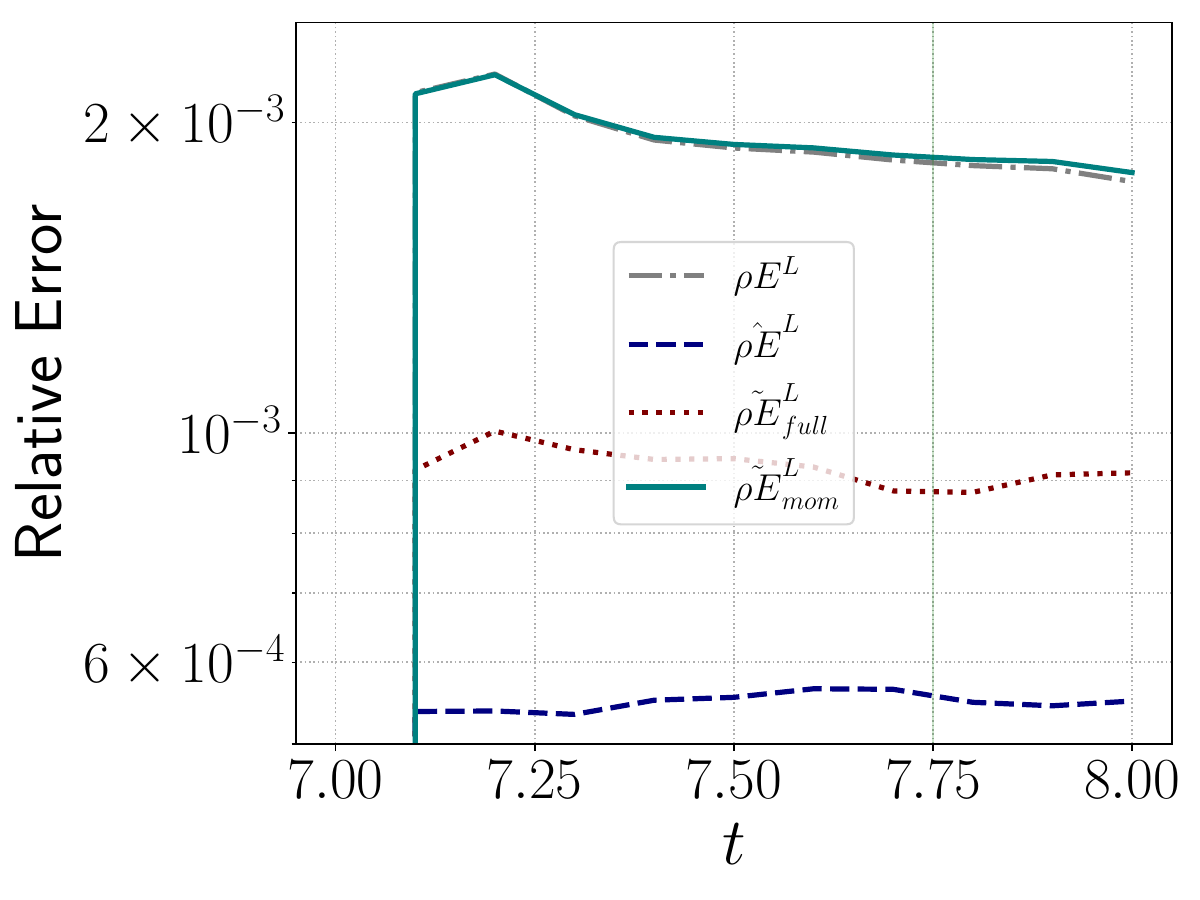}
  }
  \caption{3D Taylor--Green vortex: error histories of (a) $\rho$, (b) $\rho {\boldsymbol{\varphi}}$, and (c) $\rho E$ for $Re=1600$ with discrete forcing approaches. 
  }
  \figlab{cns3d-tgv-errhistory-lara}
\end{figure}

Compared with the discrete corrective forcing approach \cite{de2022accelerating,de2023accelerating}, 
our method involves learning a continuous corrective forcing term that interacts directly with the governing equations. Consequently, it can dynamically adjust its time step size during both the training and prediction phases. 
In our previous study \citep{kang2023learning}, numerical experiments showed that our proposed approach demonstrated enhanced robustness to variations in time step size during prediction for the one-dimensional convection-diffusion equation.
In particular, our proposed approach outperformed the discrete corrective forcing approach in terms of accuracy except when evaluated using the time step size designed especially for the discrete corrective forcing approach.
However, in the case of the three-dimensional Taylor--Green vortex example, both the discrete and continuous corrective forcing approaches showed minimal sensitivity to changes in time step size.
This is because spatial discretization errors were the dominant factor in this example.  
 Therefore, we report the relative error histories for both methods using the time step size employed during the training of the discrete forcing term.

The discrete corrective forcing term $\tilde{\Sb}_{\theta}$ for compressible Navier--Stokes equations is constructed as follows.
First, the projected high-order solution is generated by using the time step size $\dt^H$. Second, the discrete corrective forcing data  $\tilde{\Sb}_{\theta,n}$ is created for $n=0,\cdots,N_t-1$, where the $k$th component of $\tilde{\Sb}_{\theta,n}$ is calculated by using \eqnref{discrete-correction}.
%
We set $\dt^L=10\times \dt^H$ and consider two types of discrete forcing terms: $\tilde{\Sb}_{\theta,mom}$ and $\tilde{\Sb}_{\theta,full}$. 
The former focuses on correcting the momentum equations, 
defined as $\tilde{\Sb}_{\theta,mom}=\LRp{0,\tilde{\Sb}_{\rho \boldsymbol{\varphi}}, 0}^\top$, 
while the latter addresses the full set of equations, defined as $\tilde{\Sb}_{\theta,full}=\LRp{\tilde{S}_\rho,\tilde{\Sb}_{\rho \boldsymbol{\varphi}}, \tilde{S}_{\rho E}}^\top$. 
The discretely corrected solutions obtained by using $\tilde{\Sb}_{\theta,mom}$ and $\tilde{\Sb}_{\theta,full}$ are denoted as  $\tilde{\ub}_{mom}^L$ and $\tilde{\ub}_{full}^L$, respectively.
Next, we construct two nonlinear mappings using a feed-forward neural network. The input feature is 
$\LRp{\Pmat^L\rho^H,\Pmat^L \rho u^H, \Pmat^L \rho v^H, \Pmat^L \rho w^H, \Pmat^L\rho E^H}^T$, and the output features are $\tilde{\Sb}_{\theta,mom}$ and $\tilde{\Sb}_{\theta,full}$. For $\tilde{\Sb}_{\theta,mom}$, 
the neural network architecture has dimensions $\LRc{40,64,32,64,24}$, while for $\tilde{\Sb}_{\theta,full}$, the architecture is $\LRc{40,64,32,64,40}$.
Both discrete corrective forcing terms are trained by using the  
AdaBelief \cite{zhuang2020adabelief} optimizer with a learning rate of $10^{-3}$, $2$ batches and $1,000$ epochs. 

Figure \figref{cns3d-tgv-errhistory-lara} shows the relative errors for density, momentum, and total energy of the low-order (first-order) solution ($\ub^L$), the augmented solution ($\hat{\ub}^L$), and the discretely corrected solutions ($\tilde{\ub}^L_{mom}$ and $\tilde{\ub}^L_{full}$) over $t\in[7,8]$ at $Re=\LRc{100,1600}$. 
Overall, our approach (the continuous corrective forcing method) outperforms the discrete forcing approach across all variables: density, momentum, and total energy.
For density, $\hat{\rho u}^L$ exhibits errors that are half those of both $\tilde{\rho u}^L_{mom}$ and $\tilde{\rho u}^L_{full}$ at $t=8$. 
For momentum, 
the error of $\hat{\rho\boldsymbol{\varphi}}^L$ 
is approximately half that of $\tilde{\rho\boldsymbol{\varphi}}^L_{mom}$  and 1.5 times lower than that of $\tilde{\rho\boldsymbol{\varphi}}^L_{full}$ at $t=8$.
For total energy, $\hat{\rho E}^L$ achieves significantly lower errors: four times less than $\tilde{\rho E}^L_{full}$ and $6.9$ times less than $\tilde{\rho E}^L_{mom}$ at $Re=100$. 
At $Re=1600$, $\hat{\rho E}^L$ shows $1.6$ times lower errors compared with $\tilde{\rho E}^L_{full}$ and $3.2$ times lower compared with $\tilde{\rho E}^L_{mom}$. 
Among the discrete corrective forcing approaches, $\tilde{\Sb}_{\theta,full}$ provides more accurate results than $\tilde{\Sb}_{\theta,mom}$.
For momentum, $\tilde{\rho\boldsymbol{\varphi}}^L_{mom}$ has errors that are $1.7$ times larger than $\tilde{\rho\boldsymbol{\varphi}}^L_{full}$ at $Re=100$ and twice as large at $Re=1600$.

\section{Conclusions}
\seclab{Conclusion} 

Building on our previous work~\cite{kang2023learning}, this study presents an end-to-end differentiable NODE-DG framework for solving the compressible Navier--Stokes equations. We develop a differentiable DG solver for these equations on a structured mesh, integrating a standard DG solver with a neural network source term as defined in \eqnref{goveq-with-nnsource}. 
The neural network source function, consisting of a 
$D$-layer
feed-forward neural network, slides over the low-order solutions of all the elements  and generates a local source approximation for each element. 
The network parameters are trained by using NODEs, ensuring interaction with the governing equations throughout the training process. 
This hybrid model integrates traditional numerical methods with machine learning, harnessing the accuracy of numerical techniques and the efficiency of machine learning approaches. 
The NODE-DG framework with a source term offers several distinct advantages:
	(1)	improved accuracy of low-order DG approximations by effectively capturing subgrid-scale dynamics;
	(2)	robustness to nonuniform or missing temporal data;
	(3)	elimination of operator-splitting errors;
    (4) total mass conservation; and
	(5)	learning a continuous-in-time operator, enabling predictions with variable step sizes. Specifically, it allows for predictions with larger time step sizes compared with the high-order DG solvers and hence 
    significantly speeds up projected high-order DG simulations.    
 
We demonstrate the proposed methodology through two-dimensional Kelvin--Helmholtz instability and three-dimensional Taylor--Green vortex examples. 
In the Kelvin--Helmholtz instability example, we demonstrate that augmenting the low-order approximation with a neural network source term enables more accurate capture of the vortical structure compared with the low-order solution. Furthermore, to showcase the NODE-DG framework's ability to handle variable coefficients and missing data, we include the Reynolds number as an input to the neural network source term and modify the loss function to accommodate the missing data. For long-term stability, we introduce a zero-mean source term that preserves total mass. We provide theoretical proof of this mass conserving property and numerically confirm that the mass conserving source maintains the total mass, whereas the na\"{\i}ve source does not. To further enhance stability, we add $L_2$ regularization to the loss function, which shows a tradeoff between the accuracy and numerical stability: as the strength of the regularization increases, the augmented approximation gradually converges toward the low-order solution.
In the Taylor--Green vortex example, we consider 12 datasets with varying Reynolds numbers $Re=\LRc{100,200,400,1600}$ and time intervals $t\in \LRc{[0,1], [7,8], [14,15]}$ to investigate laminar, transitional, and turbulent regimes. Similar to the Kelvin--Helmholtz example, augmenting a neural network source term indeed enhances the low-order approximation's accuracy. 
Specifically, the augmented solution significantly recovers the misrepresented vortical structures in the low-order solutions, as shown in the snapshots at $t=8$ and $t=15$. 
 We also performed a sensitivity study with respect to the kernel width $k_w$ and the number of steps $m$. Results indicate that turbulent flows perform better when trained with single-kernel width, whereas laminar flows benefit from multiple-kernel width. When compared with the second- and the third-order approximation, the augmented solution (first-order) demonstrates comparable error levels to both at $Re=1600$. 
We reported the wall-clock times for the low-order approximation, the augmented low-order approximation, and the high-order approximation. 
The high-order approximation is $620$ times more computationally expensive than the low-order approximation. In comparison, the augmented solution is two orders of magnitude faster than the high-order approximation. Specifically, with a single-kernel width, the augmented approximation is $620$ times more efficient than the high-order solution.

Compared with the discrete corrective forcing approach, our proposed approach achieves smaller errors in density, momentum, and total energy for the case with $Re=100$ and $Re=1600$. Our approach can be viewed as an extension of the discrete corrective forcing approach. We demonstrated that the discrete corrective forcing approach relies on the temporal discretization of a sequential operator-splitting step, whereas our proposed method is based on the continuous form of ordinary differential equations without any operator-splitting step. 
Consequently, our approach learns the continuous source dynamics, enabling it to adapt to variable time steps during both training and prediction, without introducing operator-splitting errors. 
In practice, operator-splitting error may not be noticeable when other error sources, such as spatial discretizaiton errors, dominate, as shown in the Taylor--Green vortex example. Nevertheless, we highlight that our NODE-DG approach is free from the operator-splitting errors, is capable of handling nonuniform data, and supports end-to-end differentiability.
%
The neural network parameters are optimized by using gradient-based methods in direct interaction with the governing equations, providing improved accuracy and interpretability compared with black-box machine learning models.

This study successfully demonstrates the potential of incorporating a neural network source function with neural ordinary differential equations for solving multidimensional compressible Navier--Stokes equations. Augmenting the neural network source function not only enhances the accuracy of low-order DG approximations but also significantly accelerates projected high-order simulations.
Another significant aspect of our method is its potential application in incorporating physical parameterizations.
By integrating observational data into the model, our approach could learn unseen dynamics through supervised learning in the context of NODEs.
Differentiable hybrid models such as our NODE-DG model have the potential to become widely adopted tools in simulation-based applications because of their speed, accuracy, and efficiency. They show great promise for use in turbulent fluid dynamics, climate modeling, weather prediction, and other fields.
Additionally, our approach offers advantages for parallel training of neural network parameters. DG methods are well suited for parallel computing, and the local nature of the neural network source function makes the approach scalable and independent of problem size. We intend to investigate this aspect further in future work. 
Furthermore, we plan to explore sophisticated neural network architectures, such as graph neural networks on unstructured grids, to address more complex problems.
The current approach does not account for added noise, and as a result, statistical analysis has been omitted. Addressing this limitation will also be a priority in our future work.



\appendix





\section{Compressible Navier--Stokes Equations}
\seclab{cns-gov}

\newcommand{\velocity}{{\boldsymbol{\varphi}}}

The homogeneous compressible Navier--Stokes equations in $\Omega$ are described by
\begin{subequations}
\eqnlab{cns-gov}
\begin{align}
  \eqnlab{cns-mass}
  \dd{\rho}{t}     + \Div \LRp{\rho \velocity} &= 0,\\
  \eqnlab{cns-momentum}
  \dd{\rho \velocity}{t} + \Div \LRp{\rho \velocity \otimes \velocity + \Ical \pres} &= \Div \sigma,\\
  \eqnlab{cns-energy}
  \dd{\rho E}{t}   + \Div \LRp{\rho \velocity H} &= \Div (\sigma \velocity) - \Div \Pi,
\end{align}
\end{subequations}
where $\rho$ is the density $[\si{\kilo\gram\per\cubic\meter}]$;
$\velocity$ is the velocity vector $[\si{\meter\per\second}]$;\footnote{
}
$\pres$ is the pressure $[\si{\newton\per\meter\squared}]$;
$\rho E = \rho e + \half \rho \norm{\velocity}^2 $ is the total energy $[\si{\joule\per\cubic\meter}]$;
$e=\frac{\pres}{\rho(\gamma -1)}$ is the internal energy $[\si{\joule\per\kilogram}]$;
$H = E + \frac{\pres}{\rho} = \frac{a^2}{\gamma -1} + \frac{1}{2}\norm{\velocity}^2$ is the total specific enthalpy $[\si{\joule\per\kilogram}]$;
$\sigma = \mu \LRp{\Grad \velocity + \Grad (\velocity)^T - \frac{2}{3} \Ical \Div \velocity } $ is the viscous stress tensor; $\Pi = -\kappa \Grad T$ is the heat flux; 
$T_m$ is the temperature; 
$\kappa = \mu~c_p~Pr^{-1}$ is the heat conductivity $[\si{\watt\per\meter\per\kelvin}]$;
$\mu$ is the dynamic viscosity $[\si{\pascal\second}]$;
$Pr$ is the Prandtl number; 
$a = \LRp{\gamma\pres\rho^{-1}}^\half$ is the sound speed $[\si{\meter\per\second}]$ for an  ideal gas;
$\gamma = c_p c_v^{-1}$ is the ratio of the specific heats; 
and $c_p$ and $c_v$ are the specific heat capacities at constant pressure and at constant volume $[\si{\joule\per\kilogram \per\kelvin}]$, respectively.

By the nondimensionalized variables using the speed of sound as a
reference velocity, 
$$ 
\rho^* = \frac{\rho}{\rho_\infty}, 
\pres^* = \frac{\pres}{\rho_\infty a_\infty^2},
\velocity^* = \frac{\velocity}{a_\infty},
x^* = \frac{x}{L},
t^* = \frac{t}{L/a_\infty},
\mu^* = \frac{\mu}{\mu_\infty},
\text{ and }
T^* = \frac{T}{T_
\infty}, 
$$
\noindent
we rewrite the governing equation \eqnref{cns-gov} as  
\begin{subequations}
\eqnlab{cns-gov-nondimension}
\begin{align}
  \dd{\rho^*}{t^*}     + \Grad^*\cdot \LRp{\rho \velocity^*} &= 0,\\
  \dd{\rho \velocity^*}{t^*} + \Grad^*\cdot \LRp{\rho \velocity^* \otimes \velocity^* + \Ical \pres^*} &= \Grad^* \cdot \sigma^*,\\
  \dd{\rho E^*}{t^*}   + \Grad^*\cdot \LRp{\rho \velocity^* H^*} &= \Grad^*\cdot (\sigma^* \velocity^*) - \Grad^* \cdot \Pi^*,
\end{align}
\end{subequations}
where 
$\sigma^* = \tilde{\mu} \LRp{\Grad^* \velocity^* + \Grad^* (\velocity^*)^\top -\frac{2}{3}\Ical \Grad^* \cdot \velocity^* } $,  
$\Pi^* = - \frac{\tilde{c}_p\tilde{\mu}}{Pr} \Grad^* T^* $,  $\Grad^*=\frac{1}{L}\Grad$,
$\tilde{\mu} = \frac{\mu^* ~M_\infty}{Re_\infty}$, 
$Re_\infty = \frac{\rho_\infty u_\infty L}{\mu_\infty}$, 
$M_{\infty} = \frac{u_\infty}{a_\infty}$,
and $\tilde{c}_p := \frac{1 }{\gamma - 1 }$.
The normalized equation of state  for an ideal gas is $\pres^* = \gamma^{-1} \rho^* T^* = \rho^* e^*(\gamma-1)$.
Sutherland's formula is $\mu^*=\LRp{T^*}^{\frac{3}{2}}\frac{1+\frac{su}{T_\infty}}{T^*+\frac{su}{T_\infty}}$ with $su=110.4~\si{\kelvin}$. In this study we use the nondimensionalized form \eqnref{cns-gov-nondimension}, 
and we omit the superscript ($*$).

\section*{Acknowledgments}
This work was supported by Korea University Grants (No. K2411041, K2414071, K2425851, K2514791). 
This material is based upon work supported by the U.S. Department of Energy, Office of Science, Office of Advanced Scientific Computing Research (ASCR), Applied Mathematics program, through Competitive Portfolios Project on Energy Efficient Computing: A Holistic Methodology, and the Scientific Discovery through Advanced Computing (SciDAC) FASTMath Institute programs.
We also gratefully acknowledge the use of ThetaGPU and Polaris in the resources of the Argonne Leadership Computing Facility, which is a DOE Office of Science User Facility supported under Contract DE-AC02-06CH11357.
We also gratefully acknowledge the use of GPU in the resources of the Argonne Leadership Computing Facility, which is a DOE Office of Science User Facility supported under Contract DE-AC02-06CH11357.



\bibliographystyle{elsarticle-num}
\bibliography{main}







 \begin{center}
	\scriptsize \framebox{\parbox{4in}{Government License (will be removed at publication):
			The submitted manuscript has been created by UChicago Argonne, LLC,
			Operator of Argonne National Laboratory (``Argonne").  Argonne, a
			U.S. Department of Energy Office of Science laboratory, is operated
			under Contract No. DE-AC02-06CH11357.  The U.S. Government retains for
			itself, and others acting on its behalf, a paid-up nonexclusive,
			irrevocable worldwide license in said article to reproduce, prepare
			derivative works, distribute copies to the public, and perform
			publicly and display publicly, by or on behalf of the Government. The Department of Energy will provide public access to these results of federally sponsored research in accordance with the DOE Public Access Plan. http://energy.gov/downloads/doe-public-access-plan.
}}
	\normalsize
\end{center}

\end{document}